\documentclass[12pt]{article}
\usepackage{graphicx, amssymb, amsmath,mathabx}
\usepackage{natbib}
\usepackage{epsfig}
\usepackage{epstopdf}
\usepackage{booktabs}
\usepackage{color, colortbl,xcolor}
\usepackage{soul}
\usepackage{float}
\usepackage{arydshln}
\usepackage{multirow}
\usepackage{wrapfig} 
\usepackage{threeparttable}
\usepackage{enumitem} 
\usepackage[margin=1in]{geometry}
\usepackage[colorlinks=true,allcolors=blue]{hyperref}%

\DeclareGraphicsExtensions{.pdf,.eps.gz,.eps,.epsi.gz,.epsi,.ps,.ps.gz}

\long\def\symbolfootnote[#1]#2{\begingroup%
	\def\thefootnote{\fnsymbol{footnote}}\footnote[#1]{#2}\endgroup}

\def\wt{\widetilde}
\def\diag{\hbox{diag}}
\def\wh{\widehat}
\def\wc{\widecheck}

\def\boxit#1{\vbox{\hrule\hbox{\vrule\kern6pt
			\vbox{\kern6pt#1\kern6pt}\kern6pt\vrule}\hrule}}

\def\trace{\hbox{trace}}

\def\wh{\widehat}

\newcommand{\epsv}{\bm{\varepsilon}}

\newcommand{\bPhi}{\mathbf{\Phi}}

\newcommand{\bmu}{\boldsymbol{\mu}}

\newcommand{\0}{\mathbf{0}}
\newcommand{\E}{\mathbb{E}}
\newcommand{\Var}{\mathrm{Var}}
\newcommand{\Cov}{\mathrm{Cov}}

\newcommand{\bm}{\boldsymbol}

\newcommand{\bxi}{{\bm\xi}}
\newcommand{\bphi}{{\bm\phi}}
\newcommand{\bpsi}{{\bm\psi}}

\def\bse{\begin{eqnarray*}}
	\def\ese{\end{eqnarray*}}
\def\bsq{\begin{equation*}}
\def\esq{\end{equation*}}
\def\bq{\begin{equation}}
\def\eq{\end{equation}}

\def\trace{\hbox{trace}}
\def\wh{\widehat}
\def\wt{\widetilde}

\def\argmin{\mbox{argmin}}

\def\diag{\mbox{diag}}

\def\A{{\bf A}}
\def\a{{\bf a}}
\def\B{{\bf B}}

\def\e{{\bf e}}
\def\q{{\bf q}}

\def\b{{\bf b}}

\def\Q{{\bf Q}}

\def\1{{\bf 1}}
\def\0{{\bf 0}}

\def\diag{\hbox{diag}}

\def\trace{\hbox{trace}}

\def\wh{\widehat}
\def\wt{\widetilde}

\def\diag{\hbox{diag}}
\def\log{\hbox{log}}

\def\squarebox#1{\hbox to #1{\hfill\vbox to #1{\vfill}}}

\def\trace{\hbox{trace}}

\def\bse{\begin{eqnarray*}}
	\def\ese{\end{eqnarray*}}
\def\bsq{\begin{equation*}}
\def\esq{\end{equation*}}
\def\bq{\begin{equation}}
\def\eq{\end{equation}}

\def\wh{\widehat}
\def\wt{\widetilde}
\def\diag{\hbox{diag}}
\def\log{\hbox{log}}

\def\diag{\hbox{diag}}

\def\boxit#1{\vbox{\hrule\hbox{\vrule\kern6pt\vbox{\kern6pt#1\kern6pt}\kern6pt\vrule}\hrule}}

\newtheorem{theo}{Theorem}

\newtheorem{lem}{Lemma}

\newcommand{\be}{\begin{equation}}
\newcommand{\ee}{\end{equation}}

\newcommand{\bea}{\begin{eqnarray}}
\newcommand{\eea}{\end{eqnarray}}
\newcommand{\bsa}{\begin{eqnarray*}}
	\newcommand{\esa}{\end{eqnarray*}}

\def\bSigma{\boldsymbol \Sigma}

\def\bR{\mathbf R}

\usepackage{color}

\newcommand{\CB}{{\cal B}}

\newcommand{\CR}{{\bf R}}
\newcommand{\mCR}{{\cal R}}

\newcommand{\CT}{{\cal T}}

\def\wt{\widetilde}

\def\bse{\begin{eqnarray*}}
	\def\ese{\end{eqnarray*}}
\newcommand{\ben}{\begin{eqnarray}}
\newcommand{\een}{\end{eqnarray}}

\usepackage[normalem]{ulem}

\begin{document}

	\thispagestyle{empty} \baselineskip=30pt \vskip 5mm
	\begin{center} {\LARGE{\bf {Bias-correction} and {Test} for {Mark-point Dependence with }Replicated Marked Point Processes 
		}}
	\end{center}

	
	\baselineskip=19.5pt	
	\vskip 10mm	
	
	\begin{center}
		Ganggang Xu, Jingfei Zhang, Yehua Li and Yongtao Guan
	\end{center}
	
	\symbolfootnote[0]{
		Ganggang Xu (Email: gangxu@bus.miami.edu) is Assistant Professor, Jingfei Zhang (Email: ezhang@bus.miami.edu) is Associate Professor, and
		Yongtao Guan (Email: yguan@bus.miami.edu) is Professor,
		Department of Management Science, University of Miami, Coral
		Gables, FL 33124. Yehua Li (Email: yehuali@ucr.edu) is Professor, Department of Statistics, University of California, Riverside, CA 92521. Zhang's research is supported by NSF grant DMS-2015190 and Guan's research is supported by NSF grant DMS-1810591. For correspondence, please contact Yongtao Guan.
	}
	
	\begin{center}
		{\large{\bf Abstract}}
	\end{center}
	
	{
		Mark-point dependence plays a critical role in research problems that can be fitted into the general framework of marked point processes. In this work, we focus on adjusting for mark-point dependence when estimating the mean and covariance functions of the mark process, given independent replicates of the marked point process. We assume that the mark process is a Gaussian process and the point process is a log-Gaussian Cox process, where the mark-point dependence is generated through the dependence between two latent Gaussian processes. Under this framework, naive local linear estimators ignoring the mark-point dependence can be severely biased. We show that this bias can be corrected using a local linear estimator of the cross-covariance function and establish uniform convergence rates of the bias-corrected estimators. 
			Furthermore, we propose a test statistic based on local linear estimators for mark-point independence, which is shown to converge to an asymptotic normal distribution in a parametric $\sqrt{n}$-convergence rate. 
			Model diagnostics tools are developed for key model assumptions and a robust functional permutation test is proposed for a more general class of mark-point processes. 
		The effectiveness of the proposed methods is demonstrated using extensive simulations and applications to two real data examples.
	}

	\par\vfill\noindent
	{\bf Some key words:} {Mark-point Dependence; Marked Point Processes. }
	\par\medskip\noindent
	{\bf Short title}: {Nonparametric Estimation and Test for Marked Point Process}

	\clearpage\pagebreak\newpage \pagenumbering{arabic}
	\baselineskip=25.5 pt
	
	\allowdisplaybreaks

	\section{Introduction}
	In many scientific fields, numerical variables of interest are commonly observed at some random event times for a collection of subjects. For example, \cite{fok2012functional} considered systemic lupus erythematosus disease activity index (SLEDAI) scores of patients at times of flare episodes; \cite{gervini2020joint} studied the bid prices of Palm M515 personal digital assistants on week-long eBay auctions. In these examples, the events in turns are flare episodes and bids, while the associated numerical variables are SLEDAI score and bid price. The event times in each example are random and can be viewed as a realization from a point process, whereas the numerical variables are often referred to as marks. The random event times and the marks together form a so-called marked point process \citep{illian2008statistical}. The marks are often well defined over the entire study domain, but not just at event times. For example, SLEDAI scores could potentially be obtained at any time \citep{fok2012functional}, and it is therefore reasonable to assume a separate mark process that generated the SLEDAI scores for each patient.

	Marked point process data commonly arise in the analysis of longitudinal data with irregularly scattered observation times, where independent observations from different subjects are typically available.  Tools from functional data analysis have been used to model such data by treating the mark processes as random functions and the observed mark values as discrete observations on the functions \citep{Hsing-Eubank15}. While there has been extensive recent literature on this topic \citep[e.g.,][]{yao2005functional, chen2012modeling, ZhangWang2016AOS,   WangWongZhang2021}, most of existing work rely on a convenient but restrictive assumption stipulating that the marks and points are independent; we will refer to this assumption as \textit{mark-point independence} in this paper. 
		Potential mark-point dependence is not considered except in a few papers \citep[e.g.,][]{fok2012functional,gervini2020joint}.
	In real applications, however, the mark-point independence assumption may be invalid. For example, the SLEDAI scores are expected to be high at times of flare episodes \citep{fok2012functional} and hence the SLEDAI scores and the flare episode times may be correlated. Ignoring such mark-point dependence can lead to {biased estimation results for} the mark process. Hence, testing mark-point independence and correcting 
	any biases caused by such dependence can have a major impact on statistical practice in this area.

	In this paper, we consider the problem of estimating the mean and covariance functions of the mark process nonparametrically and testing mark-point independence, when independent replicates of the marked point process are available. To that end, we assume that the point process in each replicate is a log-Gaussian Cox process \citep[LGCP;][]{moller1998log}. Mark-point dependence can be modeled through the correlation between the latent Gaussian process defining the LGCP and the mark process which is also assumed to be Gaussian. {Similar assumptions have also been made}~in \cite{diggle2010geostat} and \cite{gervini2020joint}. Under the proposed modeling framework, we show that the naive {local linear} estimators of the {mean and covariance functions} that ignore the mark-point dependence are biased, where the {biases can be corrected by using a local linear estimator of} the cross-covariance {function} between the mark process and the latent Gaussian process for the point process; see Theorems~\ref{thm1}-\ref{thm2} for more details. The resulting bias-corrected estimators are shown to be uniformly consistent for their respective population counterparts.  

	Our proposed approach estimates the mean and covariance functions via nonparametric smoothing. Unlike the likelihood-based approaches \citep[e.g.,][]{fok2012functional,gervini2020joint}, the proposed estimators do not require fully specifying the data generating mechanism of the marked point process. {In particular, there is no need to model the point process {other than assuming it to be an LGCP} and requires no explicit assumption on how the mark-point dependence is generated. In this sense, these estimators are therefore more robust to model misspecifications than the existing {likelihood-based} methods.} A second advantage of the proposed method is computational since the local linear estimators can be efficiently computed given the selected bandwidths. In contrast, \cite{gervini2020joint} used the Karhunen-Lo\`eve expansion to {approximate} the mark process and the Gaussian process for the point process separately {with some basis functions. The resulting model parameters are estimated by} a penalized maximum likelihood approach, {which can be computationally intensive with several tuning parameters to be selected.}

	Finally, another important contribution of this work is the introduction of a new testing procedure for mark-point independence. To the best of our knowledge, the proposed test is the first formal test 
	designed for marked point processes with replicates. The limiting distribution of the proposed test statistic under mark-point independence is shown to be normal with mean $0$ and a variance that can be estimated using observed data. Surprisingly, the proposed test statistic converges to its limiting distribution at the classical parametric rate of $n^{1/2}$, even though it is constructed based on some nonparametric estimators; see Section~\ref{sec:test} and Theorem~\ref{thm3} for details. Our test relies on the assumptions that the mark process is Gaussian and the point process is an LGCP. We describe a set of diagnostic tools to check these assumptions, and propose a 
	functional permutation test
		for mark-point independence {that does not rely on distributional assumptions of the underlying marked-point process}.
		Our simulation studies demonstrate the validity and power of the proposed functional permutation test for {a variety of} mark-point process models.
	
	We note that some methods are developed to test the mark-point dependence \citep[see, e.g.,][]{schlather2004detect,guan2007test,zhang2014kolmogorov,zhang2017independence} in a single spatial mark-point process, though the majority require the marked point process to be stationary, which can be implausible in many applications. For example, both bid intensity and bid price may increase with time during an auction \citep{gervini2020joint}; as a result, neither the mark (i.e., bid price) process nor the point (i.e., bid time) process is stationary.

	The rest of the paper is organized as follows. 
	In Section \ref{sec:model}, we describe our model. In Section \ref{sec:naive_estimators}, we discuss the naive mean and covariance estimators for the mark process and their biases in the presence of mark-point dependence. 
	In Section \ref{sec:bias_correct_test}, we propose an estimator for the cross-covariance function between the mark process and 
	point processes, based on which we propose bias-corrected estimators for the mean and covariance functions of the mark process; we also propose a 
	testing procedure for the mark-point independence. Asymptotic properties of the proposed estimators and the test statistic are studied in Section \ref{sec:theory}. 
	{In Section~\ref{sec:second}, we describe a functional permutation test and some diagnostic tools to assess model assumptions.}
	Numerical performances of the proposed methods are illustrated by simulation studies in Section \ref{sec:simu} and two real datasets in Section \ref{sec:data}. Finally, some concluding remarks are provided in Section \ref{sec:conclusions}, and implementation details, additional numerical results, together with technical proofs, are collected in the supplement.
	
	%
	%
	%
	%
	%
	%
	%
	%
	%
	%
	%
	%
	%
	%
	%
	%
	%
	%
	%
	%
	
	\section{Model Specification}\label{sec:model}
	Consider a marked point process defined over a time window $\mathcal{T}\subset\mathbb{R}$. Let $\{[s,Z_i(s)]:s\in N_i, i=1,\ldots,n\}$ denote $n$ {independent} realizations of the process, where $N_i=\{s_{ij}: s_{ij}\in\mathcal{T},j=1,\ldots,n_i\}$ is the set of $n_i$ events from the $i$th point process and $Z_i(s)$ is the associated mark for an event at $s\in\CT$. We assume that the random event times in $N_i$ are generated by an LGCP, with the latent intensity function
	\be
	\label{point}
	\lambda_i(s)=\lambda_0(s)\exp\left[X_i(s)\right],
	\ee
	for $s\in\CT$. In the above, $\lambda_0(\cdot)$ is a baseline intensity function and $X_i(\cdot)$ is a latent zero-mean Gaussian process with a variance function $\sigma_X^2(\cdot)$, $i=1,\ldots,n$. Conditional on the latent intensity function, an LGCP is simply an inhomogeneous Poisson process. The first- and second-order marginal intensity functions of the point process are therefore
	\be
	\label{marginal}
	\rho(s)\equiv\E\left[\lambda_i(s)\right]=\lambda_0(s)\exp\left[\sigma_X^2(s)/2\right],
	\ee
	\be
	\label{marginal2}
	\rho_2(s,t)\equiv\E\left[\lambda_i(s)\lambda_i(t)\right]=\rho(s)\rho(t)\exp\left[C_X(s,t)\right],
	\ee
	where $C_X(s,t)=\Cov[X_1(s),X_1(t)]$, $s,t\in\CT$.

	We assume that the mark process is well defined for all $s\in\mathcal{T}$. More specifically,
	\be
	\label{mark}
	Z_i(s)=\mu(s)+Y_i(s)+e_i(s),
	\ee
	where $\mu(\cdot)$ is some {deterministic} function, $Y_i(\cdot)$ is a {zero-mean} Gaussian process with a variance {function} $\sigma_Y^2(\cdot)$, and  $e_i(\cdot)$ is a {zero-mean} Gaussian white noise process with a variance {function} $\sigma_e^2(\cdot)$, $i=1,\ldots,n$. 

	Denote $C_Y(s,t)=\Cov[Y_1(s),Y_1(t)]$ and $C_{XY}(s,t)=\Cov[X_1(s),Y_1(t)]$, for any $s,t\in\CT$. When $C_{XY}(\cdot,\cdot)\nequiv 0$, the mark process and the point process are not independent. We are interested in estimating the mean function $\mu(\cdot)$ and the covariance function $C_Y(\cdot,\cdot)$ based on the observed data. {We remark that 
		if the point process {reduces to} Poisson{, i.e., $C_{X}(\cdot,\cdot)\equiv 0$}, it always holds that $C_{XY}(\cdot,\cdot)\equiv 0$.}
	
	%
	%
	%
	%
	%
	%
	%
	%
	%
	%
	%
	%
	%
	%
	%
	%
	%
	%
	%
	%
	%
	
	%
	\section{Naive Estimation of the Mean and Covariance Functions}
	\label{sec:naive_estimators}
	When the mark process and the point process are independent, local {linear} estimators for $\mu(\cdot)$ and $C_Y(\cdot,\cdot)$ {are well studied \citep[e.g.,][]{yao2005functional}}. We refer to these estimators as the naive estimators. Define $K_{1,h_{\mu}}(s)=h_{\mu}^{-1}K_1(s/h_{\mu})$, where $K_1(\cdot)$ is a kernel function and $h_{\mu}$ is a bandwidth. {Then, the naive local linear estimator for $\mu(s)$, {$s\in\CT$,} is defined as $\wt\mu(s)=\wt\beta_{0,\mu}$, where $\wt\beta_{0,\mu}$ is obtained by minimizing 
		\[
		L_{n,\mu}(\beta_{0,\mu},\beta_{1,\mu})=\sum_{i=1}^n\sum_{u\in N_i}\left[Z_i(u)-\beta_{0,\mu}-\beta_{1,\mu}(u-s)\right]^2K_{1,h_{\mu}}(u-s)
		\]
		with respect to $\beta_{0,\mu}$ and $\beta_{1,\mu}$.} Let $\bphi_{h_{\mu}}(s)=(1,s/h_{\mu})^\top$, and denote by $\e_p$ a $p$-dimensional vector whose first entry equals $1$ and all other entries equal $0$. Then,
	\be
	\wt\mu(s)=\e_{2}^\top\left[\wh\A_{n,h_{\mu},1}(s)\right]^{-1} \wh\A_{n,h_{\mu},2}(s), {\quad s\in \CT,} \,\,\text{ where }\label{mutilde}
	\ee 
	\bea
	\wh\A_{n,h_{\mu},1}(s)&=&\frac{1}{n}\sum_{i=1}^n\sum_{u\in N_i} K_{1,h_{\mu}}(u-s)\bphi_{h_{\mu}}(u-s)\bphi_{h_{\mu}}(u-s)^\top ,\label{Ahat}\\
	\wh\A_{n,h_{\mu},2}(s)&=&\frac{1}{n}\sum_{i=1}^n\sum_{u\in N_i} K_{1,h_{\mu}}(u-s) \bphi_{h_{\mu}}(u-s) Z_i(u).\label{Ahat2}
	\eea 
	
	
	Define $K_{2,h_y}(s,t)=h_y^{-2}K_2(s/h_y,t/h_y)$, where $K_2(\cdot,\cdot)$ is a {bivariate} kernel function and $h_y$ is a bandwidth. Let $\wt W_i(u,v)=[Z_i(u)-\wt\mu(u)][Z_i(v)-\wt\mu(v)]$. {The naive local linear estimator for $C_Y(s,t)$, $s,t \in\CT$, is defined as $\wt C_Y(s,t)=\wt\beta_{0,y}$, where $\wt\beta_{0,y}$ is obtained by minimizing 
		\[
		L_{n,C_Y}(\beta_{0,y},\beta_{1,y}, \beta_{2,y})=\sum_{i=1}^n\mathop{\sum\sum}_{u,v\in N_i}^{\neq}\left[\wt W_i(u,v)-\beta_{0,y}-\beta_{1,y}(u-s)-\beta_{2,y}(v-t)\right]^2K_{2,h_y}(u-s,v-t)
		\]
		with respect to $\beta_{0,y}$,  $\beta_{1,y}$ and $\beta_{2,y}$. In the above, the $\ne$ sign indicates $u\ne v$.} Let $\bpsi_{h_y}(s,t)=(1,s/h_y,t/h_y)^\top$. The local linear estimator $\wt C_Y(s,t)$ can be written as
	\be
	\wt C_Y(s,t)=\e_{3}^\top\left[\wh\B_{n,h_y,1}(s,t)\right]^{-1}\wh\B_{n,h_y,2}(s,t), {\quad s,t \in 
		\CT}, \text{ where }\label{cytilde}
	\ee
	\bea
	\wh\B_{n,h_y,1}(s,t)&=&\frac{1}{n}\sum_{i=1}^n\mathop{\sum\sum}^{\neq}_{u,v\in N_i}K_{2,h_y}(u-s,v-t)\bpsi_{h_y}(u-s,v-t)\bpsi_{h_y}(u-s,v-t)^\top,\label{Bhat}\\
	\wh\B_{n,h_y,2}(s,t)&=&\frac{1}{n}\sum_{i=1}^n\mathop{\sum\sum}^{\ne}_{u,v\in N_i}K_{2,h_y}(u-s,v-t)\bpsi_{h_y}(u-s,v-t)\wt W_i(u,v).\label{Bhat2}
	\eea

	When the mark process and the point process are independent, $\wt\mu(\cdot)$ and $\wt C_Y(\cdot,\cdot)$ are consistent estimators for $\mu(\cdot)$ and $C_Y(\cdot,\cdot)$ under mild conditions \citep{yao2005functional, li2010uniform}. However, in the presence of mark-point dependence, both estimators will have non-negligible biases for their target parameters as we will show in the following two subsections. To facilitate the derivations, we first present a technical lemma.
	
	\begin{lem}\label{lm:stein}
		Let $X$, $Y_1$ and $Y_2$ be three normal random variables with means $0$, $\mu_1$ and $\mu_2$ and variances $\sigma^2_X$, $\sigma_1^2$ and $\sigma_2^2$, respectively. Then it holds that,
		\be
		\label{stein0}
		\E \left[Y_1\exp(X)\right]= [\mu_1+\Cov(X,Y_1)]\exp(\sigma_X^2/2), \ee
		\be
		\label{stein1}
		\E \left[Y_1^2\exp(X)\right]=\left\{\sigma_1^2 + [\mu_1+\Cov(X,Y_1)]^2\right\}\exp(\sigma_X^2/2), 
		\ee
		\be
		\label{stein2}
		\E \left[Y_1 Y_2\exp(X)\right]=\left\{\Cov(Y_1,Y_2) + \left[\Cov(X,Y_1)+\mu_1\right]\left[\Cov(X,Y_2) + \mu_2\right]\right\}\exp(\sigma_X^2/2).
		\ee
	\end{lem}
	{The proof is given in the supplement.}

	\subsection{Bias of $\wt\mu(\cdot)$}\label{biasmu}
	To derive the bias of $\wt\mu(s)$, $s\in\CT$, we first note that 
	$$
	\sum_{u\in N_i}K_{1,h_{\mu}}(u-s)\bphi_{h_{\mu}}(u-s)Z_i(u)=\int_\CT K_{1,h_{\mu}}(u-s)\bphi_{h_{\mu}}(u-s)Z_i(u) N_i(du),
	$$ 
	where $N_i(du)$ denotes the random number of events from the $i$th point process in an infinitesimal time interval $du$ at $u\in\CT$. Then, we have that {for any $s\in\CT$, }
	$$
	\A_{h_{\mu},2}(s)\equiv\E\left[\wh\A_{n,h_{\mu},2}(s)\right]=\int_\CT K_{1,h_{\mu}}(u-s)\bphi_{h_{\mu}}(u-s)\E \left[Z_i(u)N_i(du)\right],
	$$
	where $\wh\A_{n,h_{\mu},2}(s)$ is as defined in \eqref{Ahat2}. It follows from \eqref{marginal} and \eqref{mark} that
	\begin{eqnarray*}
		\A_{h_{\mu},2}(s)&=&\int_\CT \mu(u)\rho(u)K_{1,h_{\mu}}(u-s)\bphi_{h_{\mu}}(u-s)du\\
		&&+\int_\CT \lambda_0(u)\E\left\{Y_i(u)\exp[X_i(u)]\right\}K_{1,h_{\mu}}(u-s)\bphi_{h_{\mu}}(u-s)du,
	\end{eqnarray*}
	where $\rho(\cdot)$ is as defined in \eqref{marginal}. Since $X_i(u)$ and $Y_i(u)$, $u\in\CT$, are both normal random variables, it {immediately} follows from \eqref{stein0} that $\E\left\{Y_i(u)\exp[X_i(u)]\right\}=C_{XY}(u,u)\exp\left[{\sigma_X^2(u)/2}\right]$. If $\mu(\cdot)$ and $C_{XY}(\cdot,\cdot)$ are smooth functions such that $\mu(u)\approx \mu(s)$ and $C_{XY}(u,u)\approx C_{XY}(s,s)$ for {any} $u$ in a small neighborhood around $s$, then $\A_{h_{\mu},2}(s)\approx \left[\mu(s)+C_{XY}(s,s)\right]\a_{h_{\mu}}(s)$, where 
	\be\label{a}
	\a_{h_{\mu}}(s)=\int_\CT \rho(u) K_{1,h_{\mu}}(u-s)\bphi_{h_{\mu}}(u-s)du.
	\ee 
	Note that, due to the definition of $\bphi_{h_{\mu}}(\cdot)$, $\a_{h_{\mu}}(s)$ is the first column of the following matrix  
	\be
	\label{den1}
	\A_{h_{\mu},1}(s)\equiv\E\left[\wh\A_{n,h_{\mu},1}(s)\right]=\int_\CT \rho(u) K_{1,h_{\mu}}(u-s)\bphi_{h_{\mu}}(u-s)\bphi_{h_{\mu}}^\top(u-s)du, 
	\ee
	where $\wh\A_{n,h_{\mu},1}(s)$ is as defined in \eqref{Ahat}. By the law of large numbers, it holds under mild conditions that {$|\wh\A_{n,h_{\mu},1}(s)- \A_{h_{\mu},1}(s)|\xrightarrow{p} 0$ and $|\wh\A_{n,h_{\mu},2}(s)- \A_{h_{\mu},2}(s)|\xrightarrow{p} 0$, as $n\to\infty$, where $\xrightarrow{p}$ stands for convergence in probability}. It then follows from the definition of $\wt\mu(\cdot)$ in~\eqref{mutilde} {and the continuous mapping theorem} that for any $s\in\CT$,
	\be
	\label{bias-mu}
	{\wt\mu(s)} =     
	\e_{2}^\top\left[\A_{h_{\mu},1}(s)\right]^{-1}\a_{h_{\mu}}(s)\left[\mu(s)+C_{XY}(s,s)\right] \times[1+o_p(1)] \xrightarrow{p}\mu(s)+C_{XY}(s,s)\equiv \mu^*(s).
	\ee
	In other words, $\wt\mu(\cdot)$ is asymptotically unbiased for $\mu^*(\cdot)$, but not for $\mu(\cdot)$ unless $C_{XY}(\cdot,\cdot)\equiv 0$. 
	%
	%
	%
	%
	%
	%
	%
	%
	%
	%
	%
	%
	%
	%
	%
	%
	%
	%
	%
	
	\subsection{Bias of $\wt C_Y(\cdot,\cdot)$}
	
	For ease of presentation, we replace $\wt W_i(u,v)$ in \eqref{Bhat2} by $W_i(u,v)=[Z_i(u)-\mu^*(u)][Z_i(v)-\mu^*(v)]$. This is not of particular concern since $\wt\mu(\cdot)$ converges to $\mu^*(\cdot)$ uniformly in probability; see Theorem~\ref{thm1} in Section~\ref{sec:theory}. Now define $N_i^{(2)}(du,dv)=N_i(du)N_i(dv)I(u\ne v)$, where $I(\cdot)$ is an indicator function. Note that for any real function $f(\cdot,\cdot)$, it holds that
	$$
	\mathop{\sum\sum}^{\ne}_{u,v\in N_i} f(u,v)=\int_{\CT^2} f(u,v) N_i^{(2)}(du,dv).
	$$
	Then, for any $s,t\in\CT$, it holds that
	$$
	\B_{h_y,2}(s,t)\equiv\E \left[\wh\B_{n,h_y,2}(s,t)\right]=\int_{\CT^2} K_{2,h_y}(u-s,v-t)\bpsi_{h_y}(u-s,v-t) \E\left[W_{i}(u,v) N_i^{(2)}(du,dv)\right],
	$$
	where $\wh\B_{n,h_{y},2}(s)$ is as defined in \eqref{Bhat2}. Since {for any $u,v\in\CT$,}
	$$
	\E\left[W_{i}(u,v) N_i^{(2)}(du,dv)\right]=\lambda_0(u)\lambda_0(v)\E\left\{W_{i}(u,v) \exp\left[X_i(u)+X_i(v)\right]\right\}dudv,
	$$
	it follows from \eqref{marginal2} and \eqref{stein2} that for any $s,t\in\CT$,
	$$
	\B_{h_y,2}(s,t)=\int_{\CT^2} \rho_2(u,v) C_Y^*(u,v) K_{2,h_y}(u-s,v-t) \bpsi_{h_y}(u-s,v-t)du dv,
	$$ 
	where $\rho_2(u,v)$ is as defined in \eqref{marginal2} and
	\be
	\label{Cystar}
	C_Y^*(u,v)=C_Y(u,v)+C_{XY}(u,v)C_{XY}(v,u).
	\ee 
	If $C_{Y}(\cdot,\cdot)$ and $C_{XY}(\cdot,\cdot)$ are both smooth functions such that $C_{Y}(u,v)\approx C_{Y}(s,t)$ and $C_{XY}(u,v)\approx C_{XY}(s,t)$ for $(u,v)$ in a small neighborhood around $(s,t)$, then $\B_{h_y,2}(s,t)\approx C_Y^*(s,t)\b_{h_y}(s,t)$, where 
	\be\label{b}
	\b_{h_y}(s,t)=\int_{\CT^2} \rho_2(u,v) K_{2,h_y}(u-s,v-t) \bpsi_{h_y}(u-s,v-t)du dv.
	\ee
	Note that, due to the definition of $\bpsi_{h_y}(\cdot,\cdot)$, $\b_{h_y}(s,t)$ is the first column of the matrix 
	\be
	\label{den2}
	\B_{h_y,1}(s,t)\equiv \E\left[\wh\B_{n,h_y,1}(s,t)\right]=\int_{\CT^2} \rho_2(u,v) K_{2,h_y}(u-s,v-t)\bpsi_{h_y}(u-s,v-t)\bpsi_{h_y}^\top(u-s,v-t)dudv,
	\ee
	where $\wh\B_{n,h_{y},1}(s)$ is as defined in \eqref{Bhat}. By the law of large numbers, under suitable conditions, {$|\wh\B_{n,h_y,1}(s,t)- \B_{h_y,1}(s,t)|\xrightarrow{p}0$ and $|\wh\B_{n,h_y,2}(s,t)- \B_{h_y,2}(s,t)|\xrightarrow{p}0$} as $n\to\infty$. It then follows from the definition of $\wt C_Y(\cdot,\cdot)$ in~\eqref{cytilde} and the continuous mapping theorem that
	\be
	\label{bias-CY}
	{\wt C_Y(s,t)} =\e_{3}^\top\left[\B_{h_y,1}(s,t)\right]^{-1}\b_{h_y}(s,t)C_Y^*(s,t)\times[1+o_p(1)]\xrightarrow{p} C_Y^*(s,t), {\quad s,t\in\CT.}
	\ee
	This shows that the naive covariance function estimator $\wt C_Y(\cdot,\cdot)$ given in \eqref{cytilde} is an asymptotically unbiased estimator for $C_Y^*(\cdot,\cdot)$ defined in~\eqref{Cystar}, but not for $C_Y(\cdot,\cdot)$ unless $C_{XY}(\cdot,\cdot)\equiv0$.

	\section{Bias-correction and Test for Mark-point Independence}\label{sec:bias_correct_test}
	In this section, we propose a local linear estimator for the cross-covariance function $C_{XY}(\cdot,\cdot)$, based on which bias-corrected estimators for $\mu(\cdot)$ and $C_Y(\cdot,\cdot)$ are constructed. We also propose a formal testing procedure for mark-point independence. 
	The bandwidth selection procedure is detailed in Section~\ref{sec:bandwidth}.
	
	\subsection{Bias-Corrected Estimation of the Mean and Covariance Functions}\label{sec:bc-est}
	Following similar steps to obtain \eqref{cytilde}, the local linear estimator for $C_{XY}(s,t)$, $s,t\in\CT$, can be {defined as $\wh C_{XY}(s,t)=\wh\beta_{0,xy}$, where $\wh\beta_{0,xy}$ is obtained by minimizing 
		\[
		\begin{split}
		L_{n,C_{XY}}&(\beta_{0,xy},\beta_{1,xy},\beta_{2,xy})\\
		&=\sum_{i=1}^n\mathop{\sum\sum}_{u,v\in N_i}^{\neq}\left[Z_i(v)-\wt\mu(v)-\beta_{0,xy}-\beta_{1,xy}(u-s)-\beta_{2,xy}(v-t)\right]^2K_{2,h_{xy}}(u-s,v-t)
		\end{split}
		\]
		with respect to $\beta_{0,xy}$, $\beta_{1,xy}$ and $\beta_{2,xy}$, and $h_{xy}$ is a bandwidth. It can be shown that
		\be\label{cxyhat}
		\wh C_{XY}(s,t)=\e_{3}^\top\left[\wh\B_{n,h_{xy},1}(s,t)\right]^{-1}\wh\B_{n,h_{xy},3}(s,t), {\quad s,t\in\CT,}
		\ee
		where  $\wh\B_{n,h_{xy},1}(s,t)$ is as defined in \eqref{Bhat} with bandwidth $h_{xy}$, and
		\be\label{Bhat3}
		\wh\B_{n,h_{xy},3}(s,t)=\frac{1}{n}\sum_{i=1}^n\mathop{\sum\sum}^{\ne}_{u,v\in N_i}[Z_i(v)-\wt\mu(v)]K_{2,h_{xy}}(u-s,v-t)\bpsi_{h_{xy}}(u-s,v-t).
		\ee
		We will show below that $\wh C_{XY}(s,t)$ is asymptotically unbiased for $C_{XY}(s,t)$, despite the fact that $\wt\mu(\cdot)$ is a biased estimator for $\mu(\cdot)$. For ease of presentation, we replace $\wt\mu(\cdot)$ in \eqref{Bhat3} by $\mu^*(\cdot)$ as we did previously when studying the bias of $\wt C_Y(\cdot,\cdot)$. Then,
		\begin{eqnarray*}
			\B_{h_{xy},3}(s,t)&\equiv&\E\left[\wh\B_{n,h_{xy},3}(s,t)\right]\\
			&=&\int_{\CT^2} K_{2,h_{xy}}(u-s,v-t)\bpsi_{h_y}(u-s,v-t) \E\left\{[Z_i(v)-\mu^*(v)] N_i^{(2)}(du,dv)\right\}.
		\end{eqnarray*}
		Note that {for any $u,v\in\CT$,}
		$$
		\E\left\{[Z_i(v)-\mu^*(v)]N_i^{(2)}(du,dv)\right\}=\lambda_0(s) \lambda_0(s) \E\left\{[Z_i(v)-\mu^*(v)]\exp[X_i(u)+X_i(v)]\right\}dudv, 
		$$
		which is equal to $\rho_2(u,v) C_{XY}(u,v)dudv$ due to \eqref{marginal2} and \eqref{stein0}. If $C_{XY}(\cdot,\cdot)$ is a smooth function in a small neighborhood around $(s,t)$, then $\B_{h_{xy},3}(s,t)\approx C_{XY}(s,t)\b_{h_{xy}}(s,t)$, where $\b_{h_{xy}}(s,t)$ is as defined in \eqref{b} with bandwidth $h_{xy}$. Recall that $\b_{h_{xy}}(s,t)$ is the first column of the matrix $\B_{h_{xy},1}(s,t)$. It then follows from the same steps used to obtain~\eqref{bias-CY} that $\E\left[\wh C_{XY}(s,t)\right]\approx C_{XY}(s,t)$, that is, $\wh C_{XY}(s,t)$ is an asymptotically unbiased estimator for $C_{XY}(s,t)$ {for any $s,t\in\CT$}.

		{In light of~\eqref{bias-mu} and~\eqref{bias-CY} and given $\wh C_{XY}(\cdot,\cdot)$, it is natural to consider the following bias corrected estimators for $\mu(s)$ and $C_Y(s,t)$ for any $s,t\in\CT$,
			\bea
			\wh\mu(s)&=&\wt\mu(s)-\wh C_{XY}(s,s),\label{unbias-mu}\\
			\wh C_Y(s,t)&=&\wt C_Y(s,t)-\wh C_{XY}(s,t)\wh C_{XY}(t,s), \label{unbias-CY}
			\eea
			where $\wt\mu(\cdot)$ and $\wt C_Y(\cdot,\cdot)$ are as defined in~\eqref{mutilde} and~\eqref{cytilde}, respectively. As we shall show in Section~\ref{sec:theory}, $\wh\mu(\cdot)$ and $\wh C_Y(\cdot,\cdot)$ are uniformly consistent for $\mu(\cdot)$ and $C_Y(\cdot,\cdot)$, respectively.}

		\subsection{Test for Mark-Point Independence}
		\label{sec:test}
		In practice, it may be difficult to know in advance whether there exists mark-point dependence. In this section, we propose a formal procedure to test the {hypothesis} of mark-point independence, i.e., $H_0: C_{XY}(\cdot,\cdot)\equiv 0$. To motivate our test statistic, we first temporarily assume that $\mu^*(s)$ and $\sigma_Z^2(s)$, where $\sigma_Z^2(s)=\Var[Z(s)]$, are both known for any $s\in\CT$. Consider the following two random sums:
		\begin{eqnarray*}
			S_{n,1} = {1\over n} \sum_{i=1}^n\mathop{\sum\sum}^{\ne}_{u,v\in N_i} [Z_i(v)-\mu^*(v)]^2\;\;\hbox{ and }\;\; S_{n,2} = {1\over n} \sum_{i=1}^n\mathop{\sum\sum}^{\ne}_{u,v\in N_i} \sigma_Z^2(v).
		\end{eqnarray*}
		Under $H_0$, {it is straightforward to see that $\E( S_{n,1})=\E( S_{n,2})$}. More generally, note that
		$$
		\E( S_{n,1})=\int_{\CT^2} \lambda_0(u)\lambda_0(v)\E\left\{[Z_i(v)-\mu^*(v)]^2\exp[X_i(u)+X_i(v)]\right\} du dv.
		$$	
		Recall that $\mu^*(v)=\mu(v)+C_{XY}(v,v)$ {for any $v\in\CT$}. By the definition of $\rho_2(\cdot,\cdot)$ in \eqref{marginal2} and using \eqref{stein1} in Lemma 1, we can then verify that 
		$$
		\E( S_{n,1})=\int_{\CT^2} \rho_2(u,v) \left[\sigma_Z^2(v)+C_{XY}(u,v)^2\right] du dv.
		$$
		Combined with the fact that
		$
		\E( S_{n,2})=\int_{\CT^2} \rho_2(u,v)\sigma_Z^2(v) du dv,
		$	
		it holds that
		\be
		\label{test-bias}
		\E( S_{n,1})=\E( S_{n,2})+\int_{\CT^2} \rho_2(u,v)C_{XY}(u,v)^2 du dv.
		\ee
		Since the integral term in \eqref{test-bias} is strictly nonnegative and equals zero under the null, a test statistic can be formed based on the difference between $S_{n,1}$ and $ S_{n,2}$. 
		
		In practice, both $\mu^*(\cdot)$ and $\sigma_Z^2(\cdot)$ are unknown, and thus $S_{n,1}$ and $S_{n,2}$ cannot be calculated exactly. For $\mu^*(\cdot)$, we estimate it with $\wt\mu(\cdot)$ as defined in \eqref{mutilde}. For $\sigma_Z^2(\cdot)$, we consider its naive local linear estimator $\wt\sigma_Z^2(s)=\wt\beta_{0,\sigma}$, where $\wt\beta_{0,\sigma}$ is obtained by minimizing 
		\[
		L_{n,\sigma}(\beta_{0,\sigma},\beta_{1,\sigma})=\sum_{i=1}^n\sum_{u\in N_i}\left\{[Z_i(u)-\wt\mu(u)]^2-\beta_{0,\sigma}-\beta_{1,\sigma}(u-s)\right\}^2K_{1,h_{\sigma}}(u-s)
		\]
		with respect to $\beta_{0,\sigma}$ and $\beta_{1,\sigma}$, and $h_{\sigma}$ is a bandwidth. It immediately follows from the same arguments used to obtain~\eqref{mutilde} that for any $s\in\CT$,
		\be
		\label{sigmaytilde}
		\wt\sigma_Z^2(s)=\e_{2}^\top\left[\wh\A_{n,h_\sigma,1}(s)\right]^{-1} \wh\A_{n,h_\sigma,3}(s), \text{ where }
		\ee
		\[
		\wh\A_{n,h_\sigma,3}(s)=\frac{1}{n}\sum_{i=1}^n\sum_{u\in N_i} K_{1,h_\sigma}(u-s) \bphi_{h_\sigma}(u-s) [Z_i(u)-\wt\mu(u)]^2.
		\] 
		As we shall show in Section~\ref{sec:theory}, $\wt\mu(\cdot)$ and $\wt\sigma_Z^2(\cdot)$ are uniformly consistent for $\mu^*(\cdot)$ and $\sigma_Z^2(\cdot)$, respectively. {The uniform convergence of $\wt\sigma_Z^2(\cdot)$ to $\sigma_Z^2(\cdot)$ is surprising considering that $\wt C_Y(s,s)$ is biased for $C_Y(s,s)$, i.e., the variance of $Y(s)$, for $s\in\CT$, under mark-point dependence.} 
		
		{Given $\wt\mu(\cdot)$ and $\wt\sigma_Z^2(\cdot)$, we propose a test statistic of the following form}
		\be\label{test}
		T_n={1\over n} \sum_{i=1}^n\mathop{\sum\sum}^{\ne}_{u,v\in N_i} [Z_i(v)-\wt\mu(v)]^2-{1\over n} \sum_{i=1}^n\mathop{\sum\sum}^{\ne}_{u,v\in N_i} \wt\sigma_Z^2(v).
		\ee
		{Theorem~\ref{thm3} in Section~\ref{sec:theory} shows that, under some conditions, when $h_\mu=o(n^{-1/4})$ and $h_\sigma=o(n^{-1/4})$, $\sqrt{n}T_n$ converges to $N(0,\Omega_T)$ in distribution under $H_0$. The specific form of $\Omega_T$ is given in Theorem~\ref{thm3}, based on which an empirical estimator can be derived, as given in Section \ref{sec:a1} of the supplement.} 
		See also Section~\ref{sec:bandwidth} for the bandwidth selection procedure.
		
		%
		%
		%
		%
		%
		%
		%
		%
		%
		%
		%
		%
		%
		%
		%
		%
		%
		%
		%
		%
		%

		%
		%
		%

		%
		%
		%
		%
		%
		%
		%
		%
		%
		%
		%
		%
		%
		%
		%
		%
		%
		
		\section{Theoretical Properties}
		\label{sec:theory}
		\renewcommand{\labelenumi}{[C\theenumi]}
		In this section, we investigate theoretical properties of {the} proposed estimators and test statistic. We consider the asymptotic framework where the time domain $\CT$ is bounded while the number of replicates $n$ goes to infinity. 
		For any matrix (or vector) $\A$, denote $\|\A\|_{\max}=\max_{i,j}(|a_{ij}|)$ as the the max norm, and $\lambda_{\min}(\A)$ as the the smallest eigenvalues of $\A$. We use $\text{diag}\{d_1,\ldots,d_n\}$ to denote an $n\times n$ diagonal matrix with diagonal elements $d_1,\ldots,d_n$. Finally, for a function $f(\cdot)$, we denote $f^{(j)}(\cdot)$ as its $j$th derivative for some $j\geq 1$. 
		
		The following assumptions are sufficient for our theoretical investigation.
		\begin{enumerate}
			\itemsep-0.5em 
			\item Assume that $Z(s)$ is well defined for any $s\in\CT$, and $\rho(\cdot)$, $\mu(\cdot)$, $\mu^*(\cdot)$ and $\sigma_Z^2(\cdot)$ are {third-order} continuously differentiable on $\CT$ with bounded 1st-, 2nd-, and 3rd-order derivatives, and there exists a  constant $\rho_0>0$ such that $\rho(s)>\rho_0$ for all $s\in \CT$.
			\item  Assume that $K_1(\cdot)$ and $K_2(\cdot,\cdot)$ have supports $[-1,1]$ and $[-1,1]^2$, respectively.  
			\begin{enumerate}
				\itemsep-0.5em 
				\item[(a)] Define $\bm\sigma_{K_1,\bphi} = \int_{-1}^1 s^{2} K_1(s)\bphi_1(s) ds$ and $\Q_{K_1,\bphi}^{\{j\}}= \int_{-1}^1 K_1^j(s)\bphi_1(s)\bphi_1^\top(s) ds$ for $j=1,2$. Assume that $\lambda_{\min}\left(\Q_{K_1,\bphi}^{\{1\}}\right)>0$. 
				\item[(b)] Define $\bm\sigma_{K_2,\bpsi}(w) = \int_{-1}^1\int_{-1}^1 \left[(1-w)s^2+wt^{2}\right] K_2(s,t)\bpsi_1(s,t)ds dt$ and 
				$$\Q_{K_2,\bpsi}^{\{j\}}= \int_{-1}^1\int_{-1}^1 K_2^j(s,t)\bpsi_1(s,t)\bpsi_1^\top(s,t) dsdt, \;\;\hbox{for }j=1,2.$$ Assume that $\lambda_{\min}\left(\Q_{K_2,\bpsi}^{\{1\}}\right)>0$. 
			\end{enumerate}
			\item\label{assump:mu_bandwidth} Assume that {(a) $h_{\mu}\to0$ as $n\to\infty$ such that $nh_{\mu}^2/[\log(n)]^2\to\infty$; (b) $h_\sigma/h_\mu\to c$ as $n\to\infty$ for some constant $c>0$. 
			} 
			\item Assume that $\rho_2(\cdot,\cdot)$, $C_Y(\cdot,\cdot)$ and $C_{XY}(\cdot,\cdot)$ are {third-order} continuously differentiable on $\CT^2$ with bounded first-, second-, and third-order partial derivatives.
			\item\label{assump:C_XY_bandwidth} Assume that (a) $h_y\to 0$ as $n\to \infty$ such that $nh_y^2/[\log(n)]^2 \to \infty$; (b) $h_{xy}\to 0$ as $n\to \infty$ such that $nh_{xy}^2/[\log(n)]^2 \to \infty$; (c) $h_\mu^{3/2}/h_y\to 0$ and  $h_\mu^{3/2}/h_{xy}\to 0$ as $n\to\infty$. 
		\end{enumerate}
		{Assumptions C1 and C4 impose some conditions on the smoothness of the intensity functions and the mean, variance and covariance functions}. Assumption C2 specifies some desirable properties of the kernel functions used for the proposed local linear estimators and is satisfied by many popular kernel functions such as the Epanechnikov kernel. Assumption C3 {warrants} that the bandwidths $h_\mu$ and $h_\sigma$ {are }of the same order {and neither can} approach $0$ at a rate faster than $\log(n)/\sqrt{n}$ as $n\to\infty$. {In addition,} Assumption C5 requires that $h_y$ and $h_{xy}$ decay at a rate slower than $\log(n)/\sqrt{n}$ as $n\to\infty$ and $h_\mu$ cannot be too large compared to $h_y$ or $h_{xy}$.  
		Similar conditions have been commonly used in the existing literature, see, for example, \cite{yao2005functional, li2010uniform}.
		The first theorem investigates the uniform convergence of the three naive estimators when there is possible mark-point dependence. 
		
		\begin{theo}
			\label{thm1}
			{\bf (Naive estimators)} Under Assumptions C1-C5, we have that, as $n\to\infty$,
			\begin{enumerate}[noitemsep]
				\item[(a)] $	\sup_{s\in \CT} \left|\wt \mu(s) -\mu(s)+C_{XY}(s,s)\right|  = O_p\left\{h_{\mu}^{2} + \left[\log(n)/ (n h_{\mu})\right]^{1/2}\right\}$;
				\item[(b)] $\sup_{s\in \CT} \bigg |\wt {\sigma}^2_Z(s) -\sigma^2_Z(s) \bigg |  =O_p\left\{h_{\sigma}^{2} + \left[\log(n)/ (n h_{\sigma})\right]^{1/2}\right\}$;
				\item[(c)] $\sup_{s,t\in \CT} \bigg |\wt C_Y(s,t) -C_Y(s,t)-C_{XY}(s,t)C_{XY}(t,s) \bigg |  =O_p\left\{h_y^{2} + \left[\log(n)/ (n h_y^2)\right]^{1/2}\right\}$.
			\end{enumerate}
		\end{theo}
		The proof is given in the supplement. 
		
		Theorem~\ref{thm1} suggests that when there exists {mark-point} dependence (i.e. $C_{XY}(s,t)\notequiv 0$), both the naive mean function estimator $\wt\mu(\cdot)$ and the naive covariance function estimator $\wt C_Y(\cdot,\cdot)$ are biased for their respective targets.
		{The uniform convergence rates in Theorem~\ref{thm1} are comparable to the optimal rates in the literature \citep{li2010uniform},  where the non-diminishing biases caused by mark-point dependence were not considered.}
		A somewhat surprising observation is that the local linear estimator $\wt {\sigma}^2_Z(\cdot)$ is still uniformly consistent for ${\sigma}^2_Z(\cdot)$, despite the existence of mark-point {dependence}.
		The next theorem studies the asymptotic properties of the proposed bias-corrected estimators.
		\begin{theo}
			\label{thm2}
			{\bf (Bias corrected estimators)} Under Assumptions C1-C5, it holds as $n\to\infty$,
			\begin{enumerate}[noitemsep]
				\item[(a)] $\sup_{s,t\in \CT} \bigg |\wh C_{XY}(s,t) -C_{XY}(s,t)\bigg |  =O_p\left\{h_{xy}^{2} + \left[\log(n)/ (n h_{xy}^2)\right]^{1/2}\right\}$;
				\item[(b)] $\sup_{s\in \CT} \left|\wh \mu(s) -\mu(s)\right|  = O_p\left\{h_{\mu}^{2}+h_{xy}^{2} + \left[\log(n)/ (n h_{\mu})\right]^{1/2}+\left[\log(n)/ (n h_{xy}^2)\right]^{1/2}\right\}$;
				\item[(c)] $\sup_{s,t\in \CT} \bigg |\wh C_Y(s,t) -C_Y(s,t)\bigg |  =O_p\left\{h_y^{2}+h_{xy}^{2} + \left[\log(n)/ (n h_y^2)\right]^{1/2}+\left[\log(n)/ (n h_{xy}^2)\right]^{1/2}\right\}$.
			\end{enumerate}
		\end{theo}
		The proof is given in the supplement. 
		
		The key result in Theorem~\ref{thm2} is part (a), which establishes uniform consistency of the local linear estimator $\wh C_{XY}(\cdot,\cdot)$ to the cross-covariance function $C_{XY}(\cdot,\cdot)$ between the mark process and the point process. The proposed estimator achieves the same uniform convergence rate as the classical local linear estimator of the covariance function based on sparse functional data \citep{li2010uniform}. As a result of part (a), uniform consistency of the bias-corrected estimators $\wh\mu(\cdot)$ and $\wh C_Y(\cdot,\cdot)$ are subsequently established in (b) and (c). Note that the bandwidth $h_{xy}$ appears in the uniform convergence rates in parts (b)-(c) because $\wh C_{XY}(\cdot,\cdot)$ is used for bias corrections in $\wh\mu(\cdot)$ and $\wh C_Y(\cdot,\cdot)$.
		
		Finally, we give the asymptotic distribution of the test statistic proposed in Section~\ref{sec:test}. 
		Define $\rho_3(s,u,v)=\E\left[\lambda_i(s)\lambda_i(u)\lambda_i(v)\right]$, $\rho_4(s,t,u,v)=\E\left[\lambda_i(s)\lambda_i(t)\lambda_i(u)\lambda_i(v)\right]$, for $s,t,u,v\in\CT$. Expressions of $\rho_3(\cdot,\cdot,\cdot)$ and $\rho_4(\cdot,\cdot,\cdot,\cdot)$ can be derived with some algebra and be expressed in terms of $\sigma_X^2(\cdot)$ and $C_X(\cdot,\cdot)$. We omit their detailed expressions here. 
		
		\begin{theo}
			\label{thm3}
			{\bf (Test statistic)} Under Assumptions C1-C3 and assuming $nh_{\mu}^6\to 0$ and $nh_{\sigma}^6\to 0$ as $n\to\infty$, under $H_0: C_{XY}(\cdot,\cdot)\equiv 0$, we have that, 
			\[
			\sqrt{n}\left[ T_n + h_{\sigma}^2 \int_\CT \CB_{\sigma^2}(v) \rho(v)\tau(v)  dv\right]\xrightarrow{d} N(0,\Omega_T),
			\]
			where $\tau(s)=\frac{\int_\CT \rho_2(u,s)du}{\rho(s)}$ , $\CB_{\sigma^2}(s)  = {(\sigma_Z^2)^{(2)}(s)\over 2} \e_{2}^\top\left[\Q_{K_1,\bphi}^{\{1\}}\right]^{-1}\bm\sigma_{K_1,\bphi}$, {$s\in\CT$}, and 
			\[
			\begin{split}
			\Omega_T=2 &\int_{\CT^4} C_Y^2(t, v) \rho_4(s,t, u,v) ds dt du dv + 2 \int_{\CT^3}{\left\{\sigma_Z^4(v)+[3-2\tau(s)]C_Y^2(s,v)\right\}} \rho_3( s, u, v) ds du dv \\
			&{+ 2 \int_{\CT^2} \left[1-  \tau(u)\right]\left[ 1-  \tau(v)\right] C_Y^2(u,v) \rho_2( u, v) du dv} + 2 \int_\CT \left[ 1-  \tau(v)\right]\tau(v)\rho(v) \sigma_Z^4(v)  dv.
			\end{split}
			\]
			Here $\xrightarrow{d}$ denotes convergence in distribution.
		\end{theo}
		The proof is given in the supplement. 
		
		We remark that the bias term in Theorem~\ref{thm3} is negligible if $\sqrt{n}h_{\sigma}^2\to0$. Theorem~\ref{thm3} indicates that although $T_n$ is calculated using local linear estimators $\wt\mu(\cdot)$ and $\wt\sigma_Z(\cdot)$, it still achieves the parametric convergence rate $n^{-1/2}$ if $n^{1/4}h_\sigma\to 0$ as $n\to\infty$, which is a surprising result. Such a result enables us to conduct a valid hypothesis test for mark-point independence without specifying the complete data generating mechanism, provided that a consistent estimator of the variance $\Omega_T$ can be obtained. In practice, we estimate $\Omega_T$ by $\wh\Omega_{T_n}$ given in Section \ref{sec:a1} of the supplement. Validity of the resulting testing procedure and its empirical power are examined through simulations in Section~\ref{sec:simu}.
		{
			\section{Functional Permutation Test and Model Diagnostics}
			\label{sec:second}
			Our methods thus far are developed under two main assumptions, that is, the mark process is Gaussian and the point process is an LGCP. In this section, we propose a functional permutation test for testing the mark-point independence that does not reply on these two assumptions; we also devise diagnostic tools to check these two assumptions.
			
			\subsection{A Functional Permutation Test for Mark-Point Independence}
			\label{sec:perm}
			Using the Karhunen-Lo\`eve expansion, the mark process $Z_i(\cdot)$ in~\eqref{mark} can be approximated by
			\be
			\label{tpca1}
			Z_i(s)\approx \mu(s)+\sum_{k=1}^{p_Y}\xi_{ik}^Y\phi_k^Y(s)+e_i(s), \quad s\in\CT,i=1,\ldots,n,
			\ee
			where {$\xi_{ik}^Y$'s are uncorrelated random variables with mean $0$ and} variance $\eta_k$, with $\eta_k$ being the $k$th largest eigenvalue of $C_Y(\cdot,\cdot)$, and $\phi_k^Y(\cdot)$ is the associated eigenfunction, $1\le k\le p_Y$. Following similar proofs as in Section~\ref{sec:theory}, we can show that under mark-point independence, the bias-corrected estimator $\wh C_Y(\cdot,\cdot)$ in~\eqref{unbias-CY} is uniformly consistent for $C_Y(\cdot,\cdot)$, regardless of whether the mark process is Gaussian or the point process is an LGCP. 
			Denoting the $k$th eigenvalue and eigenfunction of $\wh C_Y(\cdot,\cdot)$ by $\wh\eta_k$ and $\wh\phi^{Y}_k(\cdot)$, respectively, the functional principal component (FPC) scores $\wh\bxi_i^Y=(\hat\xi^{Y}_{i1},\ldots,\hat\xi^{Y}_{ip_Y})^\top$ for the $i$th process can be obtained by
			\be
			\label{Fscore}
			\wh\bxi_i^Y=\argmin_{\bxi}\sum_{s\,\in N_i}\Big\{Z_i(s)-\wh\mu(s)-\sum_{k=1}^{p_Y}\xi^{Y}_{ik}\wh\phi^{Y}_k(s)\Big\}^2,
			\ee
			where $\wh\mu(\cdot)$ is as defined in~\eqref{unbias-mu}. 
			{In practice,} $p_Y$ can be chosen as the smallest integer such that the percentage of 
			{cumulative} variation explained by the first $k$ components {is greater than or equal to 95\%, i.e.,} $\sum_{k=1}^{p_Y}\wh\eta_k/\sum_{k=1}^{\infty}\wh\eta_k\ge 0.95$.

			{Let $c(1),\ldots,c(n)$ be a random permutation of the sequence $1,\ldots,n$, and define $\wh e_i(s)=Z_i(s)-\wh\mu(s)-\sum_{k=1}^{p_Y}\wh\xi^{Y}_{ik}\wh\phi^{Y}_k(s),s\in N_i$ for all $i$. 
				We obtain a set of permuted mark-point processes 
				$\{[s,Z_i^{c}],s\in N_i\}_{i=1,\ldots,n}$ calculated as $Z_i^c(s)=\wh\mu(s)+\sum_{k=1}^{p_Y}\wh\xi_{c(i)k}^Y\wh\phi_k^Y(s)+\wh e_i^{\text{per}}(s)$, where $\hat\xi_{c(i)k}^Y$'s are the permuted scores and $\wh e_i^{\text{per}}(s)$'s are the permuted residuals. 
					To permute the residuals, we start from a concatenation of the residual vectors from all subjects, denoted as $(\wh e_i(s), s\in N_i, i=1,\ldots,n)^\top$, and randomly permute all elements in this concatenated vector to get permuted residuals, denoted as $(\wh e_i^{\text{per}}(s), s\in N_i, i=1,\ldots,n)^\top$.
				We repeat the permutation for $B$ times and obtain the p-value as
				\be
				\label{pval}
				\text{ p-value }=\frac{1}{B}\sum_{b=1}^{B}I\Big(\Big|T_n^{c_b}\Big|>|T_n|\Big),
				\ee
				where $T_n$ and $T_n^{c_b}$ are the test statistics computed based on the observed data and the $b$th permuted data, respectively. 
				
				{In the test described above, the permutations} are designed to remove the mark-point dependence, if there is any, while preserving the marginal distributional properties of the mark process and the point process. We {shall} empirically demonstrate that the proposed functional permutation test is valid for various classes of marked point processes, and can be {implemented together with} the proposed test in Section~\ref{sec:test}. Theoretical justifications of the functional permutation test shall be deferred to future research. {More implementation details of the functional permutation test are given in Section \ref{sec:a2} of the supplement.}
				
				\subsection{Model Diagnostics}
				\label{sec:diag}
				If the mark process is Gaussian, the estimated FPC scores $\wh\bxi_{i}^Y$'s in~\eqref{Fscore} should approximately follow the normal distribution. Consequently, we can check the {normality} of $\wh\bxi_{i}^Y$'s {using, for example, the QQ plot,} to indirectly validate that the mark process is a Gaussian process.
				
				To evaluate the validity of the LGCP assumption on the point process, we first derive nonparametric estimates of $\lambda_0(\cdot)$ and $C_X(\cdot,\cdot)$ under the LGCP assumption; see Section \ref{sec:a3} of the supplement for details. Given these estimates, we simulate independent realizations from the LGCP defined by them and calculate some summary statistics from the observed and simulated point processes. A large difference between these summary statistics indicates that the LGCP assumption may be invalid. In this article, we consider the empirical nearest-neighbor distance distribution function \citep[e.g.,][]{moller2003statistical}. For the observed data, this function can be calculated as
				$$
				\wh G(d)=\frac{1}{n}\sum_{i=1}^n \frac{1}{{|N_i|}}\sum_{u\in N_i} I[d_i(u)\le d],
				$$
				where $d_i(u)$ is the distance from $u\in N_i$ to its nearest neighbor in $N_i$. The same summary statistic can be calculated from each simulated point process, 
				{and we plot $\hat G(d)$ against the average of those obtained from the simulated realizations over a range of $d$ values, along with the upper and lower simulation envelopes. If the LGCP assumption holds, the plot should be roughly linear and contained in the simulation envelopes.} {See Section~\ref{sec:data} for an illustration.}
			}			

			\section{Simulation Studies}
			\label{sec:simu}
			In this section, we evaluate the finite sample performance of the proposed methods. The mark processes are generated from $Z_i(s)=\mu(s)+Y_i(s)+e_i(s)$, $s\in [0,1]$, $i=1,\ldots,n$, with 
			{\begin{equation}\label{eq:expan} \mu(s)=\frac{1}{2} {\rm Beta}_{3,7}(s)+\frac{1}{2} {\rm Beta}_{7,3}(s),\quad Y_{i}(s)=\sum_{k=1}^{3}\xi_{ik}^Y\phi_k^Y(s),\quad e_i\sim N(0,1^2), 
				\end{equation}
				where ${\rm Beta}_{a,b}(\cdot)$ is the density function of the Beta(a,b) distribution, $\bxi_i^Y=(\xi_{i1}^Y,\xi_{i2}^Y,\xi_{i3}^Y)^\top$'s are i.i.d random vectors with mean $\0$ and covariance matrix $\bSigma_Y=\diag\{1,0.6,0.4\}$,  and 
				\begin{equation}\label{eq:y3}
				\phi_1^Y(s)=\sqrt{2}\sin(2\pi s),\quad \phi_2^Y(s)=\sqrt{2}\cos(2\pi s),\quad \phi_3^Y(s)=\sqrt{2}\sin(4\pi s).
				\end{equation}}
			The point processes are generated from an inhomogeneous Cox process according to \eqref{point}, where $X_i(\cdot)$ therein has an isotropic covariance function of the form
			\be
			\label{eq:covfun}
			C_X(s,t)=\sigma_x^2\exp\left[-{|s-t|^2}/{(2R^2)}\right],\,\sigma_x >0,R>0, \text{ for any } s,t\in[0,1].
			\ee
			For a given $\sigma_x$, we set $\lambda_0(s)=\frac{15}{8} (15+2s)\exp\left(-\frac{1}{2}\sigma_x^2\right)$, $s\in[0,1]$, such that there are on average $30$ observed events for each subject if the point process is an LGCP.
			The parameters $\sigma_x$ and $R$ jointly control the clustering strength of the point process. We fix $R=0.1$ and vary $\sigma_x$ in different scenarios. In particular, when $\sigma_x=0$, the resulting point process is an inhomogeneous Poisson process.
			
			By standard stochastic process theory, the Karhunen-Lo\`{e}ve decomposition of $X_i(\cdot)$ gives
			\[
			X_i(s)=	\sum_{k=1}^{\infty}\xi_{ik}^X\phi_k^X(s), \quad \text{ with } \E\xi_{ik}^X=0, \Var(\xi_{ik}^X)=\eta_k^X, \text{ for any } s\in[0,1],
			\]
			where $\left\{\eta_k^X,\phi_k^X(\cdot)\right\}$, $k=1, 2,\ldots$, are eigenvalue-eigenfunction pairs of the covariance function $C_X(\cdot,\cdot)$ defined in~\eqref{eq:covfun}. Denote $\bxi_i^X=(\xi_{i1}^X,\xi_{i2}^X,\xi_{i3}^X)^\top$ such that $\E \bxi_i^X=\0$ and $\bSigma_X=\Var(\bxi_i^X)=\diag\{\eta_1^X,\eta_2^X,\eta_3^X\}$. 	We let the correlation matrix between $\bxi_i^X$ and $\bxi_i^Y$ be of the form
			\be
			\label{cor}
			\bSigma_{XY}=\mathrm{Corr}\left(\bxi_i^X,\bxi_i^Y\right)=q\times \left(\begin{array}{ccc}
				1 & -0.5 & -0.25\\
				0.375 & 1 & -0.125 \\
				0.125 & 0 & 1 
			\end{array}\right),\quad \text{ for } 0\leq q\leq 0.8,
			\ee
			and {assume that the} remaining $\xi_{ik}^X$'s with $k\geq 4$ are independent of $\bxi_i^Y$. {Under such a design, $X_i(\cdot)$ and $Y_i(\cdot)$ are independent when $q=0$, and the strength of correlation between $X_i(\cdot)$ and $Y_i(\cdot)$ increases as $q$ increases. The range of $q$ is selected such that the variance-covariance matrix of the joint distribution of $(\bxi_i^X,\bxi_i^Y)$ is positive definite. 
				{When $q=0.8$, the correlation between $X(s)$ and $Z(s)$ ranges from $-0.16$ to $0.53$ for $s\in[0,1]$.}
				
				{Finally, $\bxi_i^X$'s and $\bxi_i^Y$'s are jointly simulated as follows
					\be
					\label{joint}
					\left(
					\begin{array}{c}
						\bxi_i^X\\
						\bxi_i^Y	
					\end{array}
					\right)=\bR \left(
					\begin{array}{c}
						\epsv_i^X\\
						\epsv_i^Y	
					\end{array}
					\right),\text{ with } \left(
					\begin{array}{cc}
						\bSigma_X&\bSigma_{XY}\\
						\bSigma_{XY}^\top&\bSigma_Y	
					\end{array}
					\right)=\bR^\top \bR,
					\ee
					where $\bR$ is an upper triangular matrix obtained through the Cholesky decomposition, $\epsv_i^X$ and $\epsv_i^Y$ are random vectors consisting of i.i.d random variables with mean $0$, variance $1$, and marginal distributions $\mathbb{P}_X$ and $\mathbb{P}_Y$, respectively. Three types of marginal distributions are considered for $\mathbb{P}_X$ and $\mathbb{P}_Y$, namely, $N(0,1)$, referred to as \texttt{Gaussian}, centered exponential distribution with a rate $1$, referred to as \texttt{Exp}, and scaled t-distribution with a degrees of freedom $4$, referred to as \texttt{T4}. When both $\mathbb{P}_X$ and $\mathbb{P}_Y$ are Gaussian, \eqref{joint} generates data under models~\eqref{point} and \eqref{mark}. For ease of presentation, from now on, we shall denote, for example, (\texttt{Gaussian}, \texttt{Exp}) for the setting when $\epsv_i^X$'s are Gaussian and  $\epsv_i^Y$'s are exponential.}

				\subsection{Estimation Accuracy}
				\label{sec:simacu}
				We first compare the estimation accuracy of the proposed 
				estimators in the presence of mark-point dependence. Specifically, there are four functions of interest as discussed below.
				\begin{enumerate}
					\itemsep -0.5em
					\item[(1).] $\mu(\cdot)$: the mean function of the mark process. We consider both the naive estimator $\wt\mu(\cdot)$ defined in~\eqref{mutilde} and the bias-corrected estimator $\wh\mu(\cdot)$ defined in~\eqref{unbias-mu}. 	The estimation accuracy of each estimator is evaluated through the mean absolute deviation (MAD) defined as
					$
					\text{MAD}(\wt\mu)=\int_0^1\left|\wt\mu(s)-\mu(s)\right|d s,\text{ and } \text{MAD}(\wh\mu)=
					\int_0^1\left|\wh\mu(s)-\mu(s)\right|d s.
					$
					\item[(2).] $\sigma_Z^2(\cdot)$: the variance function of the mark process. 
					{Given the naive estimator $\wt\sigma_Z^2(\cdot)$ defined in~\eqref{sigmaytilde}, we report }
					$
					\text{MAD}(\wt\sigma_Z^2)=\int_0^1\left|\wt\sigma_Z^2(s)-\sigma_Z^2(s)\right|ds.
					$
					\item[(3).] $C_Y(\cdot,\cdot)$: the covariance function of the mark process. The estimation accuracy of the naive estimator $\wt C_Y(\cdot,\cdot)$ in~(\ref{cytilde}) and the bias-corrected estimator $\wh C_Y(\cdot,\cdot)$ in \eqref{unbias-CY} are compared through
					$
					\text{MAD}(\wt C_Y)=\int_0^1\int_0^1\left|\wt C_Y(s,t)-C_Y(s,t)\right|d s d t$, and 
					$\text{MAD}(\wh C_Y)=\int_0^1\int_0^1 \left|\wh C_Y(s,t)-C_Y(s,t)\right|d s d t.
					$
					\item[(4).] $C_{XY}(\cdot,\cdot)$: the cross-covariance function between the mark process and the point process. The estimation accuracy of the estimator $\wh C_{XY}(\cdot,\cdot)$ defined in~\eqref{cxyhat} is evaluated through
					$\text{MAD}(\wh C_{XY})=\int_0^1\int_0^1\left|\wh C_{XY}(s,t)-C_{XY}(s,t)\right|d s d t.$
				\end{enumerate}
				
				{The Epanechnikov kernel is used with respective bandwidths selected following Section~\ref{sec:bandwidth} for each simulated data set}. We fix $\sigma_x=1$ for the covariance function~\eqref{eq:covfun}, and consider {varying strength of correlation} with $q=0.5,0.6,0.7,0.8$, as the sample size $n$ varies from $100$ to $600$. Individuals with more than $200$ time points are removed to enhance numerical stability.} Summary statistics based on $500$ simulation runs are illustrated in Figure~\ref{fig-1} under the (\texttt{Gaussian}, \texttt{Gaussian}) setting.
			\begin{figure}[!t]
				\centering
				\begin{tabular}{lll}
					\includegraphics[width=0.3\textwidth]{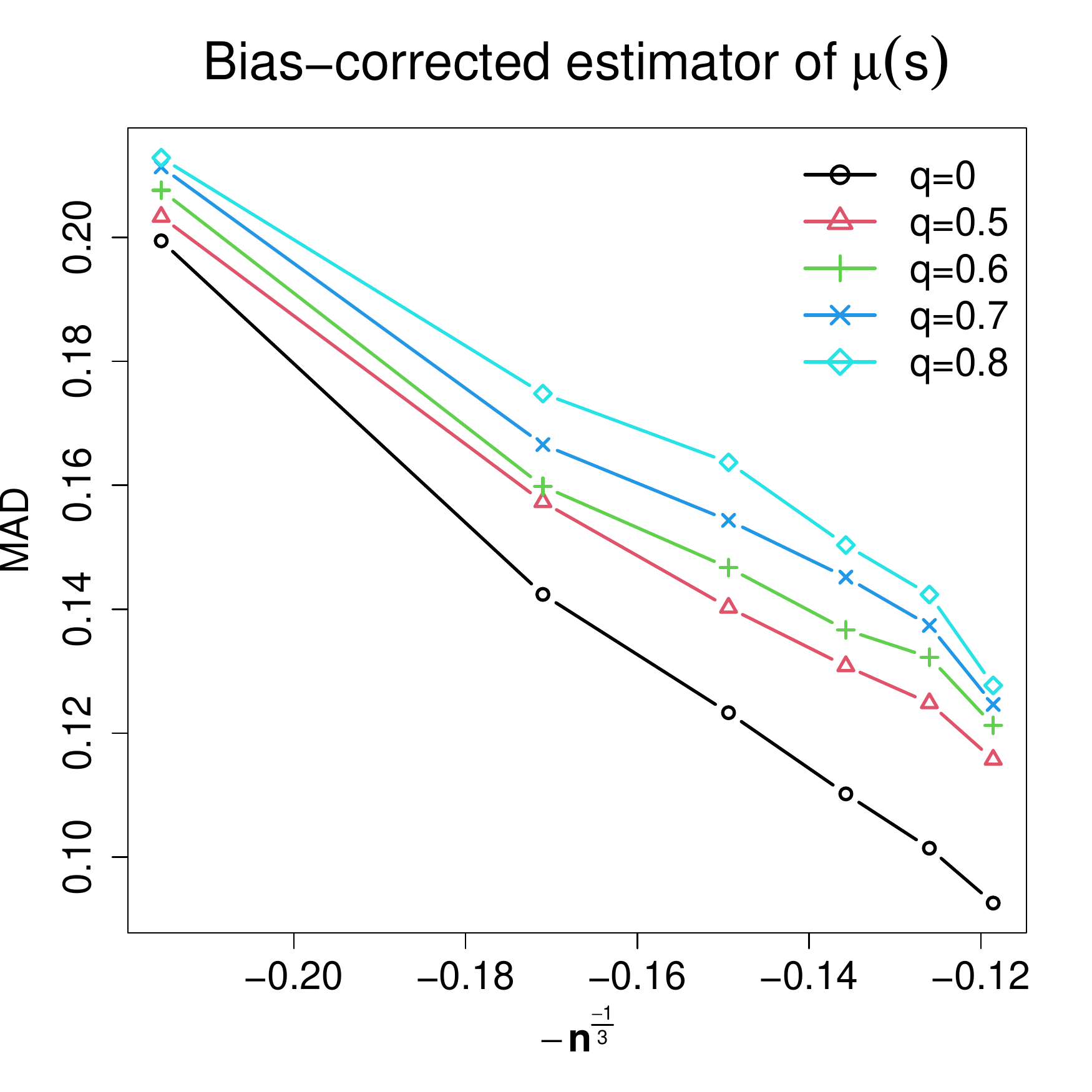} &
					\includegraphics[width=0.3\textwidth]{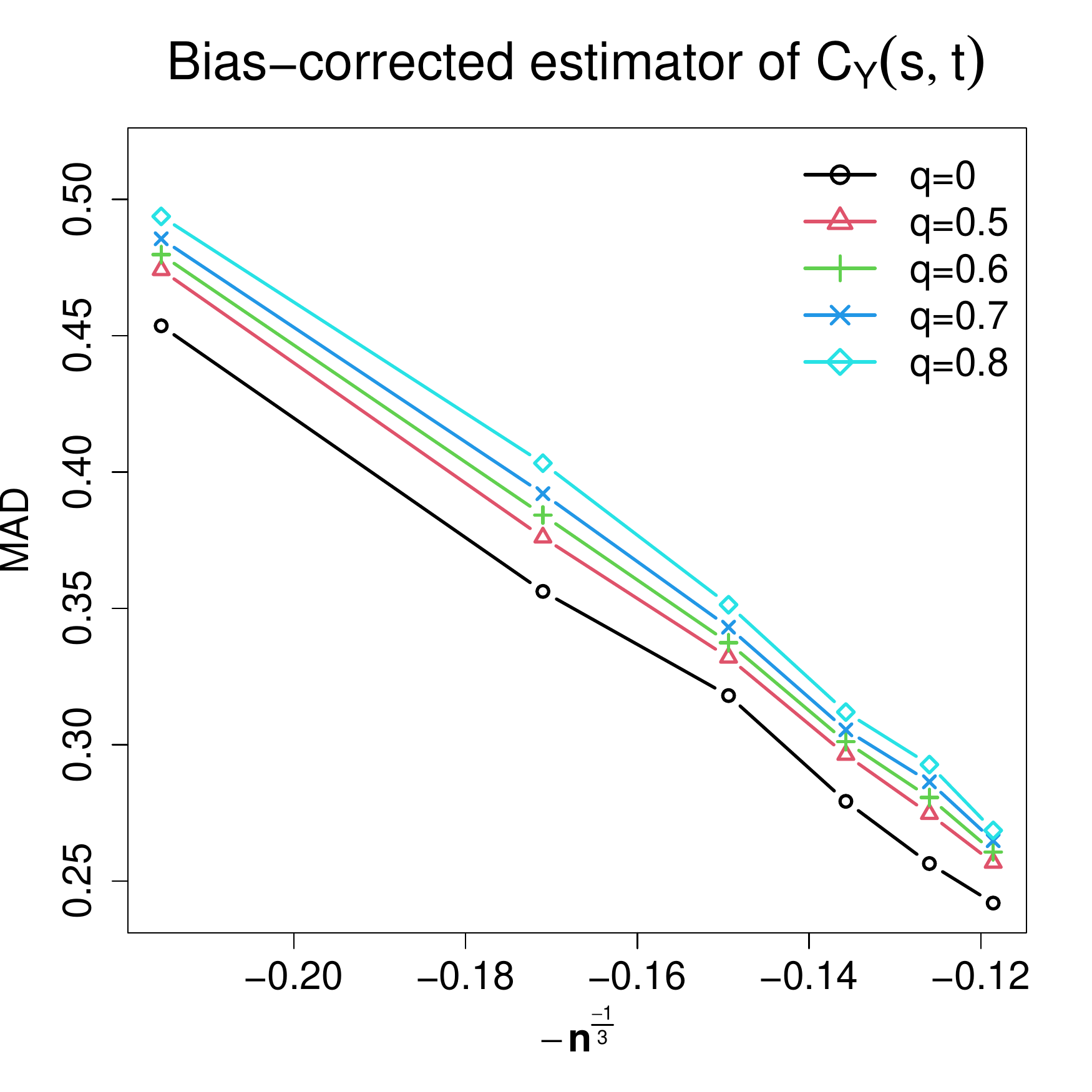}&
					\includegraphics[width=0.3\textwidth]{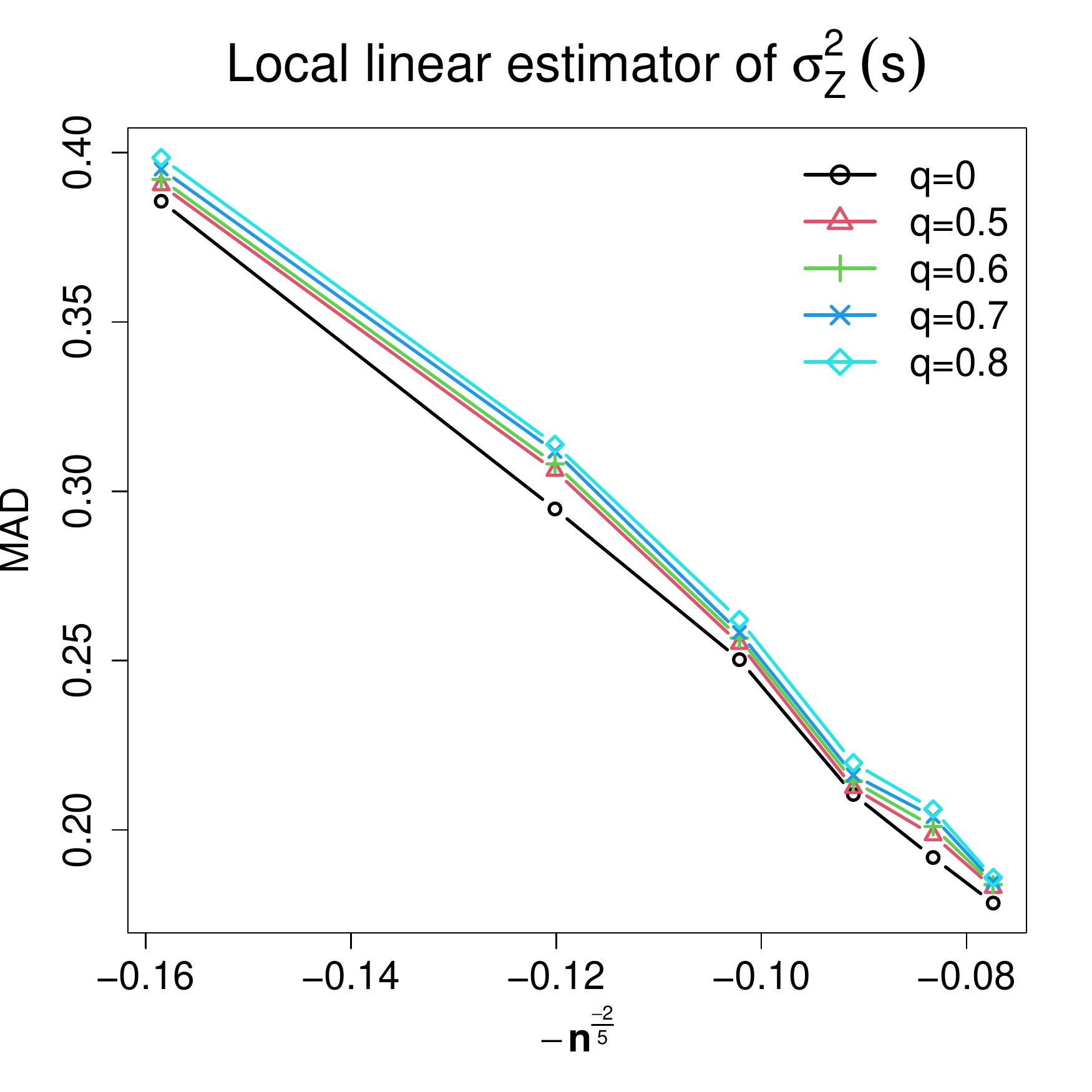}\\
					\includegraphics[width=0.3\textwidth]{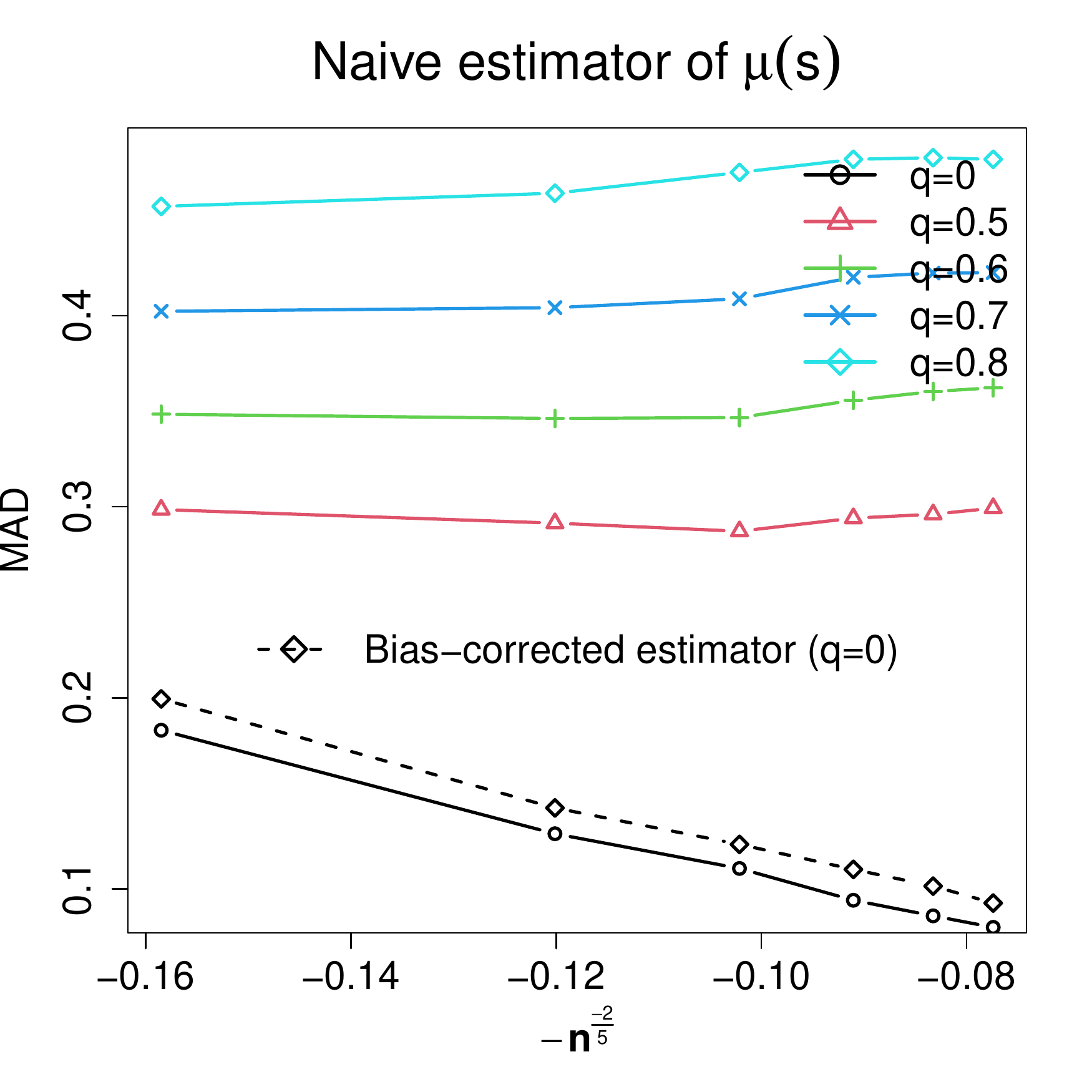} &
					\includegraphics[width=0.3\textwidth]{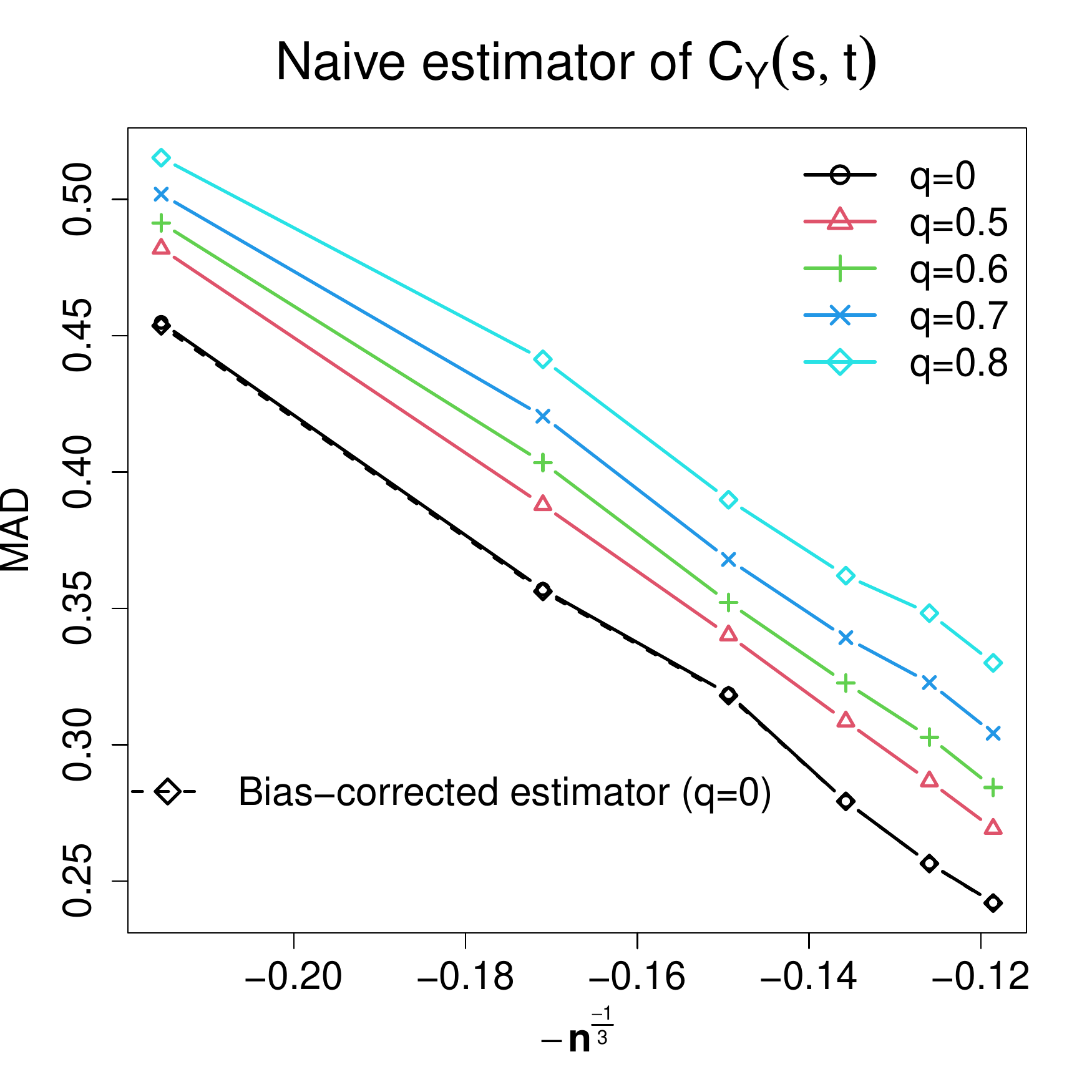}&
					\includegraphics[width=0.3\textwidth]{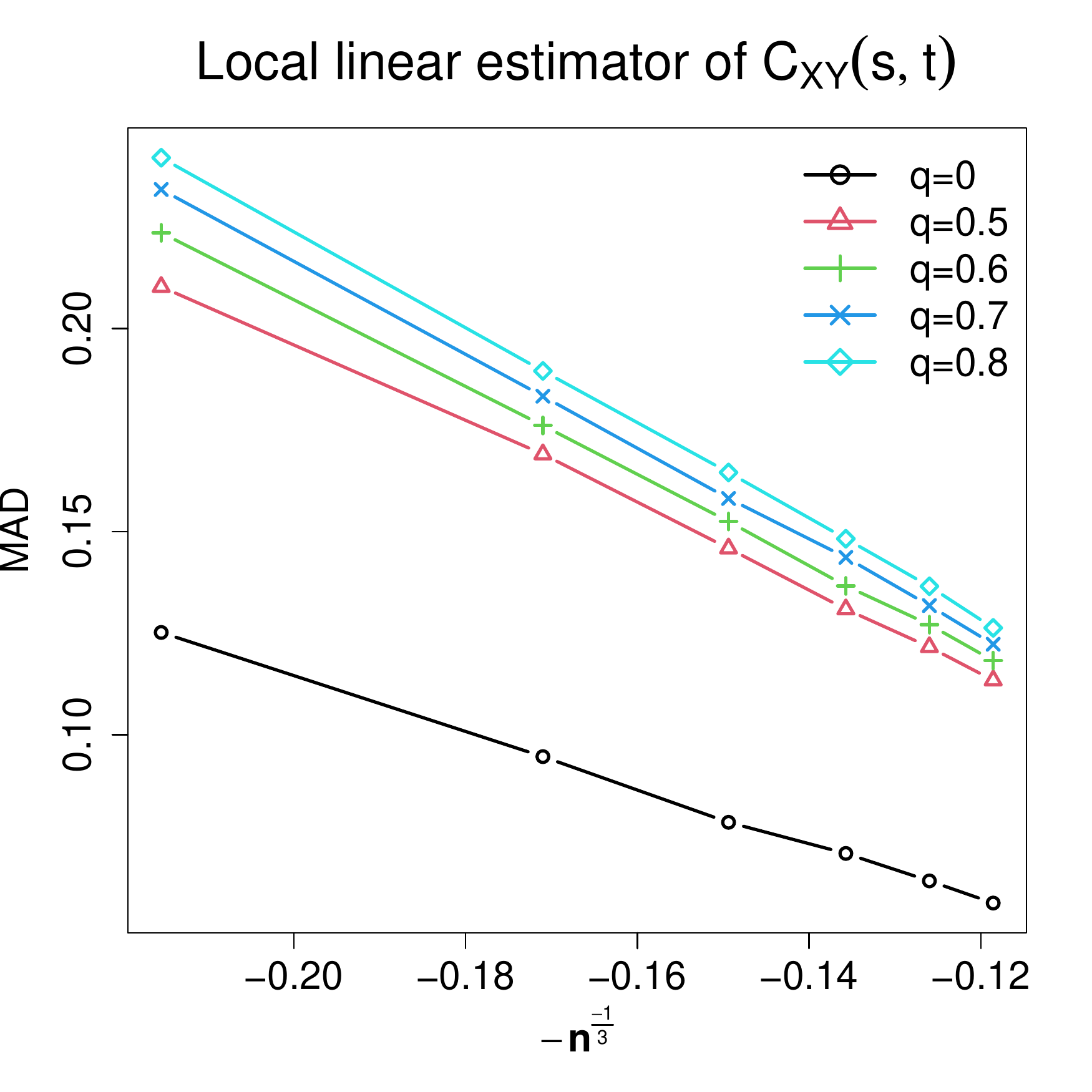}\\
				\end{tabular}
				\caption{Estimation accuracy of local linear estimators under the  (\texttt{Gaussian}, \texttt{Gaussian}) setting.}
				\label{fig-1}
			\end{figure}

			
			It is seen from Figure~\ref{fig-1} that $\text{MAD}(\wh\mu)$ and $\text{MAD}(\wt\sigma_Z^2)$ are approximately of orders $O(n^{-1/3})$ and $O(n^{-2/5})$, respectively. This agrees with our theoretical results in Theorems~\ref{thm1}-\ref{thm2}.
			Specifically, by Theorem~\ref{thm1}, the optimal bandwidth for $\wt\sigma_Z^2(\cdot)$ is roughly of the order $O(n^{-1/5})$, giving the optimal convergence rate $O(n^{-2/5})$. 
			Moreover, Theorem~\ref{thm2} indicates that the optimal convergence rate of $\wh\mu(\cdot)$ is approximately of order $O(n^{-1/3})$, due to the bandwidth $h_{xy}$ used in the bias correction term. 
			Hence, the observed rates for $\text{MAD}(\wh\mu)$ and $\text{MAD}(\wt\sigma_Z^2)$ support our theoretical findings and validate the effectiveness of the bandwidth selection criteria proposed in Section~\ref{sec:bandwidth}.

			{Similarly, Theorem~\ref{thm2} suggests that $\text{MAD}(\wh C_{XY})$ and $\text{MAD}(\wh C_{Y})$ are roughly of order $O(n^{-1/3})$ using the optimal bandwidths of order $O(n^{-1/6})$, and the graphic summaries in Figure~\ref{fig-1} strongly corroborate these conclusions and the effectiveness of the bandwidth selection criteria proposed in Section~\ref{sec:bandwidth}.}  When there is no mark-point dependence ($q=0$), it appears that the estimation efficiency loss caused by the unnecessary bias correction is minimal. We, therefore, recommend always using proposed bias corrections in practice.
			
			Next, as expected, the naive estimator of $\mu(\cdot)$, i.e. $\wt\mu(\cdot)$, suffers from considerably larger estimation errors than the bias-corrected estimators in all scenarios. To demonstrate this, in Figure~\ref{fig-2}, we give the mean and $95\%$ quantile bands of $500$ estimated functions under the (\texttt{Gaussian}, \texttt{Gaussian}) setting in the case of $q=0.8$ and $n=600$, where there is a larger bias for the naive estimator $\wt\mu(\cdot)$ than the bias-corrected estimator $\wh\mu(\cdot)$. 
			
			\begin{figure}[ht!]
				\centering
				\begin{tabular}{lll}
					\includegraphics[trim= 0 5mm 0 0, width=0.3\textwidth]{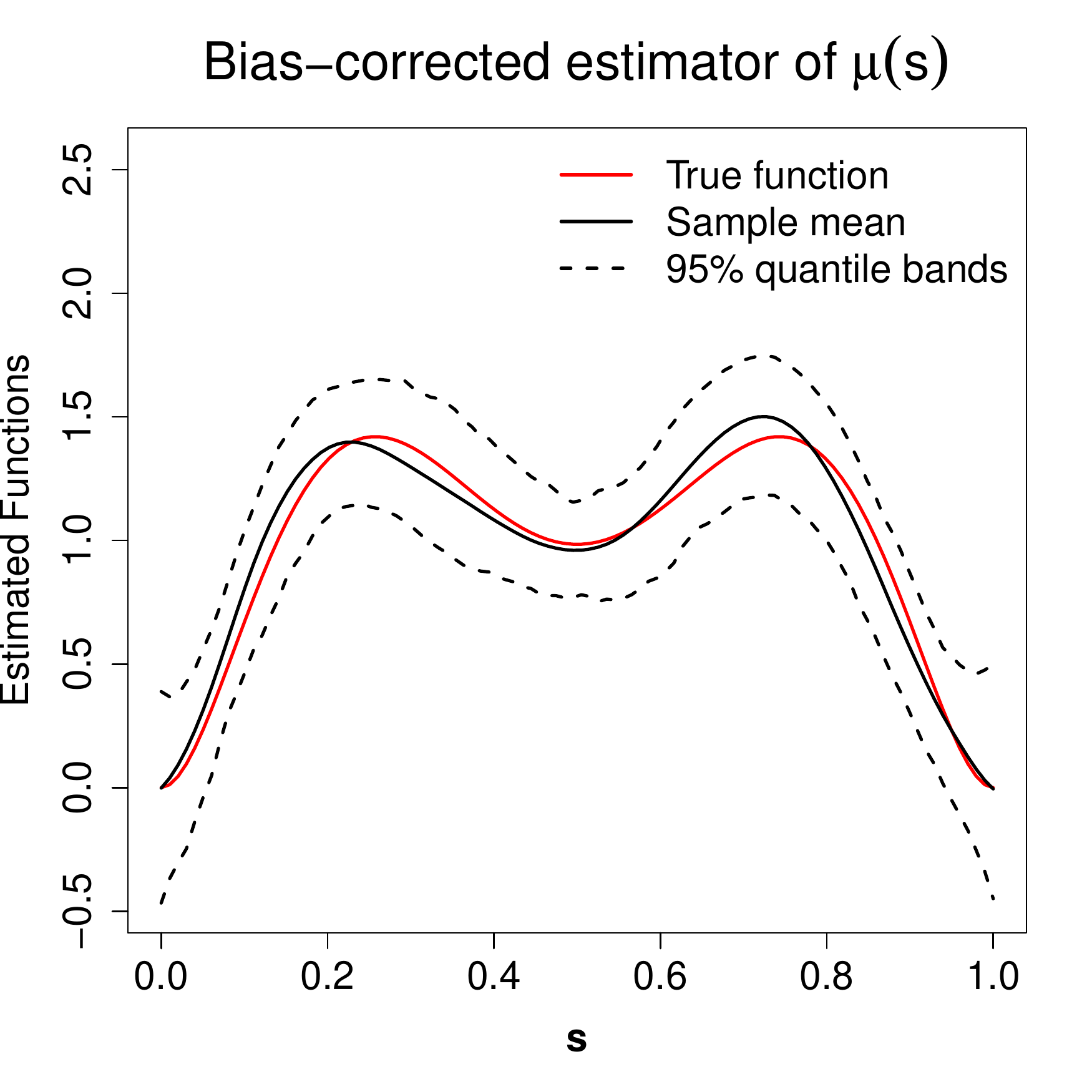} &
					\includegraphics[trim= 0 5mm 0 0, width=0.3\textwidth]{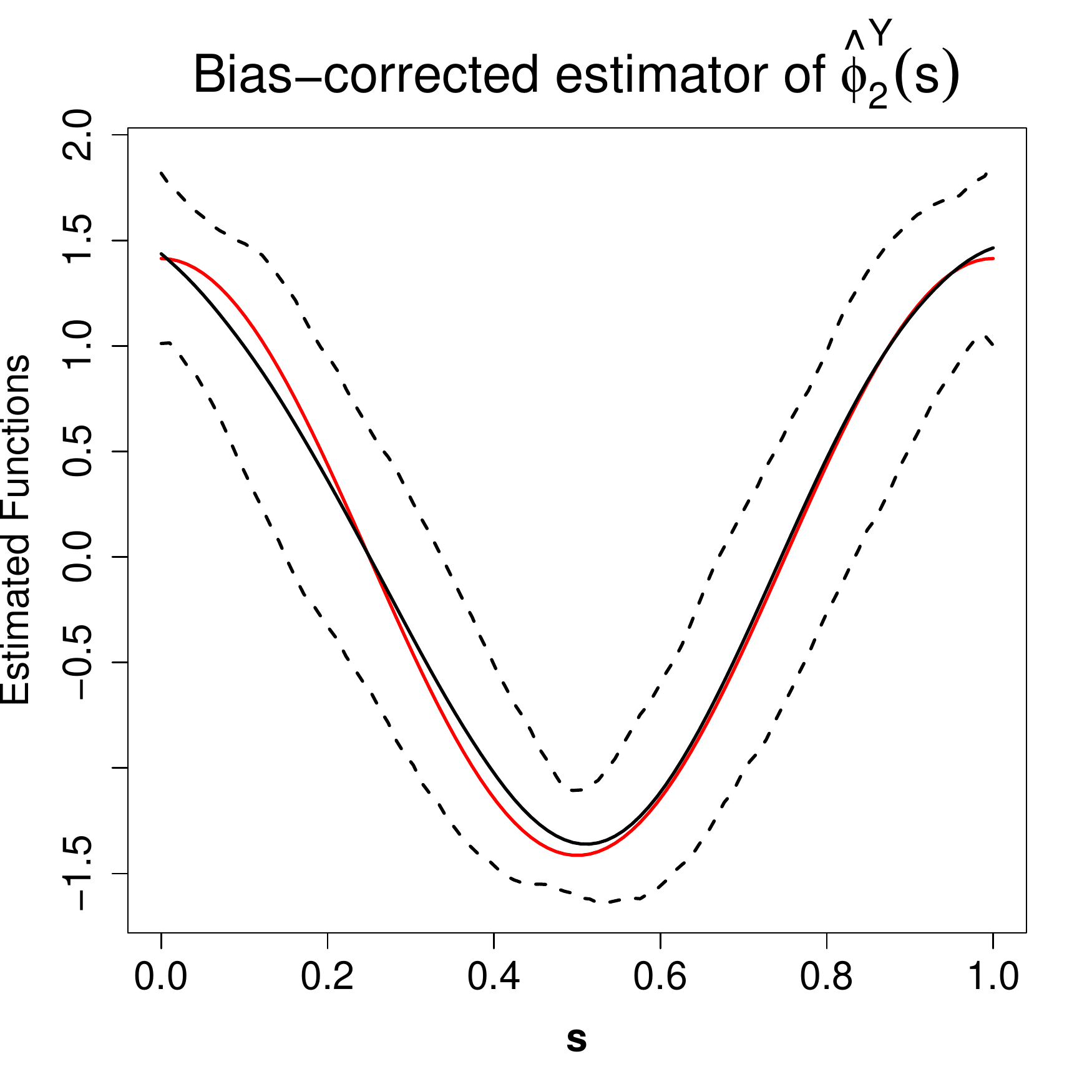}&
					\includegraphics[trim= 0 5mm 0 0, width=0.3\textwidth]{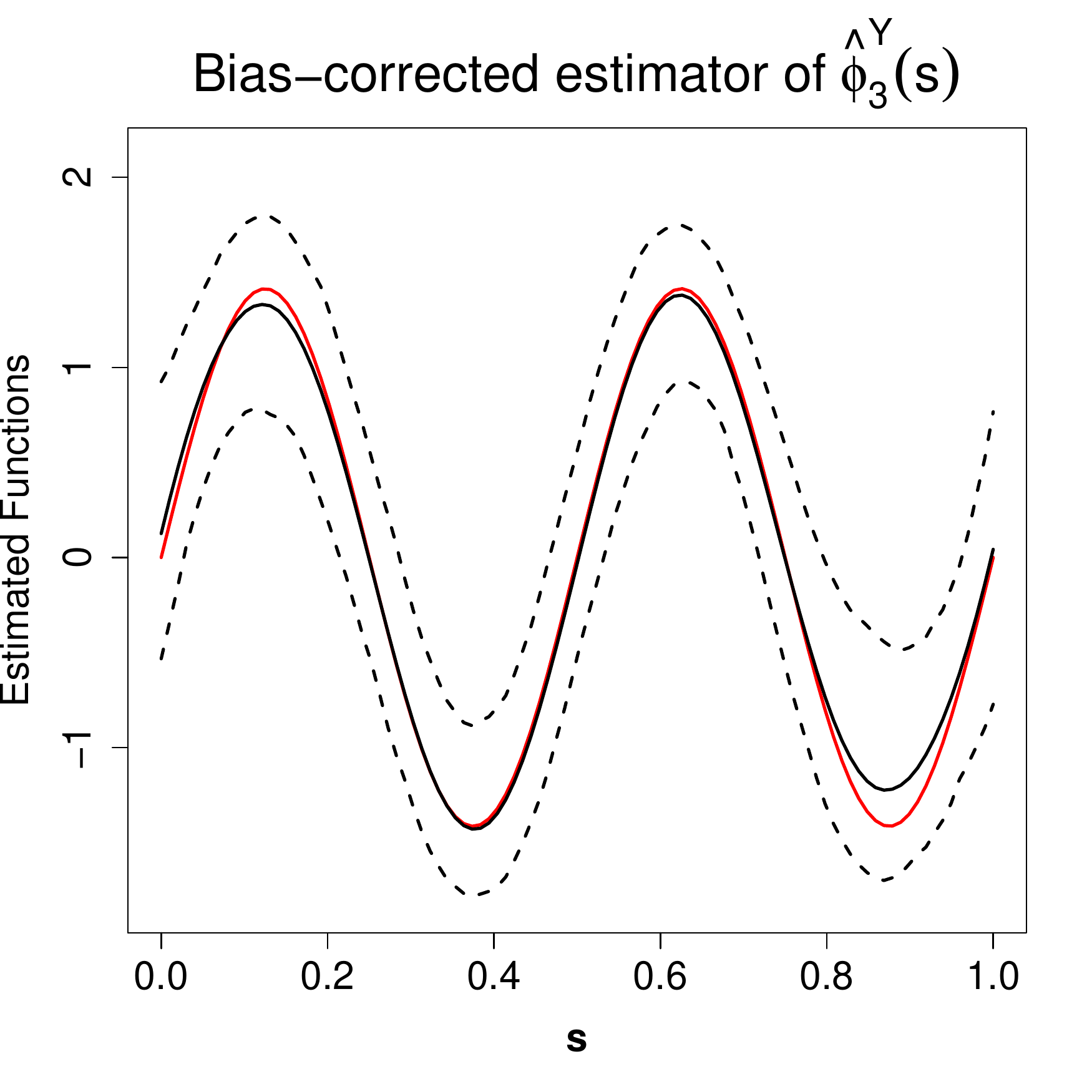}\\
					\includegraphics[trim= 0 5mm 0 0, width=0.3\textwidth]{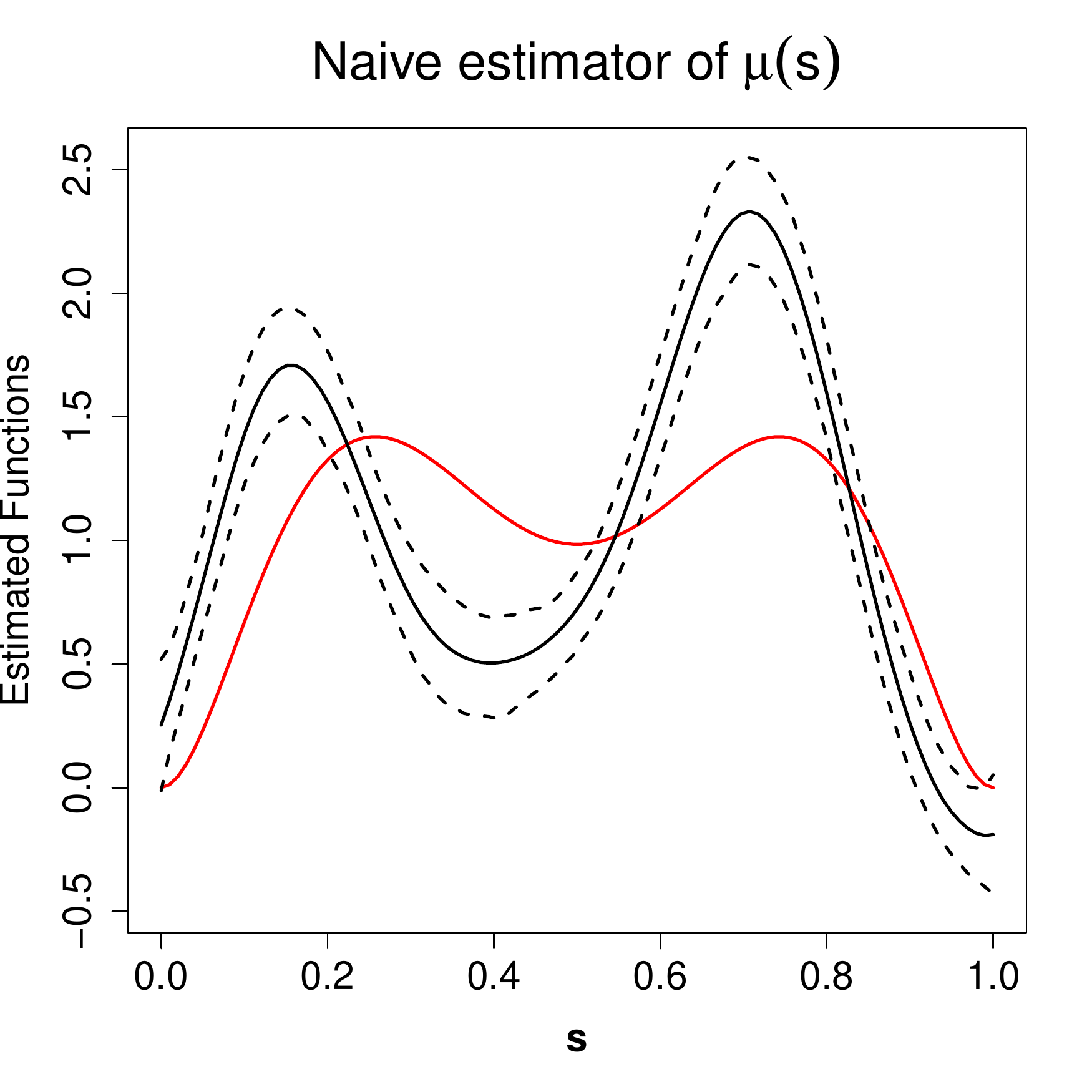} &
					\includegraphics[trim= 0 5mm 0 0, width=0.3\textwidth]{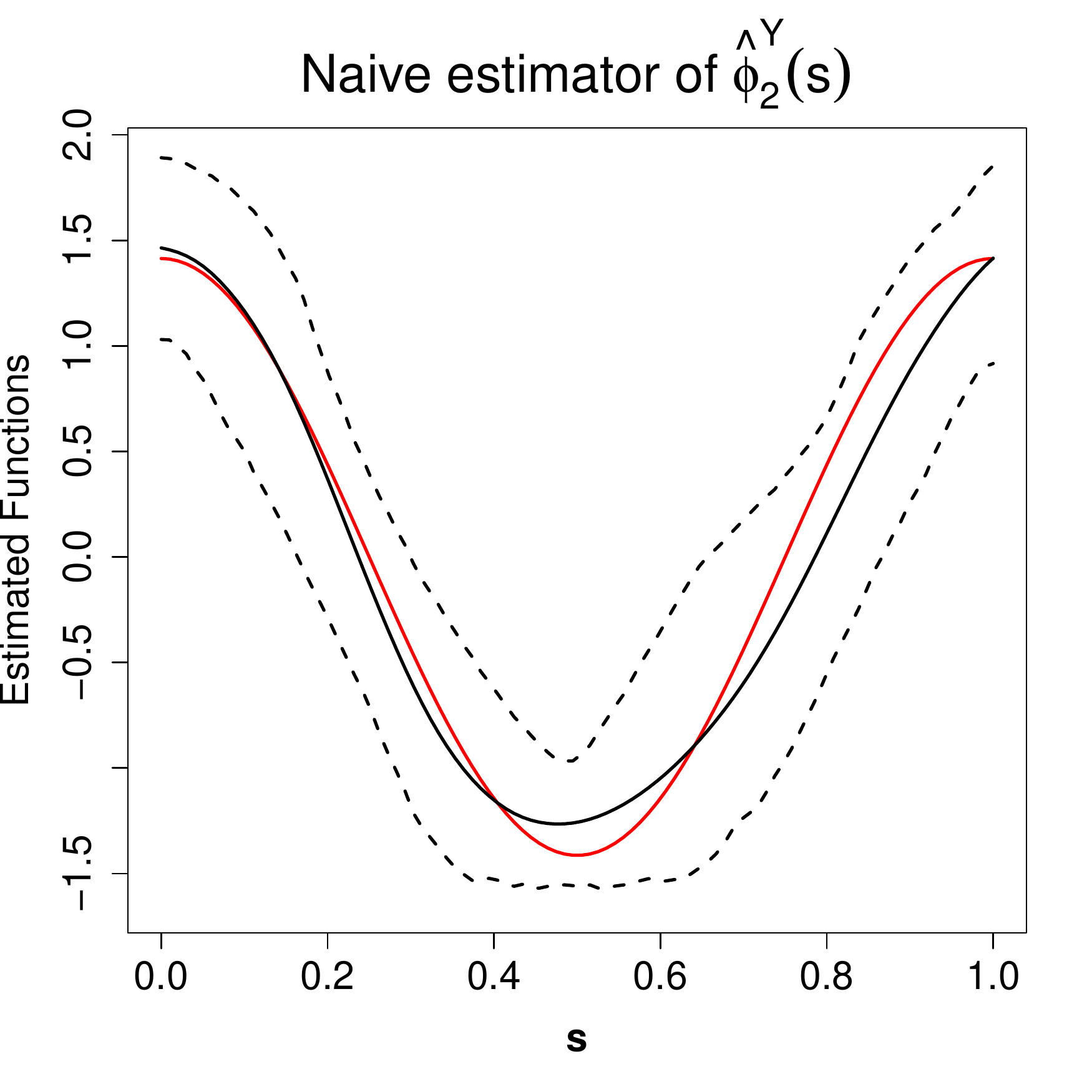}&
					\includegraphics[trim= 0 5mm 0 0, width=0.3\textwidth]{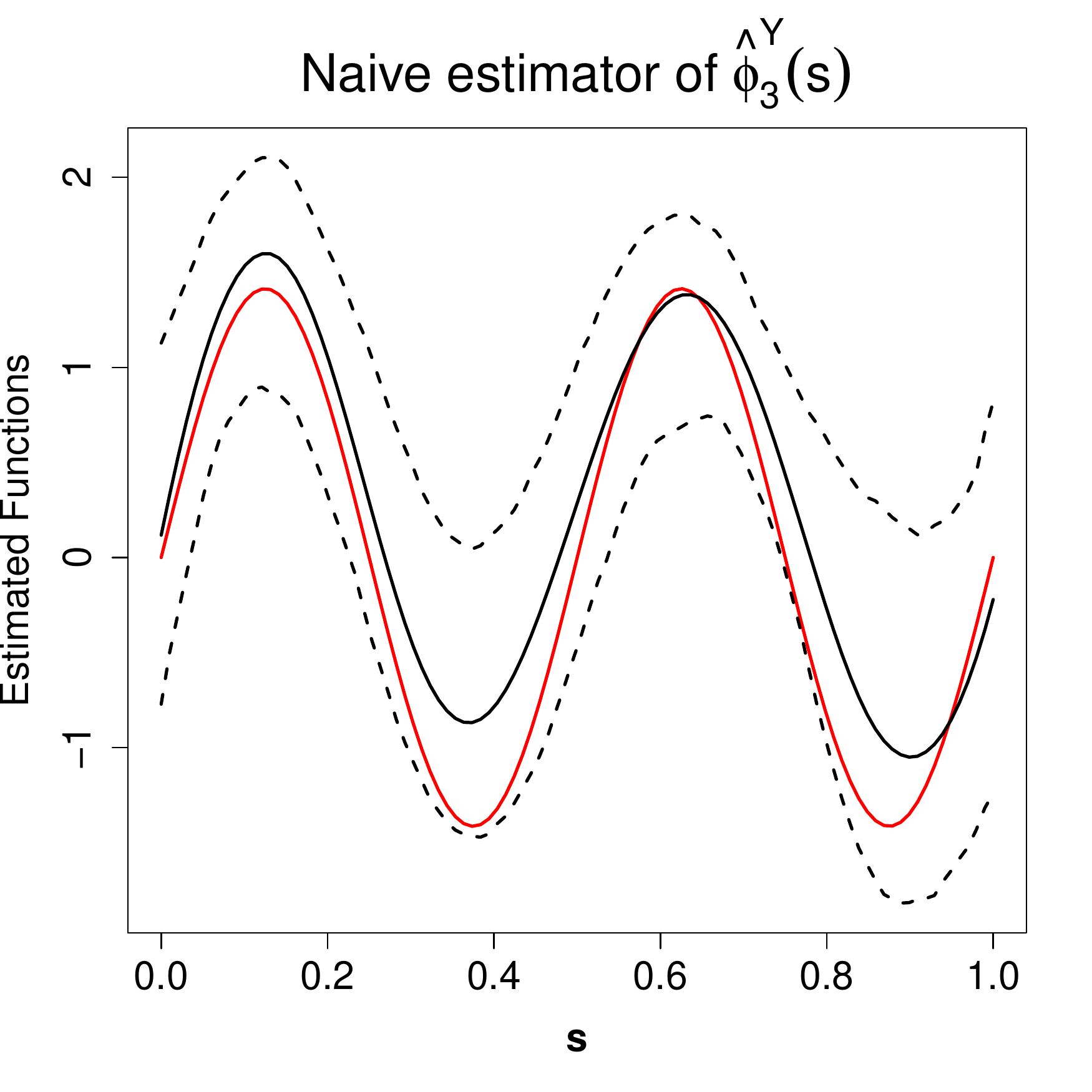}\\
				\end{tabular}
				\caption{{Estimation biases and $95\%$ quantile bands for mean functions and last two eigenfunctions of $C_Y(\cdot,\cdot)$ when $n=600$ and $q=0.8$ under the (\texttt{Gaussian}, \texttt{Gaussian}) setting.}}
				\label{fig-2}
			\end{figure}
			Figure~\ref{fig-1} suggests that the estimation errors of the bias-corrected estimator~$\wh C_Y(\cdot,\cdot)$ are generally smaller than the naive estimator~$\wt C_Y(\cdot,\cdot)$. {The improvement is moderate when the mark-point dependence is relatively small (e.g. when $q=0.5$), but becomes more pronounced when $q$ is larger. To further demonstrate this point, we summarize in Figure~\ref{fig-2} the last two eigenfunctions of $\wt C_Y(\cdot,\cdot)$ and $\wh C_Y(\cdot,\cdot)$ when $n=600$ and $q=0.8$, and compare them with those of $C_Y(\cdot,\cdot)$. The first eigenfunctions of $\wt C_Y(\cdot,\cdot)$ and $\wh C_Y(\cdot,\cdot)$ are similar and thus are not reported. It is clearly seen that $\wt C_Y(\cdot,\cdot)$ results in biased estimates of the eigenfunctions, but such biases are reduced with $\wh C_Y(\cdot,\cdot)$.} Compared to the biases observed in the estimation of $\mu(\cdot)$, the biases in the estimated eigenfunctions from the naive covariance function estimator are less pronounced, even in the case with $q=0.8$. This suggests that the  mark-point dependence may affect the estimation of the mean function $\mu(\cdot)$ more than the covariance function of the mark process.
			
			{Finally, additional simulation results for non-Gaussian $X_i(\cdot)$'s and $Y_i(\cdot)$'s are reported in \ref{sec:s1} of the supplement. In such settings, the estimation bias resulted from the mark-point dependence may not be eliminated, but it can still be effectively reduced using the proposed bias correction procedure.
			}

			\subsection{Validity of the Proposed Testing Procedures}\label{sec:val}
			In this subsection, we demonstrate the validity of the proposed testing procedures in Section~\ref{sec:test} 
			and Section~\ref{sec:perm}.
			{For each simulated dataset, the permutation test is conducted using $500$ random permutations.} In our simulation settings, {the null hypothesis of mark-point independence} corresponds to the case with $q=0$. We set $\sigma_x=0,0.5,0.7, 1$ and vary the sample size $n$ from $100$ to $600$. For each setting, we summarize test results based on $3,000$ simulated data sets. For computational efficiency, we first perform bandwidth selection using the cross-validation criteria in Section \ref{sec:bandwidth} over 50 {simulated datasets, apply the under smoothing corrections described in (\ref{eq:undersmooth}) to the average of the selected bandwidths, and hold the bandwidths fixed for all 3,000 simulated datasets.}
			
			Figures~\ref{fig-3}-\ref{fig-3-1} give empirical rejection rates of the proposed tests under $H_0$ at significance levels $0.05$ and $0.10$, respectively. Figures~\ref{fig-3} suggest that under the (\texttt{Gaussian}, \texttt{Gaussian}) setting, the empirical rejection rates of both tests {are close to} the nominal levels in all scenarios {when $n$ is greater than 200}. 
			\begin{figure}[!t]
				\centering
				\begin{tabular}{lll}
					\includegraphics[width=0.33\textwidth]{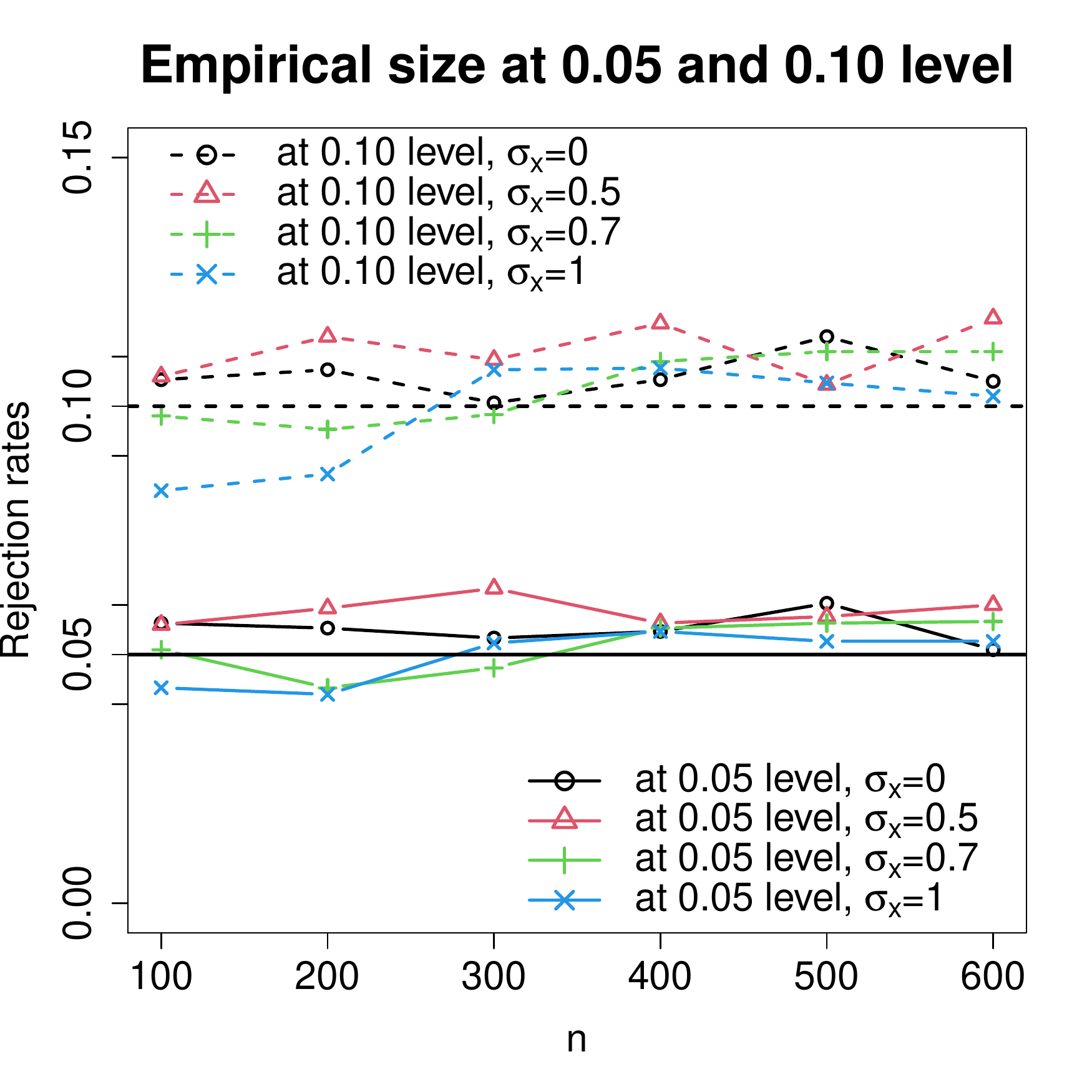} &
					\includegraphics[width=0.33\textwidth]{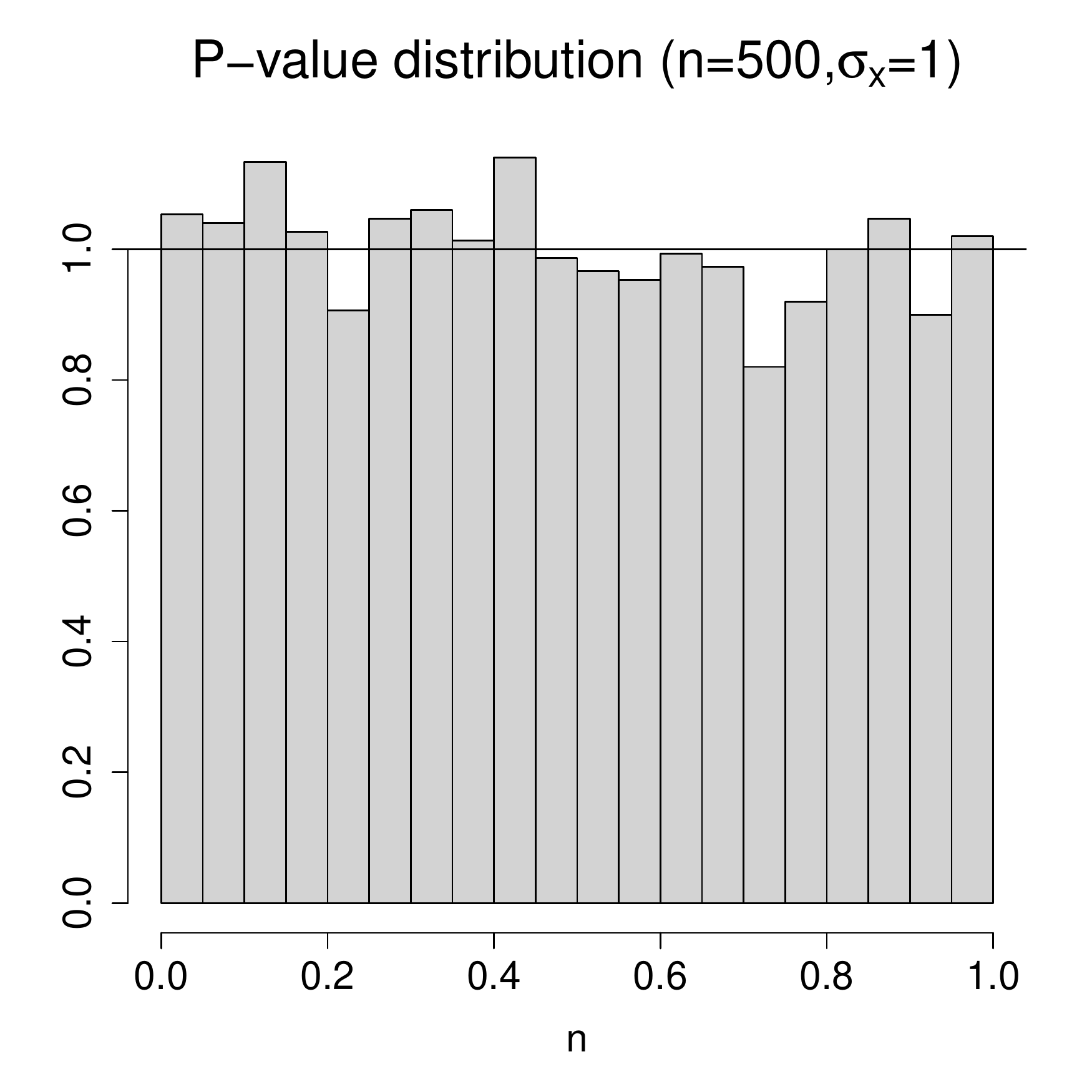}&	\includegraphics[width=0.33\textwidth]{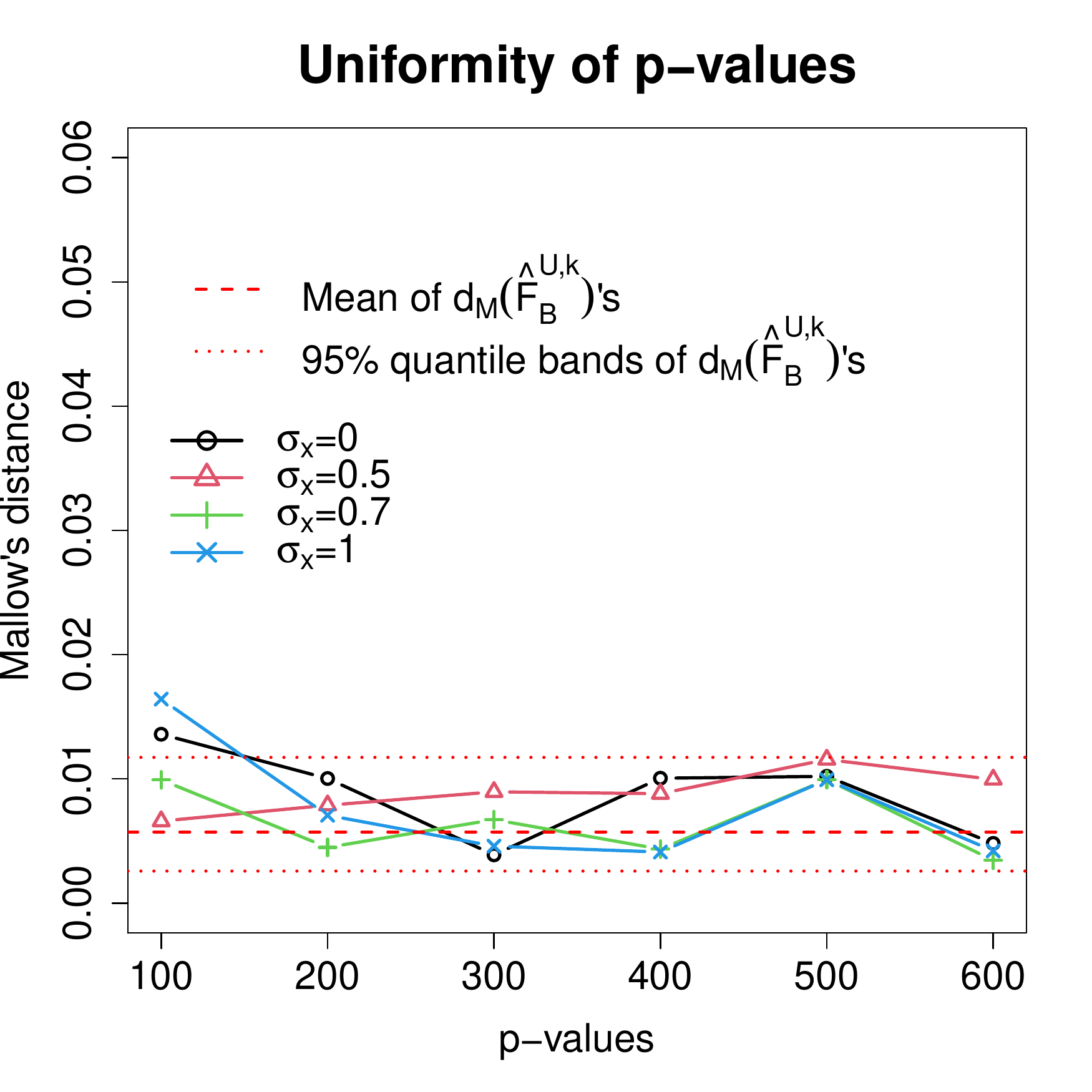}\\
					\includegraphics[trim= 0 5mm 0 0, width=0.33\textwidth]{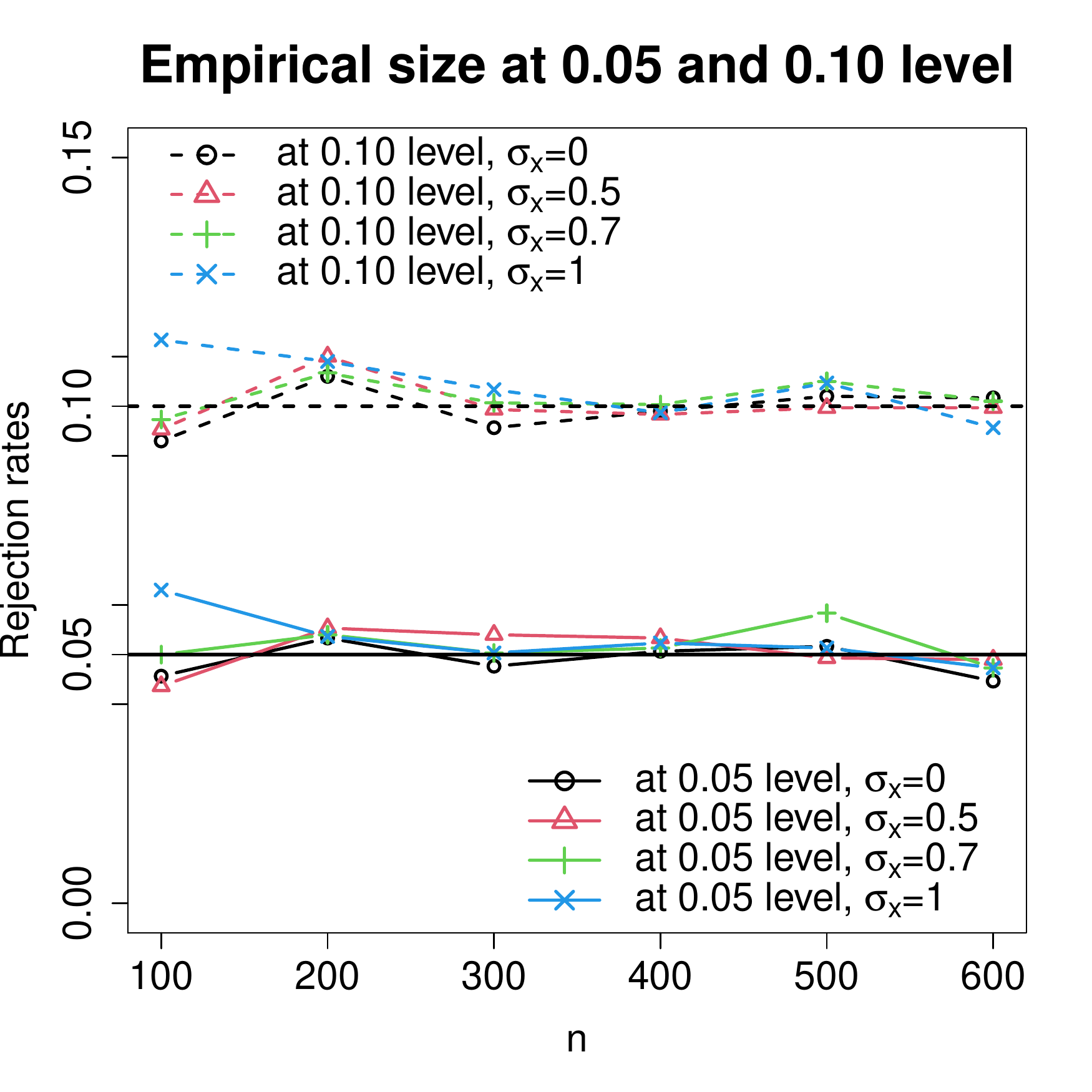} &
					\includegraphics[trim= 0 5mm 0 0, width=0.33\textwidth]{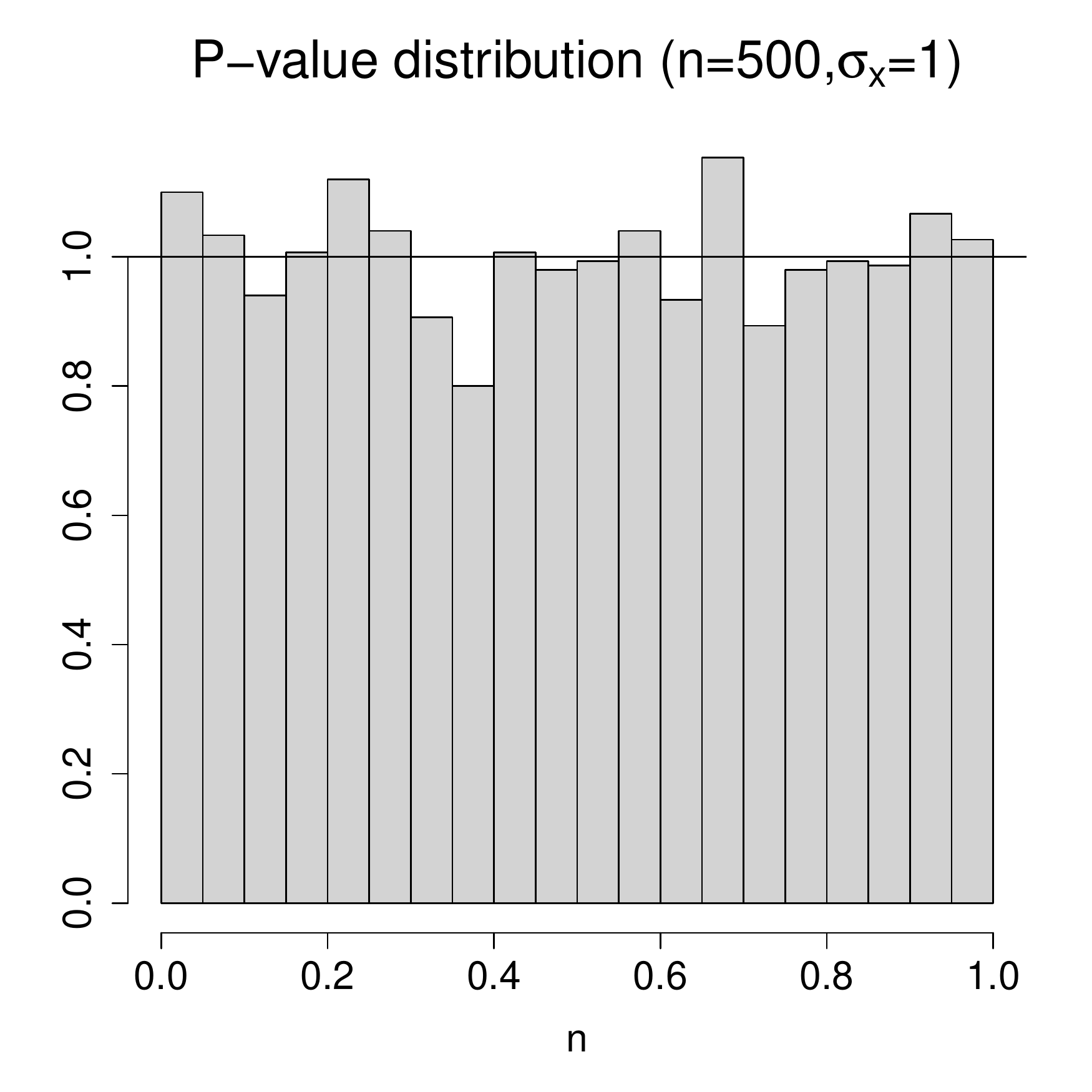}&	
					\includegraphics[trim= 0 5mm 0 0, width=0.33\textwidth]{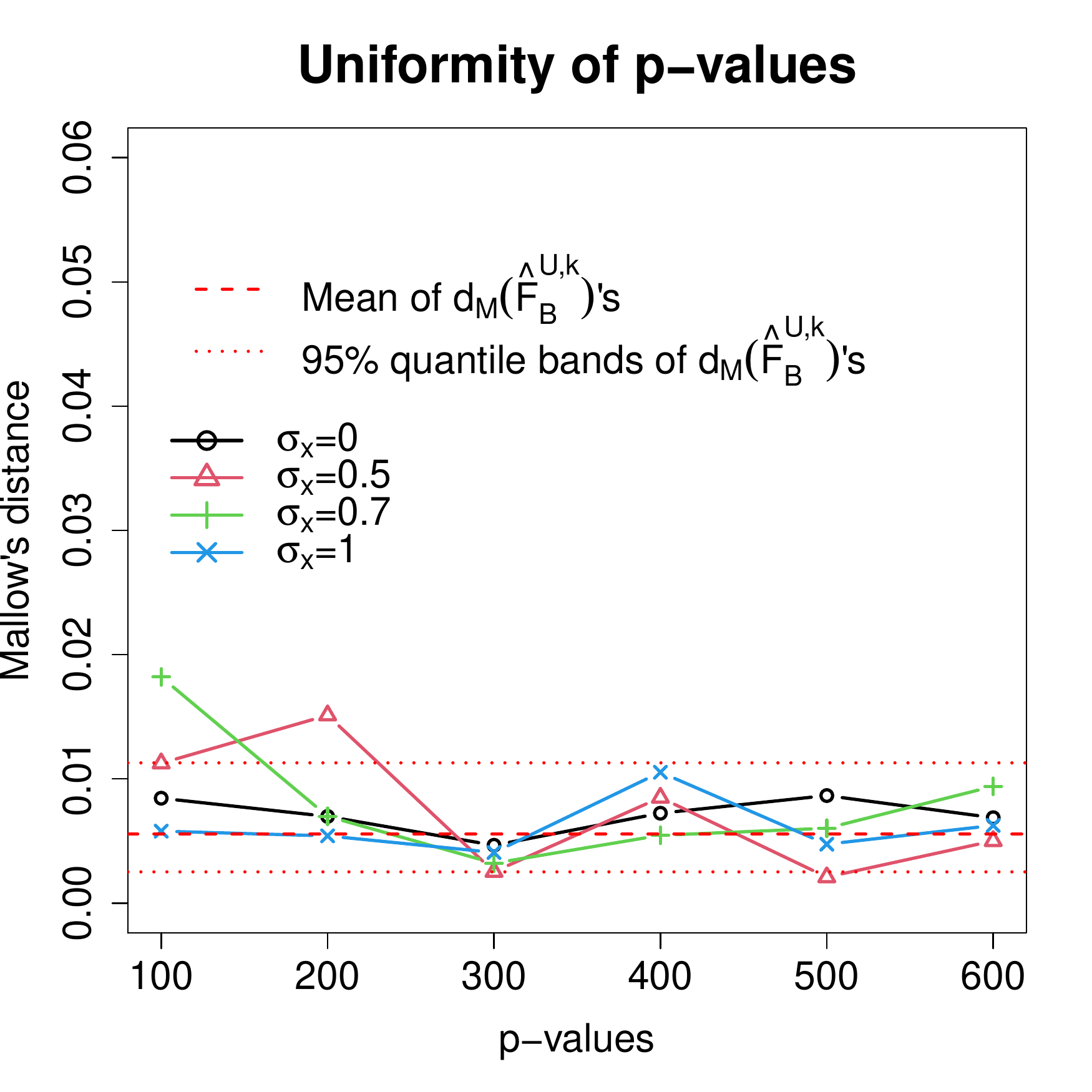}\\
				\end{tabular}
				\caption{{Validity of the proposed tests under the (\texttt{Gaussian}, \texttt{Gaussian}) setting. Top row: the asymptotic test; Bottom row: the permutation test.}}
				\label{fig-3}
			\end{figure}
			Under $H_0$, $p$-values from a valid test should follow Uniform$[0,1]$. In the middle panels of Figure~\ref{fig-3}, we provide the empirical distribution of the $p$-values under the scenario $n=500$ and $\sigma_x=1$, which indeed approximately resembles the desired uniform distribution.  To quantify the closeness between the empirical distribution $\wh F_p(\cdot)$ of the $p$-values and the uniform distribution, we also consider the Mallow's distance $d_M(\wh F_p)=\int_0^1|\wh F_p^{-1}(u)-u|d u$, where a smaller value of $d_M(\wh F_p)$ indicates that $\wh F_p$ is closer to the uniform distribution. In the right panels of Figure~\ref{fig-3}, we show the values of $d_M(\wh F_p)$ as a function of $n$ under various case scenarios. As we can see, when $n$ is sufficiently large, $d_M(\wh F_p)$ falls into the envelope computed by the empirical quantiles of $d_M(\wh F_B^{U,k})$, where $\wh F_B^{U,k}$, $k=1,\ldots,1000$, are the empirical CDFs based on $B=3,000$ random numbers drawn from the Uniform$[0,1]$. This result further supports the validity of the proposed testing procedures.

			\begin{figure}[!t]
				\centering
				\begin{tabular}{lll}
					\includegraphics[width=0.33\textwidth]{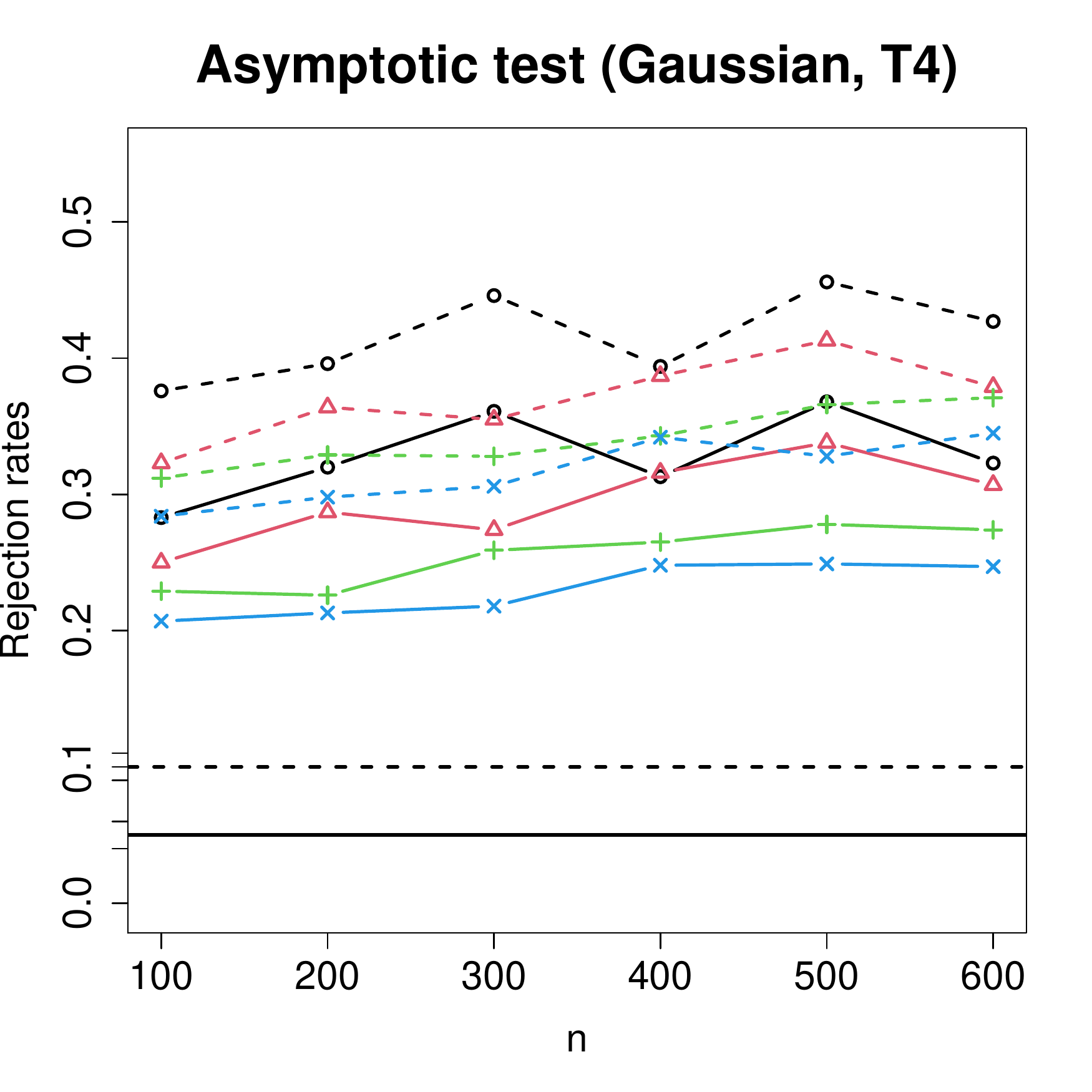}&
					\includegraphics[width=0.33\textwidth]{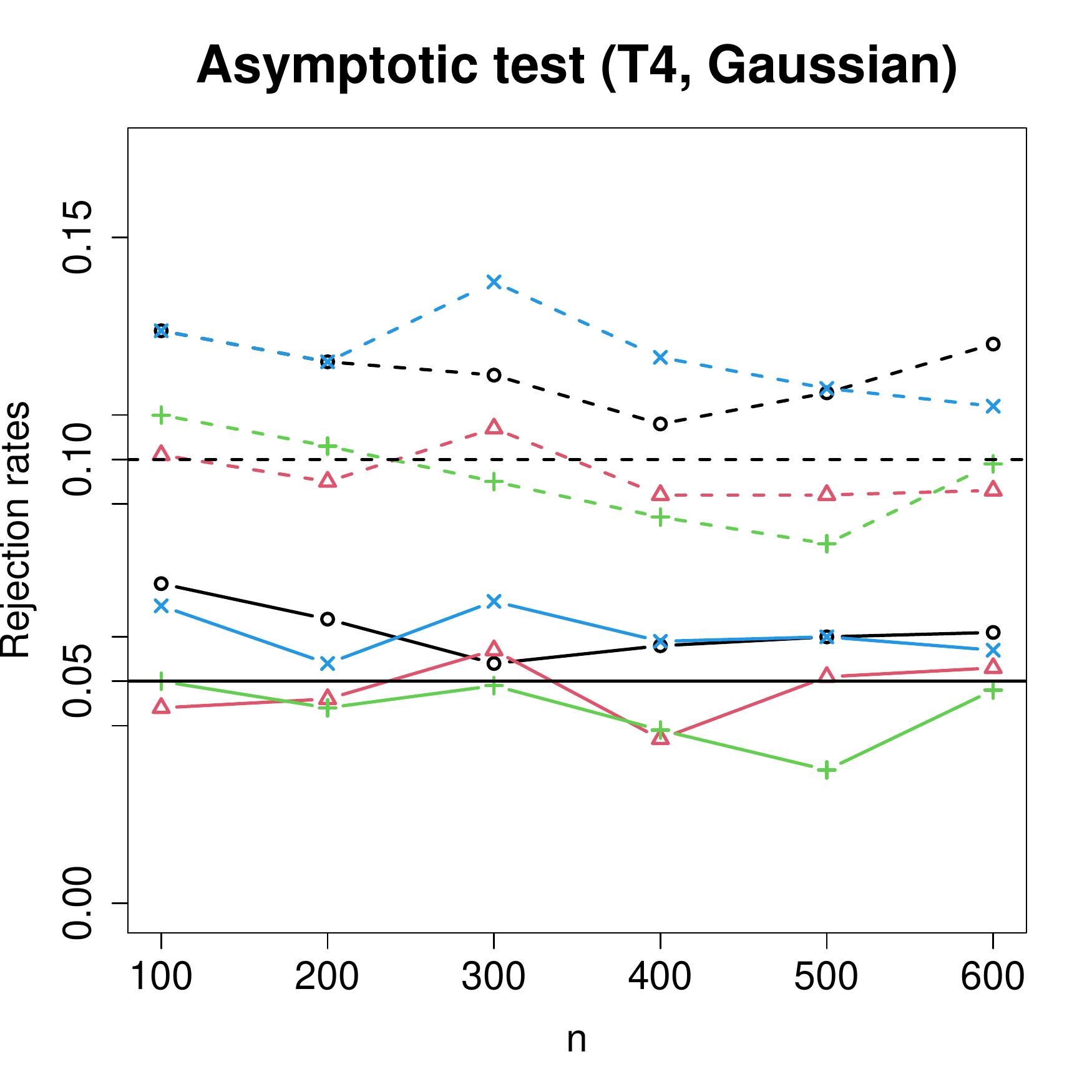}&
					\includegraphics[width=0.33\textwidth]{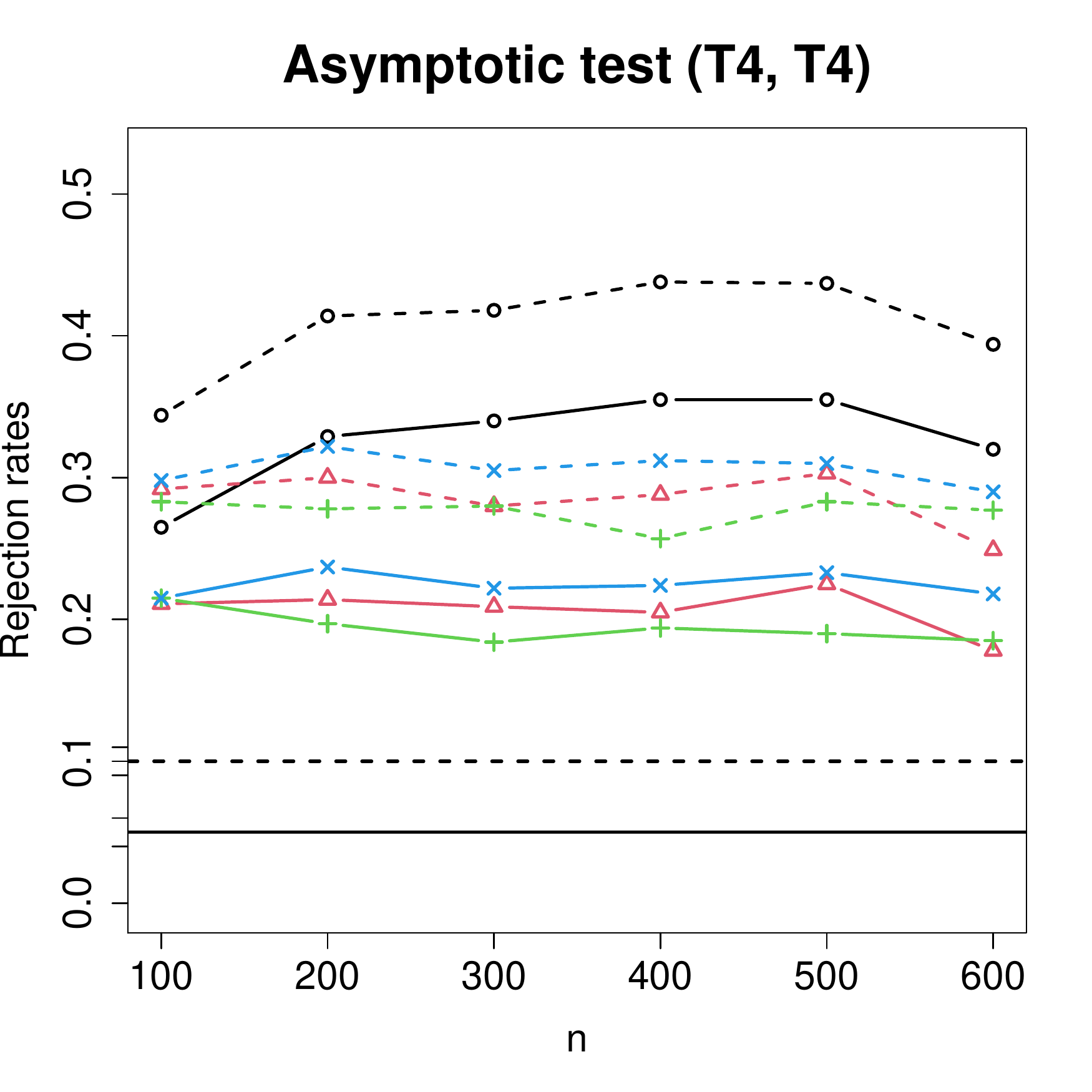}\\
					\includegraphics[trim= 0 5mm 0 0, width=0.33\textwidth]{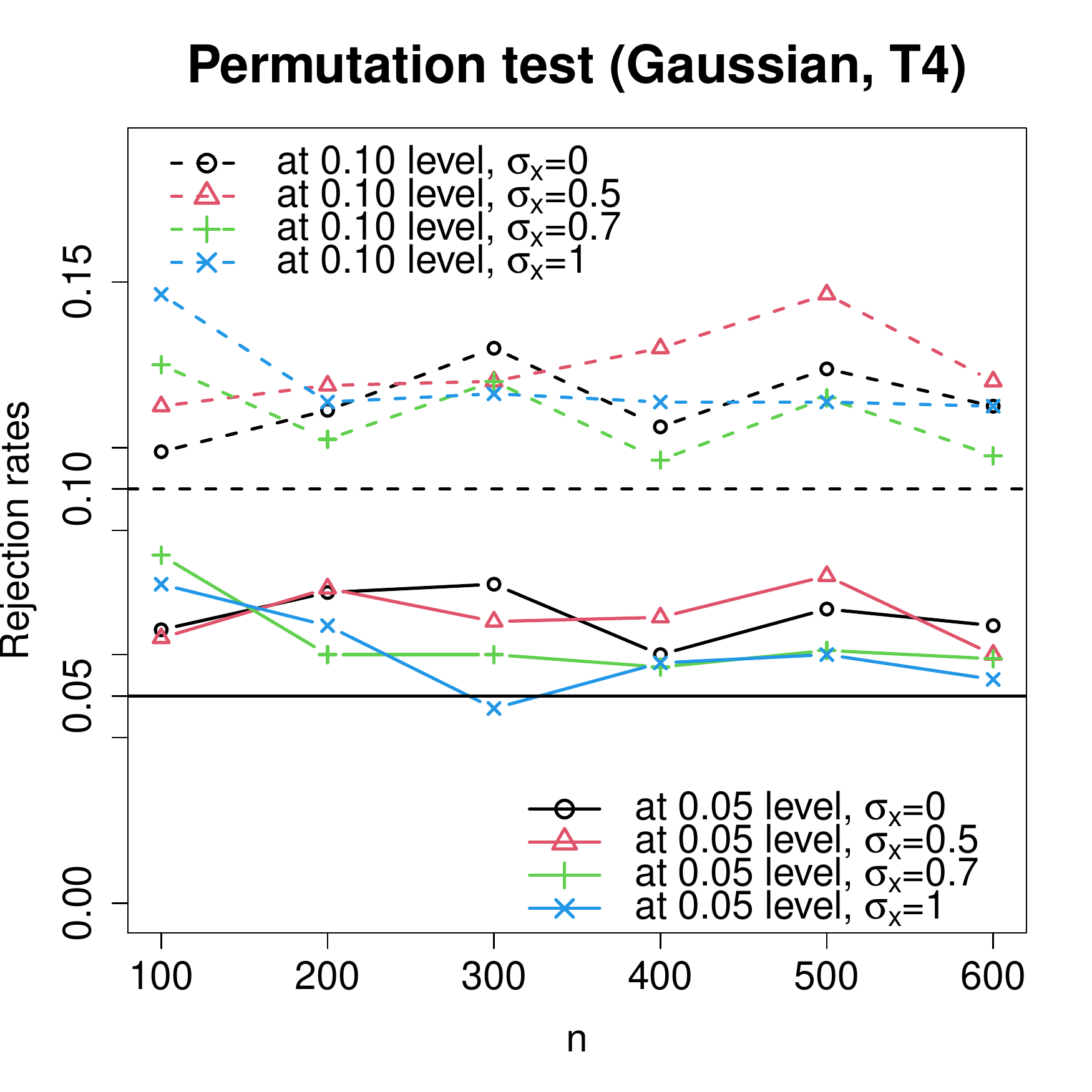}&
					\includegraphics[trim= 0 5mm 0 0, width=0.33\textwidth]{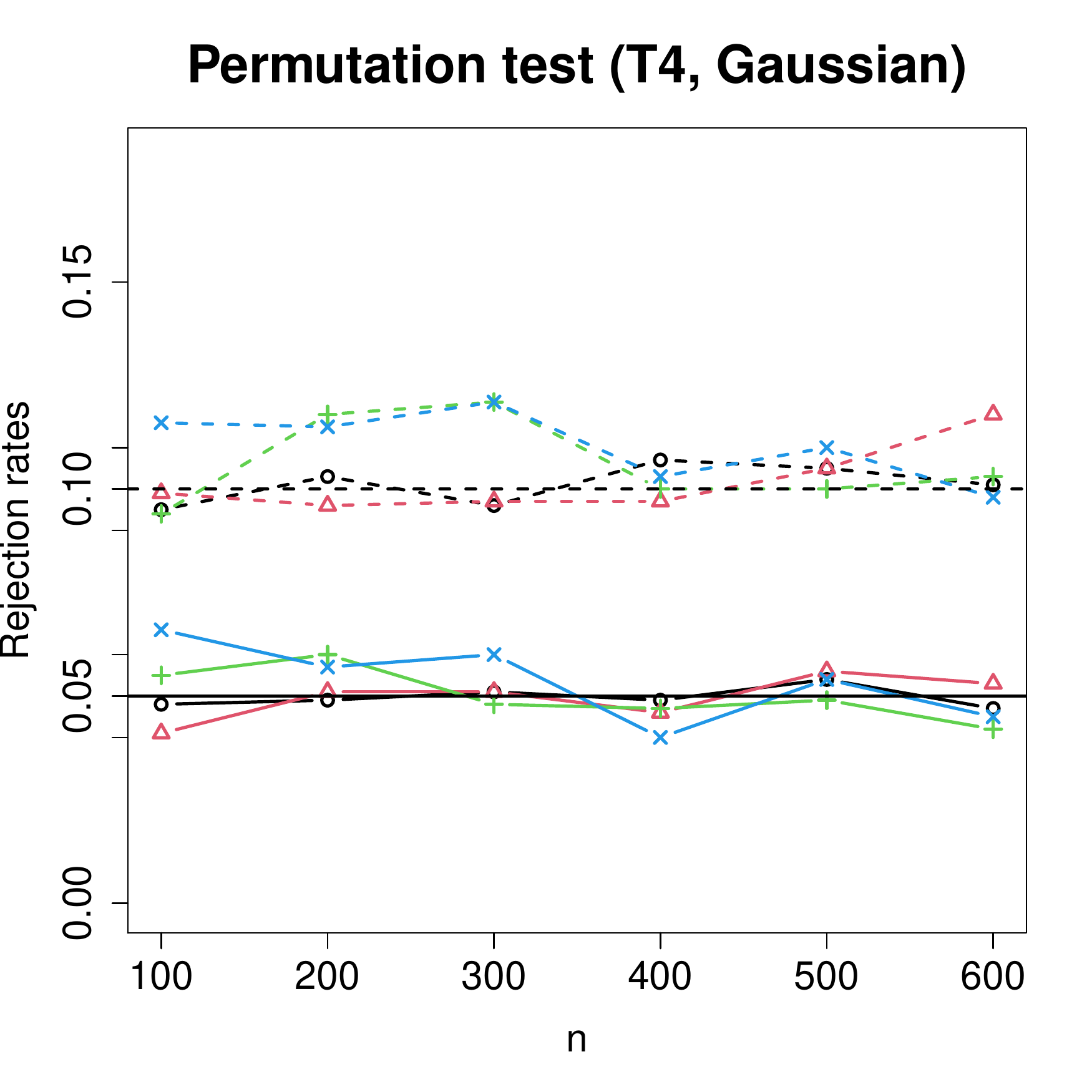}&
					\includegraphics[trim= 0 5mm 0 0, width=0.33\textwidth]{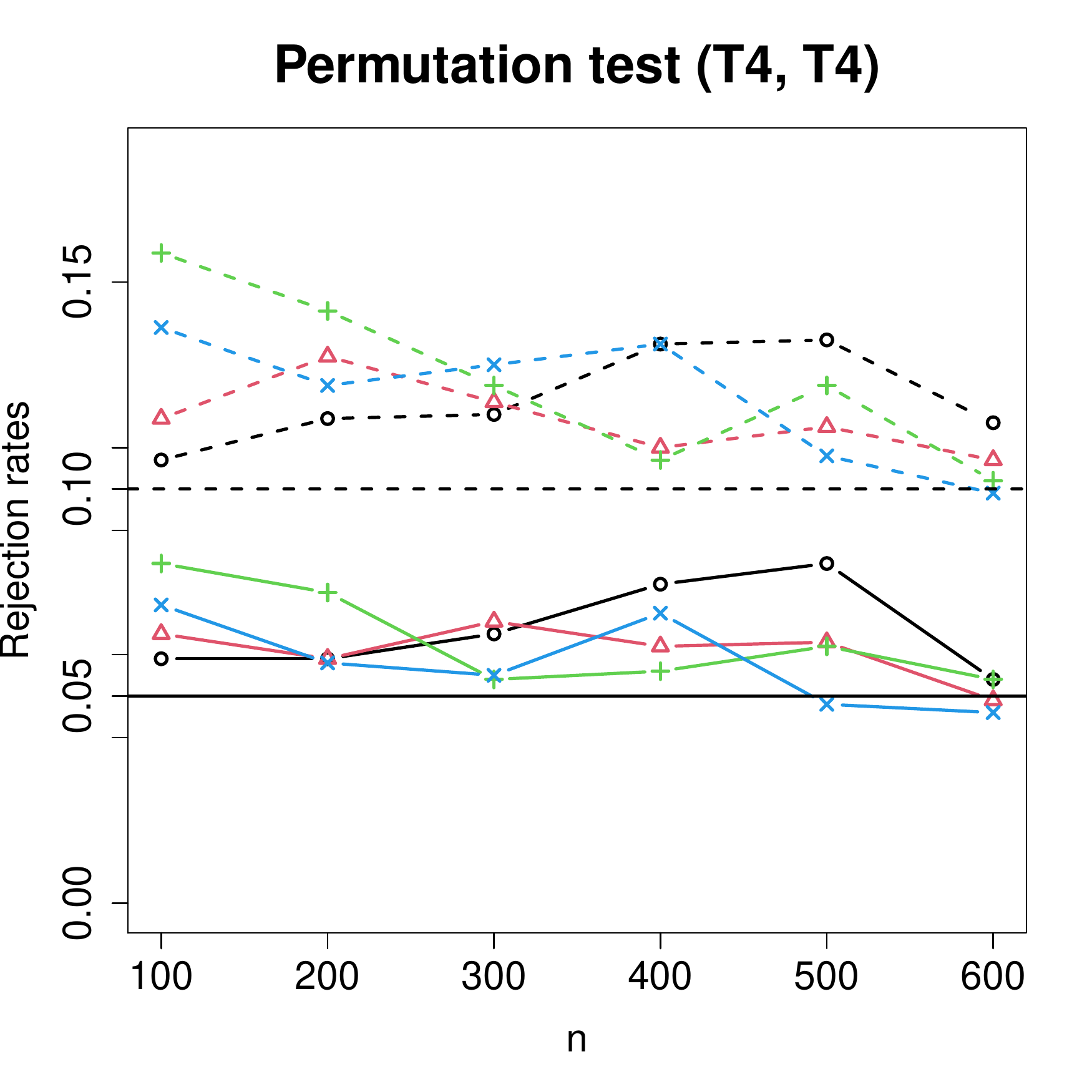}\\
				\end{tabular}
				\caption{{Rejection rates of two tests under mark-point independence in various distribution settings. Top row: the asymptotic test; Bottom row: the permutation test.}}
				\label{fig-3-1}
			\end{figure}
			
			Figure~\ref{fig-3-1} illustrates the rejection rates of the two tests when the mark process is not Gaussian and/or the point process is not an LGCP. While the asymptotic test still appears to be valid so long as the mark process is Gaussian, its rejection rates significantly exceed the nominal level when the mark process is non-Gaussian. In comparison, the permutation
			test seems to {be close to} the nominal sizes provided $n$ is sufficiently large in all settings. {Additional and similar simulation results can be found in Section~\ref{sec:s2} of the supplement.}  
			
			\subsection{{Powers Against Local Alternatives}}
			\label{sim:local}
			{
				Next, we investigate the power of the proposed tests {against local alternatives}, which is a popular approach to study the test power, see, e.g., \cite{xu2011goodness}. The asymptotic mean of $\sqrt{n}T_n$ in~\eqref{test} when $C_{XY}(\cdot,\cdot)\nequiv 0$ is 
				${\sqrt{n}}\int_{\CT}\int_{\CT} \rho_2(u,v)C_{XY}(u,v)^2 du dv$. 
				Using the simulation setup as in \eqref{cor} and {define the local alternatives} by letting $q^2=n^{-1/2}\gamma$ for some constant $\gamma>0$. {If Theorem~\ref{thm3} holds, the asymptotic mean of $\sqrt{n}T_n$ should be} proportional to $\gamma$ and is independent of $n$. We set $\gamma=3.6, 5.0,6.4$ and refer to the resulting rejection probabilities as the local power. If the limiting distribution of the test statistic is correctly specified, the local power should stay constant as $n$ increases. Data are simulated with varying values of $n$, $\sigma_x$ and $\gamma$.  Figure~\ref{fig-4} illustrates the empirical rejection rates at the $0.05$ significance level in various settings based on $1000$ data replicates. 
				\begin{figure}[ht!]
					\centering
					\begin{tabular}{lll}
						\includegraphics[width=0.33\textwidth]{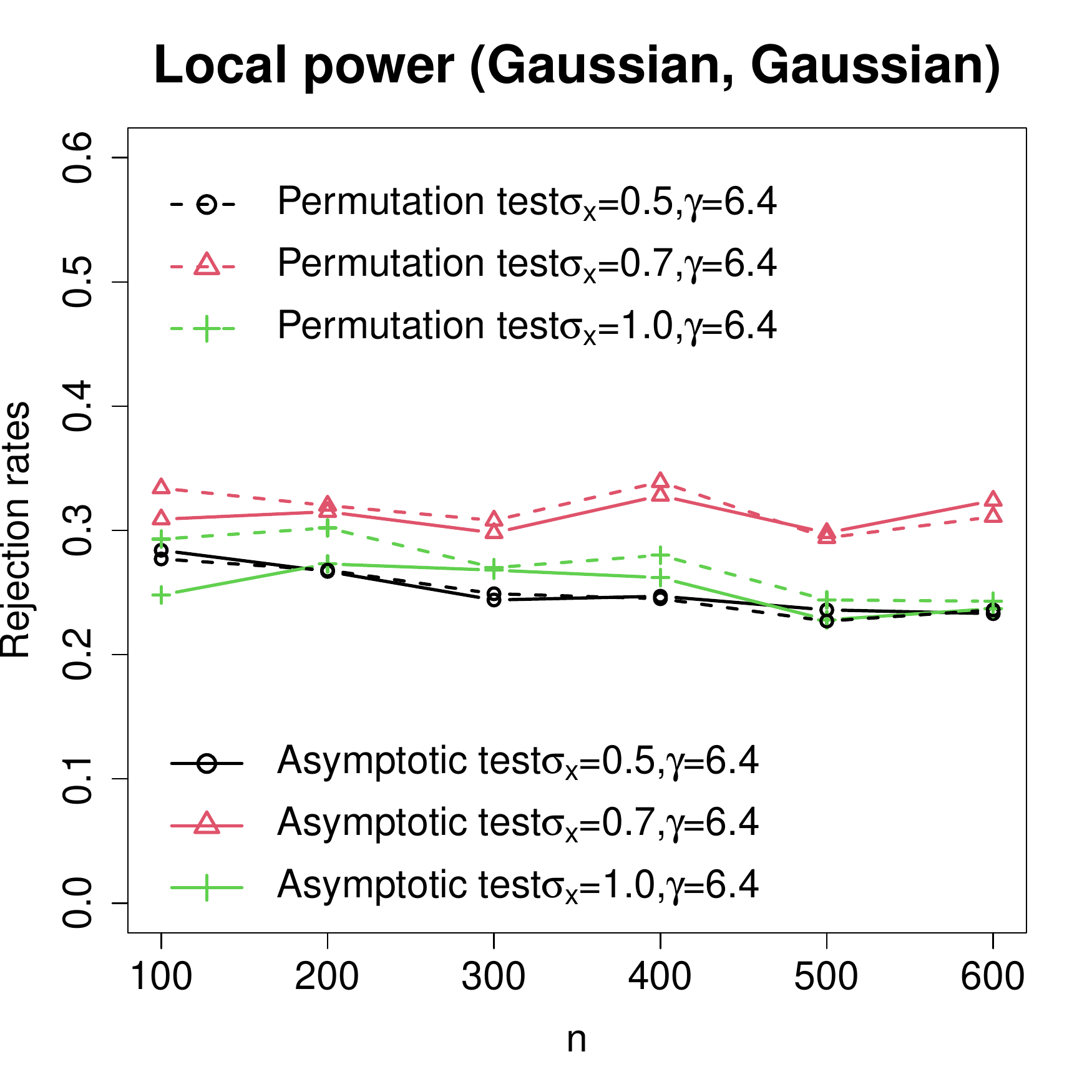} &
						\includegraphics[width=0.33\textwidth]{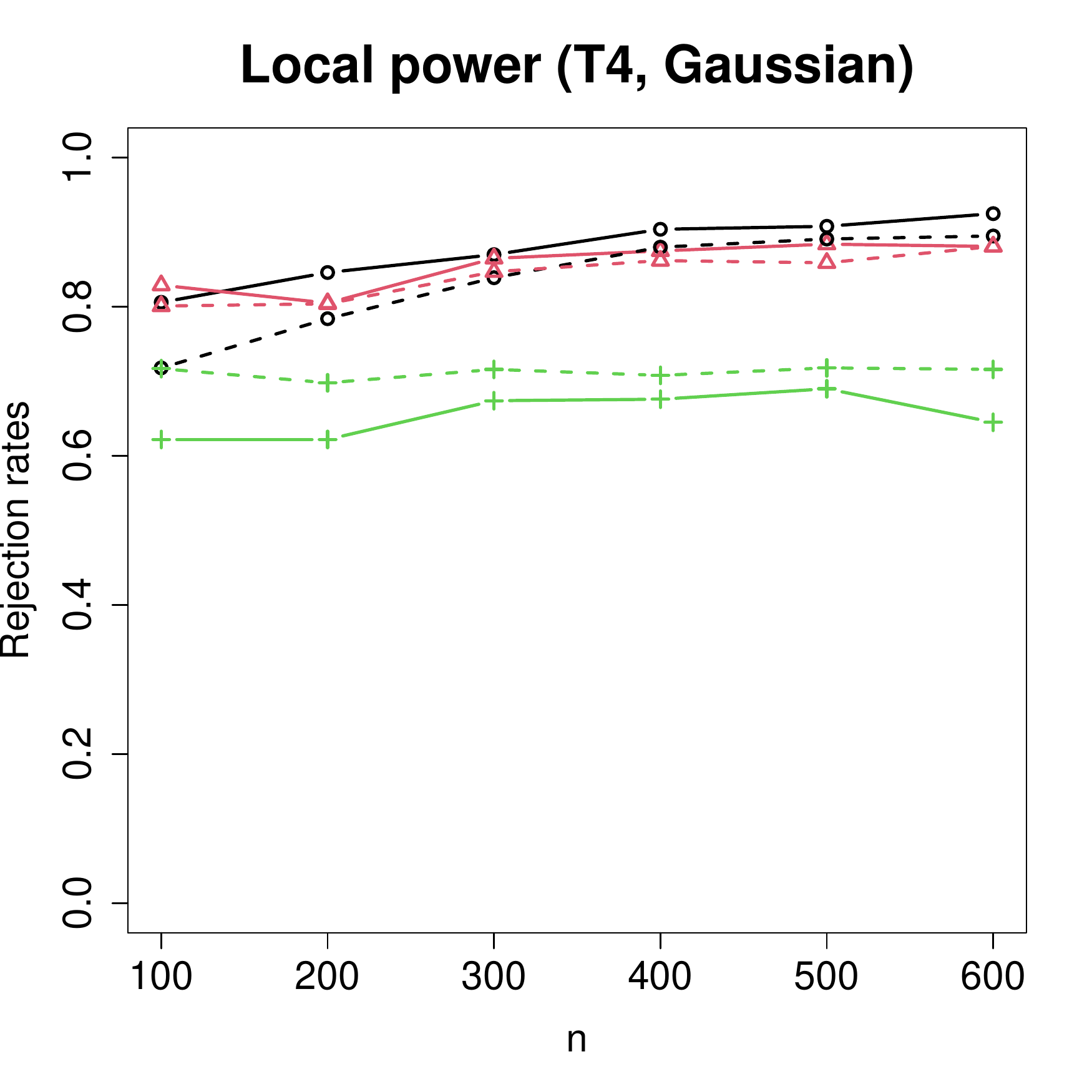} &
						\includegraphics[width=0.33\textwidth]{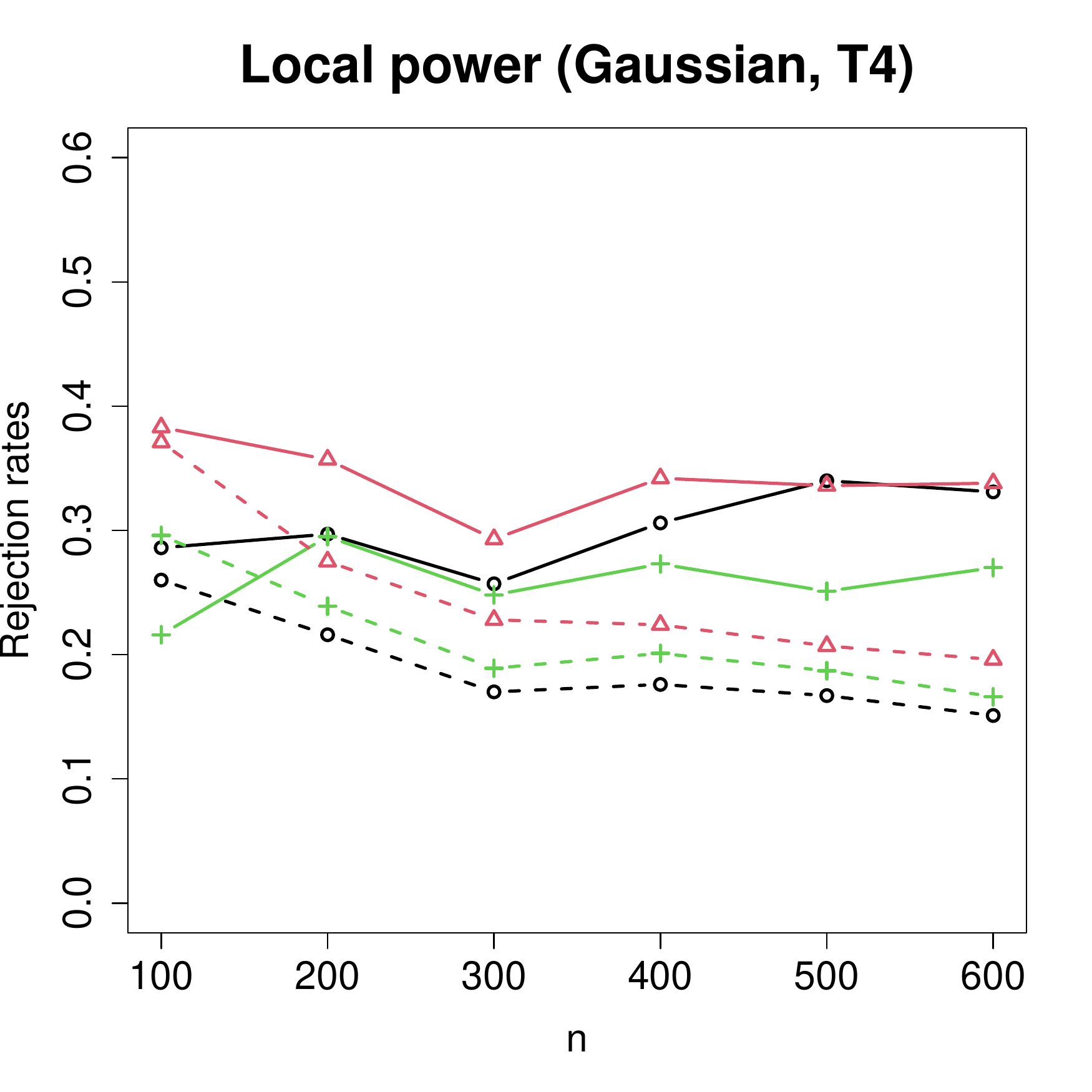} \\				\includegraphics[width=0.33\textwidth]{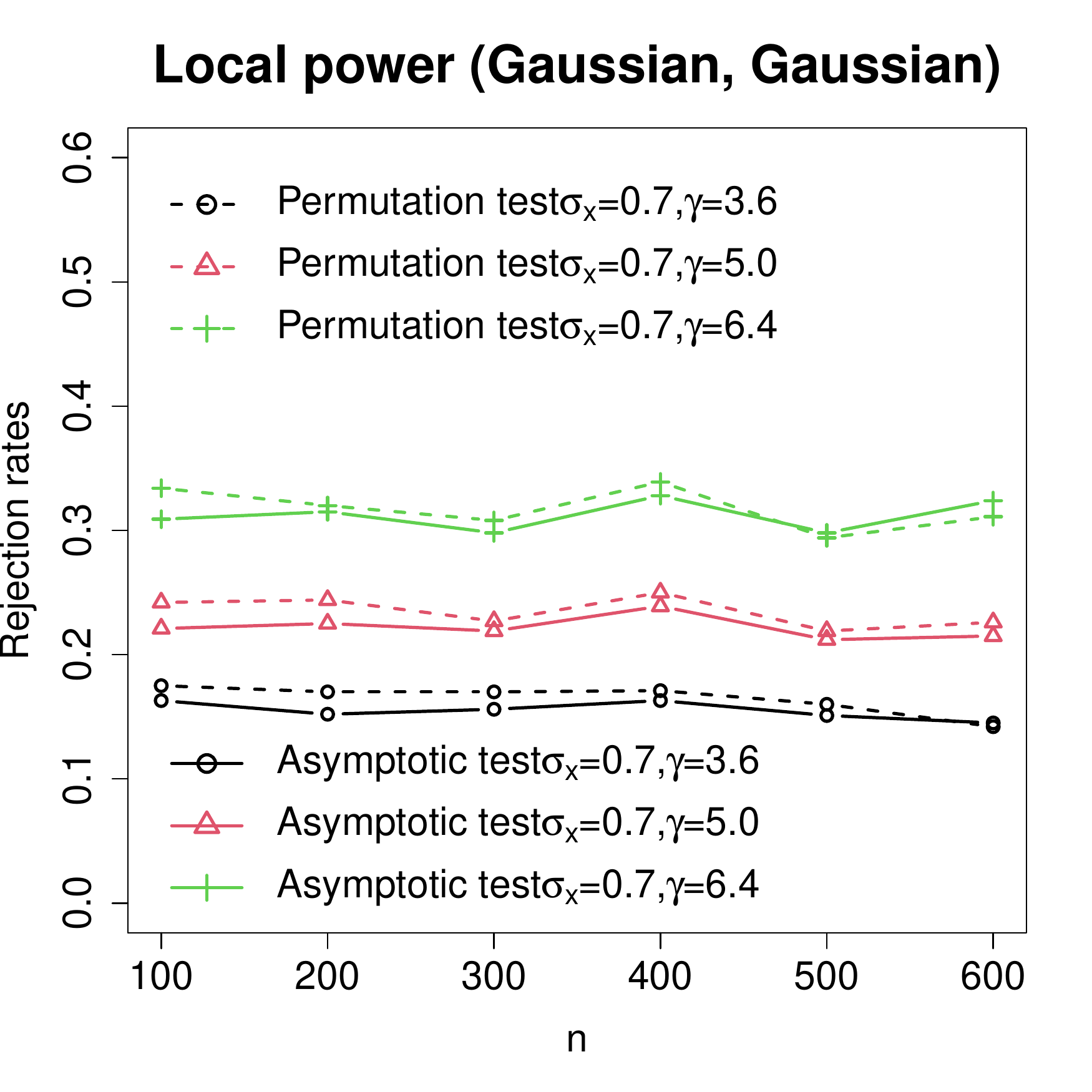} &
						\includegraphics[width=0.33\textwidth]{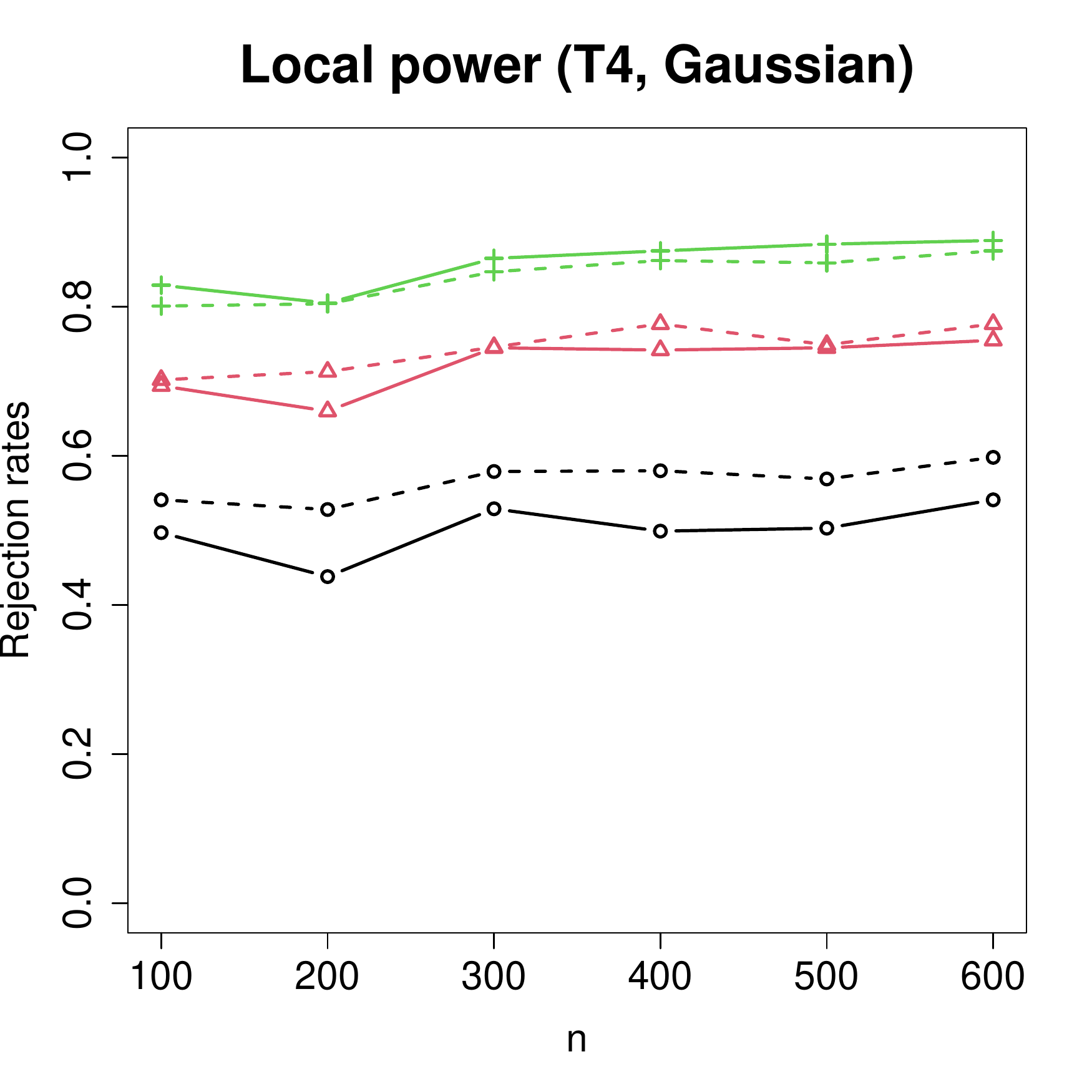} &
						\includegraphics[width=0.33\textwidth]{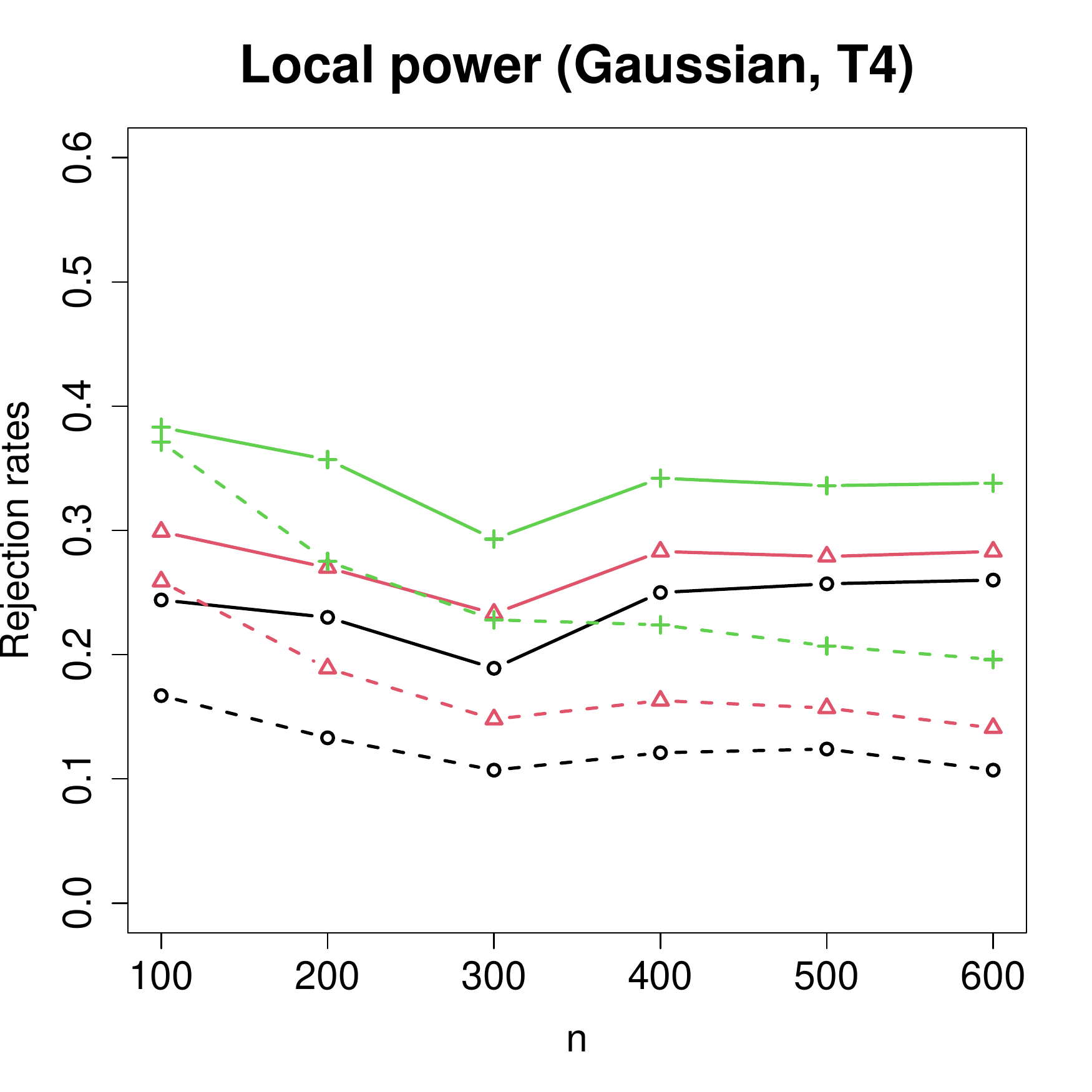}\\
					\end{tabular}
					\caption{{Rejection rates of two tests at the 0.05 level under the local alternatives with varying $\sigma_x$ (top row) and $\gamma$ (bottom row).}}
					\label{fig-4}
				\end{figure}

				The left panel of Figure~\ref{fig-4} suggests that under the (\texttt{Gaussian}, \texttt{Gaussian}) setting, the rejection rates of both tests stay roughly constant as $n$ and $\gamma$ increase. The local powers of both tests first increase as $\sigma_x$ increases. This is because the magnitude of $C_{XY}(\cdot,\cdot)$ increases with $\sigma_x$ by our simulation design. Correspondingly, the mean of the test statistic increases as suggested in~\eqref{test-bias}, leading to a greater power. However, if $\sigma_x$ continues to grow, the power of the proposed test will decrease. This can be explained by the fact that a larger $\sigma_x$ also leads to a more clustered point process {and consequently inflate} the variance of the test statistic, {which offsets} the increase in the mean of the test statistic and thus leads to a reduced power. All these observations support our theoretical findings in Theorem~\ref{thm3} under the (\texttt{Gaussian}, \texttt{Gaussian}) setting. In other settings {shown in Figure~\ref{fig-4}}, the local powers of the permutation test are roughly constant as $n$ increases, provided that $n$ is sufficiently large. This further supports the validity of the proposed functional permutation test in more general settings. Similar results are observed when the \texttt{T4} distribution is replaced by the \texttt{Exp} distribution; see Section~\ref{sec:s2} of the supplement.} 

			\section{Real Data Analysis}\label{sec:data}
			\subsection{EMA Smoking Data}\label{sec:ema_data}
			\noindent
			In this subsection, we analyze the situational associations of smoking using the ecological momentary assessment (EMA) data. Smoking is found to be cued or suppressed by immediate situational factors, such as craving, mood, and social settings, and influences of these factors are modulated by gender \citep{todd2004daily,shiffman2011point}. The data we analyze were collected in real-time from 302 smokers over 16 days \citep{shiffman2002immediate}. The participants were 43\% male and 57\% female. To be included in the study, a participant had to smoke at least 10 cigarettes per day and had been smoking for at least 2 years. Before data collection, participants were trained to use an Electronic Diary (ED) device designed to collect data in real-time. During the study period, the participants were instructed to record each cigarette in the ED, immediately before smoking. On about 4-5 randomly selected smoking occasions per day, the device administered an assessment. Each assessment provided several continuously measured mood-related ratings such as ``negative affect" and ``restless", with higher scores indicating more affective distress and stronger feelings of restlessness, respectively. See \citet{shiffman2002immediate} for more details about the data.
			
			The first few days of monitoring were designed to allow the participants to become familiar with the ED. We therefore focus on days 4-16 in our analysis as suggested in \citet{shiffman2002immediate}. 
			In each day, the observation window for events, i.e., $[0,1]$, corresponded to all waking hours and was not subject-specific, that is, participants were measured on a common time interval; see \citet{shiffman2002immediate} for more details. For each participant, we aggregate all time points observed on different days into a single day for our analysis.
			As only 1.56\% of total smoking events across all participants were within $[0,0.2]\bigcup[0.8,1]$, we focus on the time interval $[0.2,0.8]$. {See Figure~\ref{fig-skdata1} in the supplement for some sample trajectories.}
			We estimate the naive and the bias-corrected estimators of the daily mean functions $\mu(\cdot)$ and covariance functions $C_Y(\cdot,\cdot)$ for two different marks (i.e., ``negative affect" and ``restless") of male and female participants, respectively; see Figure \ref{fig-smoke} and Section~\ref{sec:sk:supp} of the supplement. The bandwidths are selected following the proposed procedures in Section \ref{sec:bandwidth}.
			
			{Following Section~\ref{sec:diag}, for the mark ``restless" of the male group, Figure \ref{fig-smoke} shows that summary statistics from the observed data fell within the $95\%$ percentile envelope based on $1,000$ simulated point processes, suggesting that the LGCP assumption for the point process is reasonable. The QQ plot of the first FPC scores of the mark process, which account for more than $95\%$ of the total variations, is reasonably close to a straight line, indicating the mark process can be considered as Gaussian. Additional diagnostic plots for other marked point processes are given in Section~\ref{sec:sk:supp} of the supplement, {supporting a similar conclusion}. 
			}
			\begin{figure}[ht!]
				\centering
				\begin{tabular}{lll}
					\includegraphics[width=0.31\textwidth]{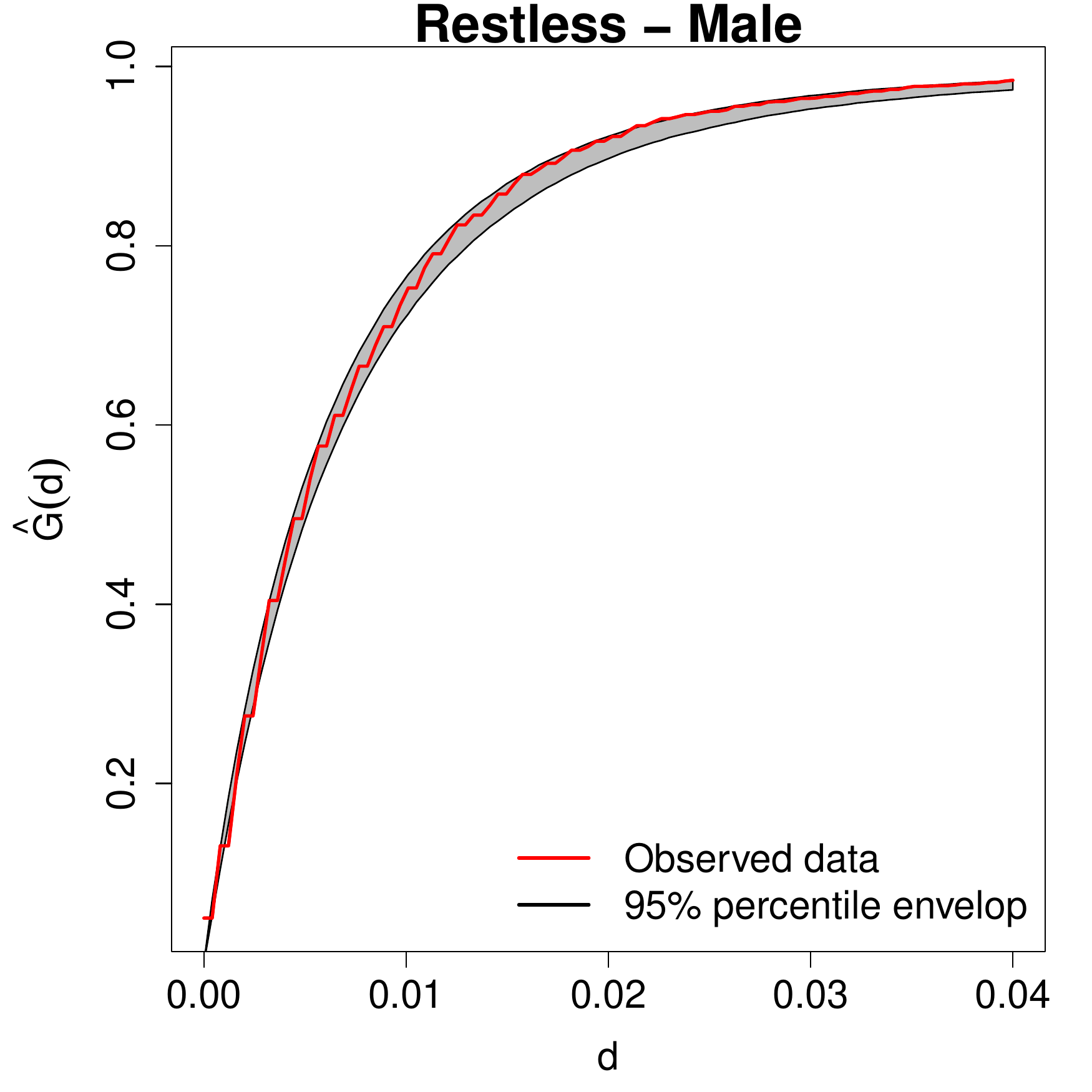} &
					\includegraphics[width=0.31\textwidth]{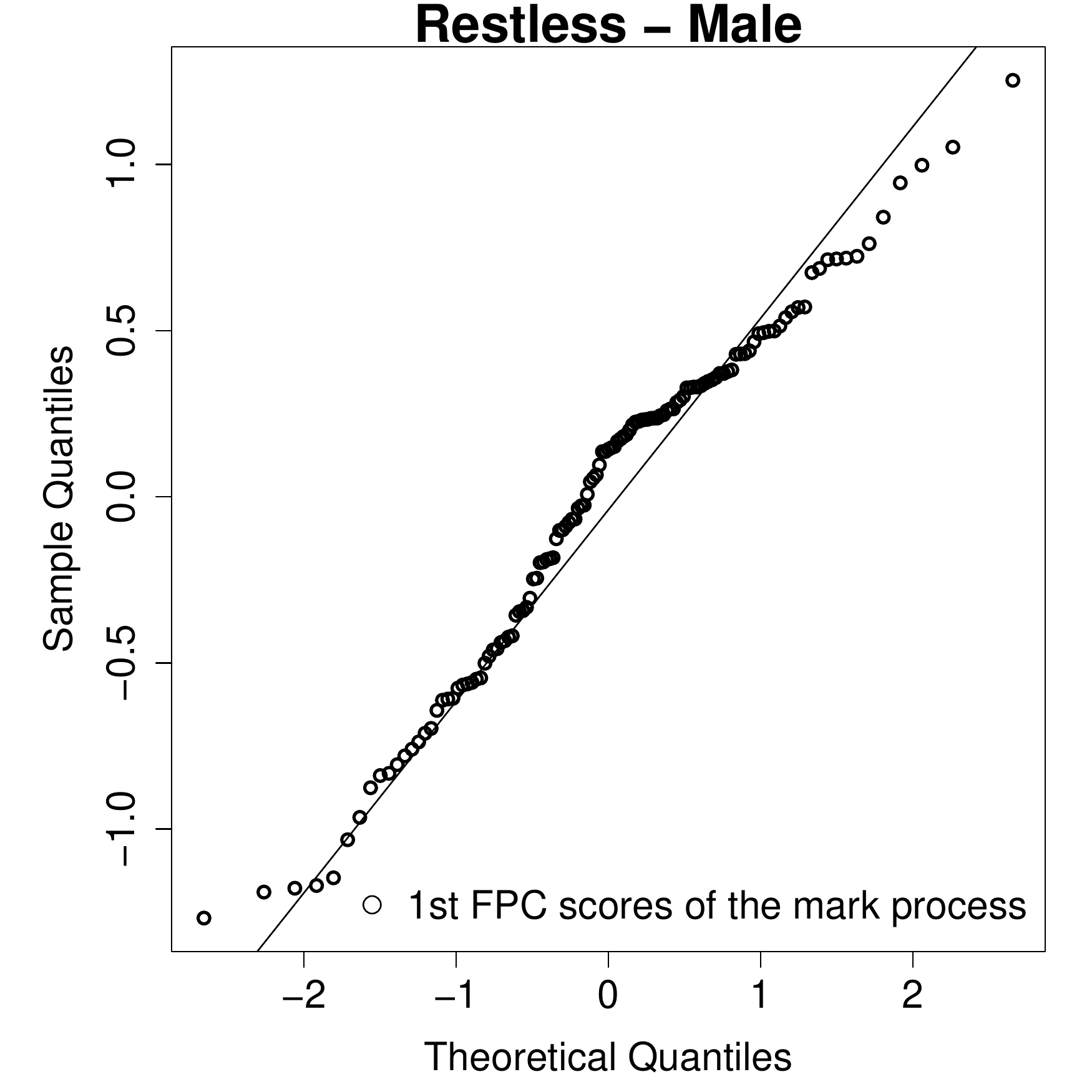} &
					\includegraphics[width=0.31\textwidth]{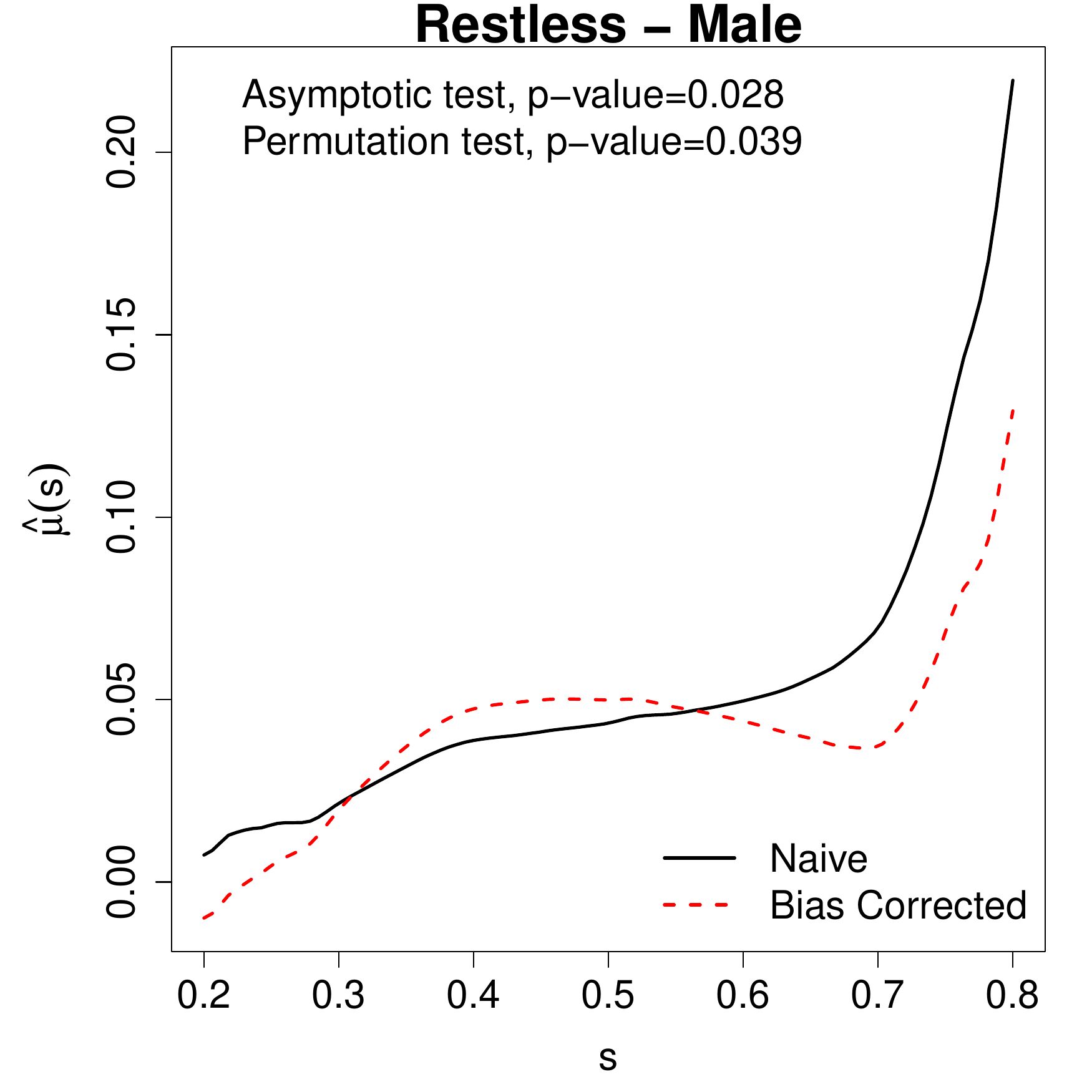} \\				\includegraphics[width=0.31\textwidth]{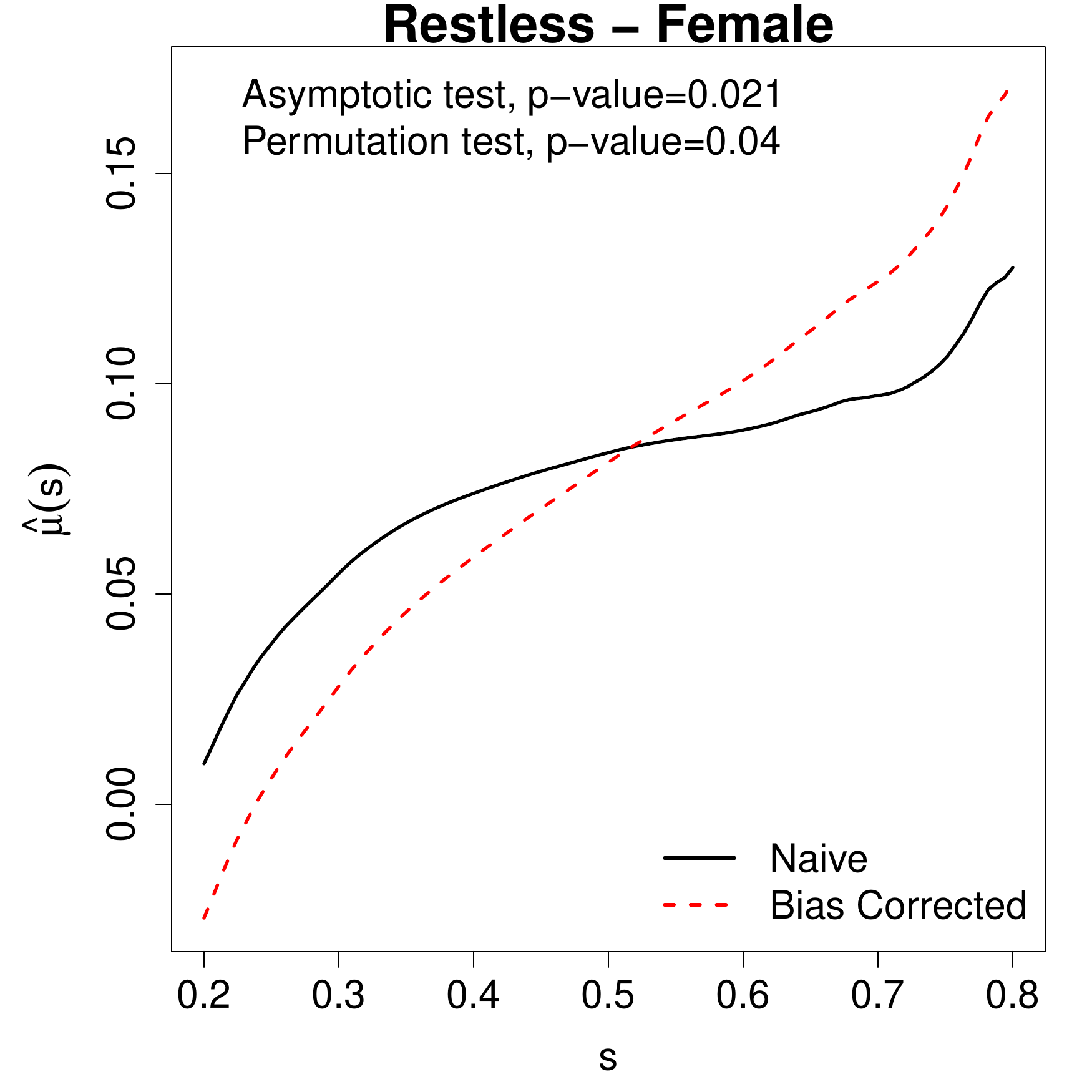} &
					\includegraphics[width=0.31\textwidth]{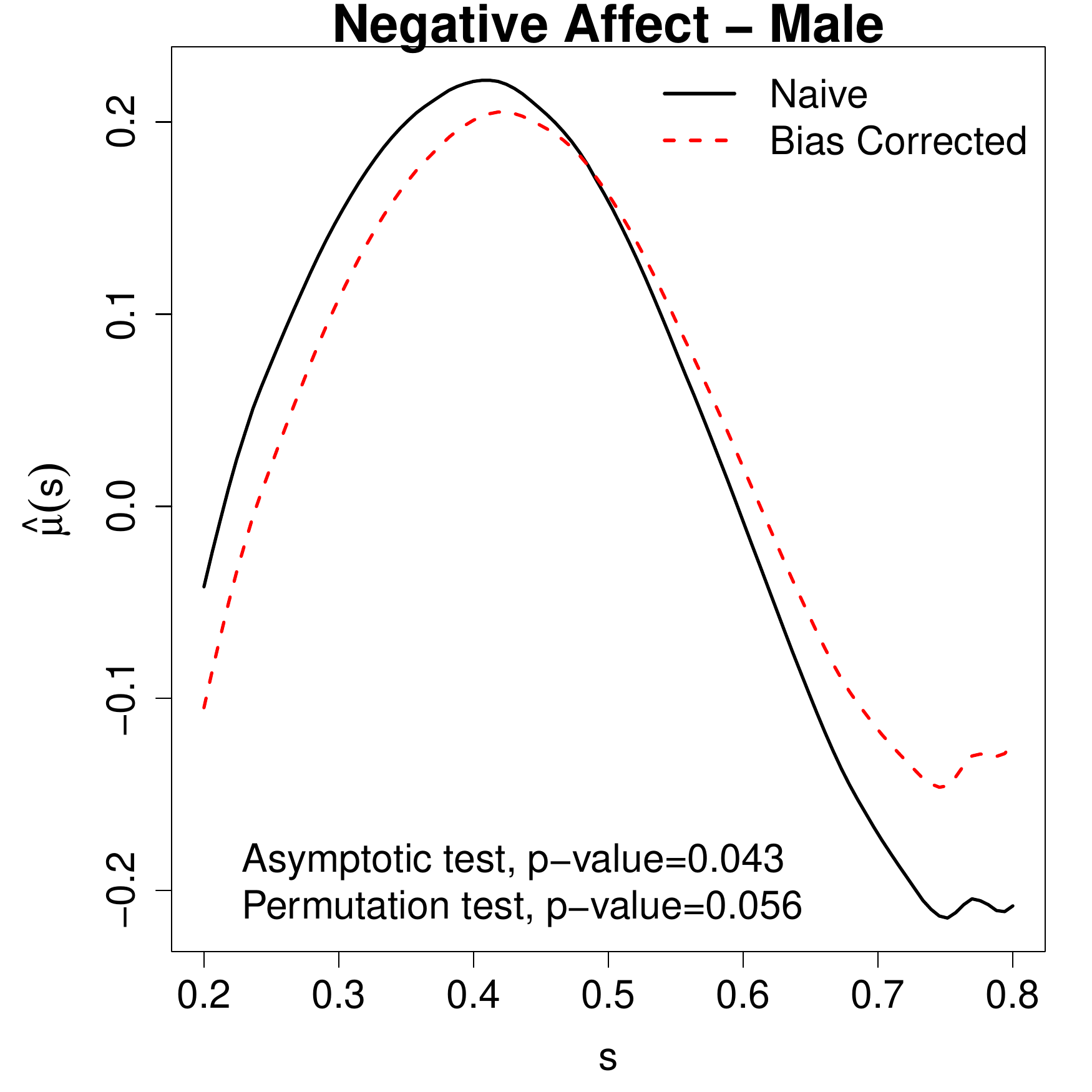} &
					\includegraphics[width=0.31\textwidth]{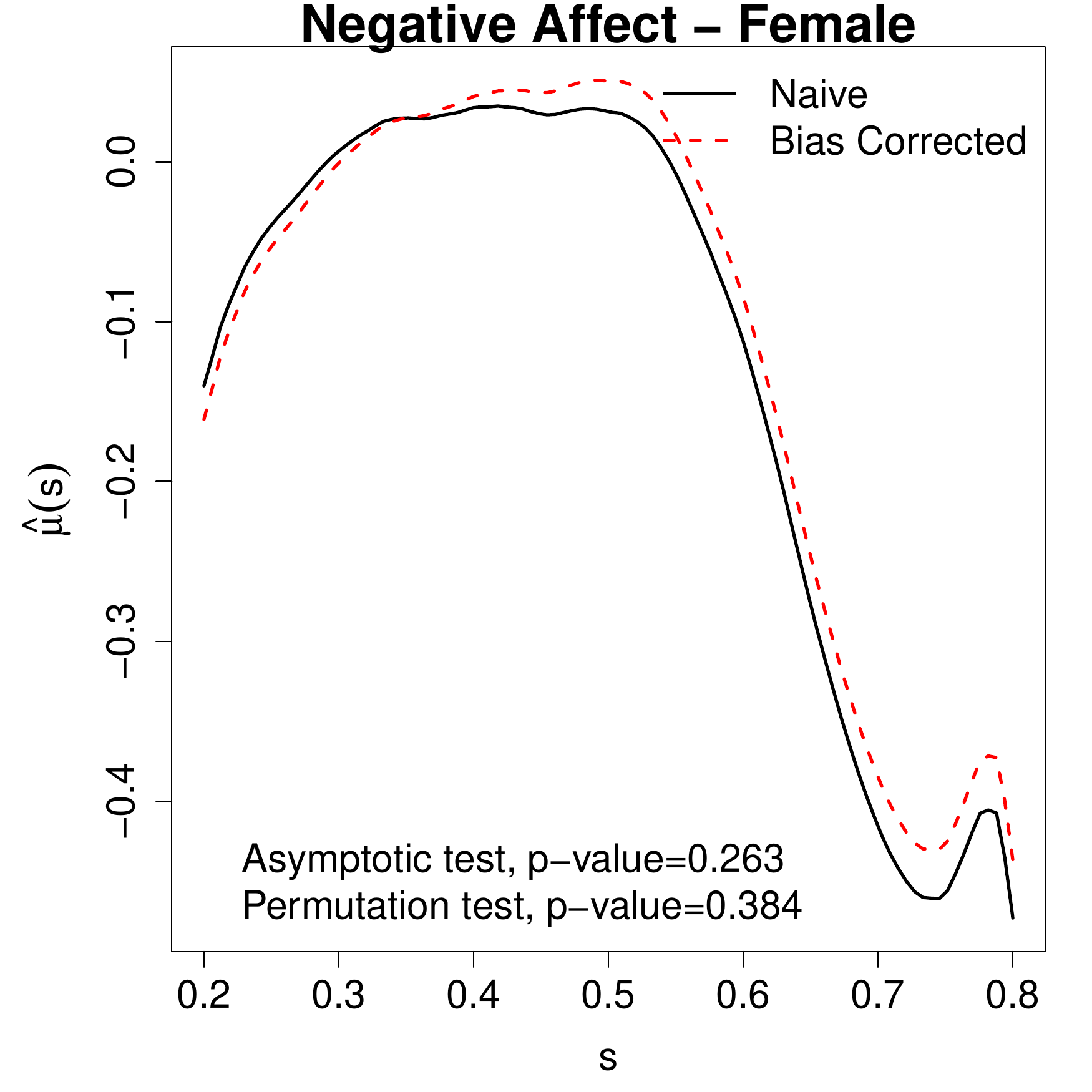} \\
				\end{tabular}
				\caption{Top left: diagnostic plot for LGCP following Section~\ref{sec:diag}; Top middle: QQ plot of the first-FPC scores following Section~\ref{sec:perm}; Top right and bottom panel:  Estimated daily mean functions of ``restless" and ``negative affect" for male and female participants, respectively.}
				\label{fig-smoke}
			\end{figure}
			
			It is seen from the reported $p$-values in Figure \ref{fig-smoke} that at the significance level of $0.10$, dependence between ``negative affect" and smoking times is significant for males, but is insignificant for females. This is consistent with the findings in \citet{shiffman2011point}. On the other hand, ``restless" and smoking times are found to have significant associations for both males and females at $0.05$ significance level. This is an interesting finding that, to our best knowledge, has not been discussed before in the literature. The estimated mean functions $\mu(\cdot)$'s are illustrated in Figure \ref{fig-smoke}. For ``negative affect", it is seen that females have a much lower level of ``negative affect" compared to males. For males, there is a notable difference between the naive estimator and the bias-corrected estimator; this is expected as the small $p$-value indicates significant mark-point dependence.

			For ``restless", there are notable differences between the naive and bias-corrected estimators for both males and females. In the bias-corrected estimates, male participants are seen to exhibit a steady level of restlessness and then an increasing trend starting around 0.7, while female participants have an increasing level of restlessness throughout the day. 
			
			From the naive and bias-corrected estimates of the covariance function $C_Y(\cdot,\cdot)$, we see that the first eigenfunction is flat for both marks and both genders (not plotted). Interestingly, the first eigenfunction accounts for more than 93\% of the variability for both marks and both genders. This suggests that the dominant mode of variation for ``negative effect" and ``restless" is the overall magnitude, and this holds for both male and female participants.
			
			\subsection{Ebay Online Auction Data}\label{sec:ebay_data}
			In this subsection, we analyze the bid prices of Palm M515 Personal Digital Assistants (PDA) on week-long eBay auctions that took place between March and May of 2003. This data set contains the bid times and prices of 194 PDA auctions, which has been analyzed by many authors. {See Figure~\ref{fig-aucdata} of the supplement for sample trajectories.}
			Earlier analyses \citep{jank2006functional} focused on dynamics of the bid price curves; 
			\citet{wu2013functional} studied the bid times as point processes; 
			\cite{gervini2020joint} modeled the bid times and bid prices jointly, since it was expected that items with lower prices tended to experience ``bid sniping" (i.e., concentrated bidding activity close to the end of the auction), {but they did not formally test for the mark-point independence.} 
			
			\begin{figure}[!t]
				\centering
				\begin{tabular}{lll}
					\includegraphics[width=0.31\textwidth]{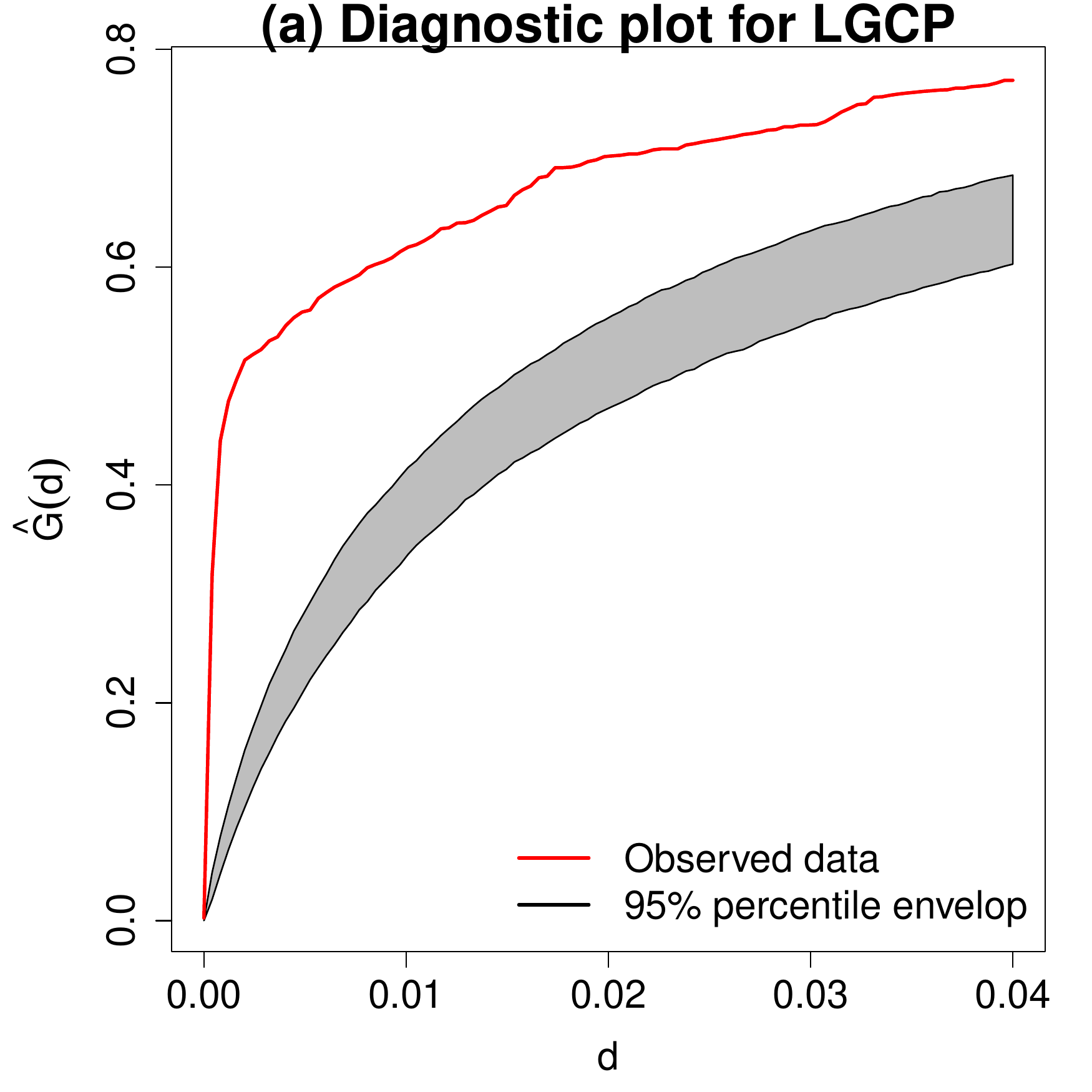} &
					\includegraphics[width=0.31\textwidth]{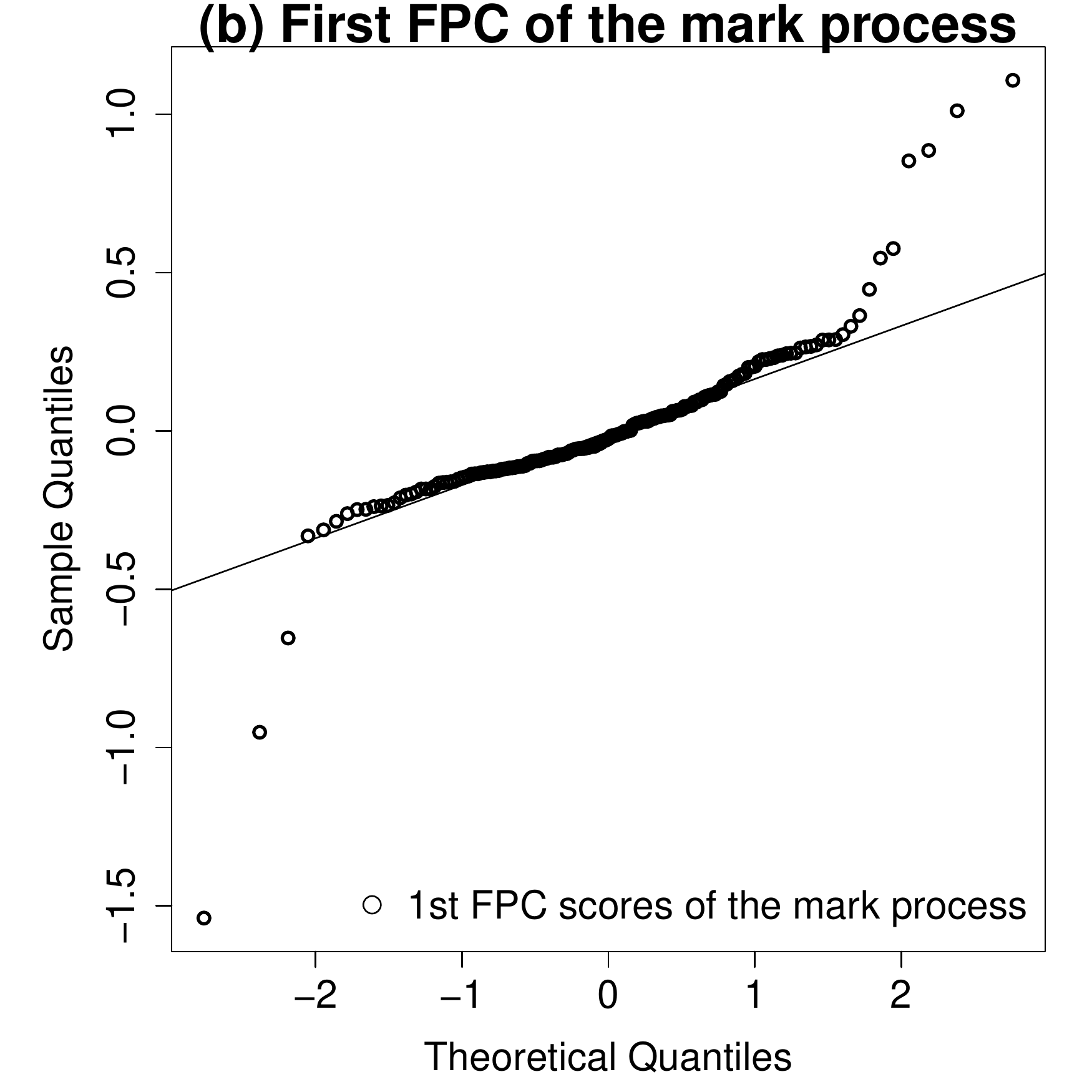}&	\includegraphics[width=0.31\textwidth]{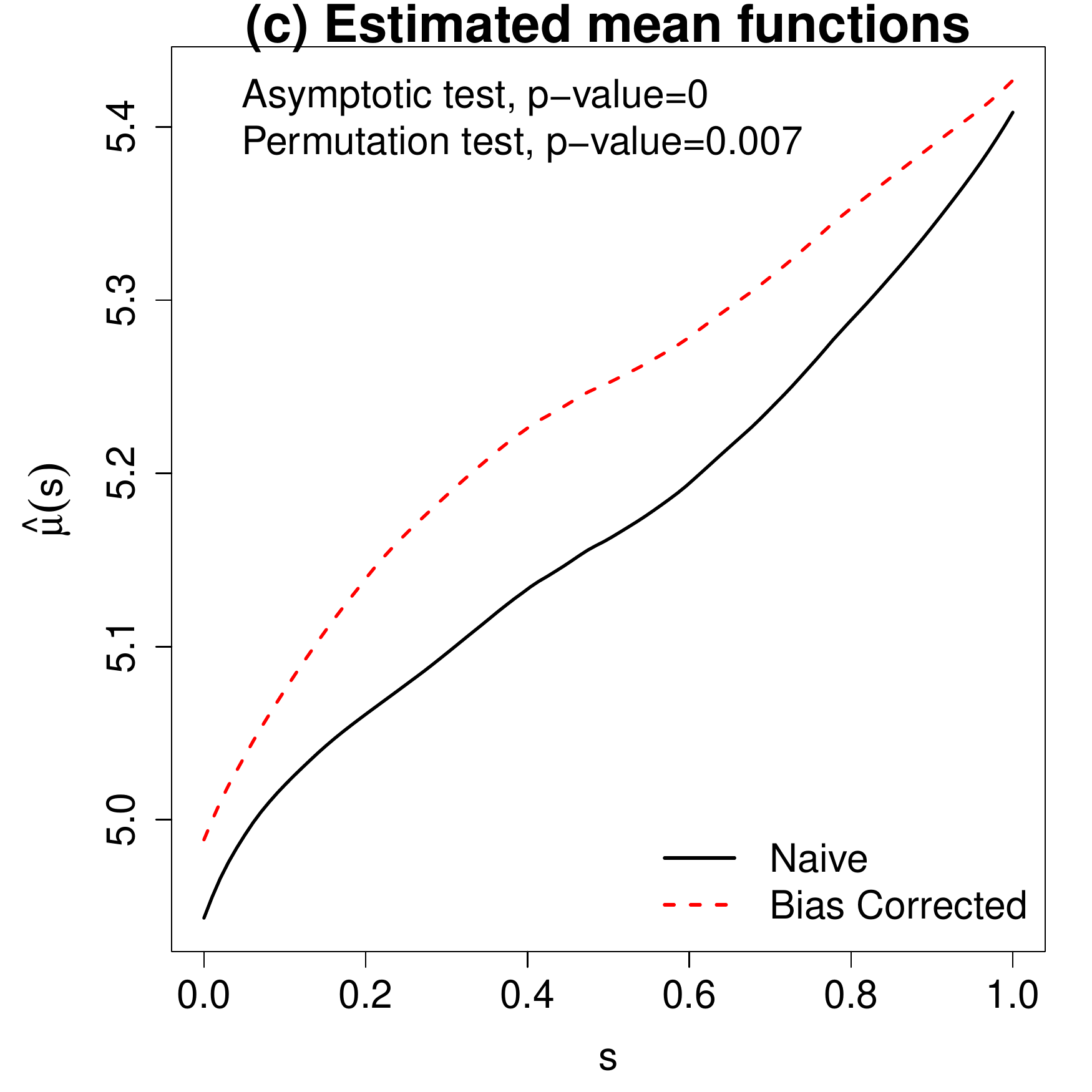}\\
				\end{tabular}\caption{{Model diagnostics and estimation for the auction data. }}
				\label{fig-ebay}
			\end{figure}
			
			
			Our analysis focuses on the last day of the seven-day auction period, which is rescaled to $[0,1]$. We remove subjects with fewer than $2$ bids in the last day, resulting in a total sample size of $n=174$. The bid prices are log-transformed as the mark process.  We {plot the diagnostic plots described in Section \ref{sec:diag}, and} apply the testing procedures for mark-point independence in Sections \ref{sec:test} and \ref{sec:perm}. {The diagnostic plots in Figure~\ref{fig-ebay}(a)-(b) suggest that the point process may not be an LGCP, as the observed nearest-neighbor distance distribution is not contained in the simulation envelope, and the mark process may not be Gaussian, as some departure from normality is seen in the QQ plot. From Figure~\ref{fig-ebay}(c), both the asymptotic and permutation tests reject the mark-point independence at the $0.05$ level. The validity of the asymptotic test is sensitive to model misspecification, but the permutation test is more robust, as numerically demonstrated in our simulation.} {On one hand, the agreement between the asymptotic test and the permutation test suggests that one can probably reject the mark-point independence at the $0.05$ level, supporting the well recognized ``bid sniping" effects toward the end of an auction. On the other hand, as demonstrated in Section~\ref{sec:s1} of the supplementary material, when the point process is not an LGCP and/or the mark process is not Gaussian, the proposed bias correction technique may not be able to eliminate the bias caused by the mark-point dependence and thus one needs to be cautious when interpreting the bias-corrected mean function.}

		}
		\section{Concluding Remarks}\label{sec:conclusions}
		In this paper, we propose {a computationally efficient} moment-based bias-correction procedure for estimating the mean and covariance functions of the mark process 
		{when there is} mark-point dependence. 
		We also propose two inferential procedures, {including an asymptotic test and a functional permutation test}, for testing the mark-point independence assumption. 
		
		While we focus on detecting mark-point dependence and correcting estimation bias caused by potential mark-point dependence, our work opens doors to a series of meaningful research directions such as bias-corrected statistical inference of functional principal component scores and bias-corrected testing of the mean functions of two or more groups (e.g., treatment v.s. control, male v.s. female).  It is also of interest to extend the current methods to mark-point process data with more complicated structures, see, e.g., \cite{xu2020semi,yin2021row}.

		Moreover, although our work has focused on temporal mark-point processes, it can be readily extended to model spatial mark-point processes if independent replicates are available, which requires using multi-dimensional kernel or product kernel functions. For future research, it is also of interest to consider the case where the marks are non-Gaussian (e.g. Poisson, Binomial). In the case of Poisson marks, we can assume that
		\be
		\label{pois}
		Z(s)\sim \text{Poisson}\left(\exp\left[\mu(s)+Y(s)\right]\right),\quad s\in\CT,
		\ee
		where $\mu(\cdot)$ and $Y(\cdot)$ are defined similarly as those in \eqref{mark}. Our proposed framework can be {extended to} derive unbiased estimators for $\mu(\cdot)$ and $C_Y(\cdot,\cdot)$, i.e., the covariance function of $Y(\cdot)$ in \eqref{pois}. {See Section~\ref{sec:s:pois} of the supplement for more details. It is also of great interest to theoretically justify the  functional permutation test for mark-point independence proposed in Section~\ref{sec:perm}.} We plan to investigate such directions in our future research.

		\baselineskip 15pt
		\bibliographystyle{asa}
		\bibliography{reference}

		\clearpage\pagebreak\newpage
		\setcounter{section}{0}
		\renewcommand{\thesection}{S.\arabic{section}} 
		\setcounter{page}{1}
		\renewcommand{\theequation}{S.\arabic{equation}}
		
		\renewcommand{\thefigure}{S.\arabic{figure}}
		\renewcommand{\thetable}{S.\arabic{table}}
		\setcounter{table}{0}
		\setcounter{equation}{0} \setcounter{figure}{0} 
		
		\renewcommand{\theenumi}{\arabic{enumi}}
		\renewcommand{\thelem}{S.\arabic{lem}}
		\setcounter{lem}{0}
		
		\begin{center}
			{\Large{\bf Supplementary Material for ``{Bias-correction} and {Test} for {Mark-point Dependence with} Replicated Marked Point Processes 
					"}}\\ \hskip 5mm \\
			\vskip1cm
		\end{center}
		\begin{center}
			Ganggang Xu, Jingfei Zhang, Yehua Li and Yongtao Guan
		\end{center}
		
	\symbolfootnote[0]{
		Ganggang Xu (Email: gangxu@bus.miami.edu) is Assistant Professor, Jingfei Zhang (Email: ezhang@bus.miami.edu) is Associate Professor, and
		Yongtao Guan (Email: yguan@bus.miami.edu) is Professor,
		Department of Management Science, University of Miami, Coral
		Gables, FL 33124. Yehua Li (Email: yehuali@ucr.edu) is Professor, Department of Statistics, University of California, Riverside, CA 92521. Zhang's research is supported by NSF grant DMS-2015190 and Guan's research is supported by NSF grant DMS-1810591. For correspondence, please contact Yongtao Guan.
	}
		
		\par\vfill\noindent
		{\bf Some key words:} {Mark-point dependence; Marked Point Processes. }
		\par\medskip\noindent
		{\bf Short title}: {Estimation and Test for Marked Point Process}

		\clearpage\pagebreak\newpage \pagenumbering{arabic}
		\baselineskip=26pt
		
		\allowdisplaybreaks
		
		\section{Technical Details}
		\subsection{Bandwidth Selection}\label{sec:bandwidth}
		{In this section, we describe a set of cross-validation scores to empirically choose bandwidths for the proposed local linear estimators. To that end, we randomly divide the sample of size $n$ into $K$ folds and denote} $\wt\mu^{(-k)}(\cdot)$, $\wt C_Y^{(-k)}(\cdot,\cdot)$, $\wt\sigma_Z^{2,(-k)}(\cdot)$ and $\wh C_{XY}^{(-k)}(\cdot,\cdot)$ {as the counterparts to} $\wt\mu(\cdot)$, $\wt C_Y(\cdot,t)$, $\wt\sigma_Z^{2}(\cdot)$ and $\wh C_{XY}(\cdot,\cdot)$ calculated by {leaving out all {data} from the $k$th fold, $k=1,\ldots,K$}. {For the purpose of function estimation,} we propose to select the bandwidths by minimizing the following {least-square type} cross-validation criteria: 
		\begin{eqnarray*}
			CV_1(h_{\mu})&=&\sum_{k=1}^K\sum_{i\in \text{Fold}_k}\sum_{u\in N_i}\left[Z_i(u)-\wt\mu^{(-k)}(u)\right]^2,\\
			CV_2(h_y)&=&\sum_{k=1}^K\sum_{i\in \text{Fold}_k}\mathop{\sum\sum}_{u,v\in N_i}^{\ne} \left\{\left[Z_i(u)-\wt\mu^{(-k)}(u)\right]\left[Z_i(v)-\wt\mu^{(-k)}(v)\right]-\wt C_Y^{(-k)}(u,v)\right\}^2,\\
			CV_3(h_\sigma)&=&\sum_{k=1}^K\sum_{i\in \text{Fold}_k}\sum_{u\in N_i}\left\{\left[Z_i(u)-\wt\mu^{(-k)}(u)\right]^2-\wt\sigma_Z^{2,(-k)}(u)\right\}^2,\\
			CV_4(h_{xy})&=&\sum_{k=1}^K\sum_{i\in \text{Fold}_k}\mathop{\sum\sum}_{u,v\in N_i}^{\ne} \left\{\left[Z_i(v)-\wt\mu^{(-k)}(v)\right]-\wh C_{XY}^{(-k)}(u,v)\right\}^2,
		\end{eqnarray*}
		{where $\text{Fold}_k$ consists of processes in the $k$th fold, $k=1,\ldots,K$. {For $CV_j(\cdot)$, $j=2,3,4$, the bandwidth $h_\mu$ used in $\wt\mu^{(-k)}(\cdot)$ is replaced by $\hat h_{\mu}=\argmin_{h_\mu}CV_1(h_\mu)$.} Our empirical studies in Section~\ref{sec:simacu} demonstrate the effectiveness of the above tuning parameter selection criteria with $K=10$.

			It is well known in the literature that the bandwidth needed for hypothesis testing is typically of a smaller order than the optimal bandwidth needed for function estimation {\citep[e.g.][]{stute2005nonparametric}.} In fact, Theorem~\ref{thm1} shows that, for estimation purpose, the optimal convergence rates of $\tilde\mu(\cdot)$ and $\tilde\sigma_Z^2(\cdot)$ {to their respective limits are attained with } $h_\mu$ and $h_\sigma$ roughly of the order $O(n^{-1/5})$. In contrast, Theorem~\ref{thm3} shows that for the proposed test to be asymptotically valid, it requires $h_\mu$ and $h_\sigma$ to be of the order $o(n^{-1/4})$. To bridge this gap, we propose to use the following re-scaled bandwidths for our test statistic $T_n$:
			\be\label{eq:undersmooth}
			\hat h_\mu^t=[n^{1/20}\log\log(n)]^{-1}\hat h_\mu, \hbox{ and } \hat h_\sigma^t=[n^{1/20}\log\log(n)]^{-1}\hat h_\sigma,
			\ee
			where $\hat h_\mu$ and $\hat h_\sigma$ are minimizers of $CV_1(\cdot)$ and $CV_3(\cdot)$, respectively. Such a choice works well for the numerical examples in our simulation studies in Section~\ref{sec:simu}, in the sense that the resulting empirical sizes are close to the nominal levels while the empirical powers remain high against various alternative hypotheses under consideration.
			
			\subsection{Empirical Variance Estimator}\label{sec:a1}
			In this section, we provide an empirical variance estimator for the asymptotic variance $\Omega_T$ given in Theorem~\ref{thm3}. We first define the following kernel estimator for the function $\tau(\cdot)$:
			$$\wh\tau(s)= \frac{\sum_{i=1}^{n}\mathop{\sum\sum}^{\ne}_{u,v\in N_i}K_{1,h_{\tau}}(v-s)}{\sum_{i=1}^{n}\sum_{v\in N_i}K_{1,h_{\tau}}(v-s)}, \quad s\in\CT,$$
			where $h_{\tau}$ is a bandwidth. It can be shown that $\wh\tau(\cdot)$ is uniformly consistent for $\tau(\cdot)$ with an appropriately chosen $h_{\tau}$, following similar steps to prove Theorem~\ref{thm1}. In practice, we empirically choose $h_{\tau}$ as the optimal bandwidth for the kernel estimator of the first-order intensity function $\rho(\cdot)$ that is selected using cross-validation.
			Next, for $i=1,\ldots,n$, define
			\[
			\begin{split}
			\wh\Omega_{T_n,i}&=2(n_i-2)(n_i-3) \mathop{\sum\sum}^{\ne}_{t,v\in N_i}[\wh C_Y^{(-i)}(t,v)]^2+2(n_i-1)(n_i-2)\sum_{v\in N_i}[\wt\sigma_Z^{2,(-i)}(v)]^2\\
			&\quad+ 2(n_i-2) \mathop{\sum\sum}^{\ne}_{s,v\in N_i}[3-2\wh\tau^{(-i)}(s)][\wh C_Y^{(-i)}(s,v)]^2+2\mathop{\sum\sum}^{\ne}_{u,v\in N_i}[1-\wh\tau^{(-i)}(u)][1-\wh\tau^{(-i)}(v)][\wh C_Y^{(-i)}(u,v)]^2\\
			&\quad+2(n_i-1)\sum_{s\in N_i}\left[1-\wh\tau^{(-i)}(s)\right][\wt\sigma_Z^{2,(-i)}(s)]^2,
			\end{split}
			\]
			where $\wh C_Y^{(-i)}(\cdot,\cdot)$, $\wt\sigma_Z^{2,(-i)}(\cdot)$ and $\wh \tau^{(-i)}(\cdot)$ are the counterparts to $\wh C_Y(\cdot,\cdot)$, $\wt\sigma_Z^{2}(\cdot)$ and $\wh \tau(\cdot)$ obtained by leaving the $i$th process out. The leave-one-out strategy is used here to to ensure that
			\[
			\begin{split}
			\E\wh\Omega_{T_n,i}=2 &\int_{\CT^4}\E[\wh C_Y^{(-i)}(t,v)]^2 \rho_4(s,t, u,v) ds dt du dv \\
			&+ 2 \int_{\CT^3}{\E\left\{[\wt\sigma_Z^{2,(-i)}(v)]^2+[3-2\wh\tau^{(-i)}(s)][\wh C_Y^{(-i)}(s,v)]^2\right\}} \rho_3( s, u, v) ds du dv \\
			&{+ 2 \int_{\CT^2} \E\left\{[1-\wh\tau^{(-i)}(u)][1-\wh\tau^{(-i)}(v)][\wh C_Y^{(-i)}(u,v)]^2\right\} \rho_2( u, v) du dv} \\
			&+ 2 \int_\CT \E\left\{\left[1-\wh\tau^{(-i)}(s)\right][\wt\sigma_Z^{2,(-i)}(s)]^2\right\}\rho(v)\tau(v) dv.
			\end{split}
			\]
			When $n$ is sufficiently large, by Theorems~\ref{thm1}-\ref{thm2} and the consistency of $\wh\tau(\cdot)$, one can show that $\E\wh\Omega_{T_n,i}\approx\Omega_T$. Consequently, it can be shown that a consistent estimator of $\Omega_T$ is defined as
			\be
			\label{emp-var}
			\wh\Omega_{T_n}=\frac{1}{n}\sum_{i=1}^{n}\wh\Omega_{T_n,i}.
			\ee
			
			\subsection{Implementation of Functional Permutation Test}\label{sec:a2}
			{ In this section, we give some details on the implementation of the functional permutation test proposed in Section~\ref{sec:perm}. It has come to our attention that the FPC estimator $\wh\bxi_i^Y$ in~\eqref{Fscore} can be problematic for some $i$'s when there are too few time points in $N_i$. For example, If $|N_i|<p_Y$, \eqref{Fscore} does not have a unique solution. In this case, we can use the best linear unbiased predictor \citep{yao2005functional} of $\bxi_i^{Y}$, denoted as  $\wh\bxi_i^{Y,blup}$, which is also valid under the mark-point independence. However, our numerical experience suggests that for smaller sample sizes, the permutation test based $\wh\bxi_i^{Y,blup}$'s tend to have overly large rejection rates under $H_0$. Given this, we use the following hybrid approach.

				If $|N_i|<p_Y$, \eqref{Fscore} does not have a unique solution and $\wh\bxi_i^Y$ needs to be replaced by $\wh\bxi_i^{Y,blup}$. When $|N_i|\ge p_Y$, we can write $\wh\bxi_i^Y=(\bPhi_i^T\bPhi_i)^{-1}\bPhi_i^\top(\bm Z_i-\wh\bmu_i)$, where $\bPhi_i$ is the design matrix of the least square problem in \eqref{Fscore}, $\bm Z_i=(Z_i(s):s\in N_i)^\top$ and $\wh\bmu_i=(\wh\mu(s):s\in N_i)^\top$. We can then roughly calculate the asymptotic variance of  $\wh\bxi_i^Y$ to be $\sigma_e^2(\bPhi_i^T\bPhi_i)^{-1}$. For each $i$, we can compute $\ell_i=\trace\left[(\bPhi_i^T\bPhi_i)^{-1}\right]$ as an indicator of the estimation variability in $\wh\bxi_i^Y$, $i=1,\ldots,n$. For all top $25\%$ values of $\ell_i$'s, we shall replace $\wh\bxi_i^Y$ with $\wh\bxi_i^{Y,blup}$. That is, if the estimation variability of $\wh\bxi_i^Y$  is relatively large, we will replace $\wh\bxi_i^Y$ with $\wh\bxi_i^{Y,blup}$.
			}
			\subsection{Kernel Smoothing of log-Gaussian Cox Process}\label{sec:a3}
			In this section, we derive the nonparametric estimates of $\lambda_0(\cdot)$ and $C_X(\cdot,\cdot)$ under the LGCP assumption. 
			Given bandwidths $h_1$ and $h_2$, we define the kernel estimator of the first and second-order intensities of the point process as
			\[
			\wh\rho(s;h_1)=\frac{1}{ne(s;h_1)}\sum_{i=1}^n\sum_{u\in N_i} K_{h_1}(u-s),\text{ for any } s\in\CT,
			\]
			\[
			\wh\rho_2(s,t;h_2)=\frac{1}{ne(s;h_2)e(t;h_2)}\sum_{i=1}^n\mathop{\sum\sum}^{\neq}_{u,v\in N_i}K_{h_2}(u-s)K_{h_2}(v-t),\text{ for any } s,t\in\CT,
			\]
			where $K_h(s)=h^{-1}K(s/h)$ and $e(s;h)=\int_\CT K_h({s-x})dx$ is an edge correction term. Under the LGCP assumption, we can utilize equations~\eqref{marginal}-\eqref{marginal2} to find kernel estimators of $\lambda_0(\cdot)$ and $C_X(s,t)$ as following
			\[
			\wh C_X(s,t)=\log[\wh\rho_2(s,t;h_2)]-\log[\wh\rho(s;h_1)] -\log[\wh\rho(t;h_1)],\text{ for any } s,t\in\CT,
			\]
			\[
			\wh \lambda_0(s)=\wh\rho(s;h_1)\exp[-\wh C_X(s,s)/2],\text{ for any } s\in\CT.
			\]
			
			{
				\section{Estimation for Poisson Marks}\label{sec:s:pois}
				In this section, we demonstrate how to correct estimation bias when the point process is an LGCP and the mark process has the following model
				\[
				Z(s)\sim \text{Poisson}\left(\exp\left[\mu(s)+Y(s)\right]\right),\quad s\in\CT,
				\]
				where $\mu(\cdot)$ and $Y(\cdot)$ are defined the same as those in \eqref{mark}.  Consider the following quantities
				\bea
				\wh A(s)&=&\frac{1}{n}\sum_{i=1}^{n}\mathop{\sum}_{u\in N_i}K_h(s-u),\nonumber\\
				\wh    B(s)&=&\frac{1}{n}\sum_{i=1}^{n}\mathop{\sum}_{u\in N_i}Z_i(u)K_h(s-u),\nonumber\\
				\wh	C(s,t)&=&\frac{1}{n}\sum_{i=1}^{n}\mathop{\sum\sum}_{u,v\in N_i}^{\neq} Z_i(u) K_h(s-u)K_h(t-v),\nonumber\\
				\wh		D(s,t)&=&\frac{1}{n}\sum_{i=1}^{n}\mathop{\sum\sum}_{u,v\in N_i}^{\neq}K_h(s-u)K_h(t-v),\nonumber\\
				\wh	E(s,t)&=&\frac{1}{n}\sum_{i=1}^{n}\mathop{\sum\sum}_{u,v\in N_i}^{\neq}\left[Z_i(u)Z_i(v)\right]K_h(s-u)K_h(t-v).\nonumber
				\eea
				Let $\rho_X(s)=\lambda_0(s)\exp\left[\sigma_X^2(s)/2\right]$, $\rho_Z(s)=\exp\left[\mu(s)+\sigma_Y^2(s)/2\right]$ and $\rho_{XZ}(s)=\rho_X(s)\rho_Z(s)$, then it is straightforward to show that
				\[
				\begin{split}
				\E\left[\wh A(s)\right]&=\int \lambda_0(u)\E\left\{\exp\left[X(u)\right] \right\}K_h(s-u)du=\int \rho_X(u)K_h(s-u)du,\\
				\end{split}
				\]
				\[
				\begin{split}
				\E\left[\wh B(s)\right]&=\int \lambda_0(u)\E\left\{\exp\left[\mu(u)+Y(u)+X(u)\right] \right\}K_h(s-u)du\\
				&=\int \rho_{XZ}(u)\exp\left[C_{XY}(u,u)\right] K_h(s-u)du,\\
				\end{split}
				\]
				\[
				\begin{split}
				\E\left[\wh C(s,t)\right]&=\int\int \lambda_0(u)\lambda_0(v)\E\left\{\exp\left[\mu(u)+Y(u)+X(u)+X(v)\right] \right\}K_h(s-u)K_h(t-v)dudv\\
				&=\int\int \rho_{XZ}(u)\rho_X(v)\exp\left[C_X(u,v)+C_{XY}(u,u)+C_{XY}(v,u)\right]K_h(s-u)K_h(t-v)dudv,\\
				\end{split}
				\]
				\[
				\begin{split}
				\E\left[\wh D(s,t)\right]&=\int\int \lambda_0(u)\lambda_0(v)\E\left\{\exp\left[X(u)+X(v)\right] \right\}K_h(s-u)K_h(t-v)dudv\\
				&=\int\int \rho_X(u)\rho_X(v)\exp\left[C_X(u,v)\right] K_h(s-u)K_h(t-v)dudv,\\
				\end{split}
				\]
				\begin{eqnarray*}
					&&\E\left[\wh E(s,t)\right]\\
					&=&\int\int \lambda_0(u)\lambda_0(v)\E\{\exp[\mu(u)+\mu(v)+Y(u)+Y(v)+X(u)+X(v)]\}K_h(s-u)K_h(t-v)dudv\\
					&=&\int\int \rho_{XZ}(u)\rho_{XZ}(v)\exp\left[C_X(u,v)+C_Y(u,v)+Q_{XY}(u,v)\right] K_h(s-u)K_h(t-v)dudv,
				\end{eqnarray*}
				where $Q_{XY}(u,v)=C_{XY}(u,v)+C_{XY}(v,u)+C_{XY}(u,u)+C_{XY}(v,v)$. Then, for an appropriately chosen bandwidth $h$, we can approximate the following quantities as
				\[
				\exp\left[C_{X}(s,t)\right]\approx\frac{\wh D(s,t)}{\wh A(s)\wh A(t)},\qquad
				\exp\left[C_{XY}(t,s)\right]\approx\frac{\wh C(s,t)\wh A(s)}{\wh B(s)\wh D(s,t)},\qquad
				\exp\left[C_{Y}(t,s)\right]\approx\frac{\wh E(s,t)\wh D(t,s)}{\wh C(s,t)\wh C(t,s)}.
				\]
				Consequently, for any $s,t\in\CT$, we can use the following estimators 
				\[
				\wh C_{X}(s,t;h)=\log\frac{\wh D(s,t)}{\wh A(s)\wh A(t)},\quad
				\wh C_{XY}(t,s;h)=\log\frac{\wh C(s,t)\wh A(s)}{\wh B(s)\wh D(s,t)},\quad
				\wh C_{Y}(t,s;h)=\log\frac{\wh E(s,t)\wh D(t,s)}{\wh C(s,t)\wh C(t,s)}.
				\]
				Similarly, the mean function of the Gaussian process $Y(\cdot)$ can then be estimated by 
				\[
				\wh\mu(s;h)=\log\frac{\wh B(s)}{\wh A(s)}-\frac{1}{2}\wh C_{Y}(s,s;h)-\wh C_{XY}(s,s;h),\text{ for any } s\in\CT.
				\]
			}
			\section{Additional Numerical Results}\label{sec:add}
			In this section, we provide additional numerical results from our simulation studies and real data analysis.
			{	
				\subsection{Estimation Accuracy}
				\label{sec:s1}
				
				In this subsection, we study the estimation accuracy of the proposed nonparametric estimators in the presence of mark-point dependence when  $X_i(\cdot)$'s and $Y_i(\cdot)$'s are not both Gaussian. From Figures~\ref{fig-1-S}-\ref{fig-2-SS}, we can see that in both (\texttt{Exp}, \texttt{Exp}) and (\texttt{T4}, \texttt{T4}) settings, the estimation bias resulting from the mark-point dependence may not be eliminated, {in the sense the bias may not decrease with sample size}, but can still be reduced using the proposed bias correction {by comparing the MADs from the naive and the bias-corrected estimators}. 
				In addition, compared to the (\texttt{Gaussian}, \texttt{Gaussian}) setting, the local linear estimator for $\sigma_z^2(\cdot)$ is no longer unbiased under the  (\texttt{Exp}, \texttt{Exp}) and (\texttt{T4}, \texttt{T4}) settings. 
				\begin{figure}[H]
					\centering
					\begin{tabular}{lll}
						\includegraphics[width=0.3\textwidth]{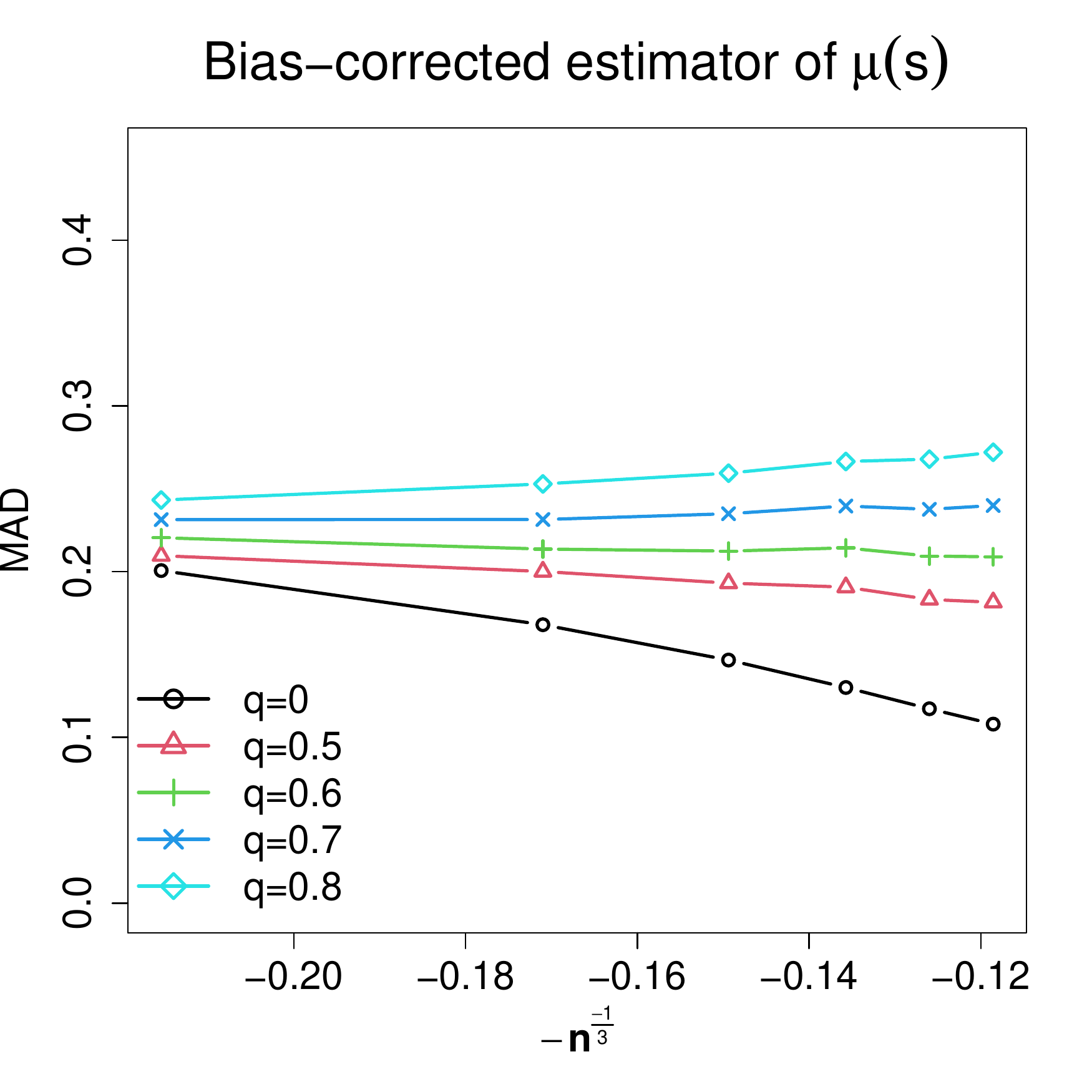} &
						\includegraphics[width=0.3\textwidth]{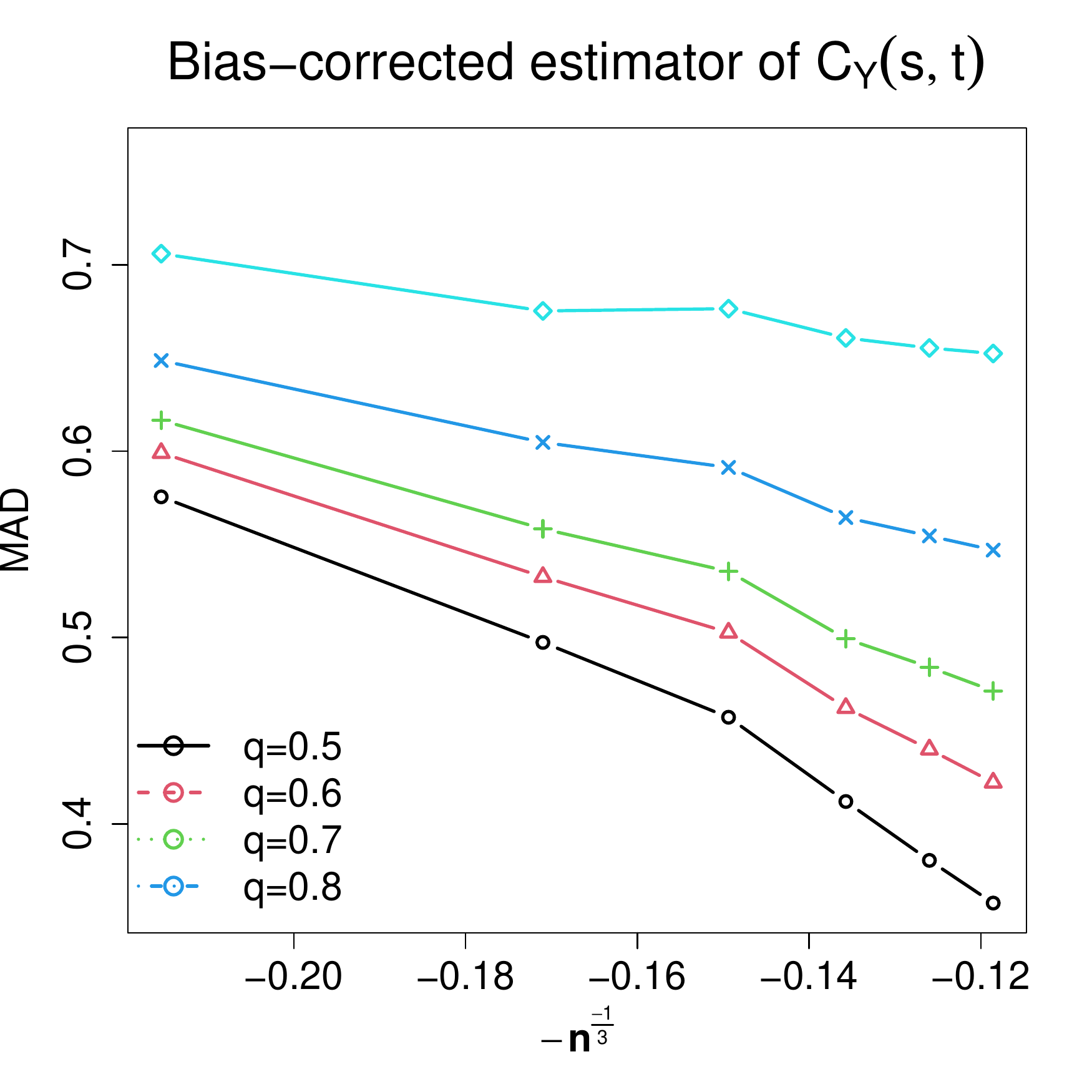}&
						\includegraphics[width=0.3\textwidth]{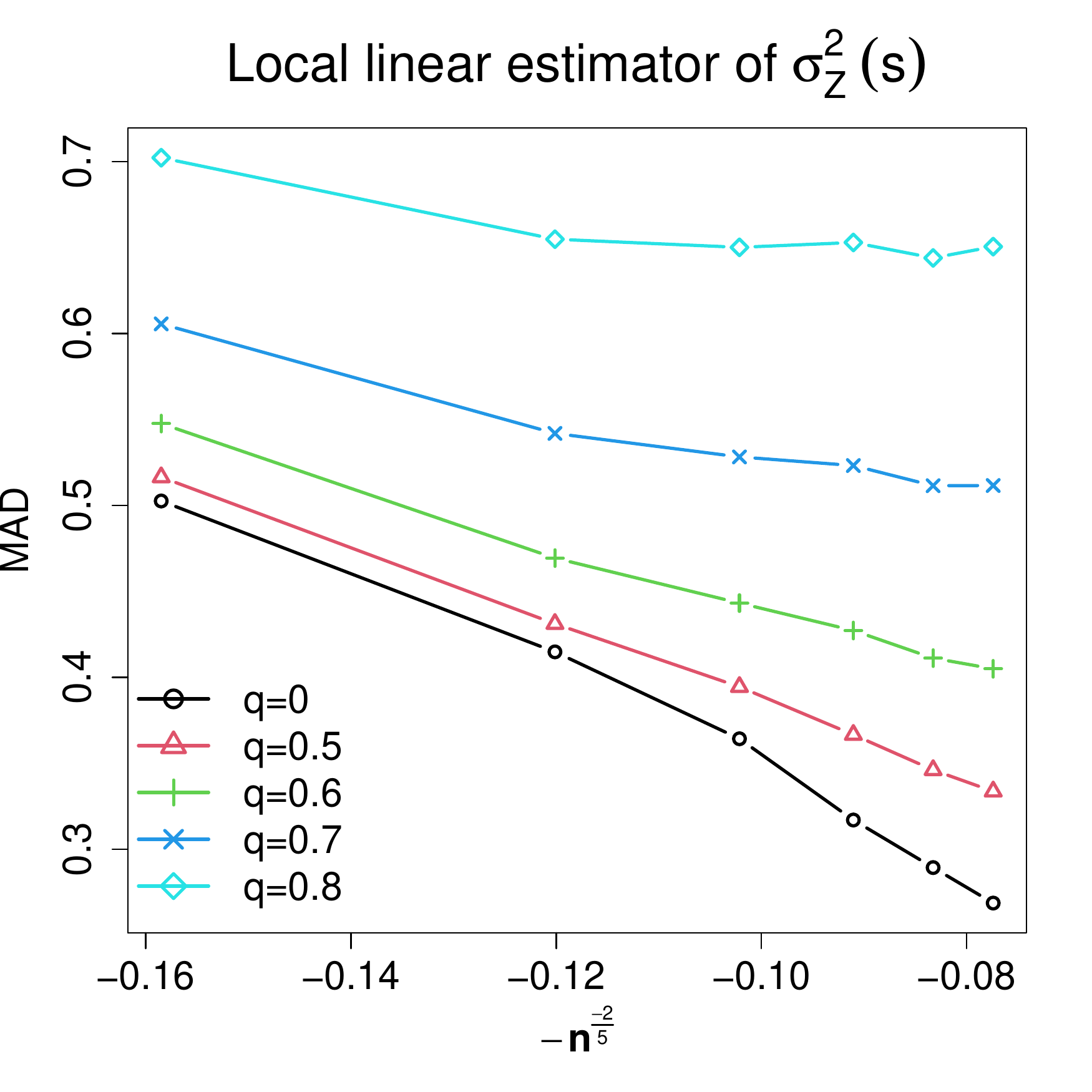}\\
						\includegraphics[width=0.3\textwidth]{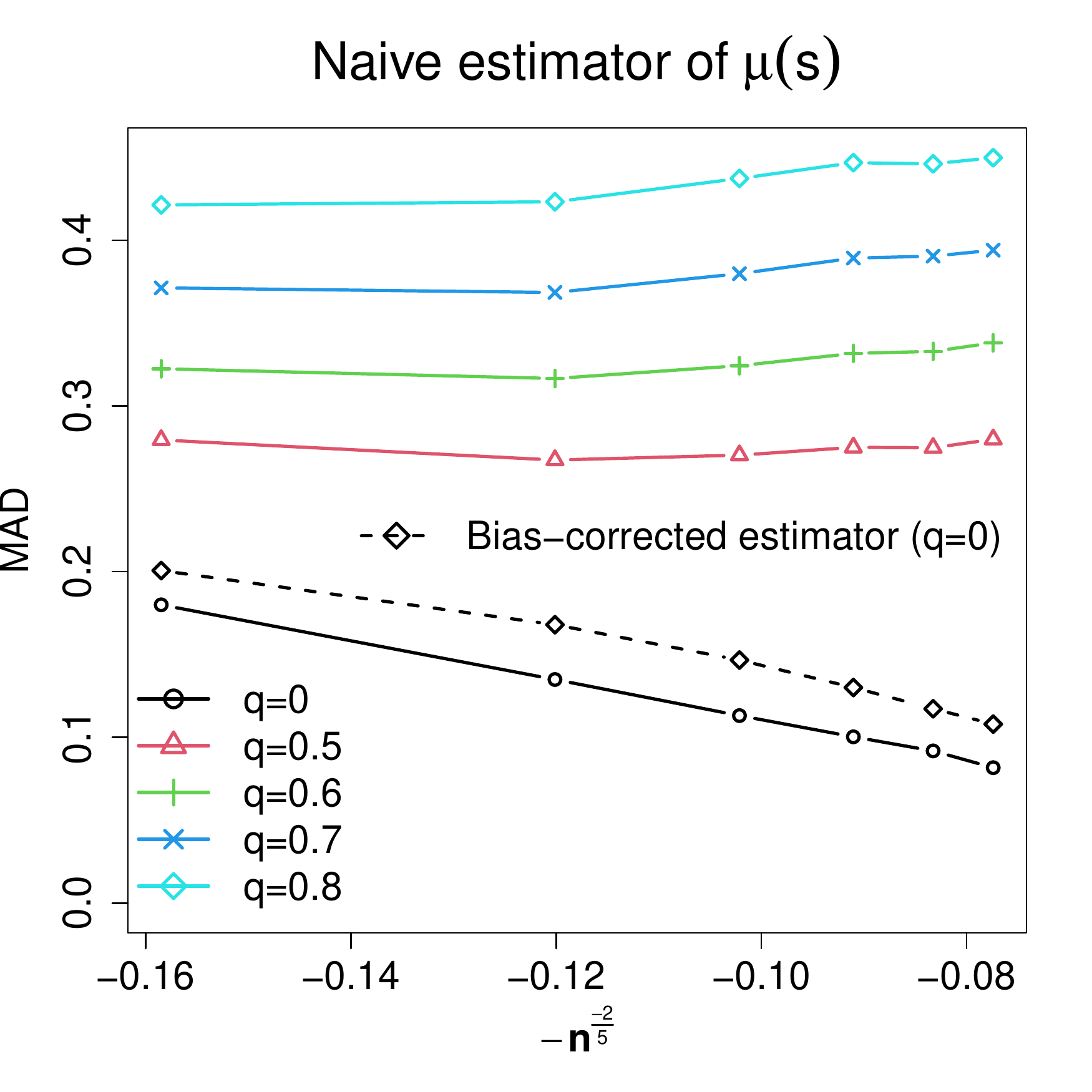} &
						\includegraphics[width=0.3\textwidth]{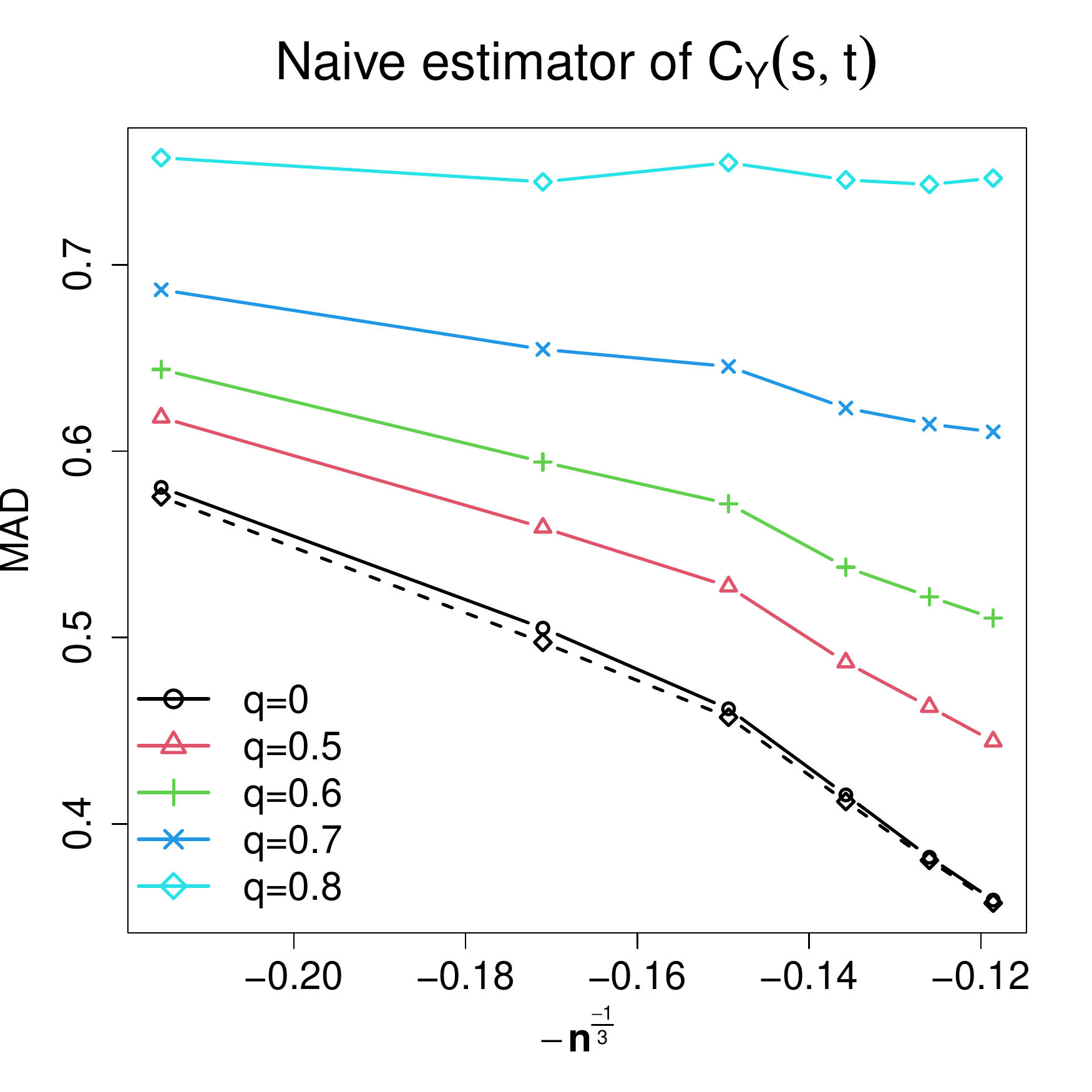}&
						\includegraphics[width=0.3\textwidth]{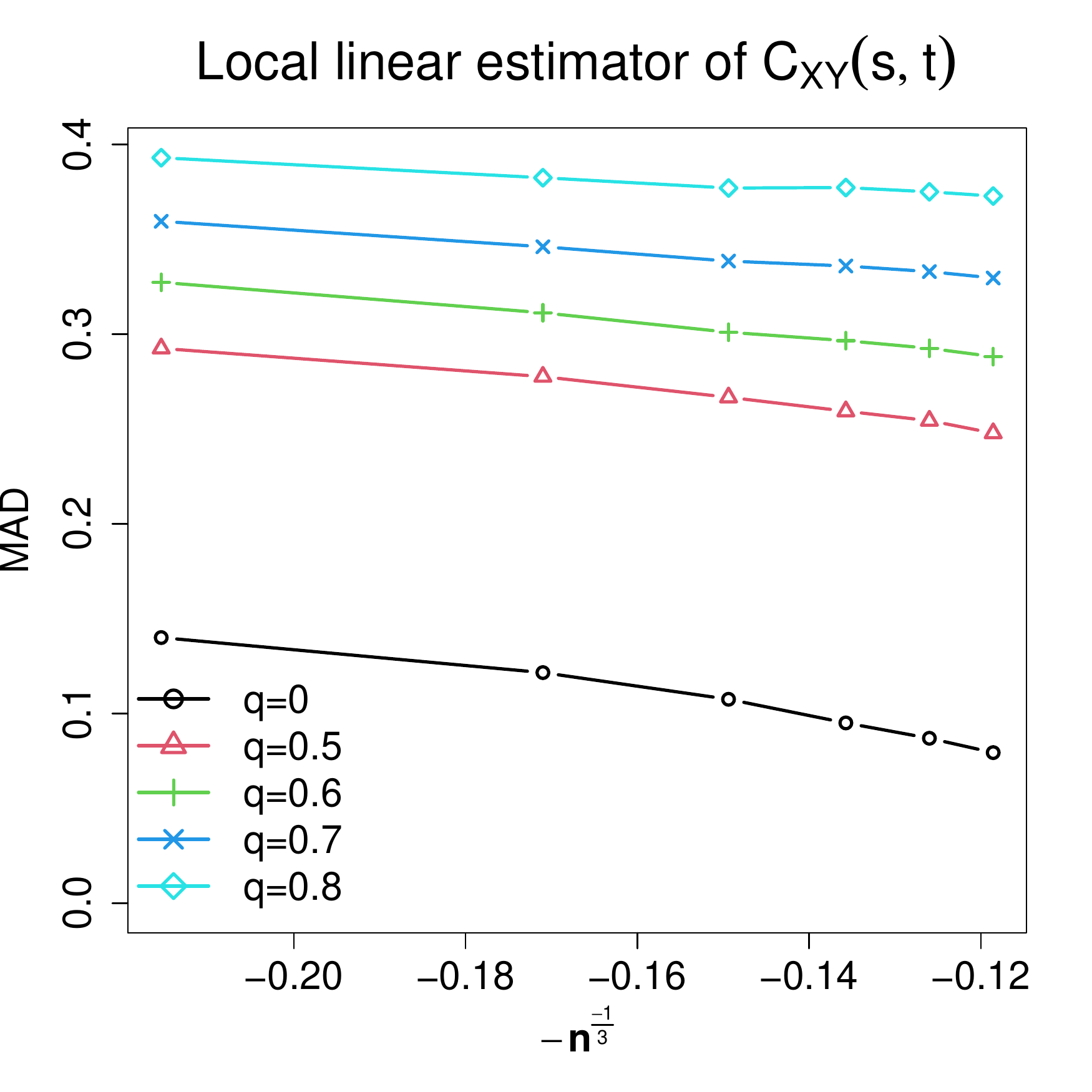}\\
					\end{tabular}
					\vskip -1em
					\caption{{Estimation accuracy of local linear estimators under the (\texttt{Exp}, \texttt{Exp}) setting.}}
					\label{fig-1-S}
				\end{figure}

				\begin{figure}[H]
					\centering
					\begin{tabular}{lll}
						\includegraphics[width=0.3\textwidth]{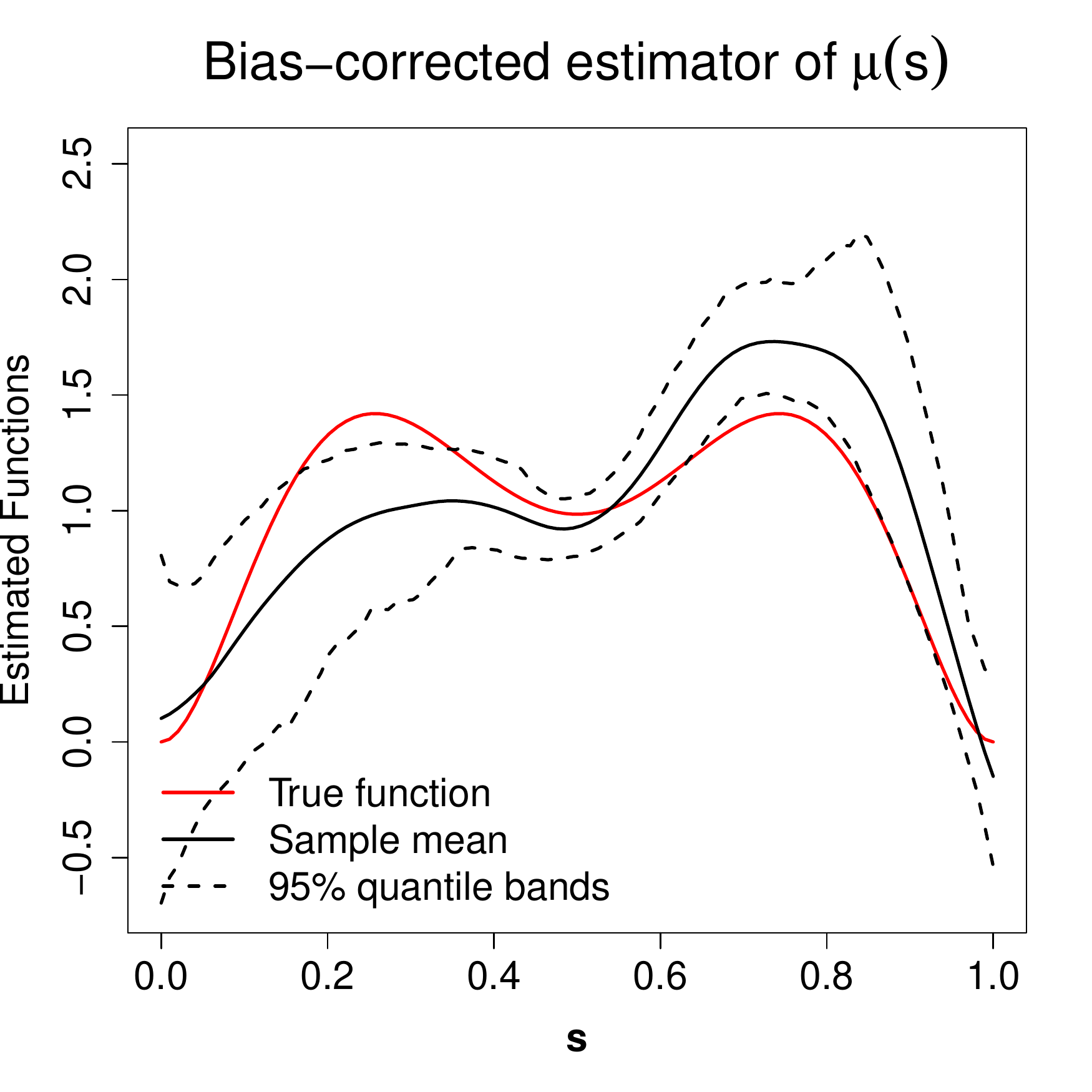} &
						\includegraphics[width=0.3\textwidth]{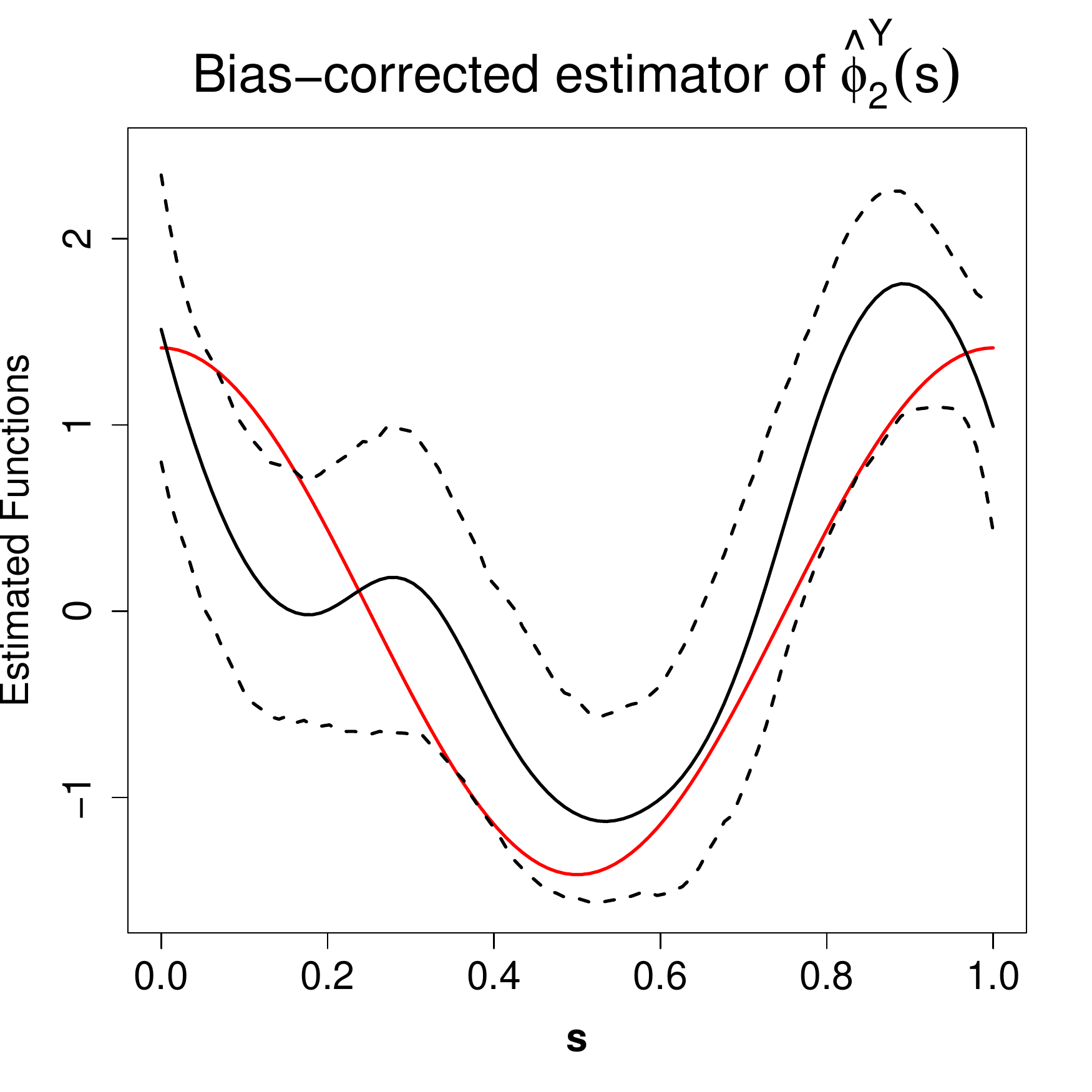}&
						\includegraphics[width=0.3\textwidth]{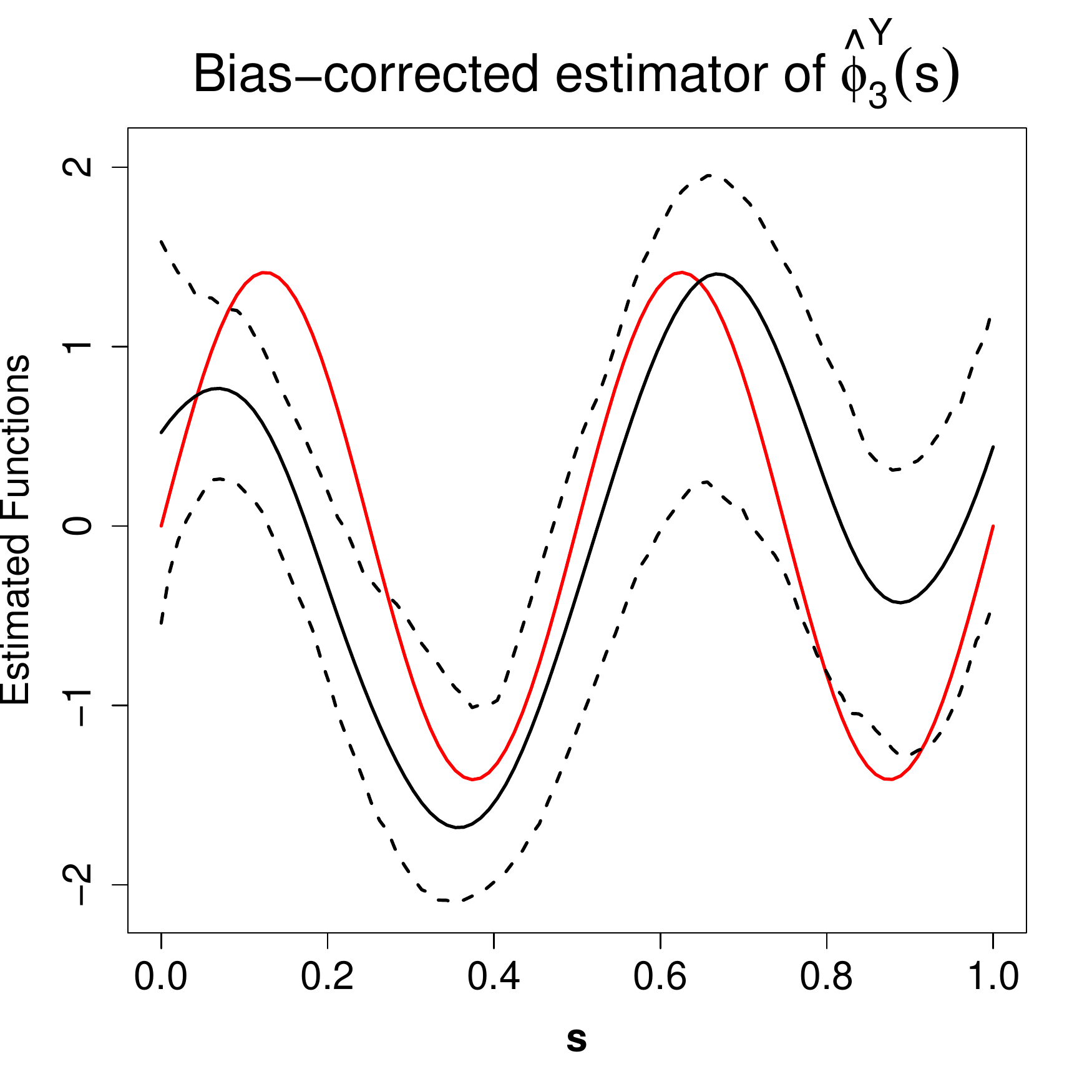}\\
						\includegraphics[width=0.3\textwidth]{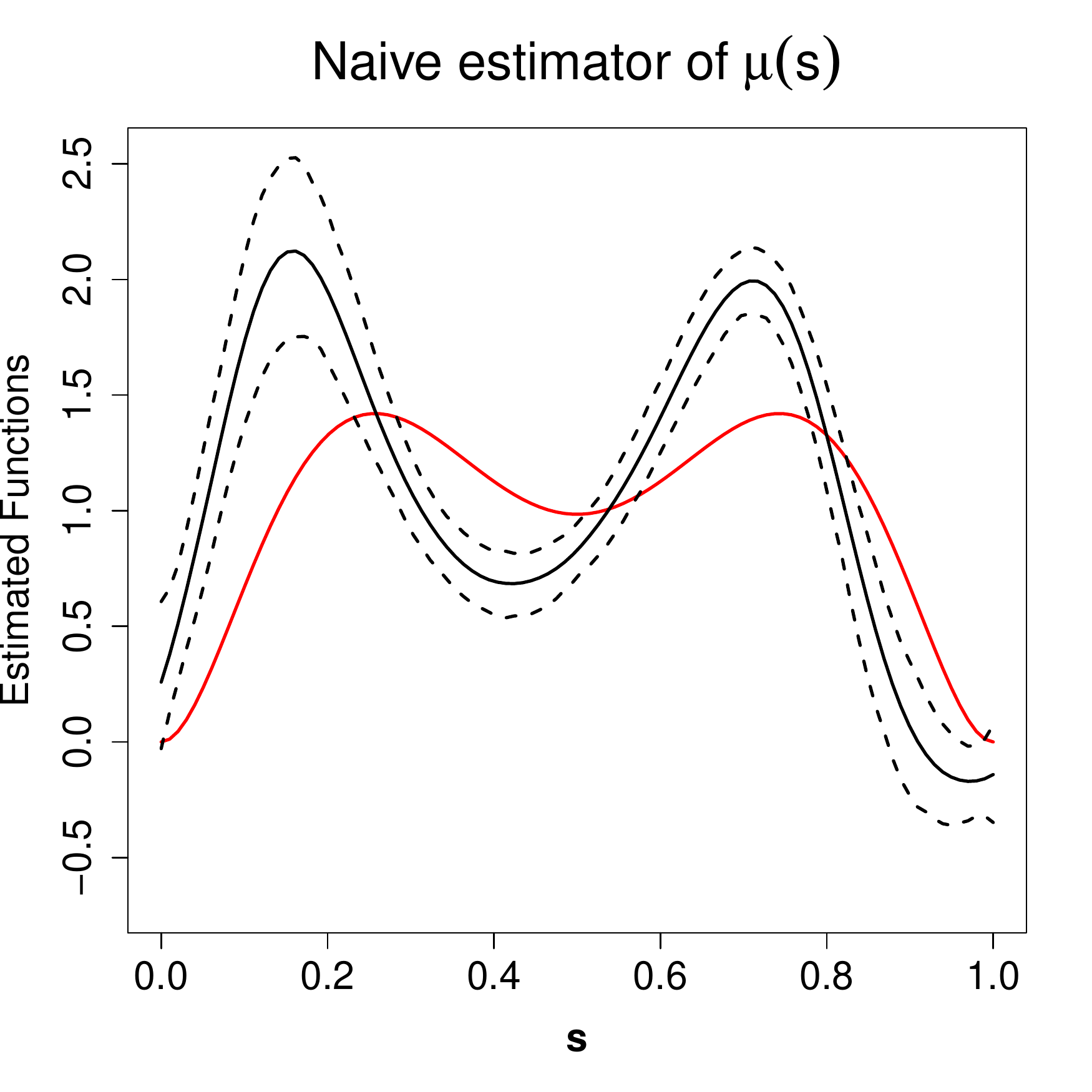} &
						\includegraphics[width=0.3\textwidth]{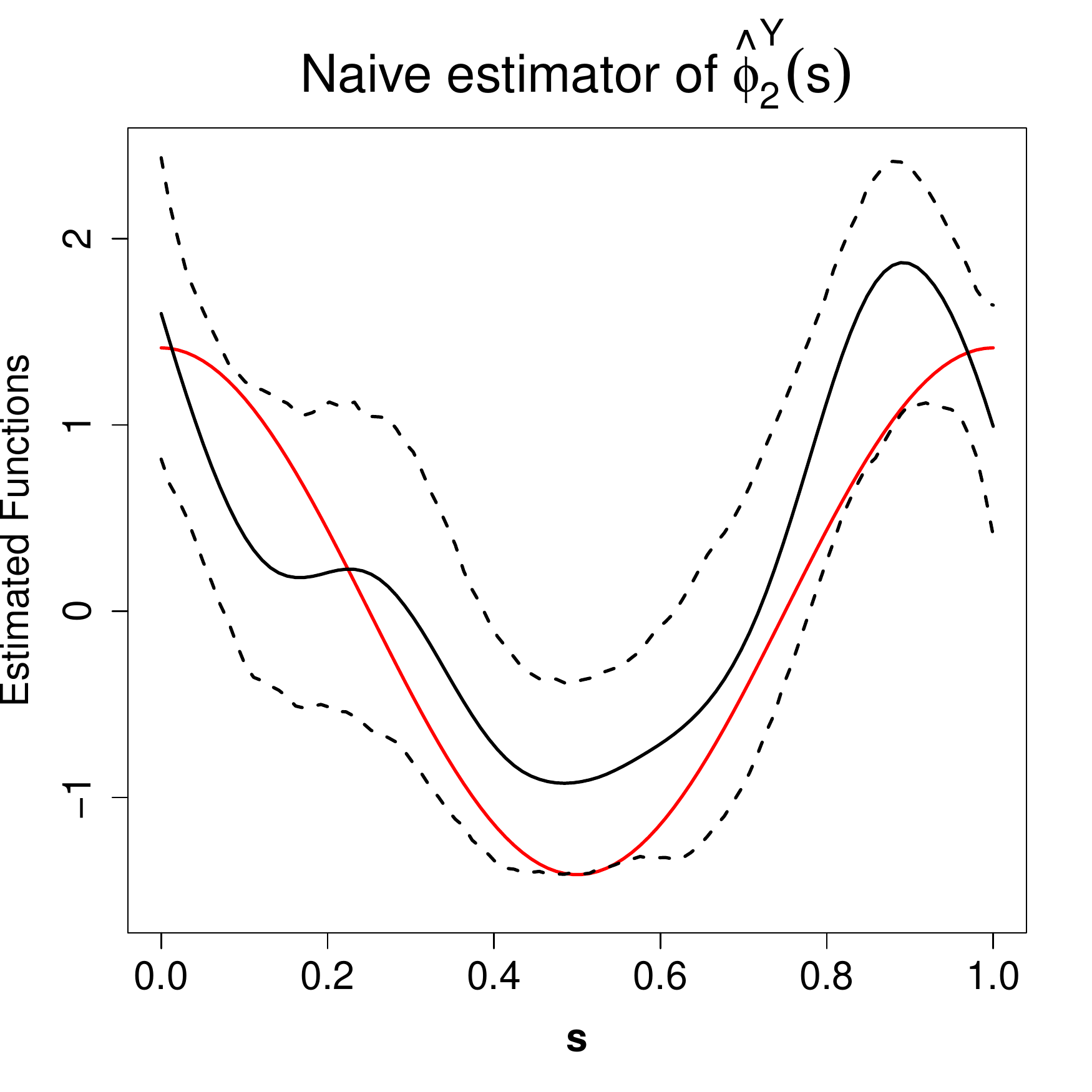}&
						\includegraphics[width=0.3\textwidth]{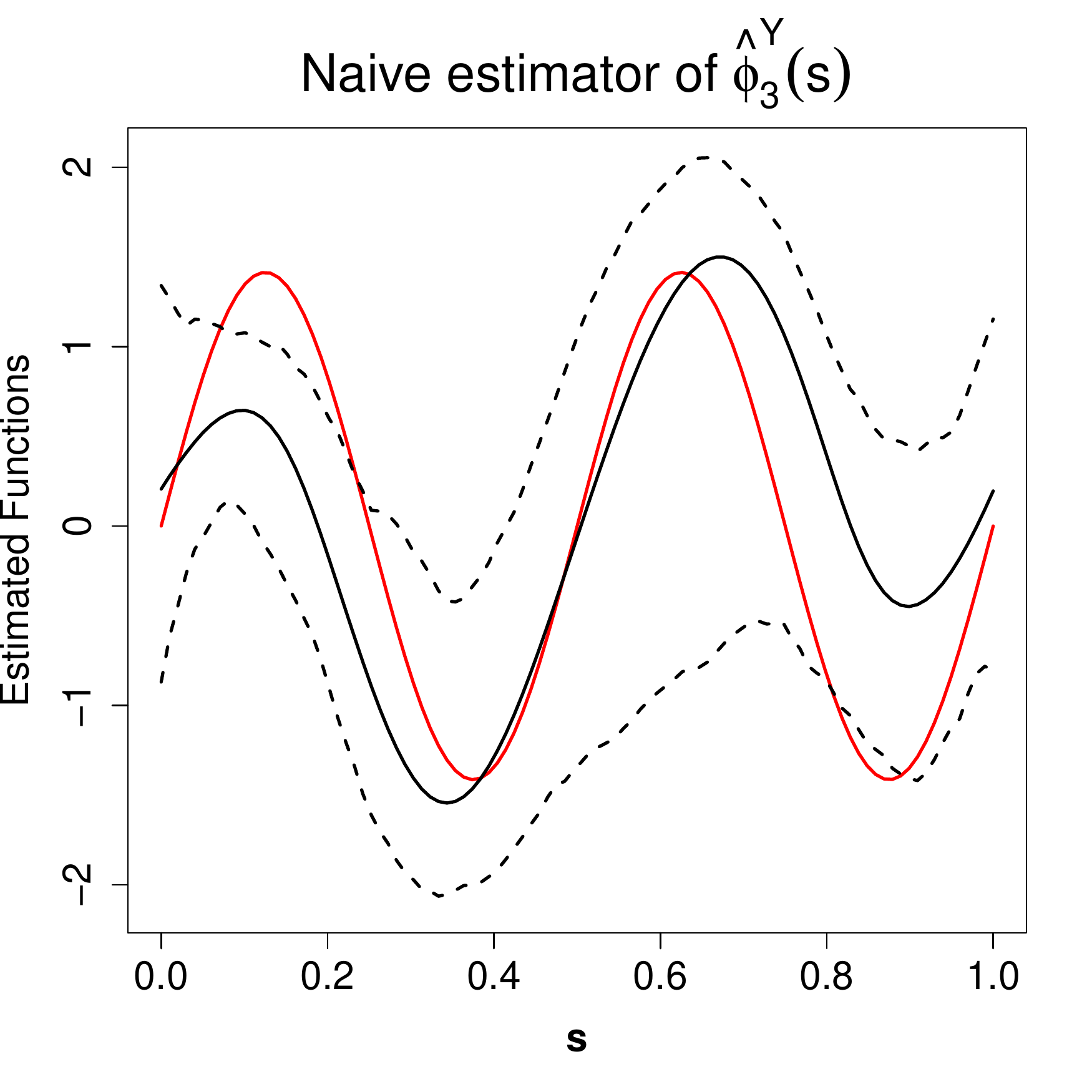}\\
					\end{tabular}
					\vskip -1em
					\caption{{Estimation biases and $95\%$ quantile bands for mean functions and last two eigenfunctions of $C_Y(\cdot,\cdot)$ when $n=600$ and $q=0.8$ under the (\texttt{Exp}, \texttt{Exp}) setting.}}
					\label{fig-2-S}
				\end{figure}
				
				\begin{figure}[H]
					\centering
					\begin{tabular}{lll}
						\includegraphics[width=0.3\textwidth]{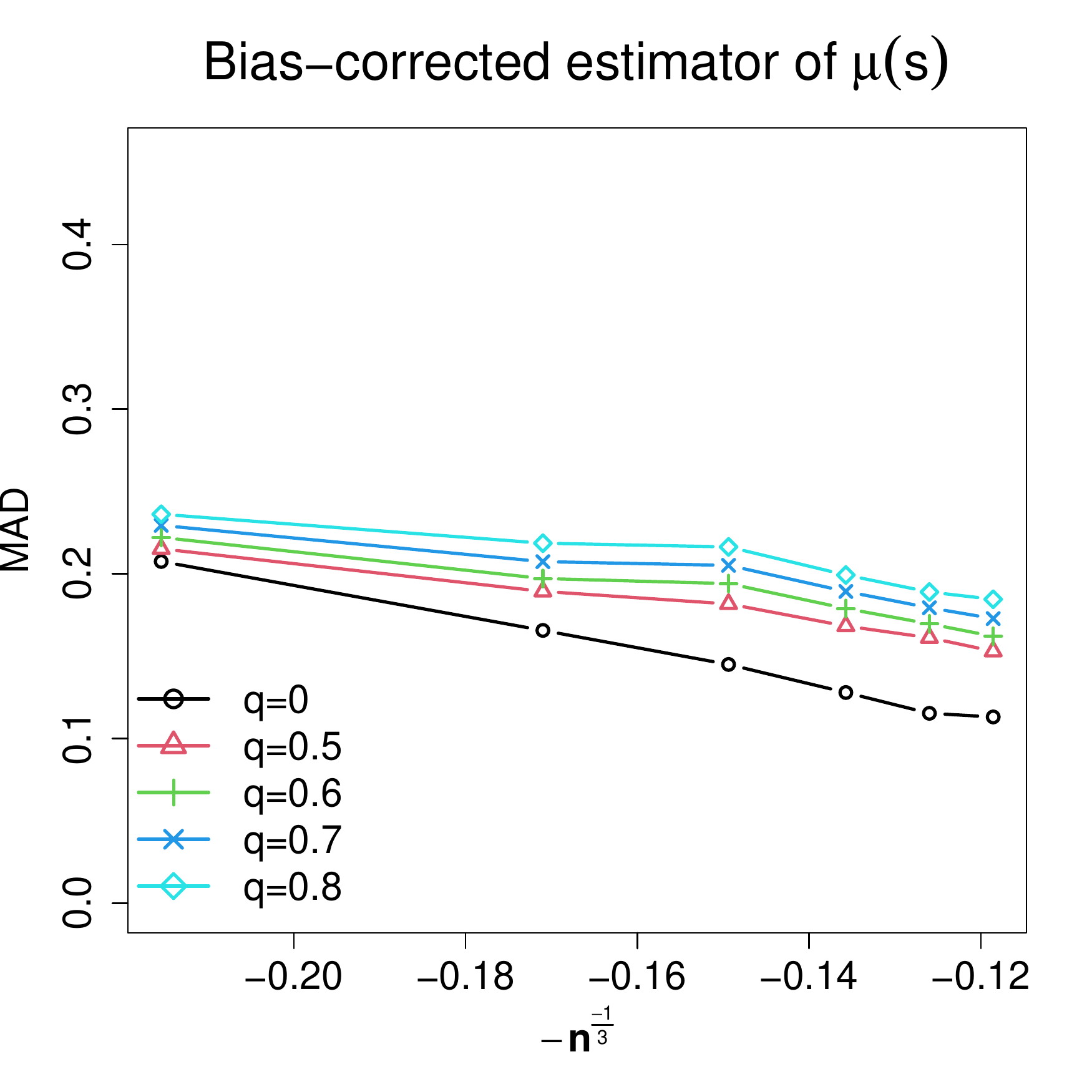} &
						\includegraphics[width=0.3\textwidth]{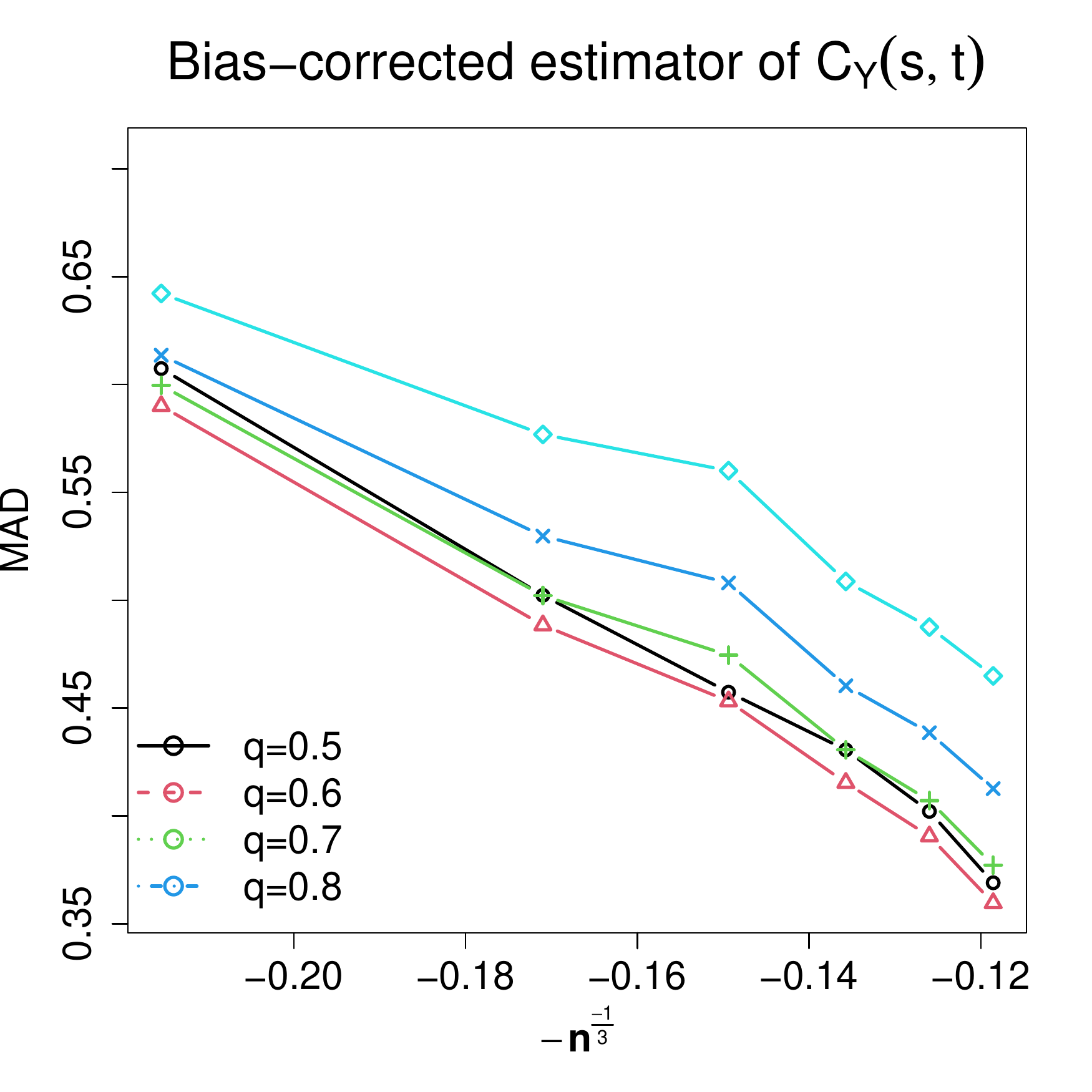}&
						\includegraphics[width=0.3\textwidth]{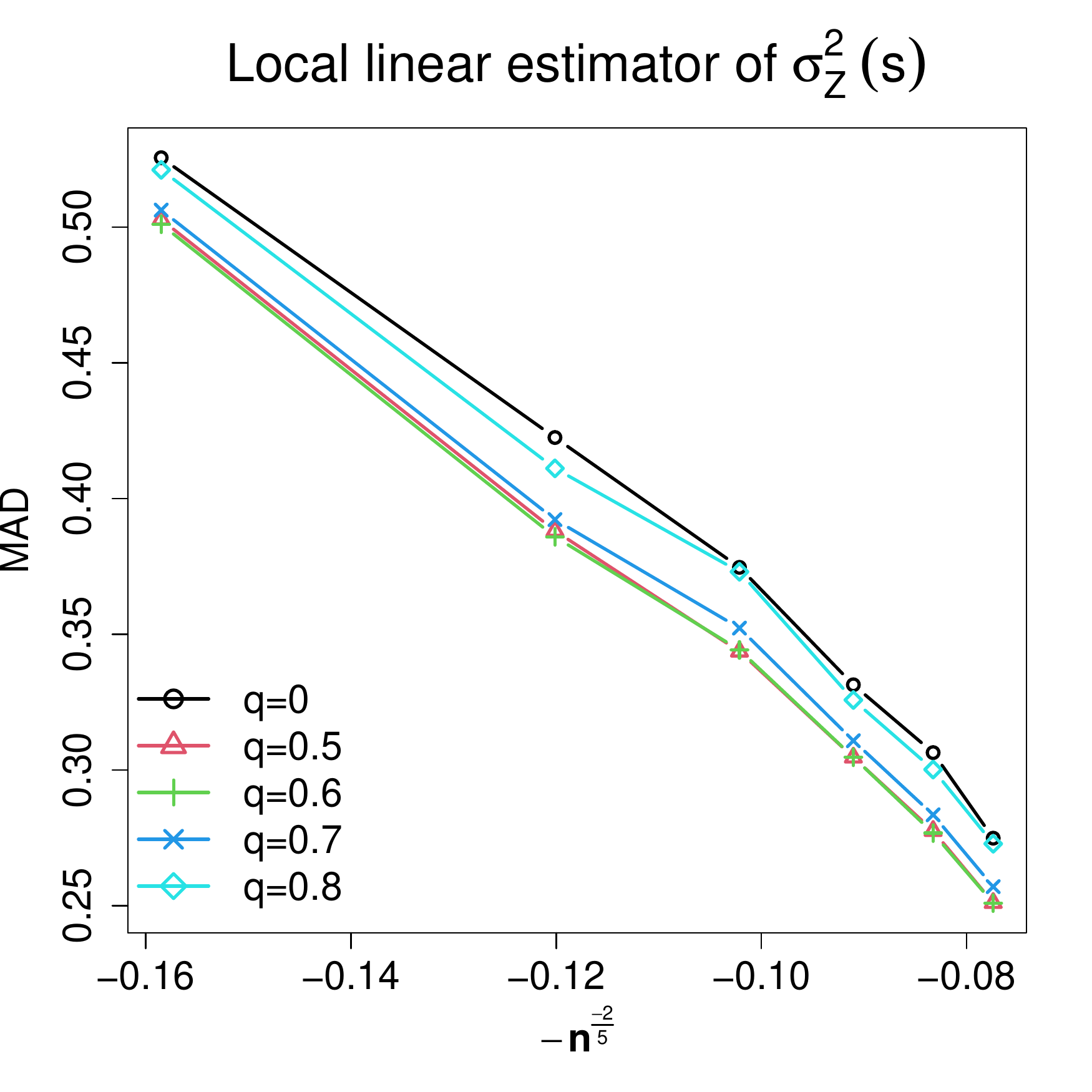}\\
						\includegraphics[width=0.3\textwidth]{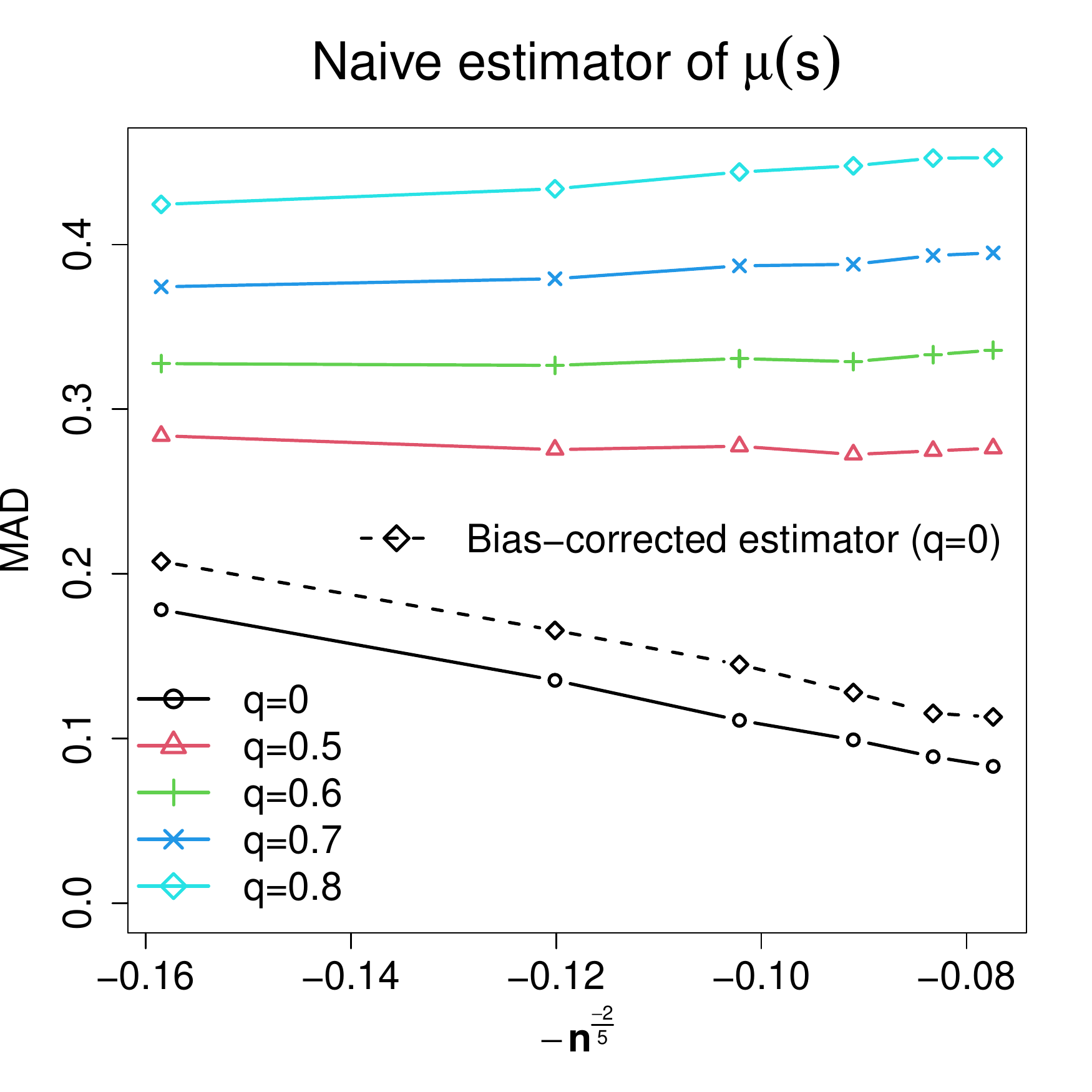} &
						\includegraphics[width=0.3\textwidth]{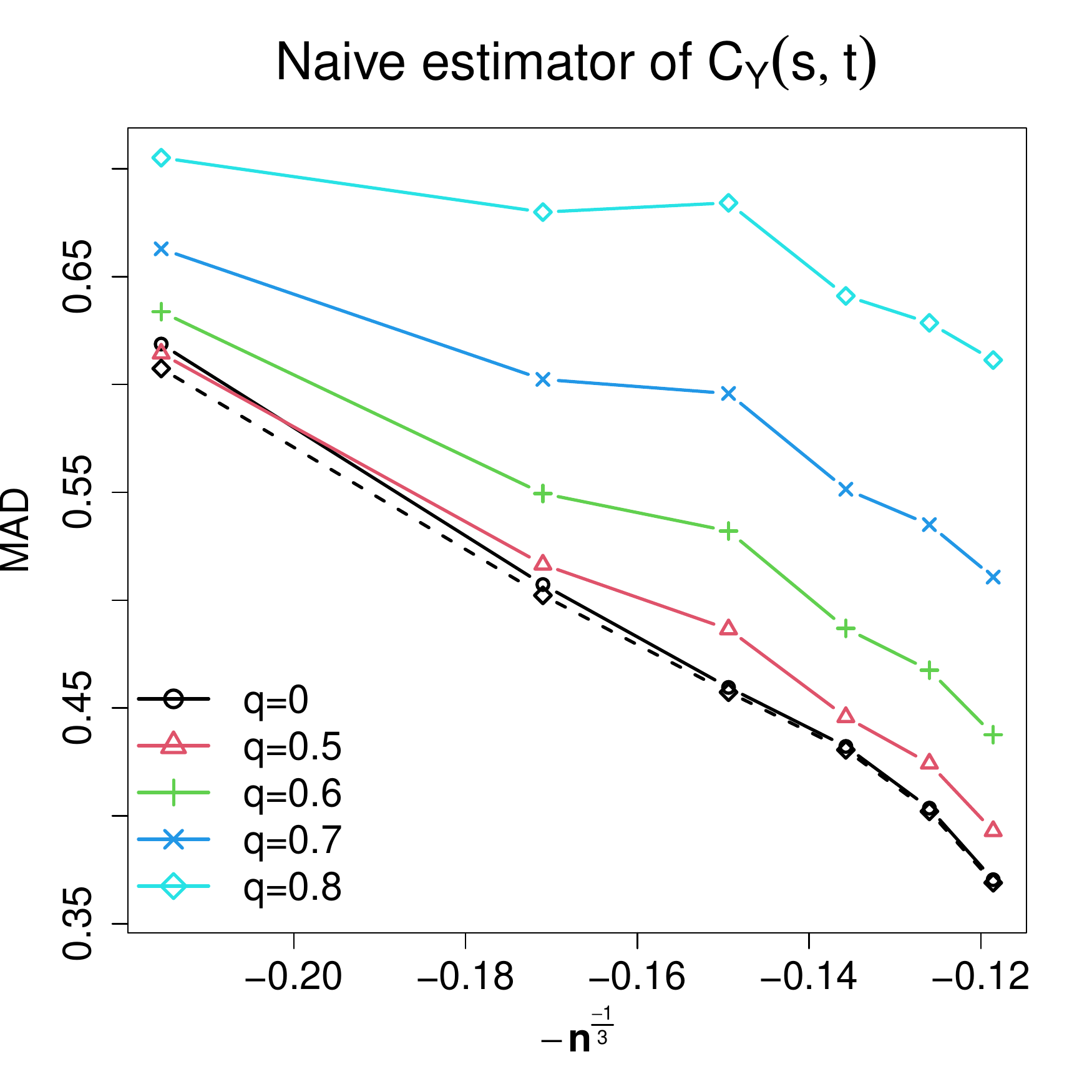}&
						\includegraphics[width=0.3\textwidth]{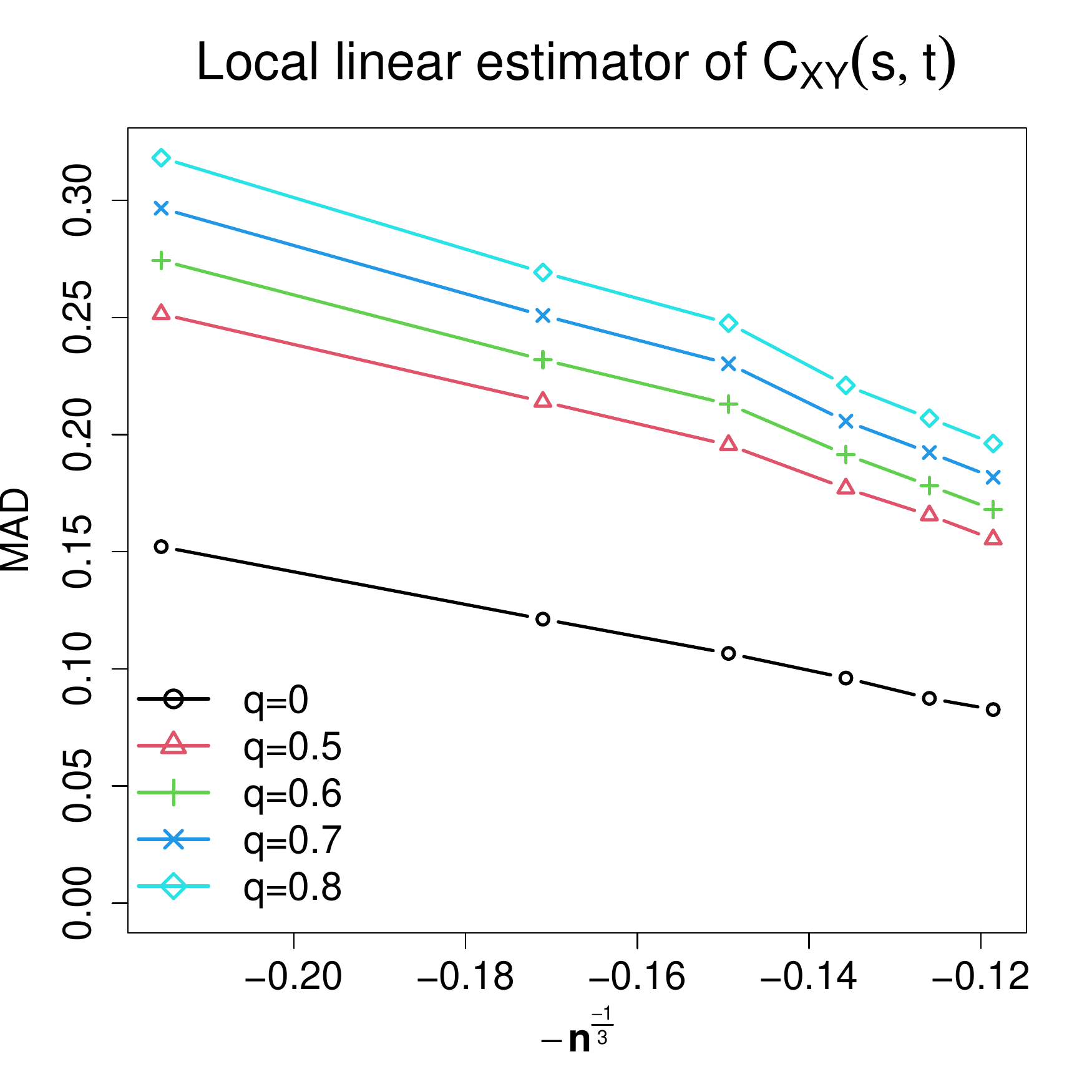}\\
					\end{tabular}
					\vskip -1em
					\caption{{Estimation accuracy of local linear estimators under the (\texttt{T4}, \texttt{T4}) setting.}}
					\label{fig-1-SS}
				\end{figure}
				
				\begin{figure}[H]
					\centering
					\begin{tabular}{lll}
						\includegraphics[width=0.3\textwidth]{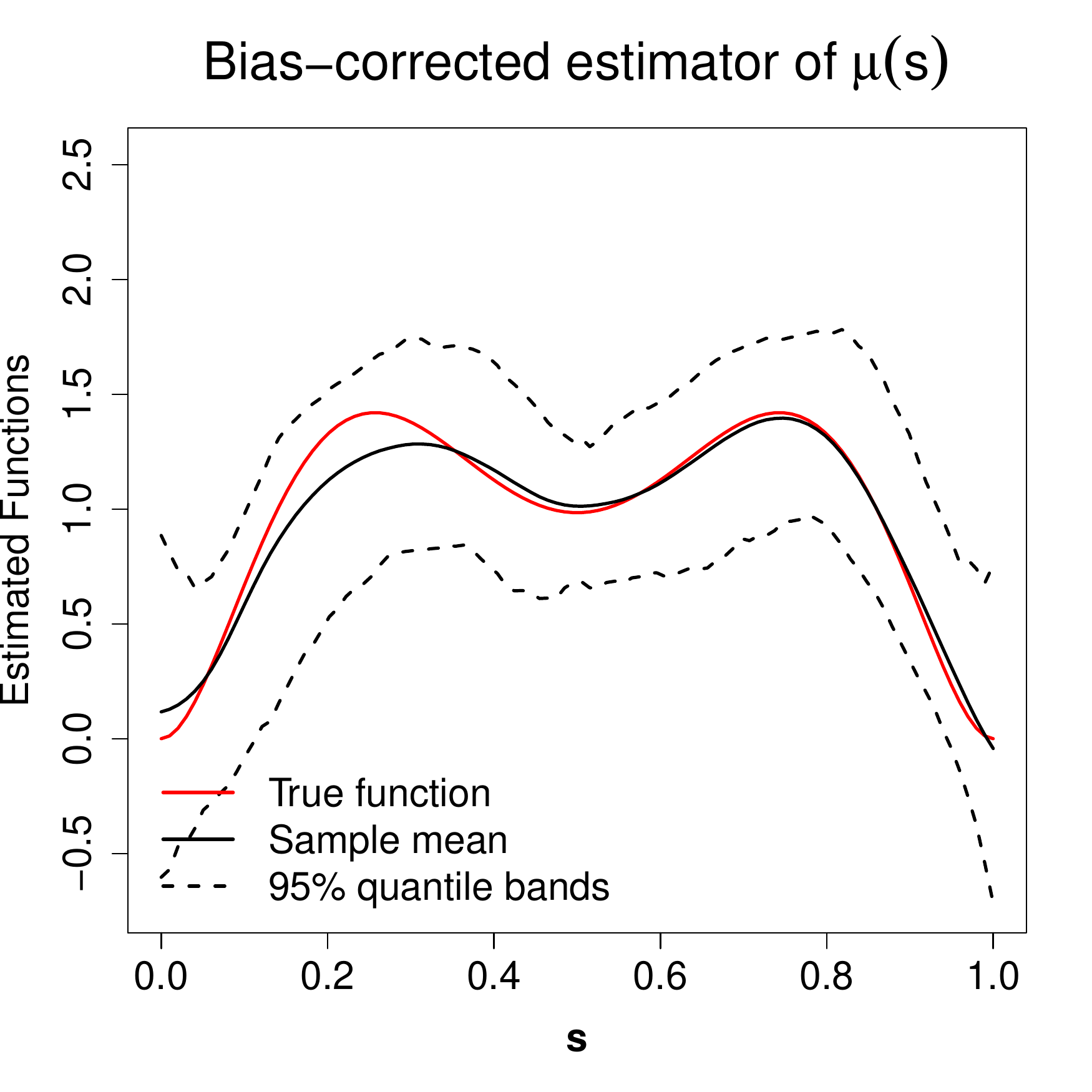} &
						\includegraphics[width=0.3\textwidth]{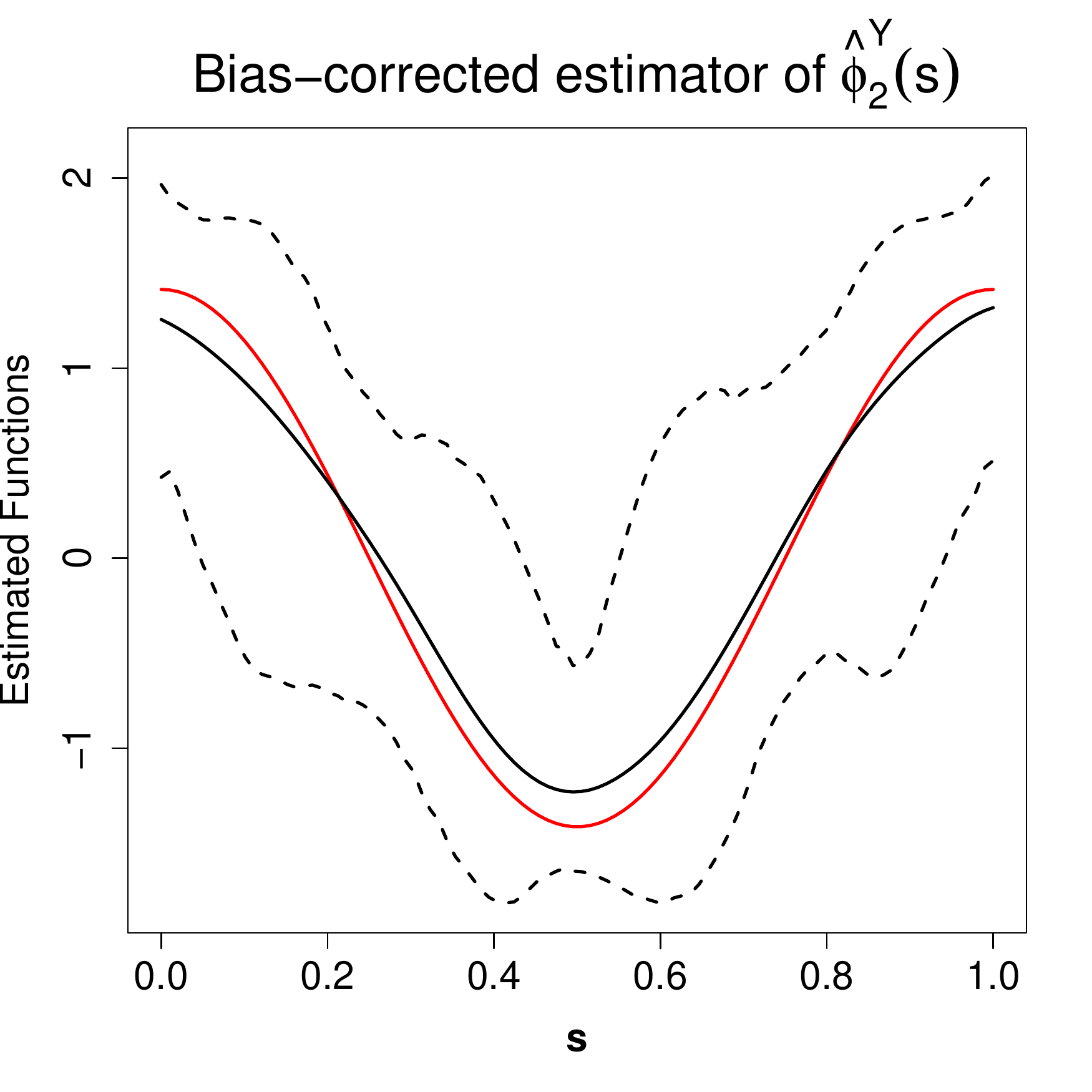}&
						\includegraphics[width=0.3\textwidth]{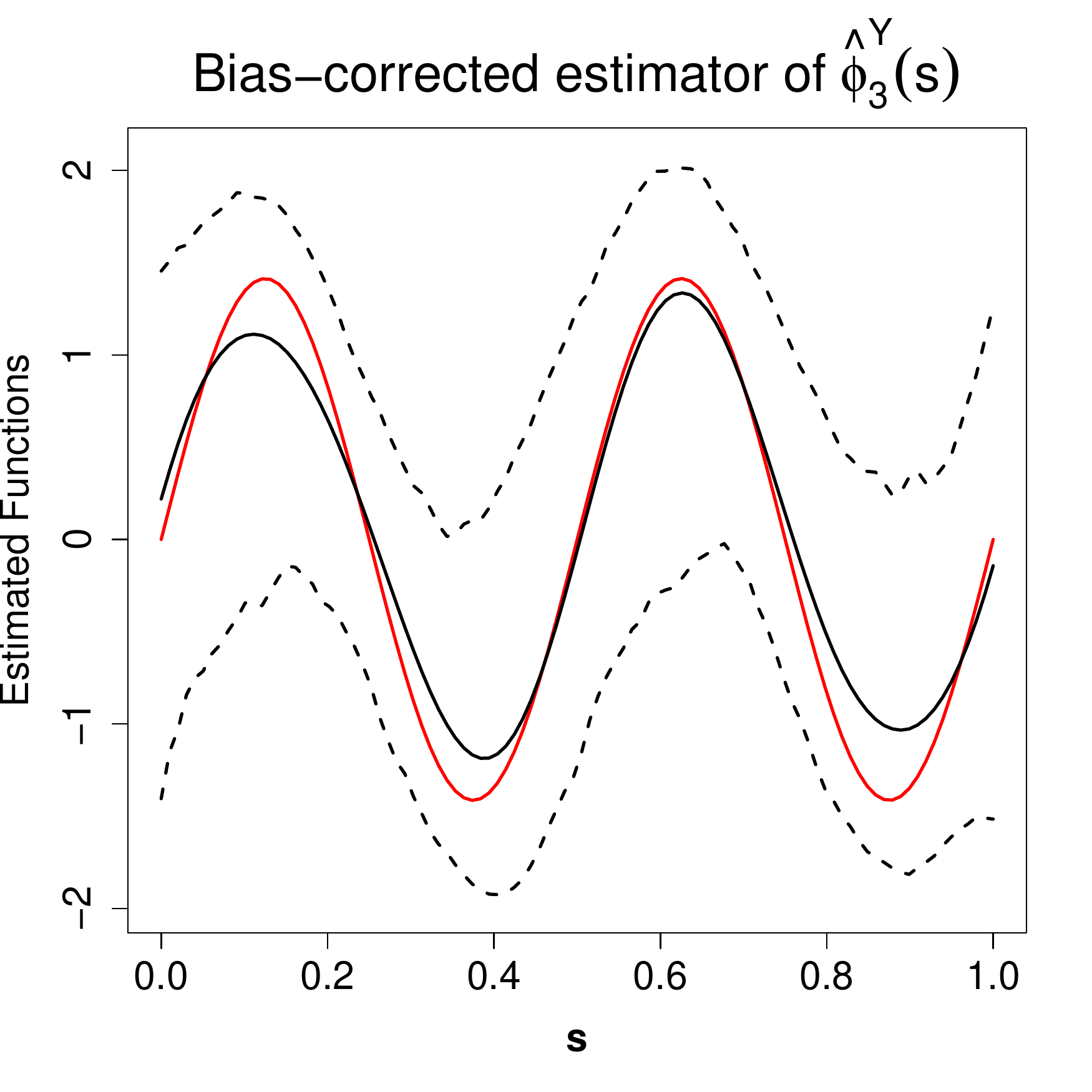}\\
						\includegraphics[width=0.3\textwidth]{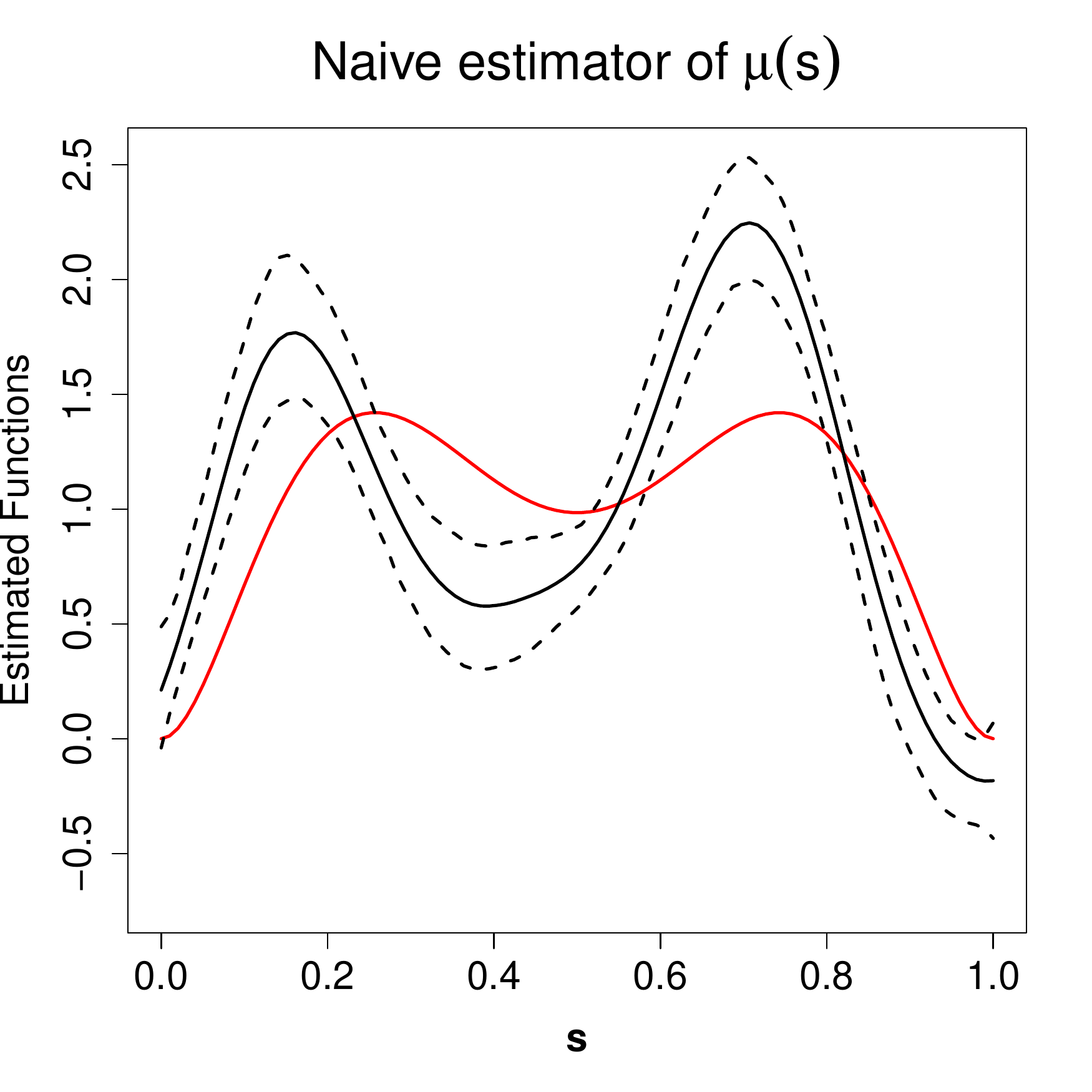} &
						\includegraphics[width=0.3\textwidth]{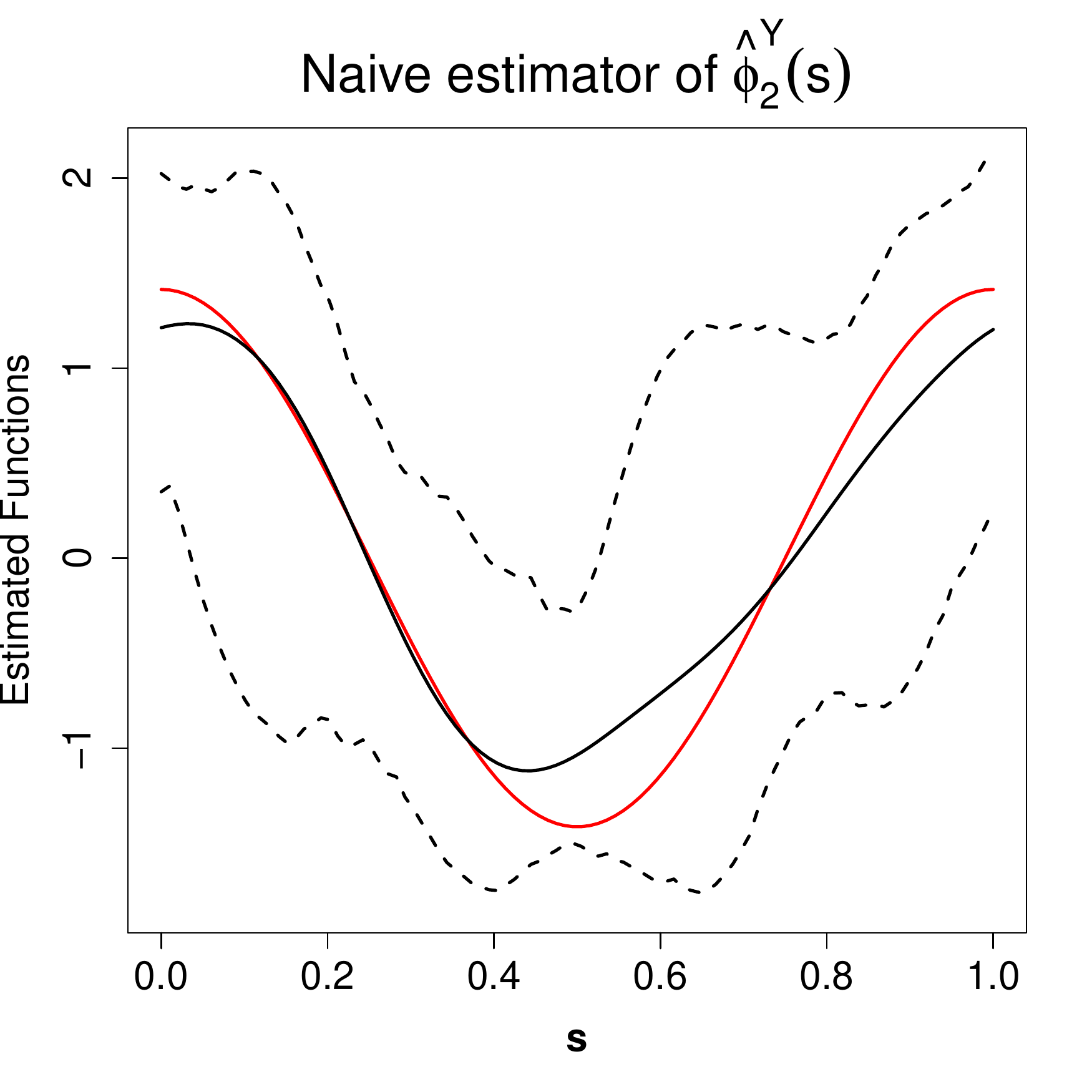}&
						\includegraphics[width=0.3\textwidth]{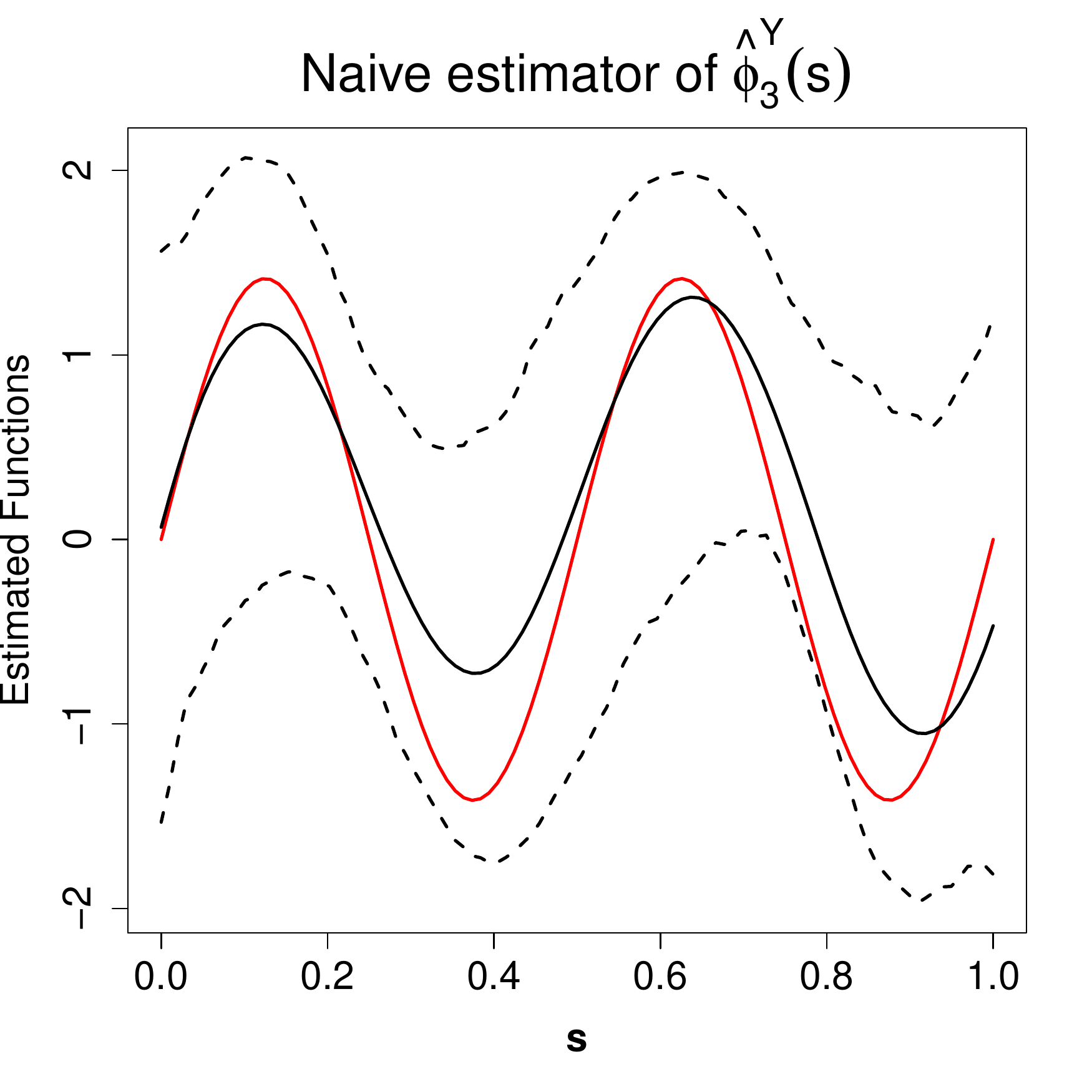}\\
					\end{tabular}
					\vskip -1em
					\caption{{Estimation biases and $95\%$ quantile bands for mean functions and last two eigenfunctions of $C_Y(\cdot,\cdot)$ when $n=600$ and $q=0.8$ under the (\texttt{T4}, \texttt{T4}) setting.}}
					\label{fig-2-SS}
				\end{figure}

				\subsection{Validity and Local Power of the Proposed Testing Procedures}
				\label{sec:s2}
				In this subsection, we provide additional simulation results of the proposed testing procedures.
				We first show the empirical rejection rates of the proposed tests under mark-point independence, summarized in Figure~\ref{fig-S3-1}. We can see that while the asymptotic test still appears to be valid 
				{so long as the mark process is Gaussian,} 
				its rejection rates significantly exceed the nominal level when the mark process is non-Gaussian. 
				The permutation test seems to achieve the nominal sizes provided $n$ is sufficiently large in all settings. 
				\begin{figure}[H]
					\centering
					\begin{tabular}{lll}
						\includegraphics[width=0.33\textwidth]{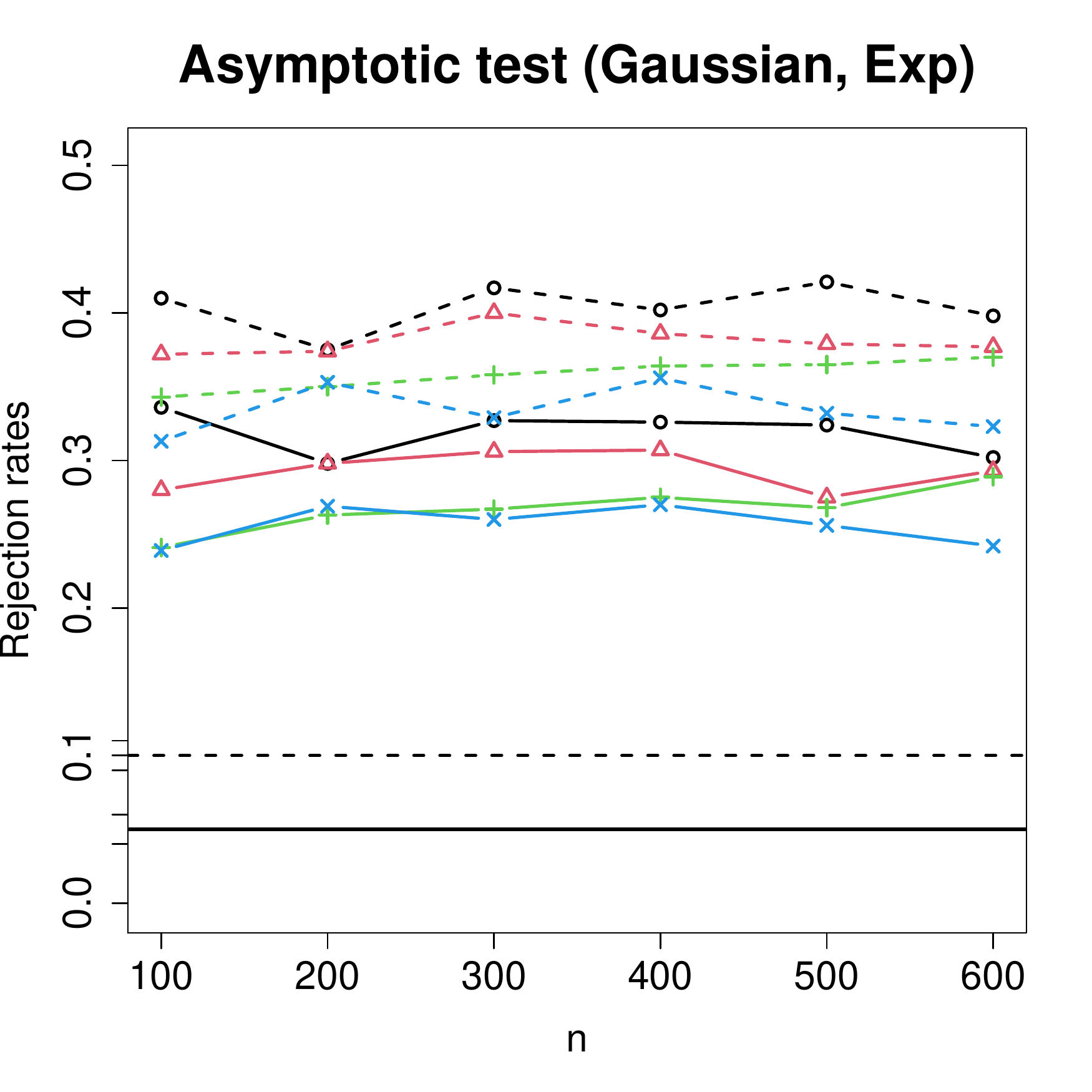}&
						\includegraphics[width=0.33\textwidth]{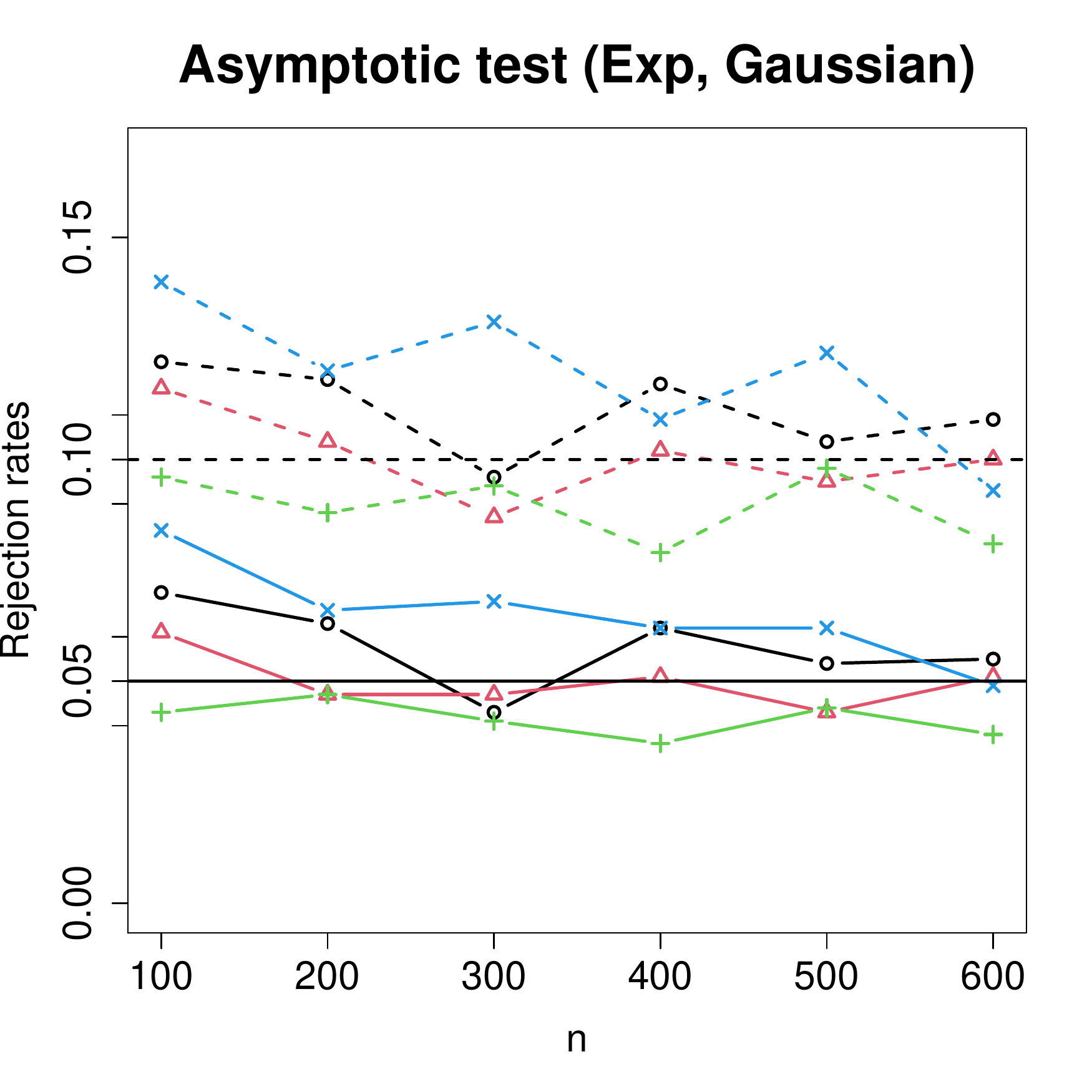}&
						\includegraphics[width=0.33\textwidth]{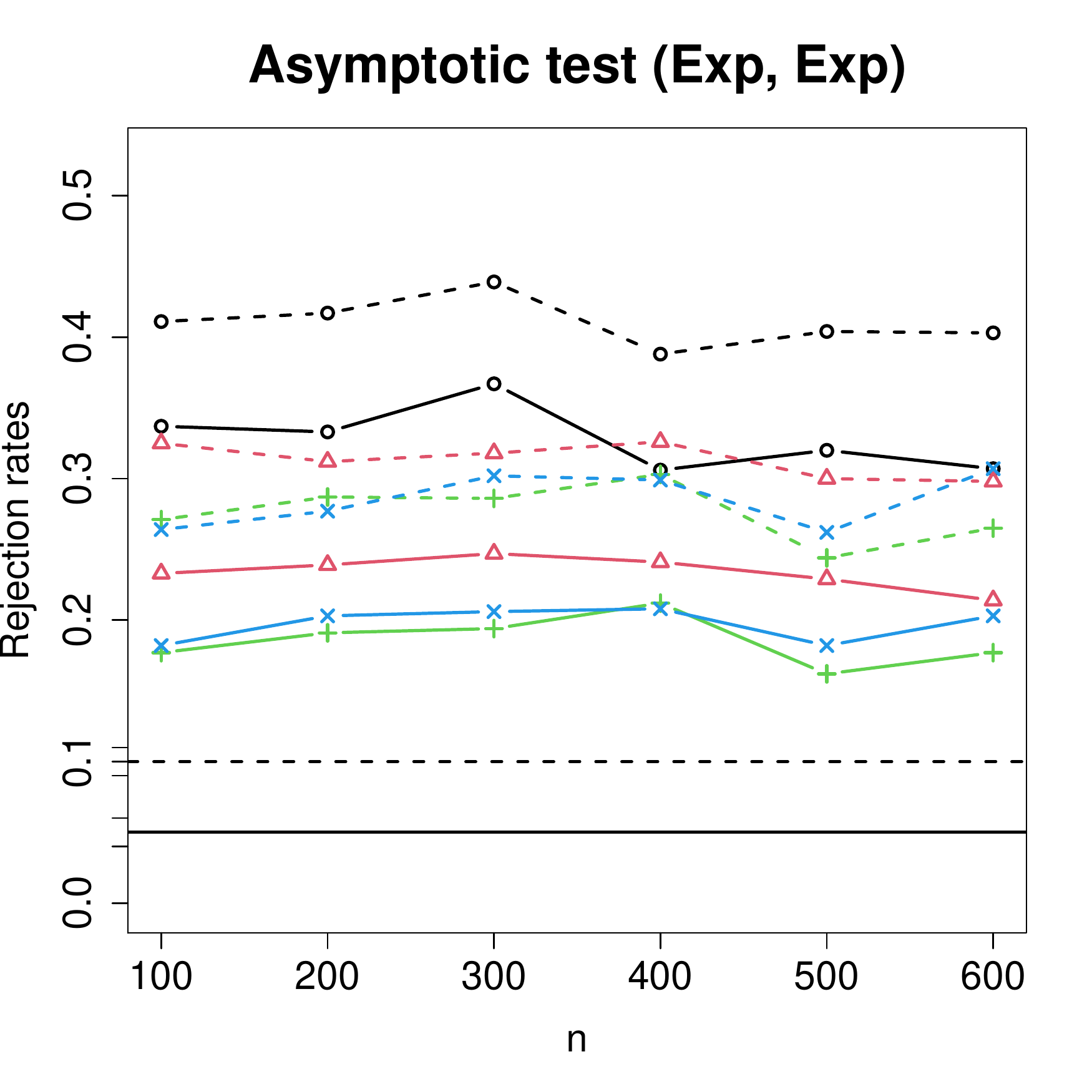}\\
						\includegraphics[width=0.33\textwidth]{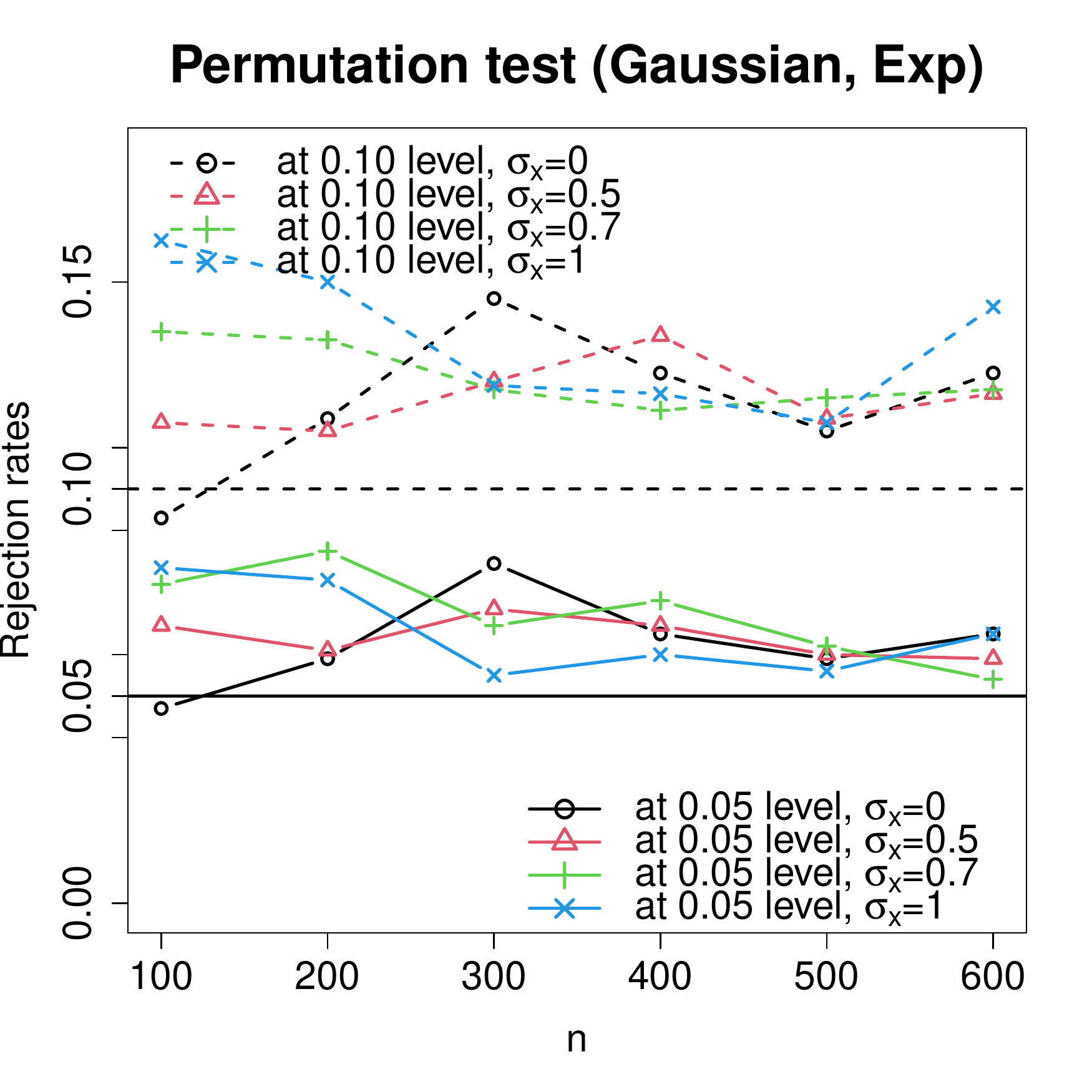}&
						\includegraphics[width=0.33\textwidth]{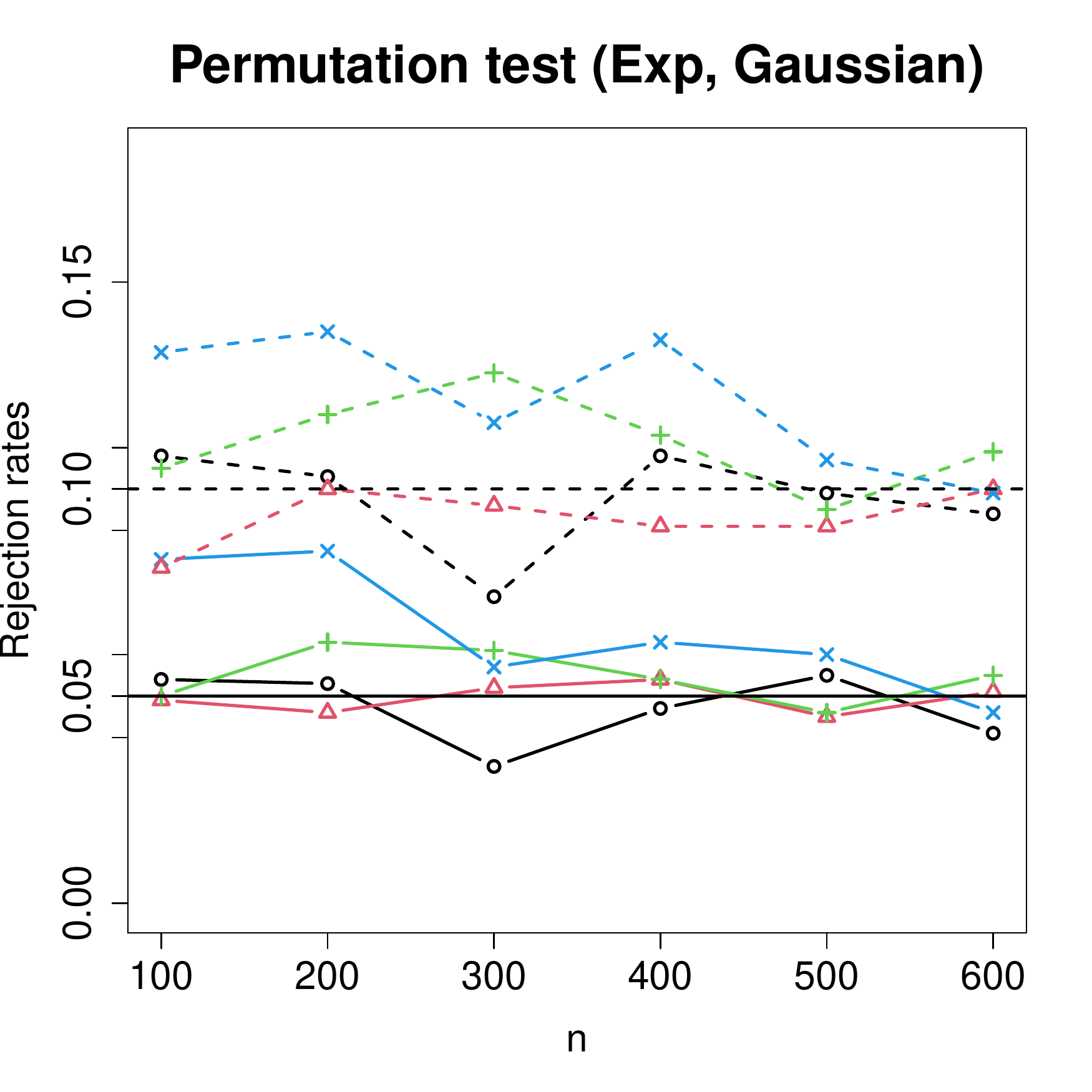}&
						\includegraphics[width=0.33\textwidth]{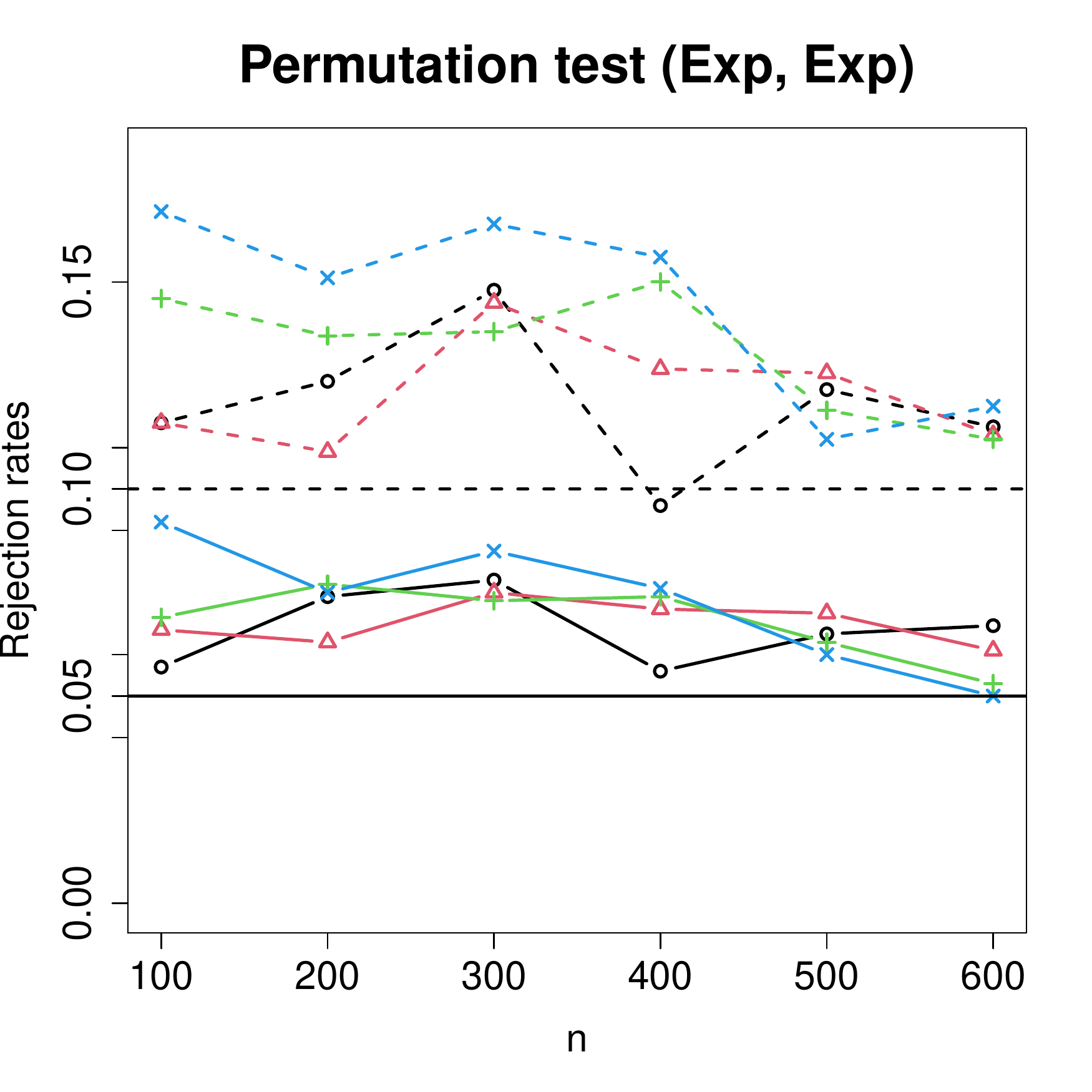}\\
					\end{tabular}
					\caption{{Rejection rates of two tests under mark-point independence in various distribution settings. Top row: the asymptotic test; Bottom row: the permutation test.}}
					\label{fig-S3-1}
				\end{figure}

				Next, we investigate the local power of the proposed tests under more non-Gaussian settings. The summary statistics are illustrated in Figure~\ref{fig-S4}. In all settings, the local powers of the permutation test are roughly constant as $n$ increases, provided that $n$ is sufficiently large. This supports the validity of the permutation test in more general settings. 
				\begin{figure}[ht!]
					\centering
					\begin{tabular}{lll}
						\includegraphics[width=0.33\textwidth]{fig4-a-1.pdf} &
						\includegraphics[width=0.33\textwidth]{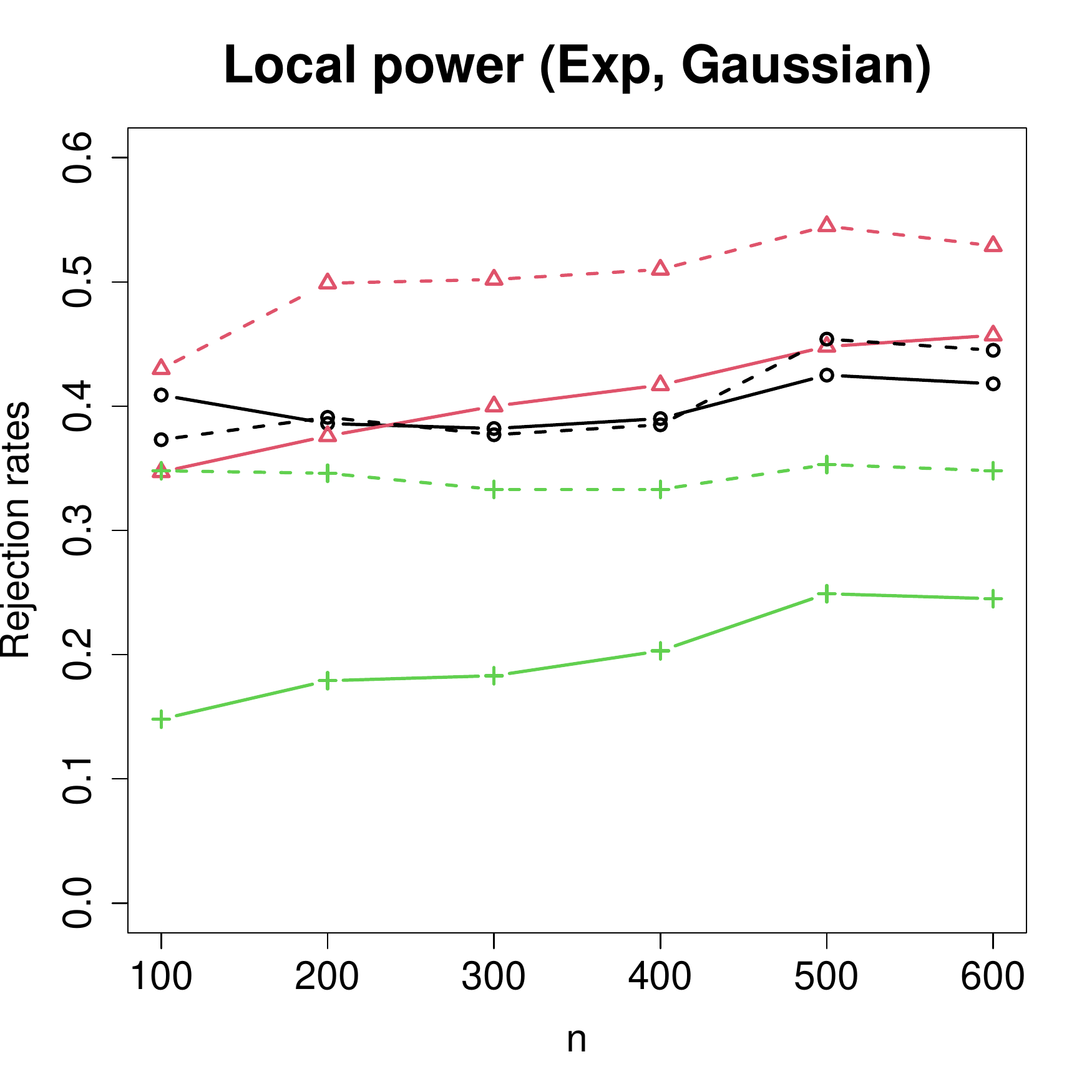} &
						\includegraphics[width=0.33\textwidth]{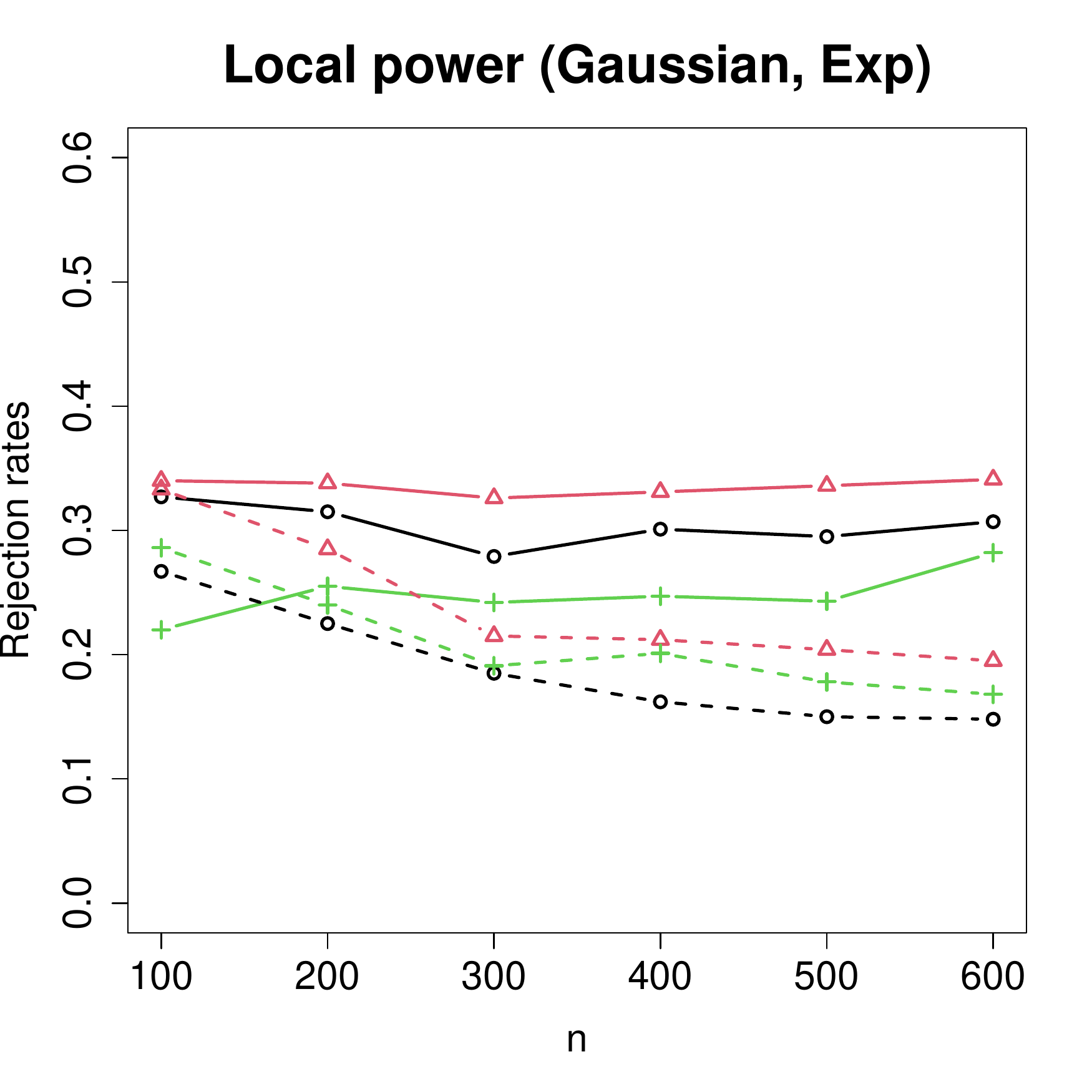} \\			
						\includegraphics[width=0.33\textwidth]{fig4-b-1.pdf} &
						\includegraphics[width=0.33\textwidth]{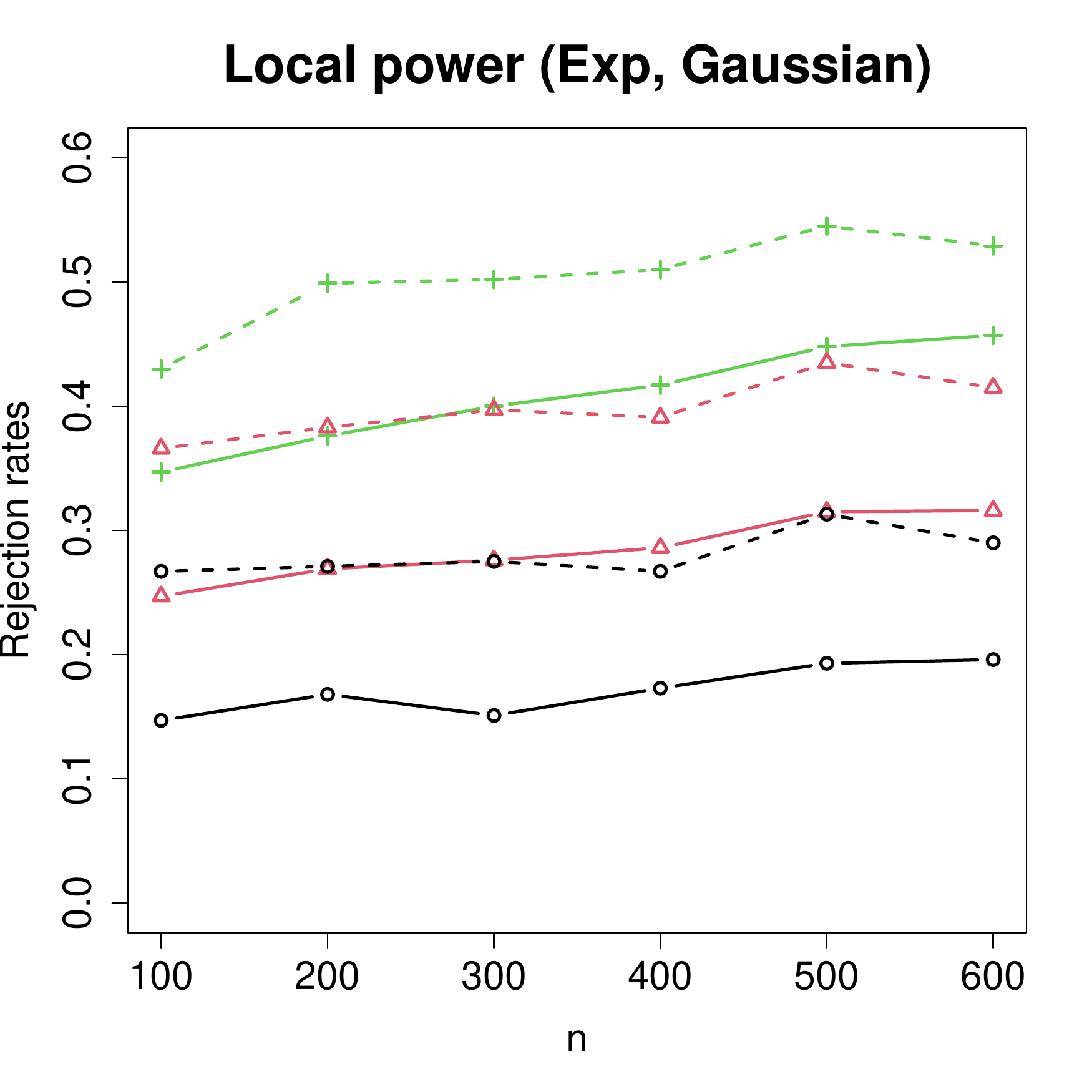} &
						\includegraphics[width=0.33\textwidth]{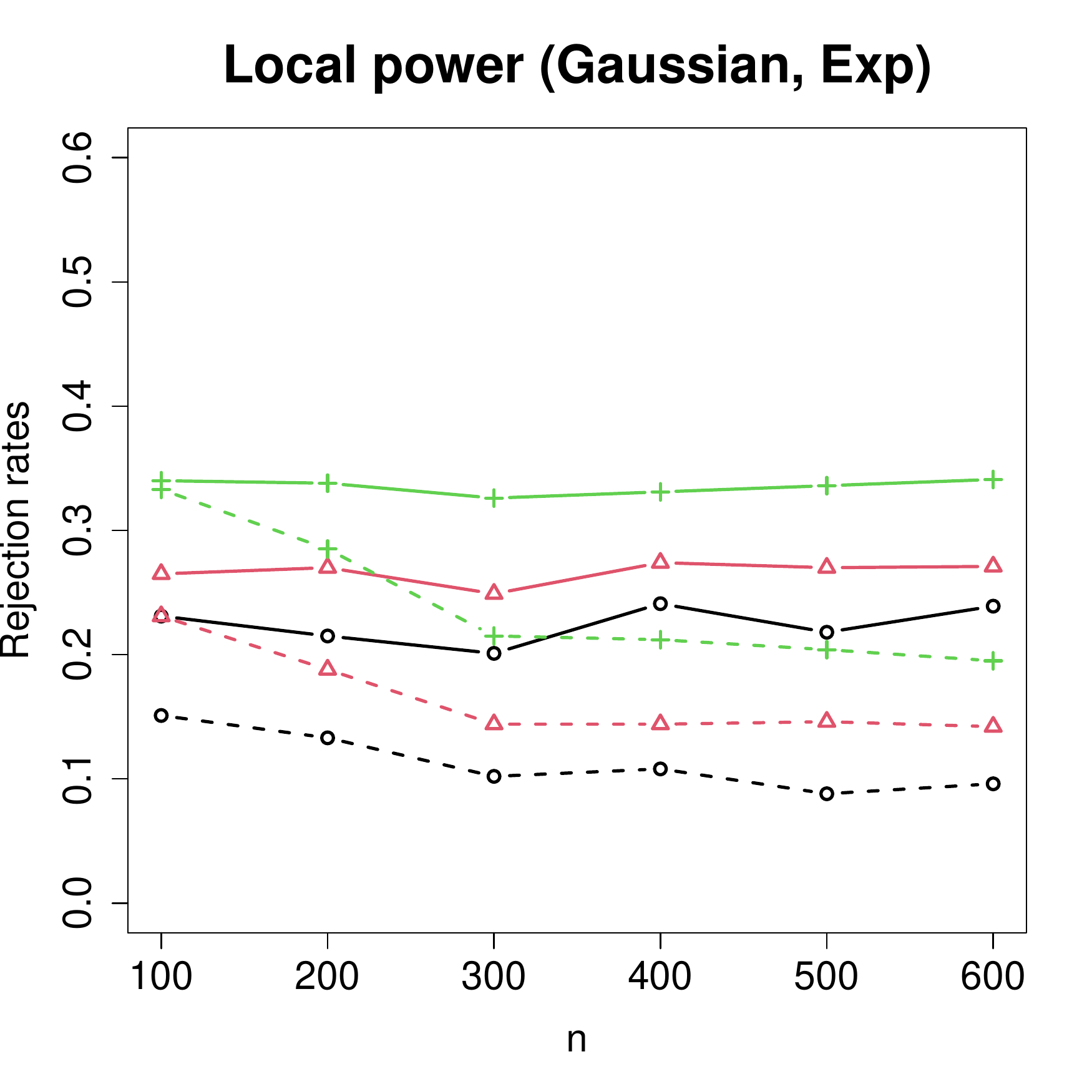}\\
					\end{tabular}
					\caption{{Rejection rates of two tests at the 0.05 level under the local alternatives with varying $\sigma_x$ (top row) and $\gamma$ (bottom row).}}
					\label{fig-S4}
				\end{figure}

				\subsection{Additional Results for Real Data Analysis }
				\label{sec:sk:supp}
				In this subsection, we first give trajectories of three sample subjects from the EMA smoking data in Figures~\ref{fig-skdata1}. 
				Next, we give all diagnostic plots for the EMA smoking data in Figures~\ref{fig-sk-S}-\ref{fig-sk-S1}. 
				Left panels of Figures~\ref{fig-sk-S}-\ref{fig-sk-S1} show that summary statistics (see Section~\ref{sec:diag}) based on the observed data fell within the $95\%$ percentile envelope based on $1,000$ simulated point processes, suggesting that the LGCP assumption for the point process is reasonable. The QQ plots of the first two (or first) FPC scores of the mark process, which account for more than $95\%$ of the total variations, are reasonably close to straight lines, indicating the mark process can be considered as a Gaussian process. We remark that, other than the Restless-Female case, the first functional principal component accounts for more than $95\%$ of the total variabilities in the respective mark process.
				\begin{figure}[ht!]
					\centering
					\begin{tabular}{ll}
						\includegraphics[width=0.415\textwidth]{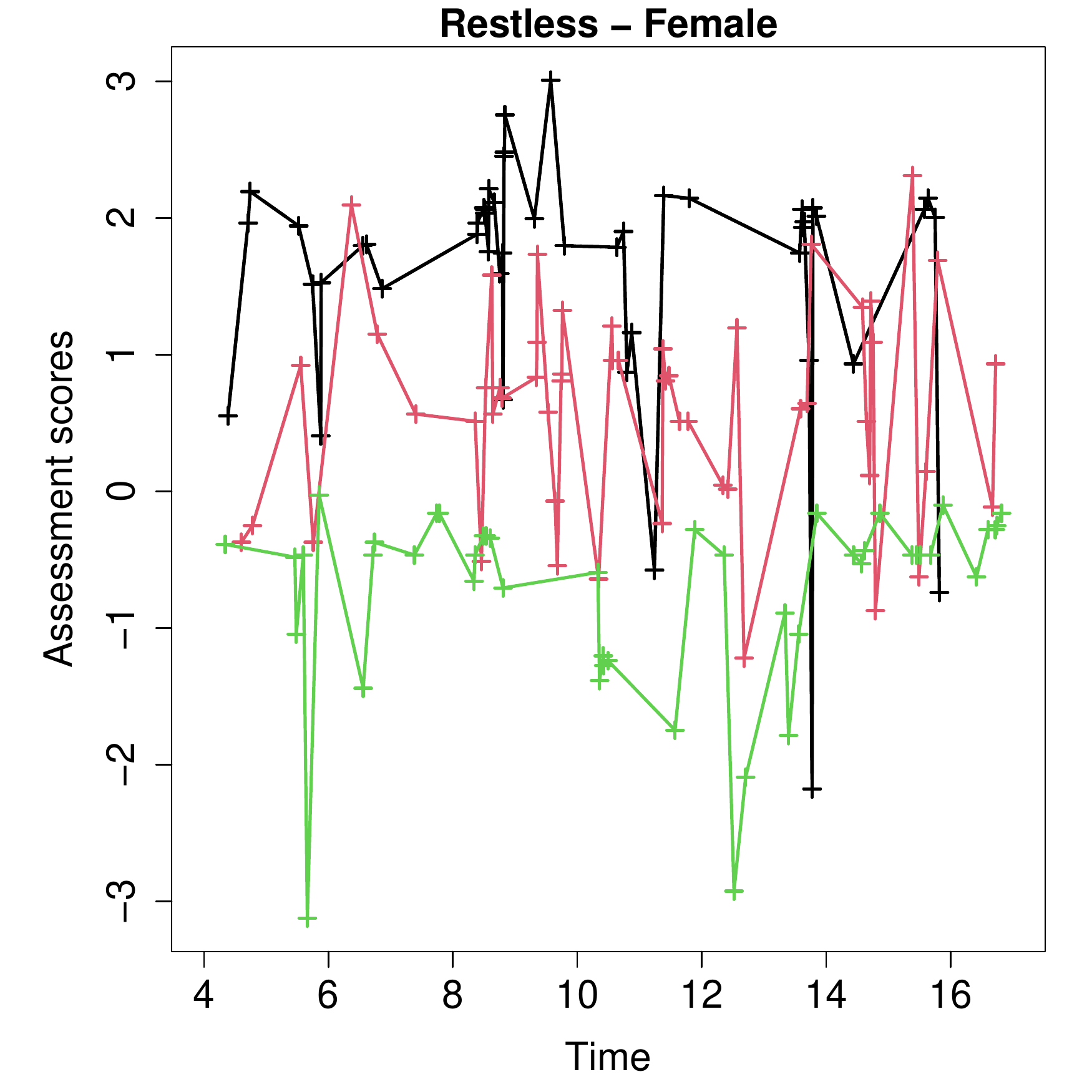} &
						\includegraphics[width=0.415\textwidth]{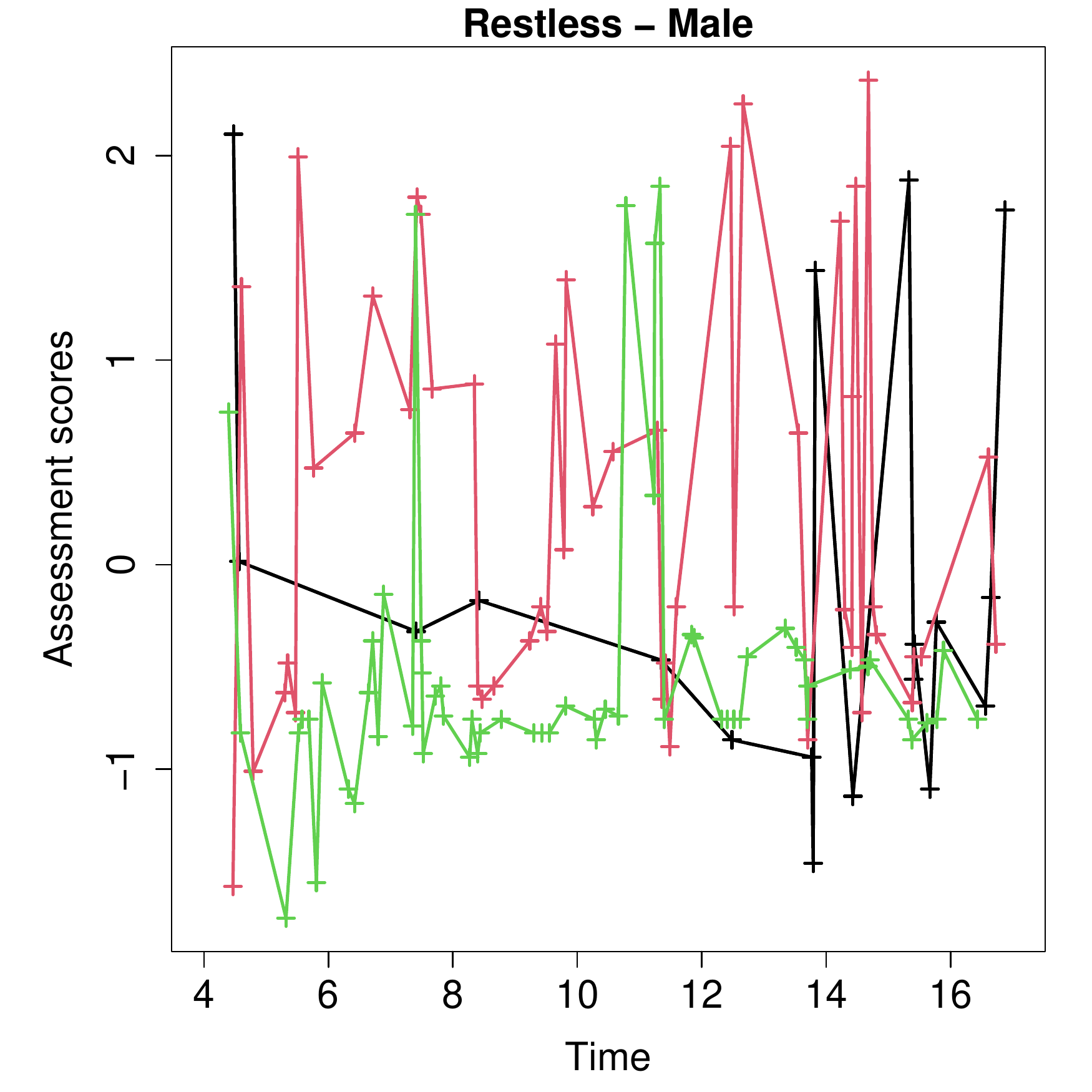} \\
						\includegraphics[width=0.415\textwidth]{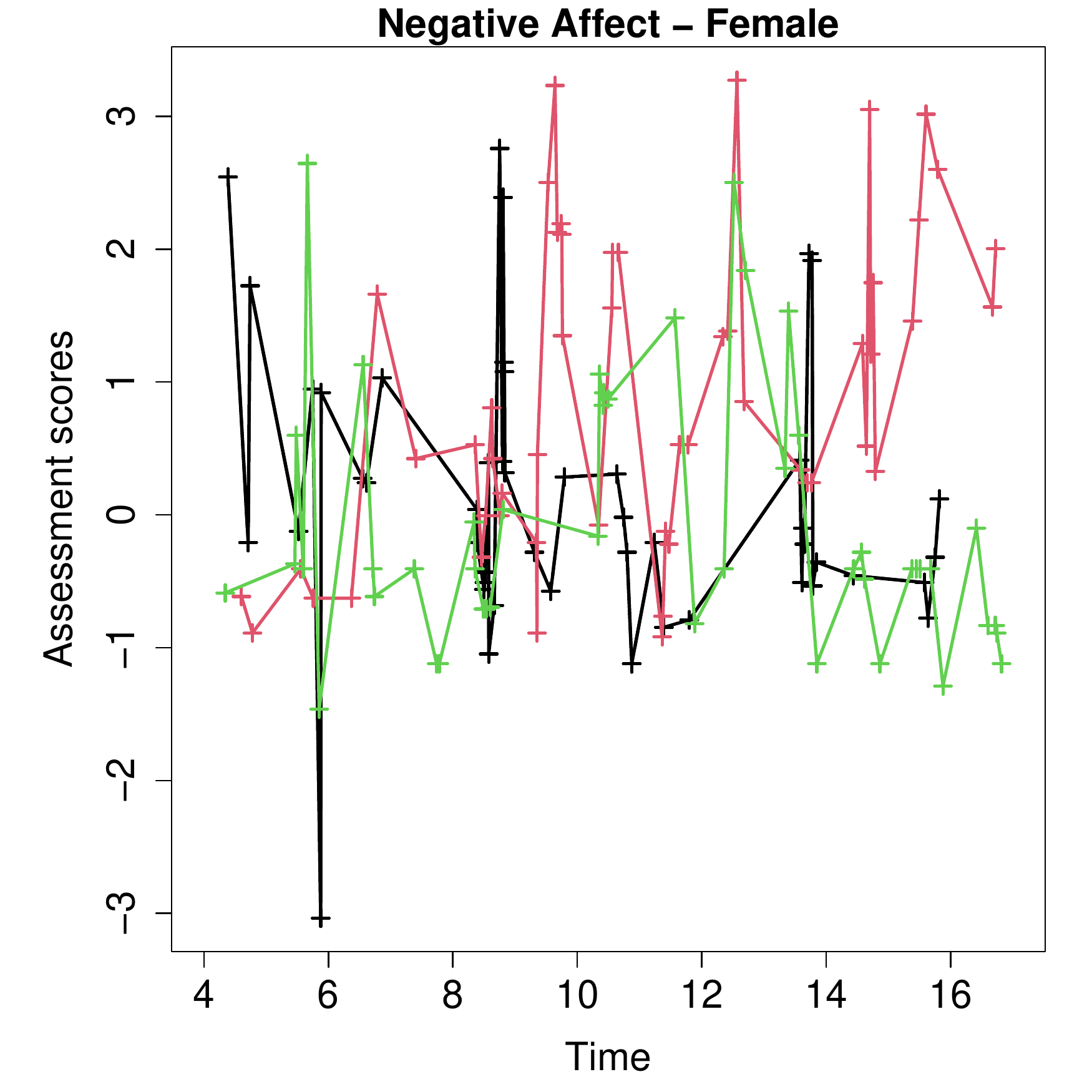} &
						\includegraphics[width=0.415\textwidth]{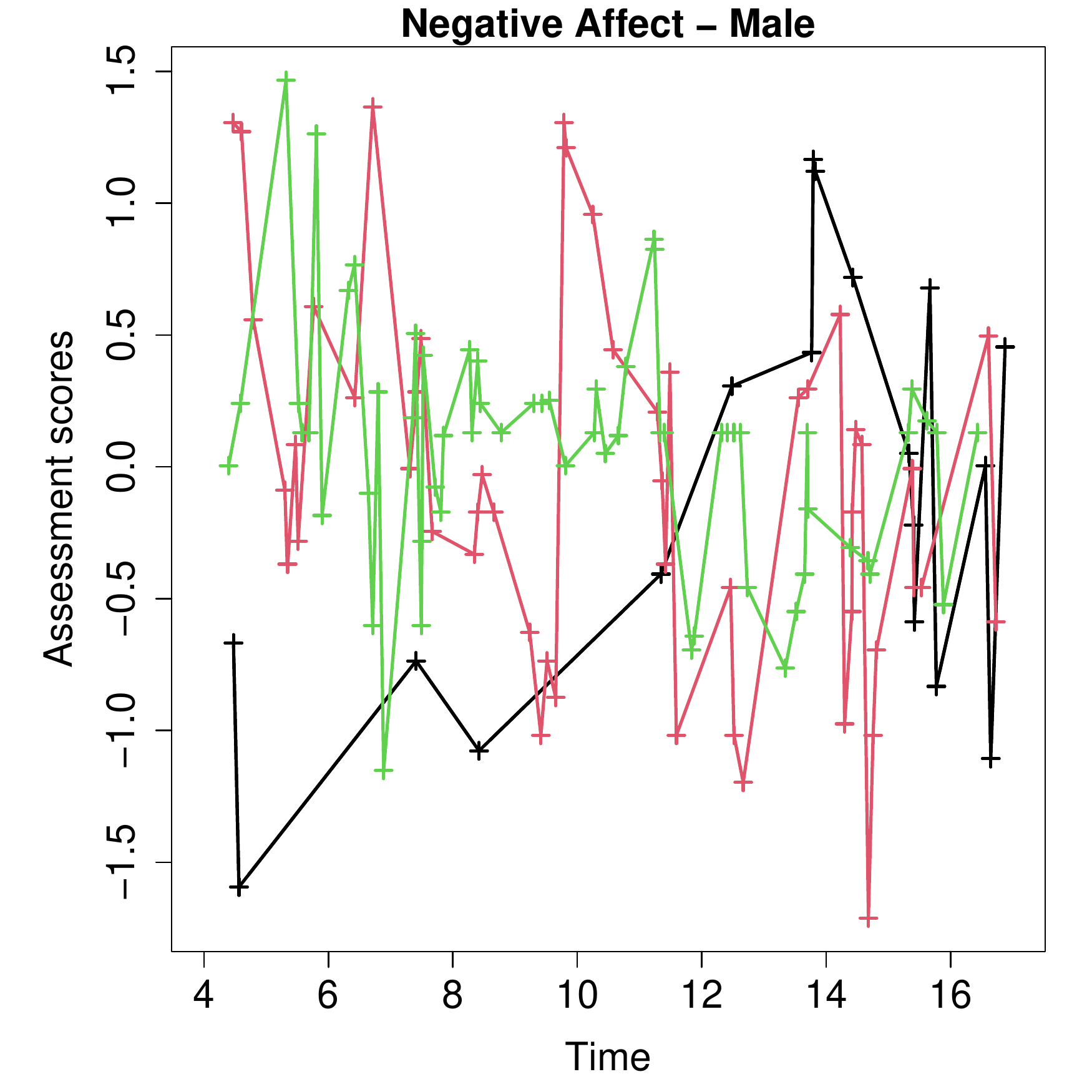} \\
					\end{tabular}
					\caption{{Trajectories of three sample subjects from the EMA smoking data.}}
					\label{fig-skdata1}
				\end{figure}
				
				\begin{figure}[ht!]
					\centering
					\begin{tabular}{lll}
						\includegraphics[width=0.3\textwidth]{fig-sk-12.pdf} &
						\includegraphics[width=0.3\textwidth]{fig-sk-22.pdf} &
						\includegraphics[width=0.3\textwidth]{fig-sk-32.pdf} \\						\includegraphics[width=0.3\textwidth]{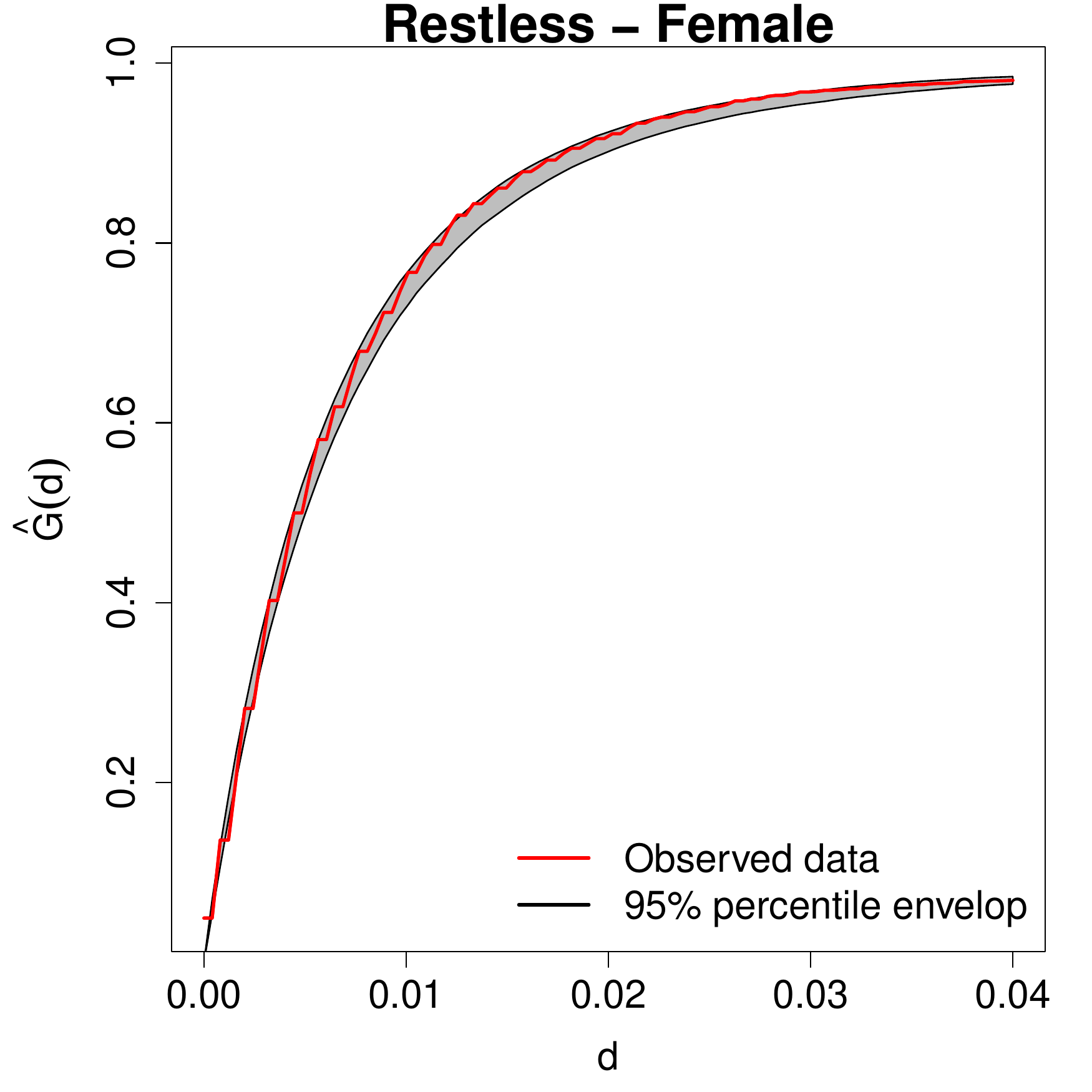} &
						\includegraphics[width=0.3\textwidth]{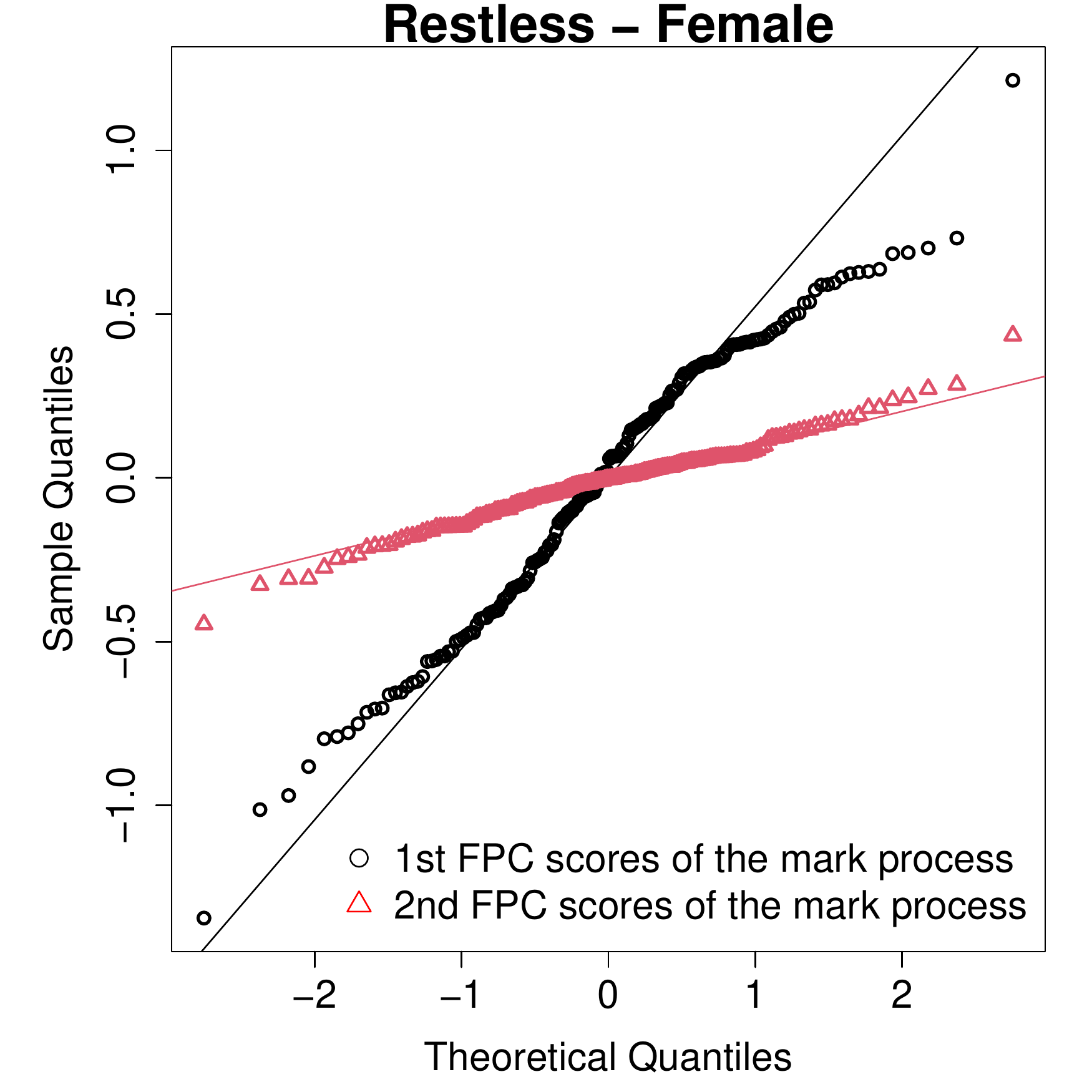} &
						\includegraphics[width=0.3\textwidth]{fig-sk-31.pdf} \\
					\end{tabular}
					\vskip -1em
					\caption{{Left panel: diagnostic plot for LGCP following Section~\ref{sec:diag}; Middle panel: QQ plots of FPC scores following Section~\ref{sec:perm}; Right panel:  Estimated daily mean functions of ``restless" for male and female participants, respectively.}}
					\label{fig-sk-S}
				\end{figure}
				
				
				\begin{figure}[ht!]
					\centering
					\begin{tabular}{lll}
						\includegraphics[trim=0 5mm 0 0, width=0.3\textwidth]{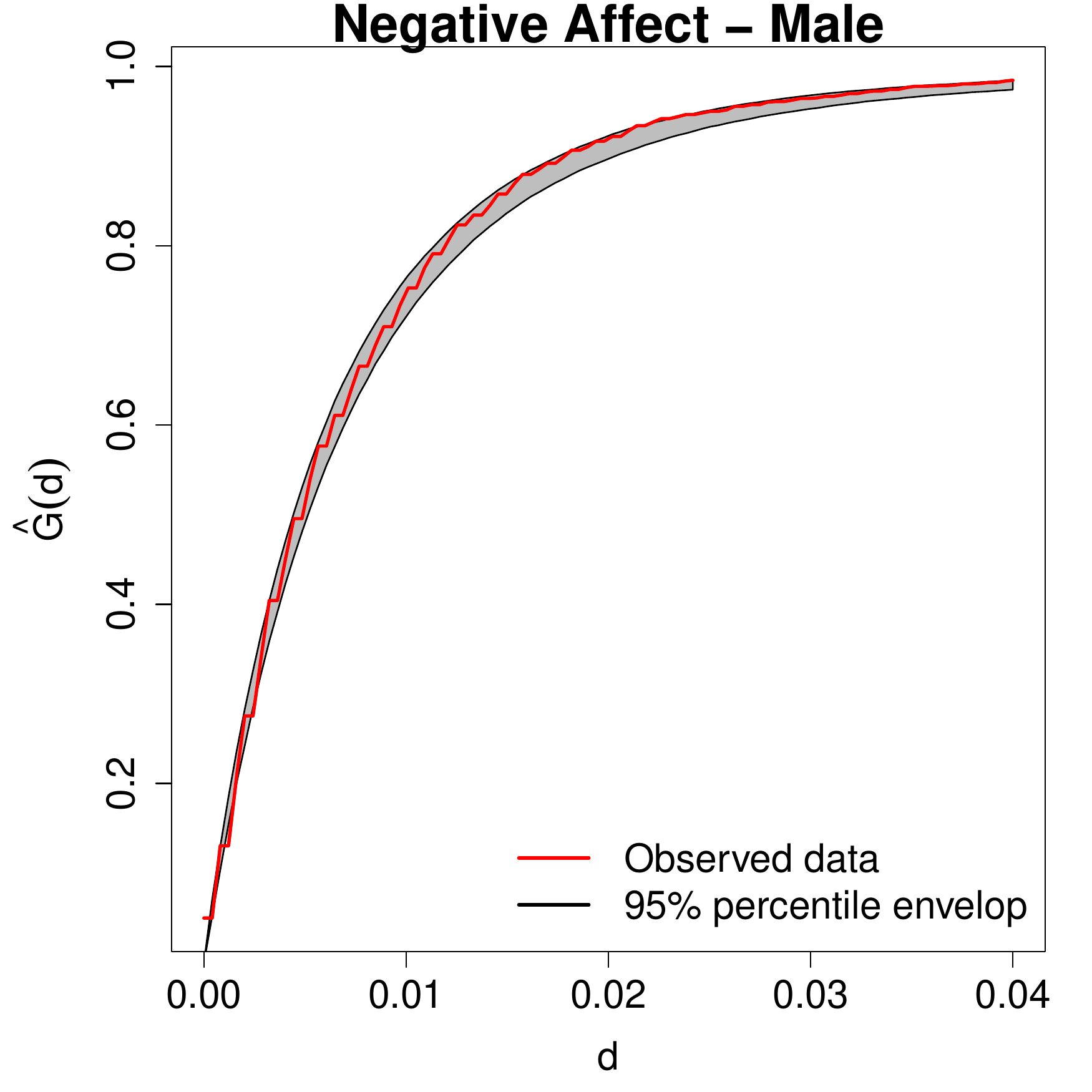} &
						\includegraphics[trim=0 5mm 0 0, width=0.3\textwidth]{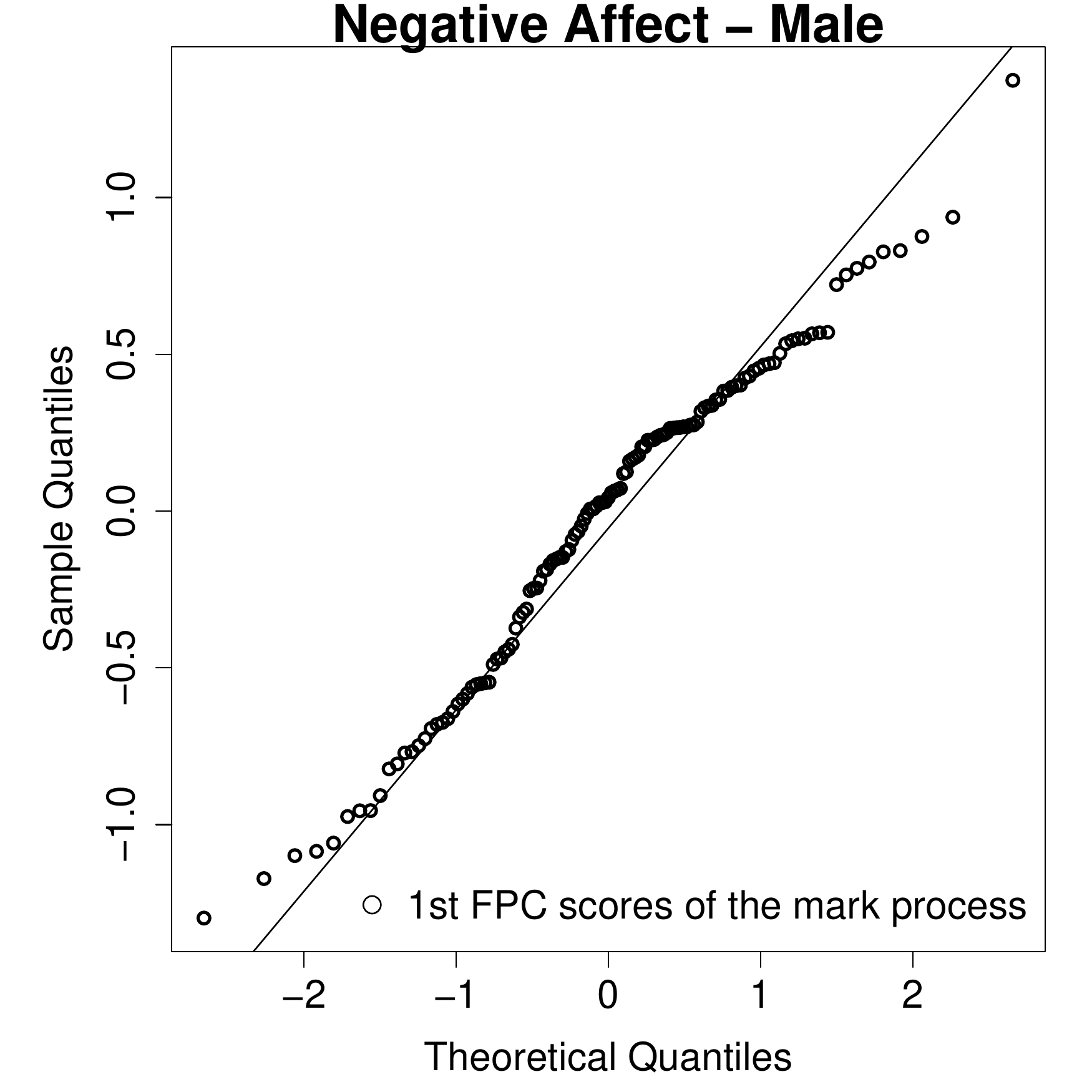} &
						\includegraphics[trim=0 5mm 0 0, width=0.3\textwidth]{fig-sk-34.pdf} \\
						\includegraphics[trim=0 5mm 0 0, width=0.3\textwidth]{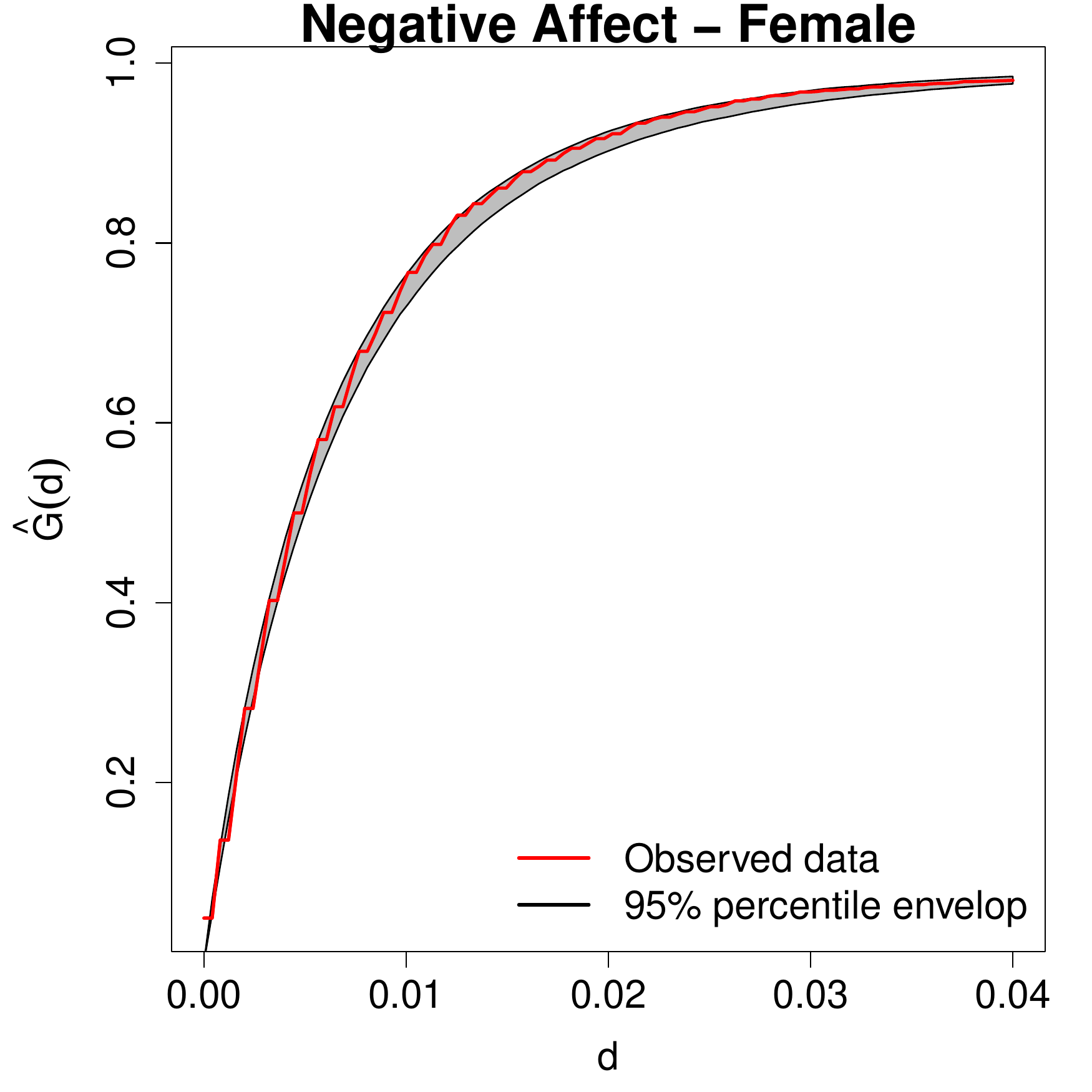} &
						\includegraphics[trim=0 5mm 0 0, width=0.3\textwidth]{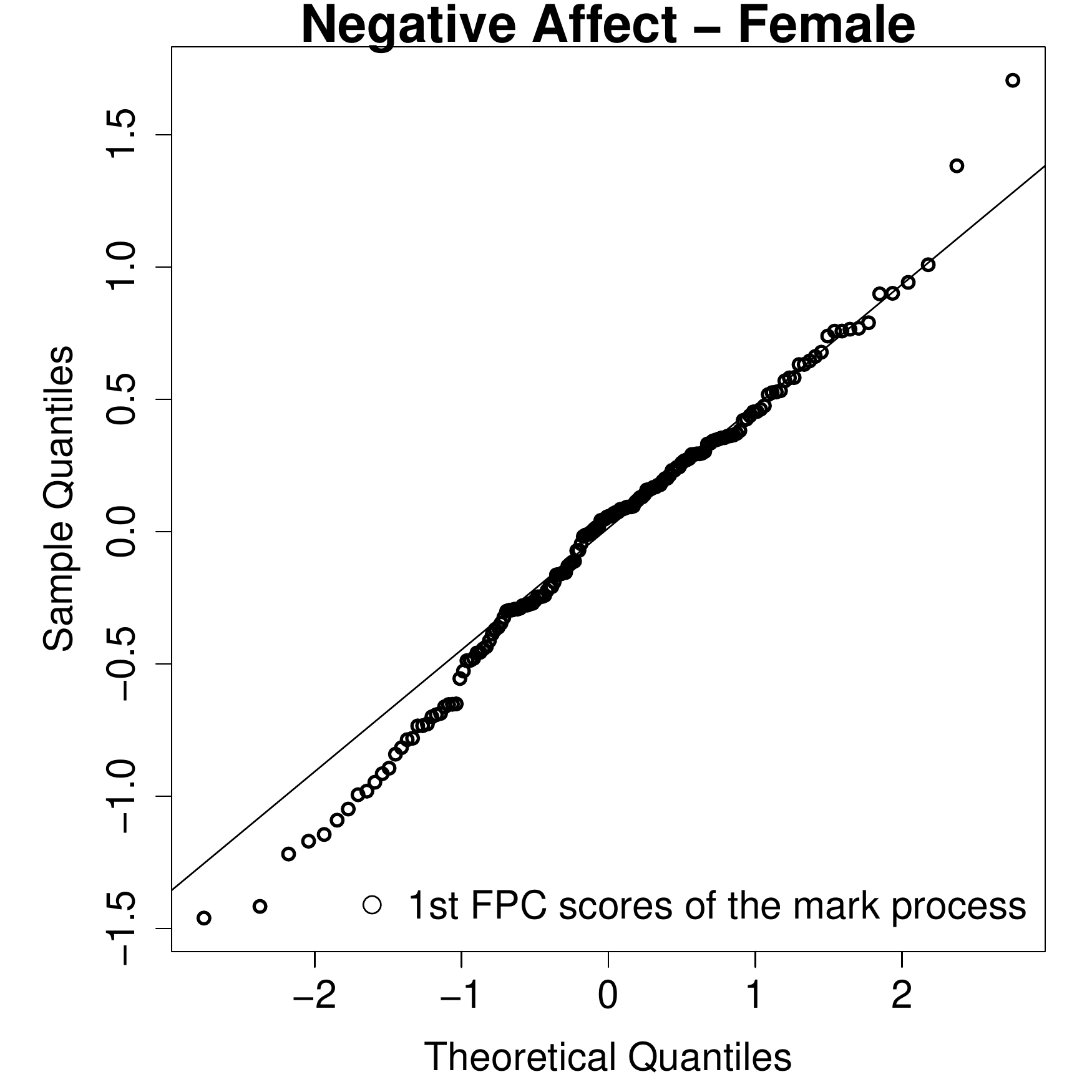} &
						\includegraphics[trim=0 5mm 0 0, width=0.3\textwidth]{fig-sk-33.pdf} \\
					\end{tabular}
					
					\caption{{Left panel: diagnostic plot for LGCP following Section~\ref{sec:diag}; Middle panel: QQ plots of FPC scores following Section~\ref{sec:perm}; Right panel:  Estimated daily mean functions of ``negative affect" for male and female participants, respectively.}}
					\label{fig-sk-S1}
				\end{figure}
				
				
				Next, we give trajectories of ten sample subjects from the eBay online auction data in Figure~\ref{fig-aucdata}.
				\begin{figure}[ht!]
					\centering
					\begin{tabular}{ll}
						\includegraphics[width=0.45\textwidth]{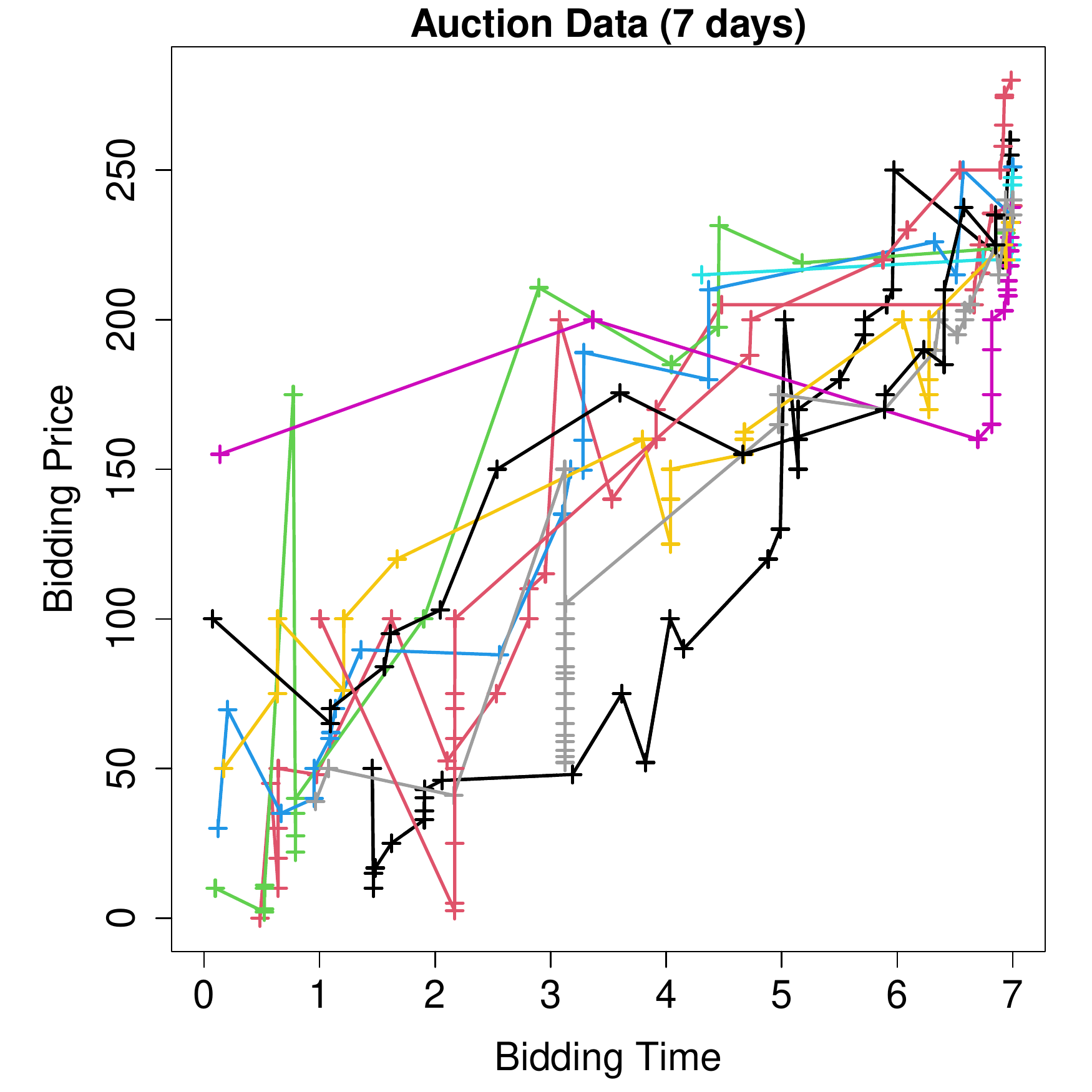} &
						\includegraphics[width=0.45\textwidth]{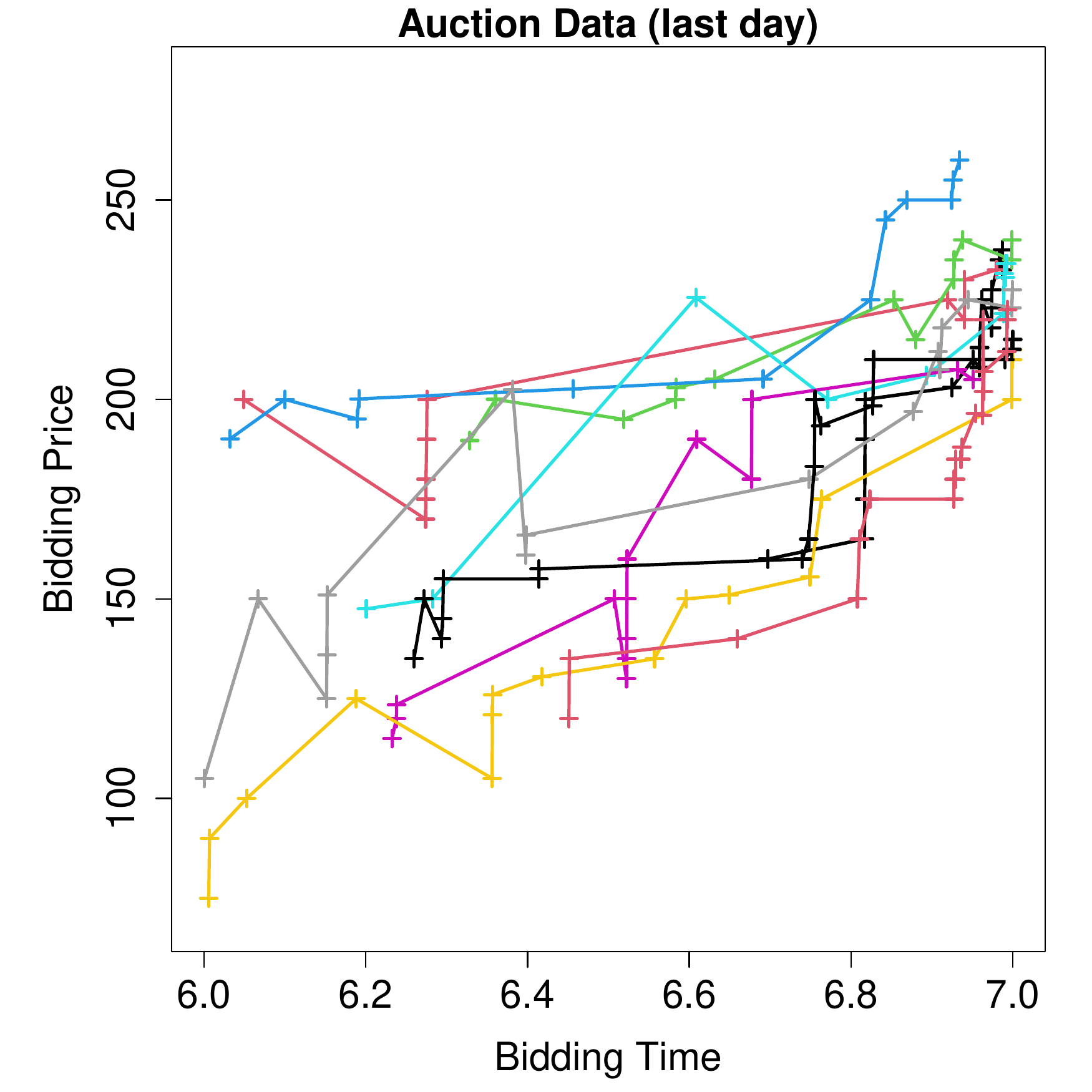} \\
					\end{tabular}
					\vskip -1em
					\caption{{Trajectories of 10 sample subjects from the eBay online auction data. Left panel: 7 day bidding period; Right panel: last bidding day.}}
					\label{fig-aucdata}
				\end{figure}
			}

			\clearpage
			\newpage
			\section{Technical Proofs}
			\label{sec:proof}
			
			\subsection{Proof of Lemma~\ref{lm:stein}}
			
			The proof of~\eqref{stein1} follows directly from~\eqref{stein2-1} of Lemma~\ref{lm:stein_extended} given in Section~\ref{sec:tech}, and hence is omitted. Next, we proceed to show~\eqref{stein2}. Since $X$ and $Y_j$, $j=1,2$, are jointly normally distributed, one has that
			\[
			\E(Y_j|X)=\mu_j+\frac{\Cov(X,Y_j)}{\sigma_X^2}X,\text{ and } \Var(Y_j|X)=\sigma_j^2-\frac{[\Cov(X,Y_j)]^2}{\sigma_X^2},\quad j=1,2.
			\]
			Define variable $Z_j=Y_j-\E(Y_j|X)$, then we have that $E(Z_j)=0$, $\Var(Z_j)=\Var(Y_j|X)$ and $\Cov(X,Z_j)=\E(XZ_j)=\E\left[X\E(Z_j|X)\right]=0$, which indicates that $X$ and $Z_j$'s are independent normal random variables and that
			\[
			Z_j\sim N\left[0,\sigma_j^2-\frac{[\Cov(X,Y_j)]^2}{\sigma_X^2}\right],\quad j=1,2.
			\]
			Denote by $a_j=\frac{\Cov(X,Y_j)}{\sigma_X^2}$, $j=1,2$. Then, we have that
			\[
			\begin{split}
			\E &\left[Y_1 Y_2\exp(X)\right]=\E \left[(Z_1+\mu_1+a_1X) (Z_2+\mu_2+a_2X)\exp(X)\right]  \\
			&=\E \left[(Z_1+\mu_1)(Z_2+\mu_2)\right]\E\exp(X)+a_2\mu_1 \E\left[X\exp(X)\right]+a_1\mu_2 \E\left[X\exp(X)\right]+a_1a_2\E \left[X^2\exp(X)\right]\\
			&=\E \left[(Z_1+\mu_1)(Z_2+\mu_2)\right] \exp(\sigma_X^2/2 )+(a_2\mu_1+a_1\mu_2)\sigma_X^2\exp(\sigma_X^2/2 )+a_1a_2\left(\sigma_X^4+\sigma_X^2\right) \exp(\sigma_X^2/2 )\\
			&=\E \left[(Z_1+\mu_1)(Z_2+\mu_2)\right] \exp(\sigma_X^2/2 )+\left[\mu_2\Cov(X,Y_1)+\mu_1\Cov(X,Y_2)\right]\\
			&\qquad+\frac{\Cov(X,Y_1)\Cov(X,Y_2)}{\sigma_X^2}\left(\sigma_X^2+1\right) \exp(\sigma_X^2/2 ).
			\end{split}
			\]
			Furthermore, it readily follows that
			\[
			\begin{split}
			\E \left[(Z_1+\mu_1)(Z_2+\mu_2)\right]&=\Cov (Z_1,Z_2)+\mu_1\mu_2=\Cov(Y_1-a_1X,Y_2-a_2X)+\mu_1\mu_2\\
			&=\Cov(Y_1,Y_2)-a_2\Cov(Y_1,X)-a_1\Cov(X,Y_2)+a_1a_2\sigma_X^2+\mu_1\mu_2\\
			&=\Cov(Y_1,Y_2)-\frac{\Cov(X,Y_1)\Cov(X,Y_2)}{\sigma_X^2}+\mu_1\mu_2.\\
			\end{split}
			\]
			Consequently, it follows that
			\[
			\begin{split}
			\E \left[Y_1 Y_2\exp(X)\right]&=\E \left[(Z_1+\mu_1)(Z_2+\mu_2)\right] \exp(\sigma_X^2/2 )+\frac{\Cov(X,Y_1)\Cov(X,Y_2)}{\sigma_X^2}\left(\sigma_X^2+1\right) \exp(\sigma_X^2/2 )\\
			&=\left[\Cov(Y_1,Y_2)-\frac{\Cov(X,Y_1)\Cov(X,Y_2)}{\sigma_X^2}+\mu_1\mu_2\right]\exp(\sigma_X^2/2 )\\
			&\qquad+\left[\mu_2\Cov(X,Y_1)+\mu_1\Cov(X,Y_2)\right]+\frac{\Cov(X,Y_1)\Cov(X,Y_2)}{\sigma_X^2}\left(\sigma_X^2+1\right) \exp(\sigma_X^2/2 )\\
			&=\left\{\Cov(Y_1,Y_2)+\left[\mu_1+\Cov(X,Y_1)\right]\left[\mu_2+\Cov(X,Y_2)\right]\right\}\exp(\sigma_X^2/2 ),
			\end{split}
			\]
			which completes the proof of Lemma~\ref{lm:stein}.
			
			\subsection{Proof of Theorem~\ref{thm1}}
			\subsubsection{ Part(a): uniform convergence of $\wt \mu(\cdot)$}
			By (\ref{mutilde}), some straightforward algebra gives that
			\[
			\begin{split}
			\wt \mu(s) - \mu^*(s)&= \e_{2}^\top\left[\wh\A_{n,h_{\mu},1}(s)\right]^{-1} \underbrace{\left[\frac{1}{n}\sum_{i=1}^n\sum_{u\in N_i} \bphi_{h_{\mu}}(u-s) K_{1,h_{\mu}}(u-s) [Z_i(u)-\mu^*(s)]\right]}_{\CR_1(s)},\\
			\end{split}
			\]
			where, {under Assumptions C1 and C3}, the term $\CR_1(s)$ can be further decomposed through the second order Taylor expansion of $\mu^*(u)$ around $s$ as $\CR_1(s) \equiv \CR_{10}(s) + \CR_{11}(s)+ \CR_{12}(s) +{O_p\left(h_{\mu}^{3}\right)}$, with 
			\bse
			\CR_{10}(s) &=& {1\over n} \sum_{i=1}^n \sum_{u\in N_i}  [Z_i(u)-\mu^*(u)] K_{1,h_{\mu}} (u-s)\bphi_{h_{\mu}}(u-s),  \\
			\CR_{11}(s) &=& {\mu^{*(1)}(s)\over n} \sum_{i=1}^n \sum_{u\in N_i}  (u-s)  K_{1,h_{\mu}} (u-s)\bphi_{h_{\mu}}(u-s)\\
			&=&\mu^{*(1)}(s)h_\mu\times\wh\a_{n,h_{\mu},2}(s), \\
			\CR_{12}(s) &=& {\mu^{*(2)}(s)\over 2n} \sum_{i=1}^n \sum_{u\in N_i}   (u-s)^2 K_{1,h_{\mu}} (u-s)\bphi_{h_{\mu}}(u-s),
			\ese
			where  $\mu^{*(j)}(\cdot)$ is the $j$th derivative of $\mu^*(\cdot)$ for $j=1,2$, and $\wh\a_{n,h_{\mu},2}(s)$ is the $2$nd column of the matrix $\wh\A_{n,h_{\mu},1}(s)$ by definition. By the definition of $\e_{2}=(1,0)^\top$, it is straightforward to see that $\e_{2}^\top\left[\wh\A_{n,h_{\mu},1}(s)\right]^{-1}\wh\a_{n,h_{\mu},2}(s)=0$, which consequently implies that
			\be
			\label{mu-dc}
			\wt \mu(s) - \mu^*(s)= \e_{2}^\top\left[\wh\A_{n,h_{\mu},1}(s)\right]^{-1} \left[\CR_{10}(s) + \CR_{12}(s) +O_p\left(h_{\mu}^{3}\right)\right].\\
			\ee
			By the Stein's Lemma, it immediately follows that
			\bse
			\E[ \CR_{10}(s)] &=& \E  \int_\CT [ Z_i(u)- \mu^*(u) ] K_{1,h_{\mu}}(u-s) N_i(du) \bphi_{h_{\mu}}(u-s)\\
			& =&    \int_\CT \E \{[Z_i(u)- \mu^*(u)] \exp[X(u)]\} \lambda_0(u) K_{1,h_{\mu}} (u-s)\bphi_{h_{\mu}}(u-s) du\\
			&=&0,
			\ese
			and by Lemma \ref{lm:stein} and Assumptions C1 and C2(a), we have that
			{\small \bse
				&&\Cov[\CR_{10}(s)] = \E [ \CR_{10}(s)\CR_{10}^\top(s)] \\
				&=& {1\over n} \E  \int_\CT [Z_1(u)- \mu^*(u) ]^2 K_{1,h_{\mu}}^2(u-s)\bphi_{h_{\mu}}(u-s)\bphi_{h_{\mu}}^\top(u-s) N_1(du)  \\
				&&  + {1\over n} \E \int_\CT\int_\CT \{ Z_1(u)- \mu^*(u) \} \{ Z_1(v)- \mu^*(v) \} K_{1,h_{\mu}}(u-s) K_{1,h_{\mu}}(v-t)\bphi_{h_{\mu}}(u-s)\bphi_{h_{\mu}}^\top(v-t) N_1^{(2)}(du, dv) \\
				&=& {1\over n}  \int_\CT  \E \bigg[ \{ Z_1(u)- \mu^*(u) \}^2 \exp\{ X(u)\} \bigg]  \lambda_0(u) K_{1,h_{\mu}}^2(u-s)\bphi_{h_{\mu}}(u-s)\bphi_{h_{\mu}}^\top(u-s) du  \\
				&& \hskip10mm + {1\over n} \int_\CT \int_\CT  \E \bigg[ \{ Z_1(u)- \mu^*(u) \} \{ Z_1(v)- \mu^*(v) \} \exp\{ X(u)+X(v)\}\bigg] \\
				&& \hskip50mm \times  \lambda_0(u) \lambda_0(v) K_{1,h_{\mu}}(u-s) K_{1,h_{\mu}}(v-t) \bphi_{h_{\mu}}(u-s)\bphi_{h_{\mu}}^\top(v-t)du  dv \\
				&=&{1\over n}  \int_\CT \sigma_Z^2(u) \rho(u) K_{1,h_{\mu}}^2(u-s) \bphi_{h_{\mu}}(u-s)\bphi_{h_{\mu}}^\top(u-s)du  \\
				&& + {1\over n} \int_\CT \int_\CT  \rho_2(u,v) \{ C_Y(u,v) + C_{XY}(u,v) C_{XY}(v,u)\} K_{1,h_{\mu}}(u-s) K_{1,h_{\mu}}(v-t)\bphi_{h_{\mu}}(u-s)\bphi_{h_{\mu}}^\top(v-t) du  dv \\
				&=& {\sigma_Z^2(s)  \rho(s) \over n h_{\mu}}\Q_{K_1,\bphi}^{\{2\}}+o\{ (n h_{\mu})^{-1}\}=O\{ (n h_{\mu})^{-1}\},
				\ese}
			where $\Q_{K_1,\bphi}^{\{2\}}$ is as defined in Assumption~C2(a). By similar calculations,  it is straightforward to show that 
			$$\E[\CR_{12}(s)]= {\mu^{*(2)}(s)h_{\mu}^{2}\over 2} \rho(s)\bm\sigma_{K_1,\bphi}    + O(h_{\mu}^{3}), \text{ and }\Var[\CR_{12}(s)]= O(h_{\mu}^{3} n^{-1}),$$ 
			where $\bm\sigma_{K_1,\bphi}$ is as defined in Assumption~C2(a) and the equation hold entry-wise. To this end, the above calculations yield that for any $s\in\CT$
			\be
			\label{mu1}
			\|\CR_{10}(s)\|_{\max}=O_p\left[(n h_{\mu})^{-1/2}\right],\text{ and } \left\|\CR_{12}(s)\right\|_{\max}=O_p\left[(n h_{\mu})^{-1/2}h_{\mu}^2+h_{\mu}^2\right].
			\ee
			
			Recall definitions of $\wh\A_{n,h_{\mu},1}(s)$ and $\A_{h_{\mu},1}(s)$ in~\eqref{Ahat} and~\eqref{den1}, and using similarly arguments as those to obtain~\eqref{mu1}, we can show that
			\[
			\|\wh\A_{n,h_{\mu},1}(s)-\A_{h_{\mu},1}(s)\|_{\max} = O_p\left[(nh_{\mu})^{-1/2}\right],\text{ and } \|\A_{h_{\mu},1}(s)-\rho(s)\Q_{K_1,\bphi}^{\{1\}}\|_{\max}=O(h_{\mu}), 
			\]
			where $\Q_{K_1,\bphi}^{\{1\}}$ is defined in Assumption~C2(a). By Assumption~C2(a), $\Q_{K_1,\bphi}^{\{1\}}$ is positive semi-definite and $\lambda_{\min}\left[\Q_{K_1,\bphi}^{\{1\}}\right]>0$, together with the Assumption C1 that $\rho(s)>0$, we have that 
			\be\label{eq:rho_hat_consist}
			\begin{split}
				&\left\|\left[\wh\A_{n,h_{\mu},1}(s)\right]^{-1}-\left[\rho(s)\Q_{K_1,\bphi}^{\{1\}}\right]^{-1}\right\|_{\max}\\&\hskip5em\leq 
				\left\|\left[\wh\A_{n,h_{\mu},1}(s)\right]^{-1}-\left[\A_{h_{\mu},1}(s)\right]^{-1}\right\|_{\max}	+\left\|\left[\A_{h_{\mu},1}(s)\right]^{-1}-\left[\rho(s)\Q_{K_1,\bphi}^{\{1\}}\right]^{-1}\right\|_{\max} \\
				&\hskip5em= O_p\left[(nh_{\mu})^{-1/2}+h_{\mu}\right].
			\end{split}
			\ee
			Therefore, combining~\eqref{mu-dc}, \eqref{mu1} and~\eqref{eq:rho_hat_consist}, we have that
			\be\label{eq:mu_tilde_expansion}
			\wt \mu(s) -\mu(s)  = \CB_\mu(s) h_{\mu}^{2}  + \mCR_{\mu}(s) +O_p\left[h_{\mu}^{3}+ h_{\mu}(nh_{\mu})^{-1/2}\right],
			\ee
			where $\CB_\mu(s) =  {\mu^{*(2)}(s)\over 2} \e_{2}^\top\left[\Q_{K_1,\bphi}^{\{1\}}\right]^{-1}\bm\sigma_{K_1,\bphi} $ and 
			\be
			\label{Rmu}
			\mCR_{\mu}(s)=\e_{2}^\top\left[\Q_{K_1,\bphi}^{\{1\}}\right]^{-1}\frac{\CR_{10}(s)}{\rho(s)}={1\over n\rho(s)} \sum_{i=1}^n \sum_{u\in N_i}  \{ Z_i(u)-\mu^*(u)\} K_{1,h_{\mu}} (u-s)w_{\bphi,h_{\mu}}(u-s),
			\ee
			with $w_{\bphi,h_{\mu}}(u-s)=\e_{2}^\top\left[\Q_{K_1,\bphi}^{\{1\}}\right]^{-1}\bphi_{h_{\mu}}(u-s)$.
			Further, some straightforward calculations give that
			\be\label{eq:mu_tilde_bias_variance}
			\E\left[\mCR_{\mu}(s)\right]=0,\text{ and } nh_\mu\Var\left[\mCR_{\mu}(s)\right]=O(1).
			\ee
			Consequently, using arguments in \cite{li2010uniform} and the asymptotic expansion (\ref{eq:mu_tilde_expansion}), the uniform convergence rate for $\wt \mu(s)$ can be established as
			\be\label{eq:mu_tilde_uniform_rate}
			\sup_{s\in \CT} |\wt \mu(s) -\mu^* (s)|  = O_p\left\{h_{\mu}^{2} + [\log(n)/ (n h_{\mu})]^{1/2}\right\} ,
			\ee
			and that
			\be\label{eq:mu_tilde_uniform_rate2}
			\sup_{s\in \CT} \bigg |\wt \mu(s) -\mu^* (s) - \CB_\mu(s) h_{\mu}^{2} - \mCR_{\mu}(s) \bigg |  = O_p\{ h_{\mu}^{3} + h_{\mu}[\log(n)/ (n h_{\mu})]^{1/2}\}.
			\ee
			\subsubsection{Part (b): uniform convergence of $\wt \sigma_Z^2(\cdot)$}
			Define $\check \sigma_Z^2(s)=\e_{2}^\top\left[\wh\A_{n,h_{\sigma},1}(s)\right]^{-1} \left\{\frac{1}{n}\sum_{i=1}^n\sum_{u\in N_i} \bphi_{h_{\sigma}}(u-s) K_{1,h_{\sigma}}(u-s) [Z_i(u)-\mu^*(u)]^2\right\}$ as the counterpart of $\wt \sigma_Z^2(s)$ when $\mu^*(\cdot)$ is known. Using similar derivations as for $\wt \mu(s)$, it can be straightforwardly shown that
			\be\label{eq:sigma2_tilde_expansion}
			\begin{split}
				\check \sigma_Z^2 (s) -\sigma_Z^2 (s)  &=\CB_{\sigma^2}(s) h_{\sigma}^2+\mCR_{\sigma^2}(s) +O_p[h_{\sigma}^{3}+ h_{\sigma}(nh_{\sigma})^{-1/2}],
			\end{split}
			\ee
			where 
			\be\label{eq:sigma2_tilde_bias}
			\begin{split}
				&		\CB_{\sigma^2}(s)  = {(\sigma_Z^2)^{(2)}(s)\over 2} \e_{2}^\top\left[\Q_{K_1,\bphi}^{\{1\}}\right]^{-1}\bm\sigma_{K_1,\bphi},\\
				&\mCR_{\sigma^2}(s)=\frac{1}{n\rho(s)}\sum_{i=1}^n\sum_{u\in N_i}\left[ \{Z_i(u)-\mu^*(u)\}^2-\sigma_Z^2(u)\right] K_{1,h_{\sigma}}(u-s)w_{\bphi,h_{\sigma}}(u-s),
			\end{split}
			\ee
			with $w_{\bphi,h_{\sigma}}(u-s)=\e_{2}^\top\left[\Q_{K_1,\bphi}^{\{1\}}\right]^{-1}\bphi_{h_{\sigma}}(u-s)$.

			Next, we bound the difference $|\check \sigma_Z^2 (s) - \wt \sigma_Z^2 (s)|$ as follows 
			\bse
			\label{sig-diff}
			%
			&&|\check \sigma_Z^2 (s) - \wt \sigma_Z^2 (s)|  \\
			&=&  \e_{2}^\top\left[\wh\A_{n,h_{\sigma},1}(s)\right]^{-1}\frac{1}{n}\sum_{i=1}^n\sum_{u\in N_i}[ \{2Z_i(u)-\mu^*(u) - \wt \mu(u)\} \{ \wt \mu(u)- \mu^*(u) \}] K_{1,h_{\sigma}}(u-s)\bphi_{h_{\sigma}}(u-s) \\
			&\leq& \left|\e_{2}^\top\left[\wh\A_{n,h_{\sigma},1}(s)\right]^{-1} \frac{2}{n}\sum_{i=1}^n\sum_{u\in N_i}[ \{Z_i(u)-\mu^*(u) \} \{ \wt \mu(u)- \mu^*(u) \}] K_{1,h_{\sigma}}(u-s)\bphi_{h_{\sigma}}(u-s)\right| \\
			&&\qquad+ \underbrace{\sup_{s\in \CT} [ \wt \mu(s)- \mu^*(s)]^2}_{O_p[h_{\mu}^4+\log(n)/(nh_{\mu})]}\times  \underbrace{\frac{1}{n}\sum_{i=1}^n\sum_{u\in N_i}\left|\e_{2}^\top\left[\wh\A_{n,h_{\sigma},1}(s)\right]^{-1} K_{1,h_{\sigma}}(u-s)\bphi_{h_{\sigma}}(u-s)\right|}_{O_p(1)} \\
			&=& \left|\e_{2}^\top\left[\wh\A_{n,h_{\sigma},1}(s)\right]^{-1} \frac{2}{n}\sum_{i=1}^n\sum_{u\in N_i}[ \{Z_i(u)-\mu^*(u) \} \{ \wt \mu(u)- \mu^*(u) \}] K_{1,h_{\sigma}}(u-s)\bphi_{h_{\sigma}}(u-s)\right| \\
			&&\qquad+ O_p[h_{\mu}^4+\log(n)/(nh_{\mu})]\times O_p(1)\\
			&=& \left|{2\over n  \rho(s)}\e_{2}^\top\left[\Q_{K_1,\bphi}^{\{1\}}(s)\right]^{-1} \sum_{i=1}^n\sum_{u\in N_i} \{Z_i(u)-\mu^*(u) \} \{ \CB_\mu(u) h_{\mu}^2 +\mCR_{\mu}(u)\} K_{1,h_{\sigma}}(u-s)\right| \\
			&& \hskip20mm \times \{1+o_p(1)\}+ O_p\{ h_{\mu}^{3} + h_{\mu}[\log(n)/ (n h_{\mu})]^{1/2}\} +O_p[h_{\mu}^4+ \log(n)/(nh_{\mu})]\\
			&=& \left|{2\over n^2  \rho(s)} \sum_{i=1}^n\sum_{u\in N_i}\sum_{i'=1}^n\sum_{v\in N_{i'}} \frac{\{Z_i(u)-\mu^*(u) \} \{Z_{i'}(v)-\mu^*(v) \}}{\rho(u)} K_{1,h_{\mu}}(u-v)w_{\bphi,h_{\mu}}(u-v)K_{1,h_{\sigma}}(u-s)\right| \\
			&& \hskip20mm \times \{1+o_p(1)\} +O_p[h_{\mu}^{2} (nh_{\mu})^{-1/2}]+O_p\left\{h_{\mu}^{3}+ h_{\mu}[\log(n)/(nh_{\mu})]^{1/2}+\log(n)/(nh_{\mu})\right\}\\
			&=&{2\over n^2  \rho^2(s)} \left|  
			\sum_{i=1}^n\sum_{u\in N_i} \{Z_i(u)-\mu^*(u) \}^2  K_{1,h_{\mu}}(0)w_{\bphi,h_{\mu}}(0)K_{1,h_{\sigma}}(u-s) \right. \\
			&& +  \sum_{i=1}^n\mathop{\sum\sum}^{\ne}_{u,v\in N_i}  \{Z_i(u)-\mu^*(u) \} \{Z_{i}(v)-\mu^*(v) \} K_{1,h_{\mu}}(u-v)w_{\bphi,h_{\mu}}(u-v)K_{1,h_{\sigma}}(u-s)\\
			&& \left. +
			\sum_{i=1}^n\sum_{u\in N_i}\sum_{i' \neq i} \sum_{v\in N_{i'}} \{Z_i(u)-\mu^*(u) \} \{Z_{i'}(v)-\mu^*(v) \}  K_{1,h_{\mu}}(u-v)w_{\bphi,h_{\mu}}(u-v)K_{1,h_{\sigma}}(u-s)\right| \\
			&& \hskip20mm \times \{1+o_p(1)\} +O_p\{h_{\mu}^{3}+ h_{\mu}[\log(n)/(nh_{\mu})]^{1/2}+\log(n)/(nh_{\mu})\}\\
			&=& O_p[(nh_{\mu})^{-1}+(nh_{\mu})^{-1/2}(nh_{\sigma})^{-1/2}]+O_p\left\{h_{\mu}^{3}+ h_{\mu}[\log(n)/(nh_{\mu})]^{1/2}+\log(n)/(nh_{\mu})\right\}\\
			&= &O_p\left\{h_{\mu}^{3}+ (nh_{\mu})^{-1/2}(nh_{\sigma})^{-1/2}+h_{\mu}[\log(n)/(nh_{\mu})]^{1/2}+\log(n)/(nh_{\mu})\right\}.
			%
			\ese
			Consequently, we have the asymptotic expansion for $\wt\sigma_Z^2 (s) -\sigma_Z^2 (s)$ as
			\[
			\label{eq:sigma2_tilde_expansion1}
			\begin{split}
			\wt\sigma_Z^2 (s) -\sigma_Z^2 (s)  &=\CB_{\sigma^2}(s) h_{\sigma}^2+\mCR_{\sigma^2}(s) \\
			&+O_p\left\{h_{\sigma}^{3}+h_{\mu}^{3}+ [h_\sigma+(nh_{\mu})^{-1/2}](nh_{\sigma})^{-1/2}+h_{\mu}[\log(n)/(nh_{\mu})]^{1/2}+\log(n)/(nh_{\mu})\right\}.
			\end{split}
			\]
			It is straightforward to see that%
			\[
			\sqrt{nh_{\sigma}} \mCR_{\sigma^2}(s)=O_p(1).
			\]
			Then by applying similar arguments in~\cite{li2010uniform} and Assumption~C3, we have that
			\be\label{eq:sigma2_tilde_uniform_rate1}
			\sup_{s\in \CT} \bigg |\wt {\sigma}^2_Z(s) -\sigma^2_Z(s) \bigg |  = O_p\left[h_{\sigma}^2+(nh_{\sigma})^{-1/2}\sqrt{\log(n)}\right],
			\ee
			and that
			\be\label{eq:sigma2_tilde_uniform_rate2}
			\sup_{s\in \CT} \bigg |\wt {\sigma}^2_Z(s) -\sigma^2_Z(s) - \CB_{\sigma^2}(s) h_{\sigma}^{2} - \mCR_{\sigma^2}(s) \bigg |  = O_p\left[h_{\sigma}^{3}+(nh_{\sigma})^{-1}\sqrt{\log(n)}+ h_{\sigma}(nh_{\sigma})^{-1/2}\log(n)\right].
			\ee

			\subsubsection{Part (c) uniform convergence rate of the naive covariance estimator $\wt C_Y(\cdot,\cdot) $}
			Define the counterparts of $\wt C_Y(s,t) $ in~\eqref{cytilde} when $\mu^*(\cdot)$ is known as follows
			\[
			\wc C_{Y}(s,t)=\e_{3}^\top\left[\wh\B_{n,h_y,1}(s,t)\right]^{-1}\left[\frac{1}{n}\sum_{i=1}^n\mathop{\sum\sum}^{\ne}_{u,v\in N_i}W_i(u,v)K_{2,h_y}(u-s,v-t)\bpsi_{h_y}(u-s,v-t)\right],
			\]
			where $W_i(u,v)=[Z_i(u)-\mu^*(u)][Z_i(v)-\mu^*(v)]$, $i=1,\ldots,n$. 
			
			Let $C_Y^*(s,t)=C_Y(s,t)+C_{XY}(s,t)C_{XY}(t,s)$, some straightforward algebra gives that
			\[
			\begin{split}
			\wc C_Y(s,t) -C_Y^*(s,t)=& \e_{3}^\top
			\left[\wh\B_{n,h_y,1}(s,t)\right]^{-1}\\
			&\qquad\times\underbrace{\left[\frac{1}{n}\sum_{i=1}^n\mathop{\sum\sum}^{\ne}_{u,v\in N_i}[W_i(u,v)-C_Y^*(s,t)]K_{2,h_y}(u-s,v-t)\bpsi_{h_y}(u-s,v-t)\right]}_{\CR_{Y}(s,t)}, 
			\end{split}
			\]
			where $\wh\B_{n,h_y,1}(s,t)$ is as defined in~\eqref{Bhat}. The term $\CR_Y(s,t)$ can be further decomposed through second order Taylor expansion of $C_Y^*(s,t)$ around $(u,v)$ as $\CR_Y(s,t) \equiv \CR_{20}(s,t) + \CR_{21}(s,t)+ \CR_{22}(s,t) +O_p\left(h_y^3+h_y^{2}/\sqrt{n}\right)$, with  
			\be
			\label{CY-dc0}
			\begin{split}
				& \CR_{20}(s,t) =\frac{1}{n}\sum_{i=1}^n\mathop{\sum\sum}^{\ne}_{u,v\in N_i}\left\{W_i(u,v)-C_Y^*(u,v)\right\}K_{2,h_y}(u-s,v-t)\bpsi_{h_y}(u-s,v-t), \\
				& \CR_{21}(s,t) ={1\over n}\sum_{i=1}^n  \mathop{\sum\sum}^{\ne}_{u,v\in N_i} \bigg \{C_Y^{*(1,0)}(s,t)(u-s)+ C_Y^{*(0,1)}(u,v) (v-t)\bigg\}
				K_{2,h_y}(u-s,v-t)\bpsi_{h_y}(u-s,v-t) \\
				&\hskip 8em=C_Y^{*(1,0)}(s,t) h_y \wh\b_{n,h_y,2}(s,t)+C_Y^{*(0,1)}(s,t) h_y \wh\b_{n,h_y,3}(s,t),\\
				& \CR_{22}(s,t) ={1\over n}\sum_{i=1}^n  \mathop{\sum\sum}^{\ne}_{u,v\in N_i} \bigg \{ C_Y^{*(1,1)}(s,t)(u-s)(v-t)+{1\over 2}  C_Y^{*(2,0)}(s,t) (u-s)^2+ {1\over 2}  C_Y^{*(0,2)}(s,t) (v-t)^2 \bigg\}\\
				& \hskip45mm  
				\times K_{2,h_y}(u-s,v-t)\bpsi_{h_y}(u-s,v-t),
			\end{split}
			\ee
			where $\wh\b_{n,h_y,j}(s,t)$, $j=2,3$, is the $j$th column of the matrix $\wh\B_{n,h_y,1}(s,t)$ by definition and $C_Y^{*(j,l)}(s,t)=\partial^{j+l} C_Y^*(s,t)/(\partial s^{j}\partial t^l)$ for $j,l=0,1,2$. By the definition of $\e_{3}$, it is straightforward to see that $\e_{3}^\top\left[\wh\B_{n,h_y,1}(s,t)\right]^{-1}\wh\b_{n,h_y,j}(s,t)=0$ for $j=2,3$, which further implies that
			\be
			\label{cy-dc}
			\wc C_Y(s,t) -C_Y^*(s,t)= \e_{3}^\top
			\left[\wh\B_{n,h_y,1}(s,t)\right]^{-1} \left[\CR_{20}(s,t) + \CR_{22}(s,t) +O_p\left(h_y^3+h_y^{2}/\sqrt{n}\right)\right].\\
			\ee
			It then follows from \eqref{marginal2} and the equation\eqref{stein2} in Lemma 1 that 
			\bse
			\E\{\CR_{20}(s,t) \}&=&  \int_\CT \int_\CT  \E\left[ \left\{W_i(u,v) -C_{Y}^*(u,v) \right\} \exp\{ X(u)+X(v)\} \right]  \\
			&&\hskip50mm  \times \lambda_0(u) \lambda_0(v) K_{2,h_y}(u-s,v-t)\bphi_{h_y}(u-s,v-t) du dv, \\
			&=&0,
			\ese
			and denote for a vector $\a$ with $\a^{dag}=\a\a^\top$, by Lemma \ref{lm:stein} and Assumption C2(b), we have that
			\bse
			\Cov \{\CR_{20}(s,t) \}&=&  {1\over n} \E \left[ \mathop{\sum\sum}^{\ne}_{u,v\in N_i} \left\{W_i(u,v)-C_Y^*(u,v)\right\}K_{2,h_y}(u-s,v-t)\bpsi_{h_y}(u-s,v-t)\right]^{dag}\\
			&=&{1\over n}  \int_\CT \int_\CT \E\left[ \left\{W_i(u,v)-C_Y^*(u,v)\right\}^2 \exp\{ X(u)+X(v)\} \right]  \\
			&&\hskip20mm  \times \lambda_0(u) \lambda_0(v) K_{2,h_y}^2(u-s,v-t)\bpsi_{h_y}^{dag}(u-s,v-t) du dv +o\{ (nh_y^2)^{-1}\} \\
			%
			&=& {O(1)\over nh_y^2} \rho_2(s,t) \Q_{K_2,\bpsi}^{\{2\}}  +o\{ (nh_y^2)^{-1}\} ,
			\ese
			where $\Q_{K_2,\bpsi}^{\{2\}}$ is defined in Assumption C2(b).
			Using similar calculations, it is straightforward to show that,
			\be
			\label{R22-dc}
			\CR_{22}(s,t) =   \frac{1}{2}\rho_2(s,t)\left\{C_{Y}^{*(2,0)}(s,t)\bm\sigma_{K_2,\bpsi}(0)+C_{Y}^{*(0,2)}(s,t)\bm\sigma_{K_2,\bpsi}(1)\right\}  h_y^2 +O_p(h_y^3+h_y/\sqrt{n}).
			\ee
			
			Using similar arguments, we can also show that
			\[
			\|\wh\B_{n,h_y,1}(s,t)-\B_{h_y}(s,t)\|_{\max} = O_p\{(nh_y^2)^{-1/2}\}\text{ and } \|\B_{h_y}(s,t)-\rho_2(s,t)\Q_{K_2,\bpsi}^{\{1\}}\|_{\max}=O(h_y), 
			\]
			together with the Assumption~C2(b) requiring $\Q_{K_2,\bpsi}^{\{1\}}$ to be positive semi-definite with $\lambda_{\min}\left[\Q_{K_s,\bpsi}^{(1)}\right]>0$, and condition~C4 that requires  $\rho_2(s,t)>0$, we have that 
			\be\label{eq:B_hat_consist}
			\begin{split}
				&\left\|\left[\B_{h_y}(s,t)\right]^{-1}-\left[\rho_2(s,t)\Q_{K_2,\bpsi}^{\{1\}}\right]^{-1}\right\|\\&
				\leq
				\left\|\left[\wh\B_{n,h_y,1}(s,t)\right]^{-1}-\left[\B_{h_y}(s,t)\right]^{-1}\right\|_{\max} + \left\|\left[\B_{h_y}(s,t)\right]^{-1}-\left[\rho_2(s,t)\Q_{K_2,\bpsi}^{\{1\}}\right]^{-1}\right\|_{\max}\\
				&= O_p[(nh_y^2)^{-1/2}]+O(h_y). 
			\end{split}
			\ee
			Therefore, \eqref{cy-dc} gives that
			\be\label{eq:cy_expansion}
			\wc C_Y(s,t) -C_Y^*(s,t) = \CB_Y(s,t) h_y^{2}  + \mCR_{Y}(s,t) +O_p\left[h_y^{3}+ h_y(nh_y^2)^{-1/2}{+(nh_y^2)^{-1}}+h_y^2/\sqrt{n}\right],
			\ee
			where $\CB_Y(s,t) =  \frac{1}{2}\e_{3}^\top\left[\Q_{K_2,\bpsi}^{\{1\}}\right]^{-1}\left[C_{Y}^{*(2,0)}(s,t)\bm\sigma_{K_2,\bpsi}(0)+C_{Y}^{*(0,2)}(s,t)\bm\sigma_{K_2,\bpsi}(1)\right] $ and that\\ $\mCR_{Y}(s,t)=\e_{3}^\top\left[\Q_{K_2,\bpsi}^{\{1\}}\right]^{-1}\frac{\CR_{20}(s,t)}{\rho_2(s,t)}$, which can be written as
			\be
			\label{RY}
			\mCR_{Y}(s,t)=\frac{1}{n\rho_2(s,t)}\sum_{i=1}^n\mathop{\sum\sum}^{\ne}_{u,v\in N_i}\left\{W_i(u,v)-C_Y^*(u,v)\right\}K_{2,h_y}(u-s,v-t)w_{\bpsi,h_y}(u-s,v-t),
			\ee
			with $w_{\bpsi,h_y}(u-s,v-t)=\e_{3}^\top\left[\Q_{K_2,\bpsi}^{\{1\}}\right]^{-1}\bpsi_{h_y}(u-s,v-t)$.

			Finally, we study the difference between $\wc C_{Y}(s,t)- \wt C_{Y}(s,t)$, which is of the form
			\be
			\label{CY-diff}
			\begin{split}
				\wc C_{Y}(s,t)&- \wt C_{Y}(s,t)=\e_{3}^\top\left[\wh\B_{n,h_y,1}(s,t)\right]^{-1}\\&\times\left[\frac{1}{n}\sum_{i=1}^n\mathop{\sum\sum}^{\ne}_{u,v\in N_i}\left\{W_i(u,v)-\wt W_i(u,v)\right\}K_{2,h_y}(u-s,v-t)\bpsi_{h_y}(u-s,v-t)\right],
			\end{split}
			\ee
			where $\wt W_i(u,v)=[Z_i(u)-\wt\mu(u)][Z_i(v)-\wt\mu(v)]$. Using the expansion~\eqref{eq:mu_tilde_uniform_rate2}, we can show that
			\be
			\label{W-dc}
			\begin{split}
				W_i(u,v)&-\wt W_i(u,v)=[\wt\mu(u)-\mu^*(u)][Z_i(v)-\mu^*(v)]+[\wt\mu(v)-\mu^*(v)][Z_i(u)-\mu^*(u)]\\&\qquad-[\wt\mu(u)-\mu^*(u)][\wt\mu(v)-\mu^*(v)]\\
				&=[\CB_\mu(u) h_{\sigma}^{2}  + \mCR_{\mu}(u) +O_p\{ h_{\mu}^{3} + h_{\mu}\{\log(n)/ (n h_{\mu})\}^{1/2}\}][Z_i(v)-\mu^*(v)]\\
				&\qquad+[\CB_\mu(v) h_{\mu}^{2}  + \mCR_{\mu}(v) +O_p\{ h_{\mu}^{3} + h_{\mu}\{\log(n)/ (n h_{\mu})\}^{1/2}\}][Z_i(u)-\mu^*(u)]\\
				&\qquad -[\wt\mu(u)-\mu^*(u)][\wt\mu(v)-\mu^*(v)].
			\end{split}
			\ee
			With the above expansion, together with the same arguments as those in~\eqref{sig-diff}, we proceed to show that for the bounded weight function $w_{\bpsi,h_y}(u-s,v-t)=\e_{3}^\top\left[\Q_{K_2,\bpsi}^{\{1\}}\right]^{-1}\bpsi_{h}(u-s,v-t)$, one has that
			{\small \bea
				&&\hskip -5em\frac{1}{n}\sum_{i=1}^n\mathop{\sum\sum}^{\ne}_{u,v\in N_i}|Z_i(u)-\mu^*(u)|K_{2,h_y}(u-s,v-t)|w_{\bpsi,h_y}(u-s,v-t)|=O_p(1),\label{e1}\\
				&&\hskip -5em\frac{1}{n}\sum_{i=1}^n\mathop{\sum\sum}^{\ne}_{u,v\in N_i}[Z_i(u)-\mu^*(u)]K_{2,h_y}(u-s,v-t)w_{\bpsi,h_y}(u-s,v-t)=O_p\left(\frac{1}{\sqrt{nh_y^2}}\right),\label{e2}\\
				&&\hskip -5em\frac{1}{n}\sum_{i=1}^n\mathop{\sum\sum}^{\ne}_{u,v\in N_i}[Z_i(u)-\mu^*(u)]\mCR_{\mu}(v)K_{2,h_y}(u-s,v-t)w_{\bpsi,h_y}(u-s,v-t)=O_p\left(\frac{1}{nh_{\mu}}+\frac{1}{\sqrt{n^2h_y^2h_{\mu}}}\right).\label{e3}
				\eea}
			The proof of~\eqref{e1} is straightforward, and the proof of~\eqref{e2} follows similarly to finding the convergence rate of $\CR_{20}(s,t)$ in~\eqref{cy-dc}, hence is omitted. Next, we focus on the proof of~\eqref{e3}.
			
			Recall the definition of $\mCR_{\mu}(\cdot)$ in~\eqref{eq:mu_tilde_expansion} and using Lemma~\ref{lm:stein} and Lemma~\ref{lm:stein_extended} repeatedly, some tedious algebra gives that
			\bse
			&&\frac{1}{n}\sum_{i=1}^n\mathop{\sum\sum}^{\ne}_{u,v\in N_i}[Z_i(u)-\mu^*(u)]\mCR_{\mu}(v)K_{2,h_y}(u-s,v-t)w_{\bpsi,h_y}(u-s,v-t)\\
			&=&\frac{1}{n^2}\sum_{i=1}^n\sum_{i'=1}^n\mathop{\sum\sum}^{\ne}_{u,v\in N_i}\sum_{u'\in N_{i'}}\frac{[Z_i(u)-\mu^*(u)]}{\rho(v)}\{ Z_{i'}(u')-\mu^*(u')\}K_{2,h_y}(u-s,v-t)w_{\bpsi,h_y}(u-s,v-t)\\
			&&
			\hskip 20em\times K_{1,h_{\mu}} (v-u')w_{\bphi,h_{\mu}}(u'-v)\\
			&=&\frac{1}{n^2}\sum_{i=1}^n\mathop{\sum\sum}^{\ne}_{u,v\in N_i}\frac{[Z_i(u)-\mu^*(u)]^2}{\rho(v)}K_{2,h_y}(u-s,v-t)w_{\bpsi,h_y}(u-s,v-t) K_{1,h_{\mu}} (v-u)w_{\bphi,h_{\mu}}(u-v)\\
			&&+\frac{1}{n^2}\sum_{i=1}^n\mathop{\sum\sum}^{\ne}_{u,v\in N_i}\frac{[Z_i(u)-\mu^*(u)]}{\rho(v)}\{ Z_{i}(v)-\mu^*(v)\}K_{2,h_y}(u-s,v-t)w_{\bpsi,h_y}(u-s,v-t)K_{1,h_{\mu}} (0)w_{\bphi,h_{\mu}}(0)\\
			&&+\frac{1}{n^2}\sum_{i=1}^n\mathop{\sum\sum\sum}^{\ne}_{u,v,u'\in N_i}\frac{[Z_i(u)-\mu^*(u)]}{\rho(v)}\{ Z_{i}(u')-\mu^*(u')\}K_{2,h_y}(u-s,v-t)w_{\bpsi,h_y}(u-s,v-t)\\
			&&
			\hskip 20em\times K_{1,h_{\mu}} (v-u')w_{\bphi,h_{\mu}}(u'-v)\}\\
			&&+\frac{1}{n^2}\sum_{i=1}^n\sum_{i'=1,i'\neq i}^n\mathop{\sum\sum}^{\ne}_{u,v\in N_i}\sum_{u'\in N_{i'}}\frac{[Z_i(u)-\mu^*(u)]}{\rho(v)}\{ Z_{i'}(u')-\mu^*(u')\}K_{2,h_y}(u-s,v-t)w_{\bpsi,h_y}(u-s,v-t)\\
			&&
			\hskip 30em\times K_{1,h_{\mu}} (v-u')w_{\bphi,h_{\mu}}(u'-v)\}\\
			&=&O_p\left(\frac{1}{nh_{\mu}}+\frac{1}{\sqrt{n^3h_y^2h_{\mu}^2}}\right)+O_p\left(\frac{1}{nh_{\mu}}+\frac{1}{\sqrt{n^3h_y^2h_{\mu}^2}}\right)+O_p\left(\frac{1}{n}+\frac{1}{\sqrt{n^3h_y^2h_{\mu}}}\right)+O_p\left(\frac{1}{\sqrt{n^2h_y^2h_{\mu}}}\right)\\
			&=&O_p\left(\frac{1}{nh_{\mu}}+\frac{1}{\sqrt{n^2h_y^2h_{\mu}}}\right).
			\ese
			Finally, using equations~\eqref{CY-diff}-\eqref{e3}, together with the same arguments in~\eqref{sig-diff}, we have that
			\[
			\begin{split}
			&\wc C_{Y}(s,t)- \wt C_{Y}(s,t)\\
			&=O_p\left(h_{\mu}^2\times\frac{1}{\sqrt{nh_y^2}}\right)+O_p\left(\frac{1}{nh_{\mu}}+\frac{1}{\sqrt{n^2h_y^2h_{\mu}}}\right)+O_p[ h_{\mu}^{3} + h_{\mu}\{\log(n)/ (n h_{\mu})\}^{1/2}]\times O_p(1)\\
			&\qquad+O_p[ h_{\mu}^{4} + \log(n)/ (n h_{\mu})]\times O_p(1)\\
			&=O_p\left( h_{\mu}^{3} + h_{\mu}\{\log(n)/ (n h_{\mu})\}^{1/2}+ \frac{h_{\mu}^2}{\sqrt{nh_y^2}}+\frac{\log (n)}{nh_{\mu}}+\frac{1}{\sqrt{n^2h_y^2h_{\mu}}}\right).
			\end{split}
			\]
			\[
			O_p\left[h_y^{3}+ h_y(nh_y^2)^{-1/2}{+(nh_y^2)^{-1}}+h_y^2/\sqrt{n}+h_{\mu}^{3} + h_{\mu}\{\log(n)/ (n h_{\mu})\}^{1/2}+ \frac{h_{\mu}^2}{\sqrt{nh_y^2}}+\frac{\log (n)}{nh_{\mu}}+\frac{1}{\sqrt{n^2h_y^2h_{\mu}}}\right]
			\]
			Therefore, under Assumptions C3 and C5, the difference  $\wc C_{Y}(s,t)- \wt C_{Y}(s,t)$ is negligible compare to $\wc C_Y(s,t) -C_Y^*(s,t)$ in~\eqref{eq:cy_expansion}, hence we have the asymptotic expansion for $\wt C_Y(s,t)$ as
			\be\label{eq:cy_expansion1}
			\wt C_Y(s,t) -C_Y^*(s,t) = \CB_Y(s,t) h_y^{2}  + \mCR_{Y}(s,t)+o_p\left[h_y^2+ (nh_y^2)^{-1/2}\right],
			\ee
			where $\CB_Y(s,t)$ and $\mCR_{Y}(s,t)$ are as defined in~\eqref{eq:cy_expansion}.
			
			Then by applying similar arguments in~\cite{li2010uniform} and Assumption C3, we have that
			\[
			\sup_{s,t\in \CT} \bigg |\wt C_Y(s,t) -C_Y^*(s,t) \bigg |  = O_p\left[h_y^2+ (nh_y^2)^{-1/2}\sqrt{\log(n)}\right],
			\]
			which completes the proof of Theorem~\ref{thm1}.
			\subsection{Proof of Theorem~\ref{thm2}}
			\subsubsection{Part (a): uniform convergence of $\wh C_{XY}(\cdot,\cdot) $}
			Define the counterparts of $\wh C_{XY}(s,t) $ in~\eqref{cxyhat} when $\mu^*(\cdot)$ is known as follows
			\[
			\wc C_{XY}(s,t)=\e_{3}^\top\left[\wh\B_{n,h_{xy},1}(s,t)\right]^{-1}\left[\frac{1}{n}\sum_{i=1}^n\mathop{\sum\sum}^{\ne}_{u,v\in N_i}[Z_i(v)-\mu^*(v)]K_{2,h_{xy}}(u-s,v-t)\bpsi_{h_{xy}}(u-s,v-t)\right]
			\]
			It is straightforward to show that
			\[
			\begin{split}
			\wc C_{XY}(s,t) -C_{XY}(s,t)=& \e_{3}^\top
			\left[\wh\B_{n,h_{xy},1}(s,t)\right]^{-1}\\&\times\underbrace{\left[\frac{1}{n}\sum_{i=1}^n\mathop{\sum\sum}^{\ne}_{u,v\in N_i}[Z_i(v)-\mu^*(v)-C_{XY}(s,t)]K_{2,h_{xy}}(u-s,v-t)\bpsi_{h_{xy}}(u-s,v-t)\right]}_{\CR_{XY}(s,t)}, 
			\end{split}
			\]
			where $\wh\B_{n,h_{xy},1}(s,t)$ is as defined in~\eqref{Bhat}, and $\CR_{XY}(s,t)=\CR_{30}(s,t) +\CR_{{31}}(s,t)+\CR_{{32}}(s,t)+O_p(h_{xy}^3+h_{xy}^2/\sqrt{n})$, {with $\CR_{{31}}(s,t)$, $\CR_{{32}}(s,t)$ similarly defined as $\CR_{21}(s,t),\CR_{22}(s,t)$ in~\eqref{CY-dc0} ($h_y$ replaced by $h_{xy}$ and partial derivatives of $C_Y(\cdot,\cdot)$ replaced by partial derivatives of $C_{XY}(\cdot,\cdot)$)}, and 
			\bse
			&& \CR_{30}(s,t) ={1\over n}\sum_{i=1}^n  \mathop{\sum\sum}^{\ne}_{u,v\in N_i} \bigg \{Z_i(v)-\mu^*(v) -C_{XY}(u,v) \bigg\} K_{2,h_{xy}}(u-s,v-t)\bpsi_{h_{xy}}(u-s,v-t).
			\ese
			Due to the same arguments to achieve~\eqref{cy-dc}, we have that
			\be
			\label{cxy-dc}
			\wc C_{XY}(s,t) -C_{XY}(s,t)= \e_{3}^\top
			\left[\wh\B_{n,h_{xy},1}(s,t)\right]^{-1} \left[\CR_{30}(s,t) + \CR_{{32}}(s,t)\right] +O_p(h_{xy}^3+h_{xy}^2/\sqrt{n}).\\
			\ee
			By Stein's lemma, it immediately follows that
			\bse
			\E\{\CR_{30}(s,t) \}&=&  \int_\CT \int_\CT \E\left[ \left \{Z_i(v)-\mu^*(v) -C_{XY}(u,v) \right\} \exp\{ X(u)+X(v)\} \right]  \\
			&&\hskip50mm  \times \lambda_0(u) \lambda_0(v) K_{2,h_{xy}}(u-s,v-t)\bpsi_{h_{xy}}(u-s,v-t) du dv, \\
			&=&0,
			\ese

			and denote for a vector $\a$ with $\a^{dag}=\a\a^\top$, by Lemma \ref{lm:stein} and Assumptions C1-C5, we have that
			\bse
			\Cov \{\CR_{30}(s,t) \}&=&  {1\over n} \E \left[ \mathop{\sum\sum}^{\ne}_{u,v\in N_i} \left \{Z_i(v)-\mu^*(v) -C_{XY}(u,v) \right\}K_{2,h_{xy}}(u-s,v-t)\bpsi_{h_{xy}}(u-s,v-t)\right]^{dag}\\
			&=&{1\over n}  \int_\CT \int_\CT \E\left[ \left \{Z_i(v)-\mu^*(v) -C_{XY}(u,v) \right\}^2 \exp\{ X(u)+X(v)\} \right]  \\
			&&\hskip20mm  \times \lambda_0(u) \lambda_0(v) K_{2,h_{xy}}^2(u-s,v-t)\bpsi_{h_{xy}}^{dag}(u-s,v-t) du dv +o\{ (nh_{xy}^2)^{-1}\} \\
			%
			&=& {\sigma_Z^2(s)+\left\{C_{XY}(s,t)-C_{XY}(t,s)\right\}^2\over nh_{xy}^2} \rho_2(s,t) \Q_{K_2,\bpsi}^{\{2\}}  +o\{ (nh_{xy}^2)^{-1}\} ,
			\ese
			where $\Q_{K_2,\bpsi}^{\{2\}}$ is defined in Assumption C2(b).
			
			{
				And similar to~\eqref{R22-dc}, we have that
				\be
				\label{R32-dc}
				\CR_{32}(s,t) =   \frac{\rho_2(s,t)}{2}\left\{C_{XY}^{*(2,0)}(s,t)\bm\sigma_{K_2,\bpsi}(0)+C_{XY}^{*(0,2)}(s,t)\bm\sigma_{K_2,\bpsi}(1)\right\}  h_{xy}^2 +O_p(h_{xy}^3+h_{xy}/\sqrt{n}).
				\ee
			}
			Using similar arguments as those to achieve decomposition~\eqref{eq:cy_expansion}, we have that
			\be\label{eq:cxy_expansion}
			\wc C_{XY}(s,t) -C_{XY}(s,t) = \CB_{XY}(s,t) h_{xy}^{2}  + \mCR_{XY}(s,t) +O_p\left[h_{xy}^3+h_{xy}(nh_{xy}^2)^{-1/2}+h_{xy}^2/\sqrt{n}\right],
			\ee
			{where $\CB_{XY}(s,t)=\frac{\rho_2(s,t)}{2}\left\{C_{XY}^{*(2,0)}(s,t)\bm\sigma_{K_2,\bpsi}(0)+C_{XY}^{*(0,2)}(s,t)\bm\sigma_{K_2,\bpsi}(1)\right\}$} and\\  $\mCR_{XY}(s,t)=\e_{3}^\top\left[\Q_{K_2,\bpsi}^{\{1\}}\right]^{-1}\frac{\CR_{30}(s,t)}{\rho_2(s,t)}$, which can be written as
			\be
			\label{RXY}
			\mCR_{XY}(s,t)=\frac{1}{n\rho_2(s,t)}\sum_{i=1}^n\mathop{\sum\sum}^{\ne}_{u,v\in N_i}\left \{Z_i(v)-\mu^*(v) -C_{XY}(u,v) \right\}K_{2,h_{xy}}(u-s,v-t)w_{\bpsi,h_{xy}}(u-s,v-t),
			\ee
			with $w_{\bpsi,h_{xy}}(u-s,v-t)=\e_{3}^\top\left[\Q_{K_2,\bpsi}^{\{1\}}\right]^{-1}\bpsi_{h_{xy}}(u-s,v-t)$.

			Next, we study the difference between $\wc C_{XY}(s,t)- \wh C_{XY}(s,t)$, which can be shown of the form
			\be
			\label{CXY-diff}
			\begin{split}
				\wc C_{XY}(s,t)&- \wh C_{XY}(s,t)=\e_{3}^\top\left[\wh\B_{n,h_{xy},1}(s,t)\right]^{-1}\\&\times\left[\frac{1}{n}\sum_{i=1}^n\mathop{\sum\sum}^{\ne}_{u,v\in N_i}\left\{\tilde{\mu}(v)-\mu^*(v)\right\}K_{2,h_{xy}}(u-s,v-t)\bpsi_{h_{xy}}(u-s,v-t)\right].
			\end{split}
			\ee
			Using the expansion~\eqref{eq:mu_tilde_expansion} together with the same arguments in~\eqref{sig-diff}, and using equations~\eqref{CY-diff}-\eqref{e3}, we have that under Assumptions C3 and C5,
			\[
			\begin{split}
			\wc C_{XY}(s,t)- \wh C_{XY}(s,t)&=O_p\left(h_{\mu}^2\times \frac{1}{\sqrt{nh_{xy}^2}}\right)+O_p\left(\frac{1}{nh_{\mu}}+\frac{1}{\sqrt{n^2h_{xy}^2h_{\mu}}}\right)+O_p[ h_{\mu}^{3}\\
			&\qquad+ h_{\mu}\{\log(n)/ (n h_{\mu})\}^{1/2}]\times O_p(1)
			\\&=O_p\left( h_{\mu}^{3} + h_{\mu}\{\log(n)/ (n h_{\mu})\}^{1/2}+ \frac{h_{\mu}^2}{\sqrt{nh_{xy}^2}}+\frac{\log(n)}{nh_{\mu}}+\frac{1}{\sqrt{n^2h_{xy}^2h_{\mu}}}\right).
			\end{split}
			\]
			
			Therefore, under Assumptions C3 and C5, the difference  $\wc C_{XY}(s,t)- \wh C_{XY}(s,t)$ is negligible compare to $\wh C_{XY}(s,t) -C_{XY}(s,t)$ in~\eqref{eq:cxy_expansion}, which yields the asymptotic expansion for $\wh C_{XY}(s,t)$ as
			\be\label{eq:cxy_expansion1}
			\wh C_{XY}(s,t) -C_{XY}(s,t) = \CB_Y(s,t) h_{xy}^{2}  + \mCR_{XY}(s,t) +o_p[h_{xy}^{2}+(nh_{xy}^2)^{-1/2}],
			\ee
			where $\CB_Y(s,t)$ and $\mCR_{XY}(s,t)$ are as defined in~\eqref{eq:cy_expansion} and~~\eqref{RXY}, respectively.
			
			Then by applying similar arguments in~\cite{li2010uniform} and Assumption C3, we have that
			\[
			\sup_{s,t\in \CT} \bigg |\wh C_{XY}(s,t) -C_{XY}(s,t) \bigg |  = O_p\left[h_{xy}^2+ (nh_{xy}^2)^{-1/2}\sqrt{\log(n)}\right].
			\]
			\subsubsection{Parts (b)-(c): uniform convergence of $\wh\mu(\cdot)$ and $\wh C_Y(\cdot,\cdot)$}
			Using the triangular inequality and Theorem~\ref{thm1} part (a), we have that
			\[
			\begin{split}
			\sup_{s\in \CT} \bigg |\wh \mu(s) -\mu(s) \bigg | &=\sup_{s\in \CT} \bigg |\wt \mu(s)-\wh C_{XY}(s,s) -\mu(s)\bigg |\\
			&\le\sup_{s\in \CT} \bigg |\wt \mu(s) -\mu(s)- C_{XY}(s,s)\bigg |+\sup_{s\in \CT} \bigg |C_{XY}(s,s)-\wh C_{XY}(s,s)\bigg |\\
			&=O_p\left[h_\mu^2+ (nh_\mu)^{-1/2}\sqrt{\log(n)}\right]+O_p\left[h_{xy}^2+ (nh_{xy}^2)^{-1/2}\sqrt{\log(n)}\right]\\
			&=O_p\left[h_\mu^2+h_{xy}^2+(nh_\mu)^{-1/2}\sqrt{\log(n)}+ (nh_{xy}^2)^{-1/2}\sqrt{\log(n)}\right].
			\end{split}
			\]
			Using the triangular inequality and Theorem~\ref{thm1} part (c), we have that
			\bse
			\sup_{s,t\in \CT} \bigg |\wh C_Y(s,t) -C_Y(s,t) \bigg | &=&\sup_{s,t\in \CT} \bigg |\wt C_{XY}(s,t) -\wh C_{XY}(s,t)\wh C_{XY}(t,s) -C_Y(s,t)\bigg |\\
			&\leq& \sup_{s,t\in \CT} \bigg |\wt C_{XY}(s,t)  -C_Y(s,t)- C_{XY}(s,t) C_{XY}(t,s)\bigg |\\
			&&\qquad+\sup_{s,t\in \CT} \bigg |\wh C_{XY}(s,t)\wh C_{XY}(t,s)- C_{XY}(s,t) C_{XY}(t,s)\bigg |\\
			&\leq& \sup_{s,t\in \CT} \bigg |\wt C_{XY}(s,t)  -C_Y(s,t)- C_{XY}(s,t) C_{XY}(t,s)\bigg |\\
			&&\qquad+\sup_{s,t\in \CT} \bigg |\left[\wh C_{XY}(s,t)- C_{XY}(s,t)\right]\left[\wh C_{XY}(t,s)- C_{XY}(t,s)\right]\bigg |\\
			&&\qquad+2\sup_{s,t\in \CT} \bigg |\left[\wh C_{XY}(s,t)- C_{XY}(s,t)\right] C_{XY}(t,s)\bigg |\\
			&=&O_p\left[h_y^2+ (nh_y^2)^{-1/2}\sqrt{\log(n)}\right]+O_p\left[h_{xy}^2+ (nh_{xy}^2)^{-1/2}\sqrt{\log(n)}\right]\\
			&=&O_p\left[h_y^2+h_{xy}^2+(nh_y^2)^{-1/2}\sqrt{\log(n)}+ (nh_{xy}^2)^{-1/2}\sqrt{\log(n)}\right],
			\ese
			which completes the proof of Theorem~\ref{thm2}.

			\subsection{Proof of Theorem~\ref{thm3}}
			Recall the definition of the test statistic
			$$
			T_n={1\over n} \sum_{i=1}^n\mathop{\sum\sum}^{\ne}_{u,v\in N_i} [Z_i(v)-\wt\mu(v)]^2-{1\over n} \sum_{i=1}^n\mathop{\sum\sum}^{\ne}_{u,v\in N_i} \wt\sigma_Z^2(v),
			$$
			which can be decomposed into four parts
			\bse
			T_n=   T_{n,1} +   T_{n,2} +  T_{n,3} +  T_{n,4},
			\ese
			where
			\bse
			&&  T_{n,1}={1\over n} \sum_{i=1}^n\mathop{\sum\sum}^{\ne}_{u,v\in N_i} [Z_i(v)-\mu^*(v)]^2-{1\over n} \sum_{i=1}^n\mathop{\sum\sum}^{\ne}_{u,v\in N_i} \sigma_Z^2(v), \\
			&&   T_{n,2}=  {1\over n} \sum_{i=1}^n\mathop{\sum\sum}^{\ne}_{u,v\in N_i} [\wt\mu(v)-\mu^*(v)]^2 ,\\
			&&  T_{n,3} = - {2\over n} \sum_{i=1}^n\mathop{\sum\sum}^{\ne}_{u,v\in N_i} [Z_i(v)-\mu^*(v)][\wt\mu(v)-\mu^*(v)],\\
			&&  T_{n,4}=-{1\over n} \sum_{i=1}^n\mathop{\sum\sum}^{\ne}_{u,v\in N_i} [\wt\sigma_Z^2(v)-\sigma_Z^2(v)].
			\ese
			As we have shown in the main text that motivates the test statistic, it holds that $\E   T_{n,1}=0$. Furthermore, some straightforward but tedious calculations yield that $\Var(  T_{n,1})=O(n^{-1})$, and we therefore conclude \textcolor{blue}{$  T_{n,1}=O_p(n^{-1/2})$}. For the second part $  T_{n,2}$, by the uniform convergence rate result (\ref{eq:mu_tilde_uniform_rate}) for $\wt \mu(\cdot)$, it is straightforward to show that $  T_{n,2}=O_p \left[h_{\mu}^4 + \log(n) / (nh_{\mu})\right]$. 
			
			Next, we proceed to derive the asymptotic rate of $  T_{n,3}$. Using the asymptotic expansion (\ref{eq:mu_tilde_expansion}), we have that
			\bse
			T_{n,3} &=&- {2\over n} \sum_{i=1}^n\mathop{\sum\sum}^{\ne}_{u,v\in N_i} \{Z_i(v)-\mu^*(v)\} \{ \CB_\mu(v) h_{\mu}^2 +\mCR_{\mu}(v)\} + O_p\{h_{\mu}^{3}+ h_{\mu}\sqrt{\log (n)/(nh_{\mu})}\}\\
			&\equiv&   T_{n,31} +  T_{n,32} +O_p\left[h_{\mu}^{3}+ h_{\mu}\sqrt{\log (n)/(nh_{\mu})}\right],
			\ese
			where $  T_{n,31}=- {2h_{\mu}^2\over n} \sum_{i=1}^n\mathop{\sum\sum}^{\ne}_{u,v\in N_i} \{Z_i(v)-\mu^*(v)\}  \CB_\mu(v) $ and $  T_{n,32}
			=- {2\over n} \sum_{i=1}^n\mathop{\sum\sum}^{\ne}_{u,v\in N_i} \{Z_i(v)-\mu^*(v)\} \mCR_{\mu}(v)$. For $  T_{n,31}$, using Lemma~\ref{lm:stein}, we have that, under $H_0$,
			\bse
			\E  T_{n,31} &=&- 2h_{\mu}^2 \int \int \E[ \{Z_i(v)-\mu^*(v)\}  \exp\{X(u)+ X(v)\} ] \CB_\mu(v) \lambda_0(u) \lambda_0(v) du dv\\
			&=&- 2h_{\mu}^2 \int \int C_{XY}(u,v) \CB_\mu(v)  \rho_2(u,v) dudv\\
			&=& 0 \quad \quad\quad \hbox{(under $H_0: C_{XY}(\cdot, \cdot)\equiv 0$)}.
			\ese
			A direct application of the central limit theorem to $  T_{n,31}$ gives that $  T_{n,31}=  O_p(h_{\mu}^2 n^{-1/2})$ under $H_0$. 
			
			For $  T_{n,32}$, we can further make the following decomposition
			\bse
			T_{n,32} &=& - {2\over  n^2 } \sum_{i=1}^n\sum_{v\in N_i}\sum_{i'=1}^n\sum_{u\in N_{i'}} \frac{\{Z_i(v)-\mu^*(v) \} \{Z_{i'}(u)-\mu^*(u) \} }{\rho(v)} K_{1,h_{\mu}}(u-v)w_{\bphi,h_{\mu}}(u-v)\\
			&=&   T_{n,321} +   T_{n,322} +  T_{n,323},
			\ese
			where
			\bse
			%
			%
			T_{n,321} &=&-{2\over n^2 } 	 \sum_{i=1}^n\sum_{u\in N_i} \frac{\{Z_i(u)-\mu^*(u) \}^2}{\rho(u)}  K_{1,h_{\mu}}(0)w_{\bphi,h_{\mu}}(0),  \\
			T_{n,322}&=& -{2\over n^2 } \sum_{i=1}^n\mathop{\sum\sum}^{\ne}_{u,v\in N_i} \frac{\{Z_{i}(v)-\mu^*(v) \}  \{Z_i(u)-\mu^*(u) \} }{\rho(v)} K_{1,h_{\mu}}(u-v)w_{\bphi,h_{\mu}}(u-v), \\
			T_{n,323}&=& -{2\over n^2 } \sum_{i=1}^n\sum_{u\in N_i}\sum_{i' \neq i} \sum_{v\in N_{i'}}\frac{\{Z_i(v)-\mu^*(v) \} \{Z_{i'}(u)-\mu^*(u) \}}{\rho(v)}  K_{1,h_{\mu}}(u-v)w_{\bphi,h_{\mu}}(u-v). 
			%
			%
			\ese
			Under Assumptions C1-C2, some straightforward calculations give that $\E(  T_{n,321})=O[(nh_\mu)^{-1}]$ and $\Var\left(  T_{n,321}\right)=O[(nh_\mu)^{-2}n^{-1}]$. Therefore, we have that $  T_{n,321}= O_p[(nh_{\mu})^{-1}]$. Similarly, we can show that 
			$
			\E   T_{n,322} = O(n^{-1}),
			$
			and $\Var(  T_{n,322})= O[ (n^3 h_{\mu})^{-1}]$, and hence $  T_{n,322}=O_p(1/n)$. Similar calculations show $\E(   T_{n,323})=0$ and $\Var(  T_{n,323})= O[(n^2 h_{\mu})^{-1}]$, which gives that $  T_{n,323}=O_p[(n^2 h_{\mu})^{-1/2}]$. We therefore conclude that 
			\[
			T_{n,3}=O_p\left[h_{\mu}^{3}+ h_{\mu}\sqrt{\log (n)/(nh_{\mu})}+(n^2 h_{\mu})^{-1/2}\right].
			\]
			
			Finally, it remains to derive the convergence rates of $  T_{n,4}$. Using the asymptotic expansion of $\wt \sigma_Z^2(\cdot)$ in \eqref{eq:sigma2_tilde_uniform_rate2}, one has that
			\bse
			T_{n,4}&=& -{1\over n} \sum_{i=1}^n\mathop{\sum\sum}^{\ne}_{u,v\in N_i} \bigg[ \CB_{\sigma^2}(v) h_{\sigma}^2  + {1\over n \rho(v)} \sum_{i'=1}^n\sum_{s\in N_{i'}}[ \{Z_{i'}(s)-\mu^*(s)\}^2-\sigma_Z^2(s)] K_{1,h_{\sigma}}(v-s)w_{\bphi,h_{\sigma}}(v-s) \bigg ]\\
			&&\qquad  +O_p\left[h_{\sigma}^{3}+(nh_{\sigma})^{-1}\sqrt{\log(n)}+ h_{\sigma}(nh_{\sigma})^{-1/2}\log(n)\right] \\
			&\equiv&   T_{n,41}+   T_{n,42}  +O_p\left[h_{\sigma}^{3}+(nh_{\sigma})^{-1}\sqrt{\log(n)}+ h_{\sigma}(nh_{\sigma})^{-1/2}\log(n)\right], 
			\ese
			where 
			$
			T_{n,41}= -{1\over n} \sum_{i=1}^n\mathop{\sum\sum}^{\ne}_{u,v\in N_i}  \CB_{\sigma^2}(v) h_{\sigma}^2 
			= - h_{\sigma}^2 \int \int \CB_{\sigma^2}(v) \rho_2(u,v) du dv +O_p(h_{\sigma}^2 n^{-1/2}).
			$
			Next, we decompose $   T_{n,42}$ as $  T_{n,42}=   T_{n,421}+   T_{n,422} +  T_{n,423}$, where
			\bse
			T_{n,421} &=& -{1\over n^2 } \sum_{i=1}^n\mathop{\sum\sum}^{\ne}_{u,v\in N_i} {1\over  \rho(v)} [ \{Z_i(v)-\mu^*(v)\}^2-\sigma_Z^2(v)] K_{1,h_{\sigma}}(0)w_{\bphi,h_{\sigma}}(0), \\
			T_{n,422} &=& -{1\over n^2 } \sum_{i=1}^n\mathop{\sum\sum}^{\ne}_{u,v\in N_i} {1\over  \rho(v)} [ \{Z_i(u)-\mu^*(u)\}^2-\sigma_Z^2(u)] K_{1,h_{\sigma}}(u-v)w_{\bphi,h_{\sigma}}(u-v), \\
			T_{n,423} &=& -{1\over n^2} \sum_{i=1}^n\mathop{\sum\sum}^{\ne}_{s, u,v\in N_i} {1\over  \rho(v)} [ \{Z_i(s)-\mu^*(s)\}^2-\sigma_Z^2(s)] K_{1,h_{\sigma}}(v-s)w_{\bphi,h_{\sigma}}(v-s), \\
			T_{n,424}&=& -{1\over n^2} \sum_{i=1}^n\mathop{\sum\sum}^{\ne}_{u,v\in N_i}  \sum_{i' \neq i }\sum_{s\in N_{i'}}{1\over \rho(v)} [ \{Z_{i'}(s)-\mu^*(s)\}^2-\sigma_Z^2(s)] K_{1,h_{\sigma}}(v-s)w_{\bphi,h_{\sigma}}(v-s).
			\ese
			It is straightforward to see that 
			$
			\E (  T_{n,421} ) =O[(nh_{\sigma})^{-1}]
			$
			and $\Var (  T_{n,421} ) =O(n^{-3} h_{\sigma}^{-2})$, therefore we conclude $  T_{n,421}= O_p[(nh_{\sigma})^{-1}]$. Similar calculations show that $  T_{n,422}= O_p(n^{-1})$ and $  T_{n,423}= O_p(n^{-1})$.

			To bound the second term~$  T_{n,424}$, define 
			\[
			T_{n,424}^*=-{1\over n} \sum_{i=1}^n\sum_{s\in N_{i}} \{[Z_{i}(s)-\mu^*(s)]^2-\sigma_Z^2(s)\}  \tau(s),
			\]
			where  $\tau(s)={\int_\CT \rho_2(u,s)du}/{\rho(s)}$. Then it immediately follows that
			\[
			T_{n,424}-  T_{n,424}^*={1\over n} \sum_{i=1}^n\sum_{s\in N_{i}}\{[Z_{i}(s)-\mu^*(s)]^2-\sigma_Z^2(s)\} S_{-i}(s),
			\]
			where $S_{-i}(s)=\tau(s)-{1\over n} \sum_{i'=1,i'\ne i}^n\mathop{\sum\sum}^{\ne}_{u,v\in N_{i'}}  K_{1,h_{\sigma}}(v-s)w_{\bphi,h_{\sigma}}(v-s)/\rho(v)$. It can be readily shown that 
			\bse
			\E\left[S_{-i}(s)\right]&=&\tau(s)-\frac{n-1}{n} \int_\CT \int_\CT \frac{\rho_2(u,v)}{\rho(v)} K_{1,h_{\sigma}}(v-s)w_{\bphi,h_{\sigma}}(v-s) du dv\\
			&=&\tau(s)-\frac{n-1}{n} \int_\CT \int_\CT \frac{\rho_2(u,s)}{\rho(s)} K_{1,h_{\sigma}}(v-s)w_{\bphi,h_{\sigma}}(v-s) du dv\\
			&&\qquad+O(1)\int_\CT \int_\CT (v-s) K_{1,h_{\sigma}}(v-s)w_{\bphi,h_{\sigma}}(v-s) dudv\\
			&=&\tau(s)-\frac{n-1}{n}\tau(s) \int_\CT K_{1,h_{\sigma}}(v-s)w_{\bphi,h_{\sigma}}(v-s)  dv\\
			&&\qquad+O(1)\int_\CT (v-s) K_{1,h_{\sigma}}(v-s)w_{\bphi,h_{\sigma}}(v-s) dv\\
			&=&\tau(s)-\frac{n-1}{n}\tau(s) \e_{2}^\top\left[\Q_{K_1,\bphi}^{\{1\}}\right]^{-1}\int_\CT K_{1,h_{\sigma}}(v-s)\bphi_{h_{\sigma}}(u-s)  dv+O(h_\sigma),\\
			\ese
			where $\Q_{K_1,\bphi}^{\{1\}}$ is as defined in condition~C2. By a simple change of variable, one can show that $\int_\CT K_{1,h_{\sigma}}(v-s)\bphi_{h_{\sigma}}(u-s)  dv=\q_2+O(h_\sigma)$, with $\q_2$ being the the first column of $\Q_{K_1,\bphi}^{\{1\}}$. Hence we have that
			\be
			\label{s1}
			\E\left[S_{-i}(s)\right]=\tau(s)-\frac{n-1}{n}\tau(s) +O(h_\sigma)=\frac{1}{n}\tau(s) +O(h_\sigma).
			\ee
			Furthermore, we an derive the bound for $\Var\left[S_{-i}(s)\right]$ as follows
			\[
			\begin{split}
			\Var\left[S_{-i}(s)\right]&=\frac{n-1}{n^2}\Var\left[\mathop{\sum\sum}^{\ne}_{u,v\in N_{i'}}  K_{1,h_{\sigma}}(v-s)w_{\bphi,h_{\sigma}}(v-s)/\rho(v)\right]\\
			&\leq \frac{n-1}{n^2}\E\left[\mathop{\sum\sum}^{\ne}_{u,v\in N_{i'}}  K_{1,h_{\sigma}}(v-s)w_{\bphi,h_{\sigma}}(v-s)/\rho(v)\right]^2=O\left((nh_\sigma)^{-1}\right).
			\end{split}
			\]
			Combining the above inequality and~\eqref{s1}, we have that
			\be
			\label{s2}
			\hskip -1em \E\left[S_{-i}(s)\right]^2=O\left[n^{-2}+h_\sigma^2+(nh_\sigma)^{-1}\right], \text{ and } |\E\left[S_{-i}(s)S_{-j}(t)\right]|=O\left[n^{-2}+h_\sigma^2+(nh_\sigma)^{-1}\right],
			\ee
			for any $s,t\in\CT$ and $i,j=1,\ldots,n$, where the second equality follows from the Cauchy-Schwartz inequality that $|\E\left[S_{-i}(s)S_{-j}(t)\right]|\leq \sqrt{\E\left[S_{-i}(s)\right]^2\E\left[S_{-j}(t)\right]^2}$.
			
			Using~\eqref{s2} and Lemma~\ref{lm:stein}, we can show that
			\bse
			&&\E\left(  T_{n,424}-  T_{n,424}^*\right)^2=\E\left\{{1\over n} \sum_{i=1}^n\sum_{s\in N_{i}}\{[Z_{i}(s)-\mu^*(s)]^2-\sigma_Z^2(s)\} S_{-i}(s)\right\}^2\\
			&=&{1\over n^2}\sum_{i=1}^n \int_\CT \lambda_0(s)\E\left[\{[Z_{i}(s)-\mu^*(s)]^2-\sigma_Z^2(s)\}^2\exp(X_i(s))\right] \E\left[S_{-i}(s)\right]^2ds\\
			&&+{1\over n^2}\sum_{i=1}^n \int_\CT \int_\CT \lambda_0(s)\lambda_0(t)\E\left[\{[Z_{i}(s)-\mu^*(s)]^2-\sigma_Z^2(s)\}\{[Z_{i}(t)-\mu^*(t)]^2-\sigma_Z^2(t)\}\exp(X_i(s)+X_i(t))\right]\\
			&&\hskip 20em\times\E\left[S_{-i}(s)S_{-i}(t)\right]dsdt\\
			&&+{1\over n^2}\sum_{i=1}^n\sum_{j=1,j\neq i}^n \int_\CT \int_\CT \lambda_0(s)\lambda_0(t)\E\bigg[\{[Z_{i}(s)-\mu^*(s)]^2-\sigma_Z^2(s)\}\{[Z_{j}(t)-\mu^*(t)]^2-\sigma_Z^2(t)\}\\
			&&\hskip 20em\times \exp(X_i(s)+X_j(t))S_{-i}(s)S_{-j}(t)\bigg]dsdt\\
			&=&{1\over n^2}\sum_{i=1}^n\sum_{j=1,j\neq i}^n \int_\CT \int_\CT \lambda_0(s)\lambda_0(t)\E\bigg\{\{[Z_{i}(s)-\mu^*(s)]^2-\sigma_Z^2(s)\}\{[Z_{j}(t)-\mu^*(t)]^2-\sigma_Z^2(t)\}\\
			&&\hskip 10em\times\exp(X_i(s)+X_j(t))\times \left[S_{-i}(s)-S_{-(i,j)}(s)\right]\left[S_{-j}(t)-S_{-(i,j)}(t)\right]\bigg\}dsdt\\
			&&\hskip 10em+O\left[n^{-3}+n^{-1}h_\sigma^2+(n^2h_\sigma)^{-1}\right]\\
			&=&O(1){1\over n^2}\sum_{i=1}^n\sum_{j=1,j\neq i}^n\sqrt{E\left\{\left[S_{-i}(s)-S_{-(i,j)}(s)\right]^2\left[S_{-j}(t)-S_{-(i,j)}(t)\right]^2\right\}}+O\left[n^{-3}+n^{-1}h_\sigma^2+(n^2h_\sigma)^{-1}\right],
			\ese
			where $S_{-(i,j)}(s)=\tau(s)-{1\over n} \sum_{i'=1,i'\ne i\text{ or }j}^n\mathop{\sum\sum}^{\ne}_{u,v\in N_{i'}}  K_{1,h_{\sigma}}(v-s)w_{\bphi,h_{\sigma}}(v-s)/\rho(v)$, for $i,j=1,\dots,n$, and the last equality follows from the Cauchy-Swartz inequality. By definition, we have that
			\[
			S_{-i}(s)-S_{-(i,j)}(s)=-{1\over n}\mathop{\sum\sum}^{\ne}_{u,v\in N_{j}}  K_{1,h_{\sigma}}(v-s)w_{\bphi,h_{\sigma}}(v-s)/\rho(v),
			\]
			which further implies that for $i\neq j$,
			\[
			\begin{split}
			E\left\{\left[S_{-i}(s)-S_{-(i,j)}(s)\right]^2\left[S_{-j}(t)-S_{-(i,j)}(t)\right]^2\right\}&=E\left[S_{-i}(s)-S_{-(i,j)}(s)\right]^2E\left[S_{-j}(t)-S_{-(i,j)}(t)\right]^2\\&=O\left[(n^2h_\sigma)^{-1}\right]\times O\left[(n^2h_\sigma)^{-1}\right]=O\left[(n^2h_\sigma)^{-2}\right].
			\end{split}
			\]
			Therefore, to this end, we have shown that
			\be
			\label{diff1}
			\begin{split}
				\E\left(  T_{n,424}-  T_{n,424}^*\right)^2&=O\left[(n^2h_\sigma)^{-1}\right]+O\left[n^{-3}+n^{-1}h_\sigma^2+(n^2h_\sigma)^{-1}\right]\\
				&=O\left[n^{-3}+n^{-1}h_\sigma^2+(n^2h_\sigma)^{-1}\right].
			\end{split}
			\ee
			Consequently, under Assumptions C3, we have that
			\[
			T_{n,424}=  T_{n,424}^*+o_p(n^{-1/2})=-{1\over n} \sum_{i=1}^n\sum_{s\in N_{i}} [\{Z_{i}(s)-\mu^*(s)\}^2-\sigma_Z^2(s)]  \tau(s)+o_p(n^{-1/2}).
			\]
			Combining results above, the leading terms in $ T_n$ are $  T_{n,1}$, $  T_{n,41}$ and $  T_{n,424}$, which yields that
			\ben
			T_n&=& {1\over n} \sum_{i=1}^n\mathop{\sum\sum}^{\ne}_{u,v\in N_i} [\{ Z_i(v)-\mu^*(v)\}^2- \sigma_Z^2(v)] - h_{\sigma}^2 \int_\CT \int_\CT \CB_{\sigma^2}(v) \rho_2(u,v) du dv \nonumber \\
			&& -{1\over n} \sum_{i =1}^n\sum_{s\in N_{i}} [\{Z_{i}(s)-\mu^*(s)\}^2-\sigma_Z^2(s)]  \tau(s)  +o_p(n^{-1/2}) \nonumber 
			\\
			&& +O_p\left\{\frac{\log(n)}{nh_{\mu}}+h_{\mu}^{3}+ \frac{h_{\mu}\log(n)}{\sqrt{nh_{\mu}}}\right\}\nonumber. \text{ (by Assumption C3, $h_\mu$ and $h_\sigma$ are of the same order.)}
			\een
			Under the Assumption C3(a), $h_{\mu}\to0$, $nh_{\mu}^2/\log(n)\to\infty$ and the additional Assumption of Theorem~\ref{thm3} that $nh_{\mu}^6\to 0$, one has that
			\ben
			T_n&=& {1\over n} \sum_{i =1}^n\sum_{s\in N_{i}} [\{Z_{i}(s)-\mu^*(s)\}^2-\sigma_Z^2(s)]  \left[n_i-1-\tau(s)\right] \nonumber \\
			&&\qquad\qquad - h_{\sigma}^2 \int_\CT \CB_{\sigma^2}(v) \rho(v)\tau(v)  dv   +o_p(n^{-1/2}).
			\een
			where $\tau(s)=\frac{\int_\CT \rho_2(u,s)du}{\rho(s)}$. Under the null hypothesis $C_{XY}(u,v)\equiv 0$, $\mu^*(u)\equiv \mu(u)$, the asymptotic variance of $ T$ is $\Omega_T /n$, and by using Lemma~\ref{lm:stein_extended}, we can show that
			\bse
			\Omega_T&=& \E \left( \mathop{\sum\sum}^{\ne}_{u,v\in N_i} [\{ Z_i(v)-\mu(v)]^2\}- \sigma_Z^2(v)] - \sum_{s\in N_{i}} [\{Z_{i}(s)-\mu(s)\}^2-\sigma_Z^2(s)]  \tau(s)  \right)^2 \\
			&=&2 \int_{\CT^4} C_Y^2(t, v) \rho_4(s,t, u,v) ds dt du dv + 2 \int_{\CT^3}{\left\{\sigma_Z^4(v)+[3-2\tau(s)]C_Y^2(s,v)\right\}} \rho_3( s, u, v) ds du dv \\
			&&{+ 2 \int_{\CT^2} \left\{ 1-  \tau(u)\right\}\left\{ 1-  \tau(v)\right\} C_Y^2(u,v) \rho_2( u, v) du dv} + 2 \int_\CT \left\{ 1-  \tau(v)\right\}\tau(v)\rho(v) \sigma_Z^4(v)  dv,%
			\ese
			where $\rho_3(s,u,v) =\lambda_0(s) \lambda_0(u) \lambda_0(v) \exp\{ \sigma_X^2(s) /2 + \sigma_X^2(u) /2+ \sigma_X^2(v) /2 +C_{X}(s,u) +C_{X}(s,v) +C_{X}(u,v) \}$, and   $\rho_4(s,t, u,v) =\lambda_0(s) \lambda_0(t) \lambda_0(u) \lambda_0(v) \exp\{ \sigma_X^2(s) /2 + \sigma_X^2(t) /2 +\sigma_X^2(u) /2+ \sigma_X^2(v) /2 +C_{X}(s,u) +C_{X}(s,v) +C_{X}(u,v) +C_{X}(s,t)+C_{X}(t,u)+C_{X}(t,v)\}$.
			
			The proof of Theorem~\ref{thm3} is now complete.

			\subsection{Technical Lemma}
			\label{sec:tech}
			In this section, we give a technical Lemma needed for the proof of Lemma~\ref{lm:stein}.
			\begin{lem}\label{lm:stein_extended}
				Let $X$ and $Y$ be two normal random variables with means equal to $0$, $\mu_Y$ and and variances equal to $\sigma^2_X$, $\sigma_Y^2$, respectively. Then, we have that $\E\{X \exp(X)\} = \sigma_X^2 \exp(\sigma_X^2/2 )$, $\E\{X^2 \exp(X)\} = (\sigma_X^2 +\sigma_X^4) \exp(\sigma_X^2/2 )$, $\E\{X^3 \exp(X)\} = (\sigma_X^6+ 3 \sigma_X^4) \exp(\sigma_X^2/2 )$, $\E\{X^4 \exp(X)\} = (\sigma_X^8 + 6\sigma_X^6 + 3\sigma_X^4) \exp(\sigma_X^2/2 )$,  and
				\ben
				&&\E \left[Y^2\exp(X)\right]=\left\{\sigma_Y^2+\left[\mu_Y+\Cov(X,Y)\right]^2\right\}\exp(\sigma_X^2/2 ), \label{stein2-1} \\
				&&\E \left[Y^3\exp(X)\right]=\left\{3\sigma_Y^2 [\mu_1+\Cov(X,Y_1)] + [\mu_Y+\Cov(X,Y)]^3\right\}\exp(\sigma_X^2/2), \label{stein3} \\
				&&\E \left[Y^4\exp(X)\right]\\\nonumber
				&&=\left\{ \left[\Cov(X,Y)+\mu_Y\right]^4 + 6 \sigma_Y^2\left[\Cov(X,Y)+\mu_Y\right]^2 + 3\sigma_Y^4 \right\}\exp(\sigma_X^2/2).\label{stein4} 
				\een
			\end{lem}
			\noindent {\bf Proof of Lemma~\ref{lm:stein_extended}.}
			A direct application of the Stein's Lemma gives that
			\[
			\begin{split}
			\E\left[X\exp(X)\right]&=\sigma_X^2\E\left[\exp(X)\right]=\sigma_X^2 \exp(\sigma_X^2/2 ),\\
			\E\left[X^2\exp(X)\right]&=\sigma_X^2\E\left[\exp(X)+X\exp(X)\right]=\left(\sigma_X^4+\sigma_X^2\right) \exp(\sigma_X^2/2 ),\\
			\E\left[X^3\exp(X)\right]&=\sigma_X^2\E\left[2X\exp(X)+X^2\exp(X)\right]=\left(\sigma_X^6+3\sigma_X^4\right) \exp(\sigma_X^2/2 ),\\
			\E\left[X^4\exp(X)\right]&=\sigma_X^2\E\left[3X^2\exp(X)+X^3\exp(X)\right]=\left(\sigma_X^8+6\sigma_X^6+3\sigma_X^4\right) \exp(\sigma_X^2/2 ).\\
			\end{split}
			\]
			Since $X$ and $Y$ are jointly normally distributed, one has that
			\[
			\E(Y|X)=\mu_Y+\frac{\Cov(X,Y)}{\sigma_X^2}X,\text{ and } \Var(Y|X)=\sigma_Y^2-\frac{[\Cov(X,Y)]^2}{\sigma_X^2},
			\]
			Define variable $Z=Y-\E(Y|X)$, then we have that $E(Z)=0$, $\Var(Z)=\Var(Y|X)$ and $\Cov(X,Z)=\E(XZ)=\E\left[X\E(Z|X)\right]=0$, which indicates that $X$ and $Z$ are independent normal random variables and that
			\[
			Z\sim N\left[0,\sigma_Y^2-\frac{[\Cov(X,Y)]^2}{\sigma_X^2}\right].
			\]
			As a result, one has that, for $j\geq 1$,
			\[
			\begin{split}
			\E \left[Y^j\exp(X)\right]&=\E\left\{\left[Z+\E(Y|X)\right]^j\exp(X)\right\}=\sum_{k=0}^{j}{j\choose k}\E\left[(Z+\mu_Y)^ka^{j-k}X^{j-k}\exp(X)\right]\\
			&=\sum_{k=0}^{j}{j\choose k}a^{j-k}\E\left[(Z+\mu_Y)^k\right]\E\left[X^{j-k}\exp(X)\right],
			\end{split}
			\]
			where $a={\Cov(X,Y)}/{\sigma_X^2}$. Consequently, we have the following three equations.
			\[
			\begin{split}
			\E \left[Y^2\exp(X)\right]&=a^2\E[X^2\exp(X)]+2a\mu_Y\E[X\exp(X)]+\left(\mu_Y^2+\sigma_Y^2-\frac{[\Cov(X,Y)]^2}{\sigma_X^2}\right)\E[\exp(X)]\\
			&=\frac{[\Cov(X,Y)]^2}{\sigma_X^2}\left(\sigma_X^2+1\right) \exp(\sigma_X^2/2 )+2\mu_Y \Cov(X,Y)\exp(\sigma_X^2/2 )\\
			&\qquad +\left(\mu_Y^2+\sigma_Y^2-\frac{[\Cov(X,Y)]^2}{\sigma_X^2}\right)\exp(\sigma_X^2/2 )\\
			&=\left\{\sigma_Y^2+\left[\mu_Y+\Cov(X,Y)\right]^2\right\}\exp(\sigma_X^2/2 ).
			\end{split}
			\]
			\[
			\begin{split}
			\E \left[Y^3\exp(X)\right]&=a^3\E[X^3\exp(X)]+3a^2\mu_Y\E[X^2\exp(X)]+3a\left(\mu_Y^2+\sigma_Y^2-\frac{[\Cov(X,Y)]^2}{\sigma_X^2}\right)\E[X\exp(X)]\\
			&\qquad+\left(\mu_Y^3+3\mu_Y\sigma_Y^2-3\mu_Y\frac{[\Cov(X,Y)]^2}{\sigma_X^2}\right)\E[\exp(X)]\\
			&=\frac{[\Cov(X,Y)]^3}{\sigma_X^2}\left(\sigma_X^2+3\right) \exp(\sigma_X^2/2 )+3\mu_Y\frac{[\Cov(X,Y)]^2}{\sigma_X^2}\left(\sigma_X^2+1\right) \exp(\sigma_X^2/2 )\\
			&\qquad+3\Cov(X,Y)\left(\mu_Y^2+\sigma_Y^2-\frac{[\Cov(X,Y)]^2}{\sigma_X^2}\right)\exp(\sigma_X^2/2 )\\
			&\qquad+\left(\mu_Y^3+3\mu_Y\sigma_Y^2-3\mu_Y\frac{[\Cov(X,Y)]^2}{\sigma_X^2}\right)\exp(\sigma_X^2/2 )\\
			&=\left\{3\sigma_Y^2\left[\mu_Y+\Cov(X,Y)\right]+\left[\mu_Y+\Cov(X,Y)\right]^3\right\}\exp(\sigma_X^2/2 ).
			\end{split}
			\]
			\[
			\begin{split}
			\E \left[Y^4\exp(X)\right]&=a^4\E[X^4\exp(X)]+4a^3\mu_Y\E[X^3\exp(X)]+6a^2\left(\mu_Y^2+\sigma_Y^2-\frac{[\Cov(X,Y)]^2}{\sigma_X^2}\right)\E[X^2\exp(X)]\\
			&+4a\left(\mu_Y^3+3\mu_Y\sigma_Y^2-3\mu_Y\frac{[\Cov(X,Y)]^2}{\sigma_X^2}\right)\E[X\exp(X)]\\
			&\qquad+\left(\mu_Y^4+6\mu_Y^2\sigma_Y^2-6\mu_Y^2\frac{[\Cov(X,Y)]^2}{\sigma_X^2}+3\left(\sigma_Y^2-\frac{[\Cov(X,Y)]^2}{\sigma_X^2}\right)^2\right)\E[\exp(X)]\\
			&=\frac{[\Cov(X,Y)]^4}{\sigma_X^4}\left(\sigma_X^4+6\sigma_X^2+3\right) \exp(\sigma_X^2/2 )+4\mu_Y\frac{[\Cov(X,Y)]^3}{\sigma_X^2}\left(\sigma_X^2+3\right) \exp(\sigma_X^2/2 )\\
			&\qquad+6\frac{[\Cov(X,Y)]^2}{\sigma_X^2}\left(\sigma_X^2+1\right)\left(\mu_Y^2+\sigma_Y^2-\frac{[\Cov(X,Y)]^2}{\sigma_X^2}\right) \exp(\sigma_X^2/2 )\\
			&\qquad+4\Cov(X,Y)\left(\mu_Y^3+3\mu_Y\sigma_Y^2-3\mu_Y\frac{[\Cov(X,Y)]^2}{\sigma_X^2}\right) \exp(\sigma_X^2/2 )\\
			&\qquad+\left(\mu_Y^4+6\mu_Y^2\sigma_Y^2-6\mu_Y^2\frac{[\Cov(X,Y)]^2}{\sigma_X^2}+3\left(\sigma_Y^2-\frac{[\Cov(X,Y)]^2}{\sigma_X^2}\right)^2\right)\exp(\sigma_X^2/2 )\\
			&=\left\{3\sigma_Y^4+6\sigma_Y^2\left[\mu_Y+\Cov(X,Y)\right]^2+\left[\mu_Y+\Cov(X,Y)\right]^4\right\}\exp(\sigma_X^2/2 ).
			\end{split}
			\]
			The proof of Lemma~\ref{lm:stein_extended} is complete.

		\end{document}